\documentclass[11pt,a4paper]{article}
\pdfoutput=1
\usepackage{jheppub,hyperref}
\usepackage{multirow, graphicx,amssymb,url,mathrsfs,amsmath}
\usepackage{wrapfig,boxedminipage,setspace,subfigure,epsfig}
\usepackage{amsxtra,amstext,latexsym,dsfont,amsfonts}
\usepackage{color,eucal}
\usepackage[section]{placeins}
\usepackage[colorlinks=true,linktocpage=true,linkcolor=blue,citecolor=blue]{hyperref}
\graphicspath{{Figures/}}
\newcommand{\mt}[1]{\textrm{\tiny #1}}
\newcommand{\be}{\begin{equation}}
\newcommand{\ee}{\end{equation}}
\newcommand{\bea}{\begin{eqnarray}}
\newcommand{\eea}{\end{eqnarray}}
\newcommand{\rh}{r_\mt{H}}

\title{  New black holes with hyperscaling violation for the   transports  of  quantum critical points with magnetic impurity} 
\author[a]{Xian-Hui Ge,}
\author[b]{Yunseok Seo,}
\author[c]{Sang-Jin Sin,}
\author[c]{Geunho Song}
\emailAdd{gexh@shu.edu.cn}
\emailAdd{yseo@gist.ac.kr}
\emailAdd{sjsin@hanyang.ac.kr}
\emailAdd{sgh8774@gmail.com}

\affiliation[a]{ Department of Physics, Shanghai University, Shanghai 200444,  China}
\affiliation[b]{ School of Physics and Chemistry, Gwangju Institute of Science and Technology, Gwangju 61005 Korea}
\affiliation[c]{ Department of Physics, Hanyang University, Seoul 04763, Korea }

 \abstract{ We consider the magneto-transports of quantum matters doped with magnetic impurities  near the quantum critical points(QCP). For this, we first find new black hole solution with  hyper-scaling violation which is 
dual to such system. By considering the fluctuation near this exact solution,  we calculated all transport coefficients using the holographic method.  
 We applied our result to  the surface state of the topological insulator with magnetic doping and found two QCP's, one bosonic and the other fermionic.  It turns out that      doped Bi$_2$Se$_3$  and   Bi$_2$Te$_3$ correspond to different QCP's. 
We also investigated transports of QCP's  as  functions of physical parameters and found that there are phase transitions  as well as  crossovers from weak localization to weak anti-localization.   
}
\begin{document}
\maketitle
 
\section{Introduction}
When many particles  are  strongly interacting, entire system is entangled and single particle picture  can not be used for system property at   low energy. 
In such case, reducing the degrees of freedom is the key issue of condensed matter physics and one such idea is to look at the quantum critical point (QCP), where 
almost all informations are lost except a few critical exponents and symmetries.   
The information loss leads  to the universality, which is parallel to the black hole (BH) physics. 
This parallelism  between  QCP and BH is the underlying principle of applying holographic method to strongly correlated condensed matter\cite{Hartnoll:2016apf,Zaanen:2015oix}. 
If a QCP is characterized by $z, \theta$ defined by the dispersion relation $\omega\sim k^{z}$  and the entropy density $s\sim T^{(d-\theta)/z}$, there exist a  metric with the same scaling symmetry.  
 \be
 ds^{2}=r^{-\theta}\bigg(-r^{2z}f(r)dt^{2}+\frac{dr^{2}}{r^2 f(r)}+r^{2}(dx^{2}+dy^{2})\bigg),
 \ee
under  $(t,r,x) \to (\lambda^z t,\lambda^{-1}r, \lambda  x)$ and 
the method developed for the exact AdS/CFT \cite{Maldacena:1997re,Witten:1998qj,Gubser:1998bc} is assumed  to hold for this case too. 
 This class of metric is called    hyper-scaling violation (HSV) metric and it has been  constructed and used     in various context \cite{Charmousis:2010zz,Gouteraux:2011ce,Huijse:2011ef,Iizuka:2012pn,Dong:2012se,Narayan:2012hk,Perlmutter:2012he,Cadoni:2012uf,Alishahiha:2012cm,Dey:2012fi,Cadoni:2012ea,Kim:2012pd} after flurry of activity on Lifshitz metrics\cite{Kachru:2008yh,Sin:2009wi,Hartnoll:2009kk,Goldstein:2009cv,Goldstein:2010aw,Keranen:2012mx,Zhao:2013pva,Lu:2013tza,Bu:2012zzb,Brynjolfsson:2009ct,Harrison:2012vy}.

The picture above suggests that the holographic principle works only at or near a QCP, while the most well charted regime in holographic research is the calculatiion of DC transport coefficients for a given gravity \cite{Zaanen:2015oix,Hartnoll:2016apf,Donos:2014uba,Blake:2015ina}.  Then 
 the most urgent  question in holography  would be how  a QCP can be characterized by the transports? Or can we distinguish the QCP by looking its  transport data? Therefore, much efforts have been given to the  transport calculation in the context of the  HSV geometries\cite{Davison:2014lua,Gouteraux:2014hca,Giataganas:2014hma,Fonda:2014ula,Papadimitriou:2014lia,Dehghani:2015gza,Liu:2016njg,Ge:2016sel,Ge:2017fix,Chen:2017gsl,Cremonini:2018jrx,Mukhopadhyay:2019dxn,Ahn:2019lrh}  which seem to be in 1-1 correspondence with the QCP's. However  the magneto-thermal conductivity had not been calculated  even after a few years of  the discovery of   exact solution that allows the presence of magnetic field in the context of HSV\cite{Ge:2017fix}.   The first result on the magneto-heat conductivity was obtained   in very recent paper \cite{Mukhopadhyay:2019dxn} but the result in the zero magnetic field limit  does not seem to be reduced to the known result by a parameter dependent  factor.   
It is also important to extract out  the  characters of QCP's, that is to examine the differences in behaviors of the transports  for different   QCP's, which is deeply hidden in the complexity of the transport formula.

In this paper we will  re-calculate the magneto-transports with different methods  in a generalized context,  motivated from  the earlier work \cite{Seo:2017yux} of some of us \cite{Seo:2017oyh,Seo:2017yux} where   the surface state of Topological insulator (TI) with magnetic doping \cite{liu2012crossover,zhang2012interplay,bao2013quantum} was studied. In \cite{Seo:2017oyh,Seo:2017yux}, 
we treated the surface of TI as a  Dirac material \cite{pkim,Lucas:2015sya,Seo:2016vks,Seo:2017oyh,Seo:2017yux,kkss},     a system with  $z=1,\theta=0$, because the topological insulator's surface has a Dirac cone as a part of its definition. 
However,  when the surface   is doped with the magnetic impurities, 
the surface gap is open. The system is strongly interacting when  the surface band  touches the Fermi level  so that the Fermi sea is small. See the red curve in the    Figure 1(a). 
\begin{figure}[ht!] 
\centering  
   \subfigure[surface gap of TI]
   {\includegraphics[width=45mm]{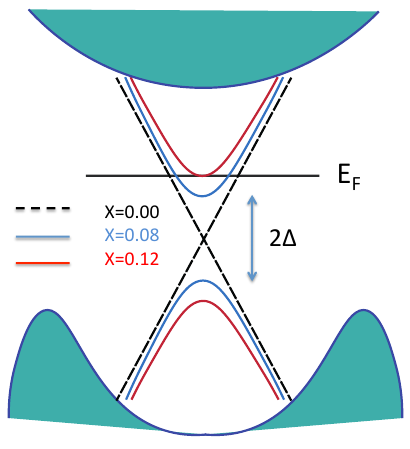} }     \subfigure[Mn doped Bi$_2$Se$_3$]
     {\includegraphics[width=45mm]{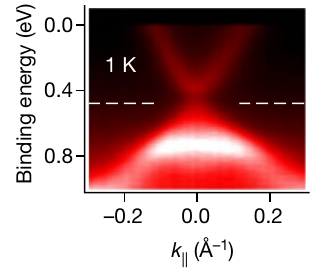} }            \subfigure[Mn doped Bi$_2$Te$_3$]
         {\includegraphics[width=37mm]{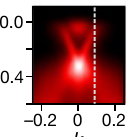} }
               \caption{(a) Evolution of density of state  as the doping changes. As we increase the doping   the surface gap increases, 
and      the fermi surface gets smaller.  At some point, the surface band touches FS and the system become strongly interacting. The green colored region represent the the bulk bands and the lines are for surface bands. 
 The figure is from \cite{Seo:2017yux}.
(b,c) The actual ARPES data for  for Mn-doped Bi$_{2}$Se$_{3}$ and Bi$_{2}$Te$_{3}$.  
Notice that for the latter, it is hard to see the surface gap. 
  The figure is from  the ref.\cite{rienks2019large}, https://rdcu.be/b3App
 }   \label{fig:111} 
\end{figure}
Then   one  may immediately ask  whether  
we can still treat the system as a Dirac material, because   strictly speaking, there is no more a Dirac cone. 
Therefore we have a  pressing reason to  re-examine the system  with the hyper-scaling violating geometry. 
We will see that comparing our theoretical result to the data, the relevant QCP for this system 
should be identified as $(z,\theta)=(1.5,1)$ for Mn doped Bi$_2$Se$_3$ and $(z,\theta)=(1,0)$ for the Cr doped Bi$_2$Te$_3$. 

Our analysis will also be useful for multilayer graphene system where the excitations are known to have dispersion relation  $\omega=k^{n}$ \cite{PhysRevB.78.121401}. 

We will see that  the behavior of the transports can be   qualitatively different,  depending on the dynamical exponent of QCP.   
 However, such difference can be recognized only by plotting the transports for various $(z,\theta)$, therefore we will put much efforts for explicitly plotting the result.  
We  checked  that our result in the    limits of zero magnetic field and in the absence of the new coupling  agrees with known result.  

The rest of the paper consists of as follows. 
In section 2, we introduce the hyperscaling violation  gravity  model with a magnetic coupling and present its   solution. 
 In section 3, we calculate various transport coefficients. 
 In section 4, we apply our result to the surface state of topological insulator with its charge carrier's fermionic nature. 
In section 5, we conclude with summary. 
In appendices A and B, we give detailed study of the long and complicated formula by plotting the magneto transports and density dependence of the various QCP's in order to explicitly show the 
presence of phase transitions across the four regions of    critical exponents  and the appearance of cross over from the weak localization to weak anti-localization. 
We also compared  the magneto-transports without and with magnetic impurity.  These are by themselves important since they are   predictions that can be compared with future experimental data, but to avoid the too many figures in the main text, we put it at the appendices. 
At the end, we explain the null energy condition for our theory. 
 
 \section{The model and its black hole solution}
To get the solution  with the dynamical exponent $z$ and the hyperscaling violating factor $\theta$, we   consider a 4 dimensional action.
\bea\label{action1}
S_{tot}&=&\int_{\mathcal{M}}d^4x\left(\mathcal{L}_0+\mathcal{L}_{int}\right)\nonumber\\
\mathcal{L}_0&=&\sqrt{-g}\left(R+\sum_{i=1}^2 V_ie^{\gamma_i\phi}-\frac{1}{2}(\partial\phi)^2-\frac{1}{4}\sum_{i=1}^2 Z_i(\phi) F_{(i)}^2-\frac{1}{2}Y(\phi)\sum\limits_{i,I}^{2}(\partial \chi_I^i)^2\right)\nonumber\\
\mathcal{L}_{int}&=&-\frac{q_{\chi}}{16}\sum_{I=1,2}(\partial\chi^{(2)}_{I})^{2}\epsilon^{\mu\nu\rho\sigma}F^{(2)}_{\mu\nu}F^{(2)}_{\rho\sigma},
\eea
where we  will use the convention  $2\kappa^2=16\pi G=1$. 
We use  ansatz  
\be
Z_i= e^{\lambda_i \phi}, \quad Y(\phi)= e^{-\lambda_2\phi},\quad \chi_i^{I}=\beta^{I} \delta_{Ii} x^i,
\ee
where  $\chi_i$'s are  $d$-massless linear axions introduced to break the translational symmetry and $(\beta^{1},\beta^{2})=(\alpha,\lambda)$ denote the strength of momentum relaxation. 
 The action consists of Einstein gravity, axion fields, and $U(1)$ gauge fields and  a dilaton field. For simplicity, we only consider two $U(1)$ gauge $F^{(1)}_{rt}$ and $F^{(2)}_{rt}$ in which the first gauge field plays the role of an auxiliary field, making the geometry asymptotic Lifshitz, and the second gauge field is the exact Maxwell field making the black hole charged.


 The equations of motion are given by
\begin{eqnarray}
&&\partial_{\mu}(\sqrt{-g}g^{\mu\nu}Y(\phi)\sum_{a}\partial_{\nu}\chi^{(i)}_{I})+\frac{q_{\chi}}{8}\partial_{\mu}(\delta^{i2}\epsilon^{\rho\sigma\lambda\gamma}F^{(2)}_{\rho\sigma}F^{(2)}_{\lambda\gamma}g^{\mu\nu}\partial_{\nu}\chi^{(2)}_{I})=0, \\
&&\partial_{\mu}(\sqrt{-g}Z_i(\phi)F^{\mu\nu}_{(i)}+\frac{q_{\chi}}{4}\delta^{i2}g^{\rho\sigma}\sum_{I}\partial_{\rho}\chi^{(2)}_{I}\partial_{\sigma}\chi^{(2)}_{I}\epsilon^{\alpha\beta\mu\nu}F^{(2)}_{\alpha\beta})=0,\\
&&R_{\mu\nu}=\frac{1}{2\sqrt{-g}}g_{\mu\nu}\mathcal{L}_0+\frac{1}{2}\partial_{\mu}\phi\partial_{\nu}\phi+\frac{Y}{2}\sum_i \partial_{\mu}\chi^{(i)}_I\partial_{\nu}\chi^{(i)}_I+\frac{1}{2}\sum_i Z_i F_{(i)\mu}^{\rho}F_{\mu \rho}^{(i)}
\nonumber\\&&+\frac{q_{\chi}}{16} \sum_{I}\frac{1}{\sqrt{-g}}(\partial_{\mu}\chi^{(2)}_{I})(\partial_{\nu}\chi^{(2)}_{I})\epsilon^{\rho\sigma\lambda\gamma}F^{(2)}_{\rho\sigma}F^{(2)}_{\lambda\gamma}.\\
&& \Box\phi+\sum_i V_{i}\gamma_ie^{\gamma_i\phi}-\frac{1}{4}\sum_i Z'_i(\phi) F^2_{(i)}-\frac{1}{2}Y'(\phi)\sum\limits_{i,I}^{2}(\partial \chi_I^i)^2=0.
\end{eqnarray}
The solution for the dilaton field is given as
\be
\phi(r)=  \nu\ln r, \quad \hbox{with } \nu=\sqrt{(2-\theta)(2z-2-\theta)}.
\ee
The gauge couplings $Z_1$ and $Z_2$ can be solved to give 
\bea
Z_1(\phi)=e^{\lambda_1\phi}= r^{\theta-4},\quad 
Z_2(\phi)=e^{\lambda_2\phi}=r^{2z-\theta-2}, \quad 
Y(\phi)=1/Z_2,
\eea
where $\lambda_1=(\theta-4)/\nu $ and $\lambda_2=\nu /(2-\theta)$.
Other exponentials and   potentials are 
\be
\gamma_1=\frac{\theta}{\nu },~~~\gamma_2=\frac{\theta+2z-6}{\nu }, \quad
V_1=\frac{z-\theta+1}{2(z-1)}(q_1)^2,~~~V_2=\frac{H^2(2z-\theta-2)}{4(z-2)},
\ee
where $H$ is a constant  magnetic field.
Finally, the  solution is given by 
\begin{eqnarray}\label{bgsol}
&&A_1=a_1(r)dt, A_2=a_2(r)dt+\frac{1}{2}H(xdy-ydx),   \\
&&\quad \chi^{(1)}_{I}=(\alpha x, \alpha y), \quad \chi^{(2)}_{I}=(\lambda x, \lambda y),  \\
&&ds^{2}=r^{-\theta}\bigg(-r^{2z}f(r)dt^{2}+\frac{dr^{2}}{r^2 f(r)}+r^{2}(dx^{2}+dy^{2})\bigg), \\
&&f(r)=1-m r^{\theta-z-2}-\frac{\beta^{2}}{(\theta-2)(z-2)}r^{\theta-2z}+\frac{q_2^{2}(\theta-z)r^{2\theta-2z-2}}{2(\theta-2)}\nonumber\\&&
\qquad\qquad+\frac{H^2 r^{2z-6}}{4(z-2)(3z-\theta-4)}+\frac{\lambda^{4}H^{2}q^{2}_{\chi}c_3}{r^{6+2z-4\theta}}-\frac{\lambda^{2}Hq_2q_{\chi}c_2}{r^{4+2z-3\theta}}, \\
&&a_1(r)=\frac{-q_1}{2+z-\theta}(\rh^{2+z-\theta}-r^{2+z-\theta}), \;\; a_2(r)= \big(\mu-q_2 r^{\theta-z} \big)-\frac{\lambda^2 H q_{\chi} c_4}{r^{z-2\theta+2}}, 
\end{eqnarray}
where $\beta^2=\alpha^2+\lambda^2$ and  $c_2$, $c_3$,  $c_4$ can be expressed in terms of  $\theta$ and $z$
\bea
&&c_2=\frac{ (z-\theta)}{(\theta-2)(2\theta-z-2)}, \quad c_3=\frac{1}{2}\frac{1}{(2-\theta)(4+z-3\theta)}, \quad c_4=\frac{1}{2\theta-z-2}.
\eea
The charge density is determined by the condition $a_i(\rh)=0$ at the event horizon:
\bea\label{muq}
q_1=\sqrt{(2z-2)(2+z-\theta)}, \quad 
q_2=\mu\rh^{z-\theta}-c_4 \Theta  H, \eea
where $  \Theta = {\lambda^2  q_{\chi}}/{\rh^{2-\theta}}.
$
The entropy density and the Hawking temperature read
\bea
s=&&4\pi r^{2-\theta}_H, \\
{4\pi}T=&& (z+2-\theta)\rh^z -\frac{\beta^2 \rh^{\theta-z}}{2-\theta}
-\frac12\frac{\rh^{2\theta-2-z}}{(2-\theta)} \Big( \Theta H -(z-\theta)q_2 \Big)^2 -\frac{H^2 \rh^{3z-6}}{4(2-z)} .
\label{r0Trelation}
\eea
Notice that all three terms in eq. (\ref{r0Trelation}) contribute negatively and  if both $z,\theta<2$ and  
at zero temperature, the solution near the horizon  becomes $AdS_2 \times R^2$: 
\be
ds^2=\frac{-d\tau^2+du^2}{l^2_{eff}u^2}+\rh^{2-\theta}(dx^2+dy^2).
\ee

\begin{figure}[ht!]
\centering
      \subfigure[Classification of $(z,\theta)$ space ]
   {\includegraphics[width=45mm]{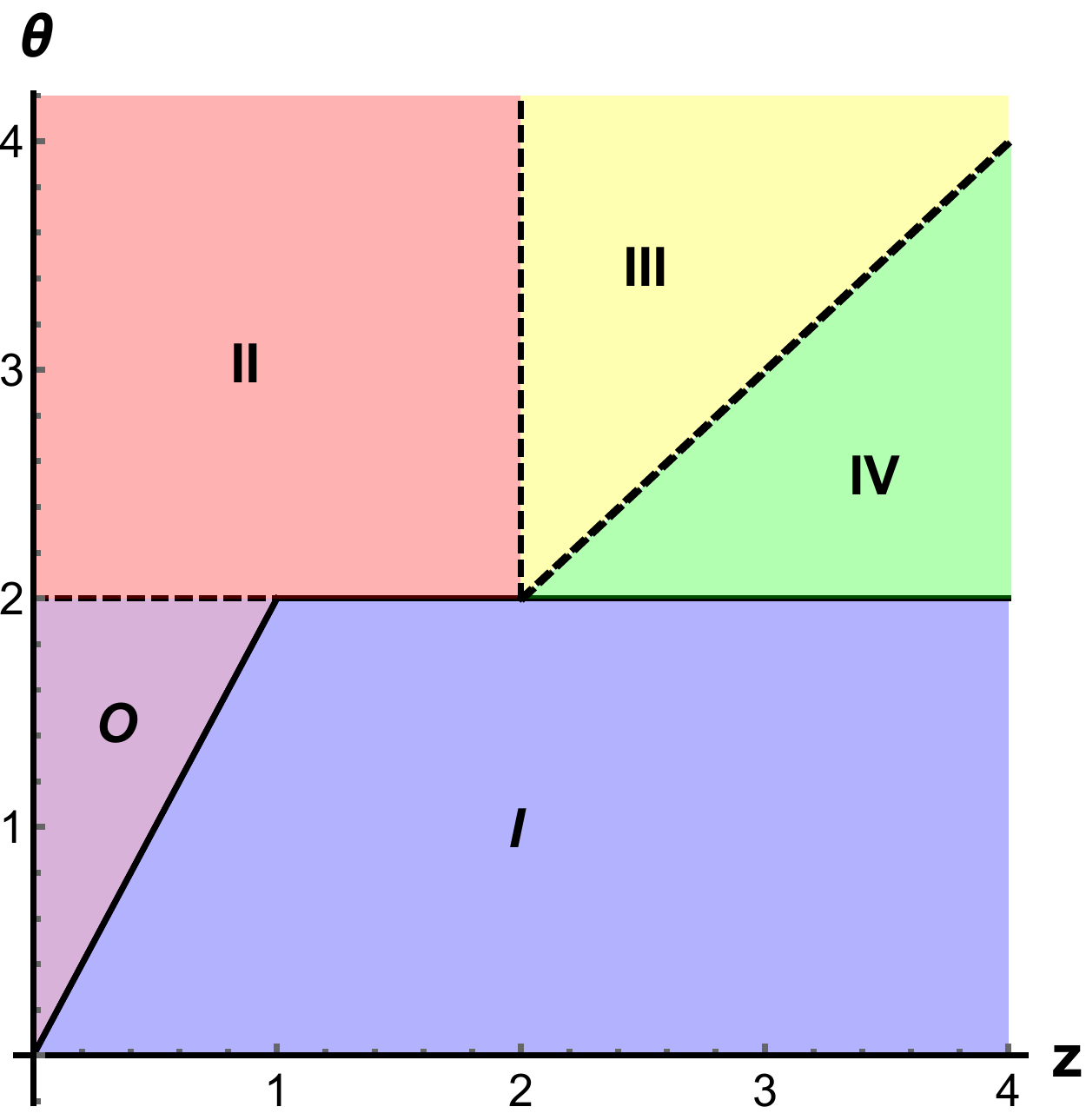} }
    \subfigure[$r_0$ $vs$ $T$ ]
   {\includegraphics[width=70mm]{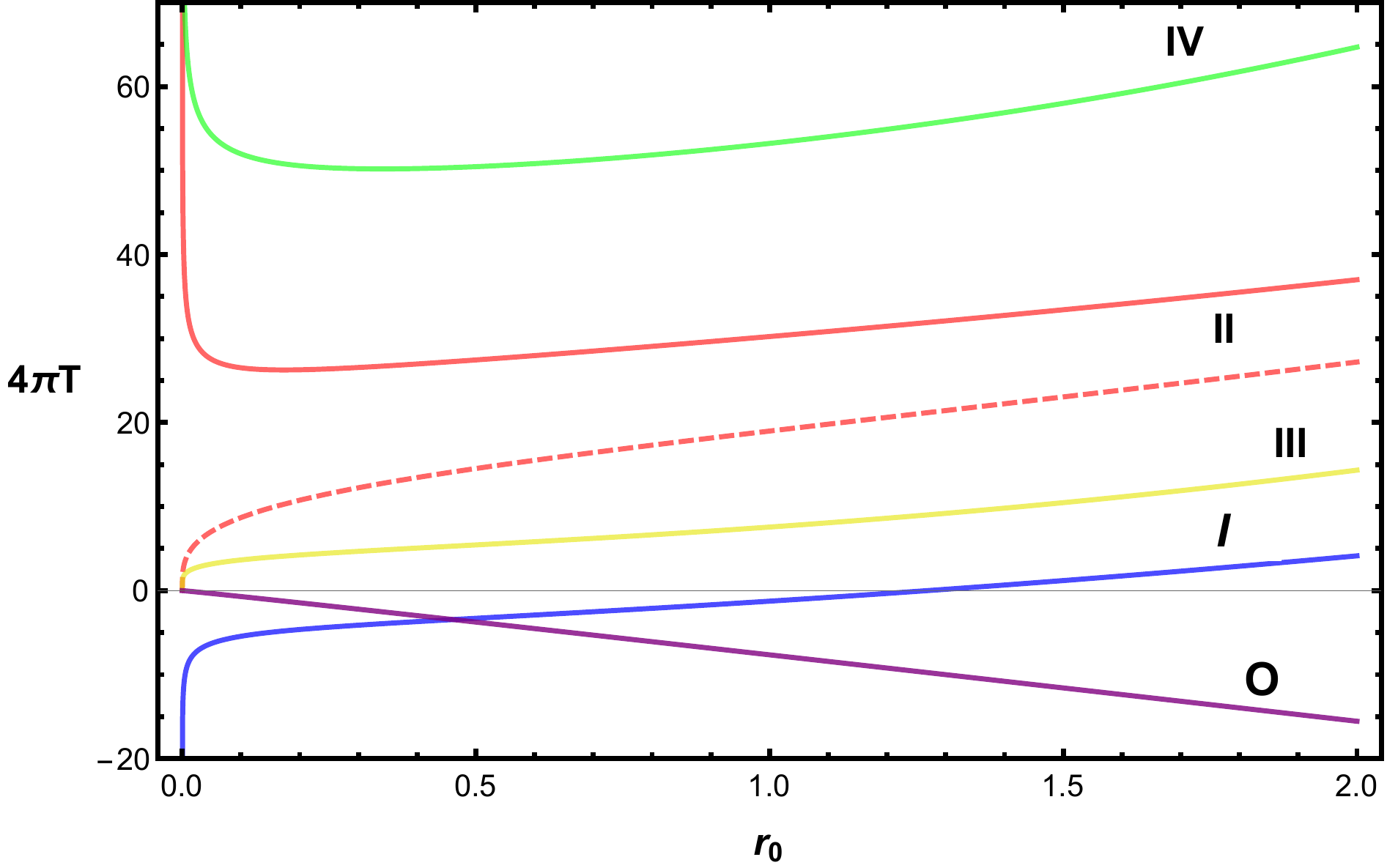} }
             \caption{Classification of  $(z,\theta)$ space by $T$-range: Each curves in (b) represent a typical curve of each region with the corresponding color of (a).  }    \label{fig:r0T} 
\end{figure}

We classify the plane of $(z,\theta)$ into several classes $O,I,II,III,IV$ depending on the behavior of the temperature as a function of $r_H$, the radius of the black hole which again is a complicated function of parameters of the theory.  See Figure \ref{fig:r0T}. 
I is  the region which allows negative $T$.  The region of $\theta>2$ has non-negative temperature, which can be further classified to three types II, III, IV.
For III, temperature is monotonic function of $r_H$ and cover the  entire positive temperature. 
For IV, there is a minimum temperature regardless of the chemical potential $\mu$.  
For II, depending on  $\mu$, temperature has minimum like IV (real red curve in (b)) or monotonic like III (dotted red curve in (b)).   
we have two subclasses which are represented by solid (dashed) red curve for the case with the finite (vanishing) $\mu$. We  do not have  positive $T$ in region O.

 \section{Calculation of magneto-transport coefficients}
We calculate the transport following the idea of linear response theory. We consider the following perturbations
\bea
&&\delta g_{tx}=h_{tx}(r)+tf_{3x}(r),~~ \delta g_{ty}=h_{ty}(r)+tf_{3y}(r),~~ \delta g_{rx}=h_{rx}(r), ~~\delta g_{ry}=h_{ry}(r),\\
&&\delta A_{ax}=b_{ax}-tf_{ax},~~ \delta A_{ay}=b_{ay}-tf_{ay}, \delta\chi_1=\varphi_{x}(r), \delta\chi_2=\varphi_{y}(r).
\eea
Requiring the linearised Einstein equations to be time-independent, we can get the explicit form of $f_{1i}$, $f_{2i}$, and $f_{3i}$ as
\bea
	&&f_{1i}=-E_{1i}+\zeta_ia_1(r),\nonumber\\
	&&f_{2i}=-E_{2i}+\zeta_ia_2(r),\nonumber\\
	&&f_{3i}=-\zeta_i U(r)
\eea
The thermo-electric transport can be calculated at the event horizon. One can take the Eddington-Finkelstein coordinates $(v,r)$ such that the background metric is manifestly regular at the horizon, 
\be
ds^2=-Udv^2-2\sqrt{UV}dv dr+W dx^2+W dy^2,
\ee
where  $v=t+\int dr \sqrt{V/U}$. 
 In this coordinate, the perturbed metric becomes
\be
\delta g_{\mu\nu}dx^\mu dx^\nu=h_{tx}dv dx+h_{ty}d v dy+\bigg(h_{rx}-\sqrt{\frac{V}{U}}h_{tx}\bigg)dr dx+\bigg(h_{ry}-\sqrt{\frac{V}{U}}h_{ty}\bigg)dr dy.
\ee

In order to ensure the regularity of the perturbed metric at the horizon, we ask the vanishing of last two terms  at the horizon so that 
\be
h_{ri}\sim \sqrt{\frac{V}{U}}h_{ti}, \hbox{ with } i=x,y. 
\ee
Similarly, we can expand the gauge fields in the Eddington-Finkelstein coordinates to get regularity condition at the horizon:
\bea
\delta A_{ai}&\sim &b_{ai}+E_{ai}v-E_{ai}\int dr \sqrt{\frac{V}{U}},
\eea
where $E_{ai}$ is constant probe electric fields with $a=1,2$ and $i=x,y$. 
The full gauge-field perturbation will have the regular expansion $\delta A_{ai}$ $\sim$ $E_{ai}v$ $+\cdots$ in the Eddington-Finkelstein coordinates, provided we demand  
\be
b'_{ai}\sim \sqrt{\frac{V}{U}}E_{ai}. 
\ee

To fix the perturbations, we need the linearized Einstein equations. The $(r,x)$ and $(r,y)$-components of the Einstein equation are given by
\bea
&&-\left(\frac{YV\beta^2}{W}+\frac{Z_2VH^2}{W^2}+\frac{q_{\chi}\lambda^2H\sqrt{UV}a_2'}{UW^2}\right)h_{rx} +\frac{V}{U}\left(\sum_{i=1,2}\left(Z_ia_i'f_{ix}\right)+f'_{3x}\right) \nonumber\\
&&\frac{Z_2 V Ha_2'}{UW}h_{ty}-\frac{VW'}{UW}f_{3x}+\frac{Z_2VH}{W}b_{2y}'=0\\
&&-\left(\frac{YV\beta^2}{W}+\frac{Z_2VH^2}{W^2}+\frac{q_{\chi}\lambda^2H\sqrt{UV}a_2'}{UW^2}\right)h_{ry} +\frac{V}{U}\left(\sum_{i=1,2}\left(Z_ia_i'f_{iy}\right)+f'_{3y}\right) \nonumber\\
&&\frac{Z_2 V Ha_2'}{UW}h_{tx}-\frac{VW'}{UW}f_{3y}+\frac{Z_2VH}{W}b_{2x}'=0
\eea
For simplification, it is fruitful to define following functions:
\bea \label{FG}
\mathcal{F}&&=W Y \beta^2+Z_2 H^2-q_2 Y \Theta H +Y\Theta^2 H^2\nonumber\\
\mathcal{G}&&=q_2-\Theta H
\eea
where all the quantities are computed at $r=r_H$. For example, $W=W(r_H)=r_H^{2-\theta}$. 
Regularity at the event horizon yields
\bea
&&h_{tx}|_{r_H}=-W\bigg(q_1(\mathcal{F}E_{1x}+\mathcal{G}E_{1y})+(\mathcal{F}-Z_2 H^2)\mathcal{G}E_{2x}+H(Z_2 \mathcal{F}+\mathcal{G}^2)E_{2y}\nonumber\\
&&\qquad\quad+s T (\mathcal{F}\zeta_x+\mathcal{G}H\zeta_y)\bigg)\bigg/(\mathcal{F}^2+H^2\mathcal{G}^2)\\
&&h_{ty}|_{r_H}=-W\bigg(q_1(\mathcal{F}E_{1y}-\mathcal{G}E_{1x})+(\mathcal{F}-Z_2 H^2)\mathcal{G}E_{2y}-H(Z_2 \mathcal{F}+\mathcal{G}^2)E_{2x}\nonumber\\
&&\qquad\quad+s T (\mathcal{F}\zeta_y-\mathcal{G}H\zeta_x)\bigg)\bigg/(\mathcal{F}^2+H^2\mathcal{G}^2)
\eea
The radially conserved electric current   ${J}^{\mu}=(J^t,J^x,J^y)$ can be defined by 
\be\label{concurrent}
{J}_a^{\mu}=\sqrt{-g}Z_{(a)} F^{\mu r}_{(a)}+\frac{q_{\chi}}{4} \delta^{2a}g^{\rho\sigma}\sum_{I}\partial_{\rho}\chi^{(2)}_{I}\partial_{\sigma}\chi^{(2)}_{I}\varepsilon^{\alpha\beta \mu r}F^{(2)}_{\alpha\beta},
\ee
where the index $a=1,2$ denotes the two currents which are dual to the two gauge fields. 
One can define the 'conserved' electric current ${\mathcal J}^i$ satisfying $\partial_r {\mathcal J}^i=0$  using 
the magnetization electric current $ {\mathcal M}_a^i= \sqrt{-g}\partial_j (Z_a F_{(a)}^{ij}+\frac{q_{\chi}}{4} \delta^{2a}g^{\rho\sigma}\sum_{I}\partial_{\rho}\chi^{(2)}_{I}\partial_{\sigma}\chi^{(2)}_{I}\varepsilon^{\alpha\beta ij}F^{(2)}_{\alpha\beta})$ as follows,  
\be
{\mathcal J}_a^i(r) =J_a^i(r) -  \int_{r_H}^r dr {\mathcal M_a}^i(r).
\ee
Notice that ${\mathcal J}$ coincides with the local boundary current and it can be evaluated at the horizon. Then we get  
\bea
&& {\mathcal J}_{1i}=Z_1 E_{1i}-\frac{q_1}{W}h_{ti}\bigg|_{r=\rh},\\
&& {\mathcal J}_{2i}=Z_2 E_{2i}-\frac{Z_2 a'_{2}}{UV}h_{ti}-\epsilon_{ij}\left(\frac{Z_2 H }{W}h_{tj}-\frac{q_{\chi}\lambda^2 E_{2j}}{\sqrt{UV}W^2}\right)\bigg|_{r=\rh},
\eea
For the transport coefficients, we will take only $J_{2i}$ and $E_{2i}$ as physical, because $A_{1i}$  is  introduced just for constructing HSV geometry.

For the heat current, following  \cite{Donos:2014cya}, 
we introduce  the total heat current 
 $Q^i$ by 
\be
Q^i=2\sqrt{-g}G^{ri}
\ee
where the two form $G^{\mu\nu}$ is associated with    the Killing vector $k=\partial_t$ by 
 \be
G^{\mu\nu}= \nabla^{\mu}k^{\nu} +\frac12 \sum_{I=1,2}Z_I
 k^{[\mu}F_I^{\nu]\sigma}A_{I\sigma}  +
 \frac14 \sum_{I=1,2}(Ex+2a_I)Z_IF_I^{\mu\nu}. 
\ee
It  satisfies 
\be
\nabla_\nu G^{\mu\nu}=-Vk^\mu/2. \label{Gcons}
\ee
Then   the  heat current is given  by   \cite{Donos:2014cya},
\bea
&& {Q}^{i}=2\sqrt{-g}\nabla^{r}k^{i}-\sum_{I=1,2}a_IJ_I^i =\frac{U^2}{\sqrt{UV}}(\frac{h_{ti}}{U})'-\sum_{I=1,2}a_IJ_I^i .
\eea
In the presence of the magnetic field, $\partial_r Q^i\neq 0$. 
In fact by taking $i$-th component of eq(\ref{Gcons}), 
\be 
 \partial_r Q^i = \partial_jG^{ij}:={\mathcal M}_Q^i . 
 \ee
Then  the 'conserved'  heat current $\mathcal{Q}^i$ is defined 
by 
\be
\mathcal{Q}^i(r)= Q^i(r) - \int_{r_H}^r dr {\mathcal M}_Q^i(r),
\ee
It turns out that near the boundary, $\mathcal{Q}^i(r)\sim T^{ti}-\mu J^i$ which is the expected boundary heat current and $ {\mathcal M}_Q^i$ gives the magnetization heat current for the  charge and energy. 
 Since $\mathcal{Q}^i$ is radially conserved, $\partial_r \mathcal{Q}^i=0$,   it is enough to compute $\mathcal{Q}^i$ at the horizon.
\bea
\mathcal{Q}_i=-\frac{U'h_{ti}}{\sqrt{UV}}({r_H}) .
\eea

Finally we can read off the transport coefficients from the following matrix:
\bea
\label{transport}
\left(\begin{array}{c} {\mathcal J}^{i} \\  {\mathcal Q}^{i} \end{array}  \right)=\left(\begin{array}{cc} \sigma_{ij} & \alpha_{ij} T \\ \bar{\alpha}_{ij} T& \bar{\kappa}_{ij} T \end{array}\right) \left( \begin{array}{c} E_j \\ \zeta_j \end{array}   \right),
\eea
where $\zeta_j=-(\nabla_jT)/T$ is the temperature gradient that drive thermal motion.   The transport coefficients are given by
\bea \label{transp}
&&\sigma_{ii}=\frac{Z_2\left(\mathcal{F}+Y \mathcal{G}^2\right)\left(\mathcal{F}-Z_2 H^2\right)}{\left(\mathcal{F}^2+H^2\mathcal{G}^2\right)},\\
&&\sigma_{ij}=\epsilon_{ij}\left(\Theta +\frac{Z_2 H\mathcal{G}\left(2\mathcal{F}+Y \mathcal{G}^2-Z_2H^2\right)}{\left(\mathcal{F}^2+H^2\mathcal{G}^2\right)}\right),\\
&&\alpha_{ii}=\bar{\alpha}_{ii}=\frac{s \mathcal{G}\left(\mathcal{F}-Z_2 H^2\right)}{\left(\mathcal{F}^2+H^2\mathcal{G}^2\right)},\\
&&\alpha_{ij}=\bar{\alpha}_{ij}=\epsilon_{ij}\frac{s H\left(Z_2 \mathcal{F}+\mathcal{G}^2\right)}{\left(\mathcal{F}^2+H^2\mathcal{G}^2\right)},\\
&&\bar{\kappa}_{ii}=\frac{s^2 T \mathcal{F}}{\left(\mathcal{F}^2+H^2\mathcal{G}^2\right)},\\
&&\bar{\kappa}_{ij}=\epsilon_{ij}\frac{s^2 T \mathcal{G}}{\left(\mathcal{F}^2+H^2\mathcal{G}^2\right)},\eea
The resitivity  is defined as the  inversion of the conductivity matrix:   
\bea
	\rho_{xx}&=&\frac{\sigma_{xx}}{\sigma_{xx}^2+\sigma_{xy}^2}=\mathcal{R}_{xx}/\mathcal{D}, 
 \nonumber \\
 	 \rho_{yx}&=&\frac{\sigma_{xy}}{\sigma_{xx}^2+\sigma_{xy}^2}=\mathcal{R}_{yx}/\mathcal{D}, 
\eea
where
\bea
	\mathcal{R}_{xx}&=&Z_2(\mathcal{F}+Y\mathcal{G}^2)(\mathcal{F}-Z_2H^2),\nonumber\\
	\mathcal{R}_{yx}&=&(\mathcal{F}^2+\mathcal{G}^2H^2)\Theta+Z_2H\mathcal{G}(2\mathcal{F}+Y\mathcal{G}^2-Z_2H^2),\nonumber\\
	\mathcal{D}&=&(Z_2\mathcal{F}+\mathcal{G}^2)^2+(\mathcal{F}^2+\mathcal{G}^2H^2)\Theta^2+H(Z_2^2H -2\mathcal{G}\Theta)(Z_2^2H^2-2Z_2\mathcal{F}-\mathcal{G}^2) .
\eea
The thermal conductivity when electric currents are turned off is given by
\bea\label{eq:kxxH}
	\kappa_{ij}&=&\bar{\kappa}_{ij}-T(\bar{\alpha}\cdot\sigma^{-1}\alpha)_{ij}  ,     \\
	\kappa_{xx}&=&\mathcal{K}_{xx}/\mathcal{D}, \qquad \kappa_{yx}=\mathcal{K}_{yx}/\mathcal{D} ,
\eea
where
\bea
	\mathcal{K}_{xx}&=&s^2T(Z_2^2\mathcal{F}+Z_2\mathcal{G}^2-Z_2^3H+\mathcal{F}\Theta^2+2Z_2H\mathcal{G}\Theta) , \nonumber\\
	\mathcal{K}_{yx}&=&s^2T(Z_2\mathcal{F}+\mathcal{G}^2)(\mathcal{G}+H\Theta).
\eea
From these quantities,  the Seebeck coefficient, $S$,  and the Nernst signal, $N$, are given by 
\bea
	S&&=(\sigma^{-1}\cdot \alpha)_{xx}=\mathcal{S}/\mathcal{D} ,\\
	N&&=(\sigma^{-1}\cdot\alpha)_{yx}=\mathcal{N}/\mathcal{D} ,
\eea
where
\bea
	\mathcal{S}&=&s(Z_2\mathcal{F}+\mathcal{G}^2)(\mathcal{G}+H\Theta) , \\
	\mathcal{N}&=&s(\mathcal{F}-Z_2H^2)(\mathcal{G}\Theta-Z_2^2H) .
\eea
The Hall angle is given by
\be
\cot \theta_H=\frac{\sigma_{xx}}{\sigma_{xy}}=\frac{\mathcal{R}_{xx}}{\mathcal{R}_{yx}}=\frac{Z_2(\mathcal{F}+Y\mathcal{G}^2)(\mathcal{F}-Z_2H^2)}{(\mathcal{F}^2+\mathcal{G}^2H^2)\Theta+Z_2H\mathcal{G}(2\mathcal{F}+Y\mathcal{G}^2-Z_2H^2)} .
\ee
Our result resembles the result of Dirac material because we could find  $\mathcal F$ and $\mathcal G$ given by eq(\ref{FG}). 
However, transport behavior is expected to depend on the class of exponent. In fact,  for $z=1$, the system becomes strongly interacting due to the small fermi surface, while higher exponents
cases are strongly interacting because the band is flat. 
Since the mechanisms for the strong correlation are  different, we do have reason to expect the difference between the Dirac and non-Dirac materials in transport coefficients as well as the band structure. 

\begin{figure}[h]
\centering \label{fig:phase}
      \subfigure[Phases of transports]
   {\includegraphics[width=50mm]{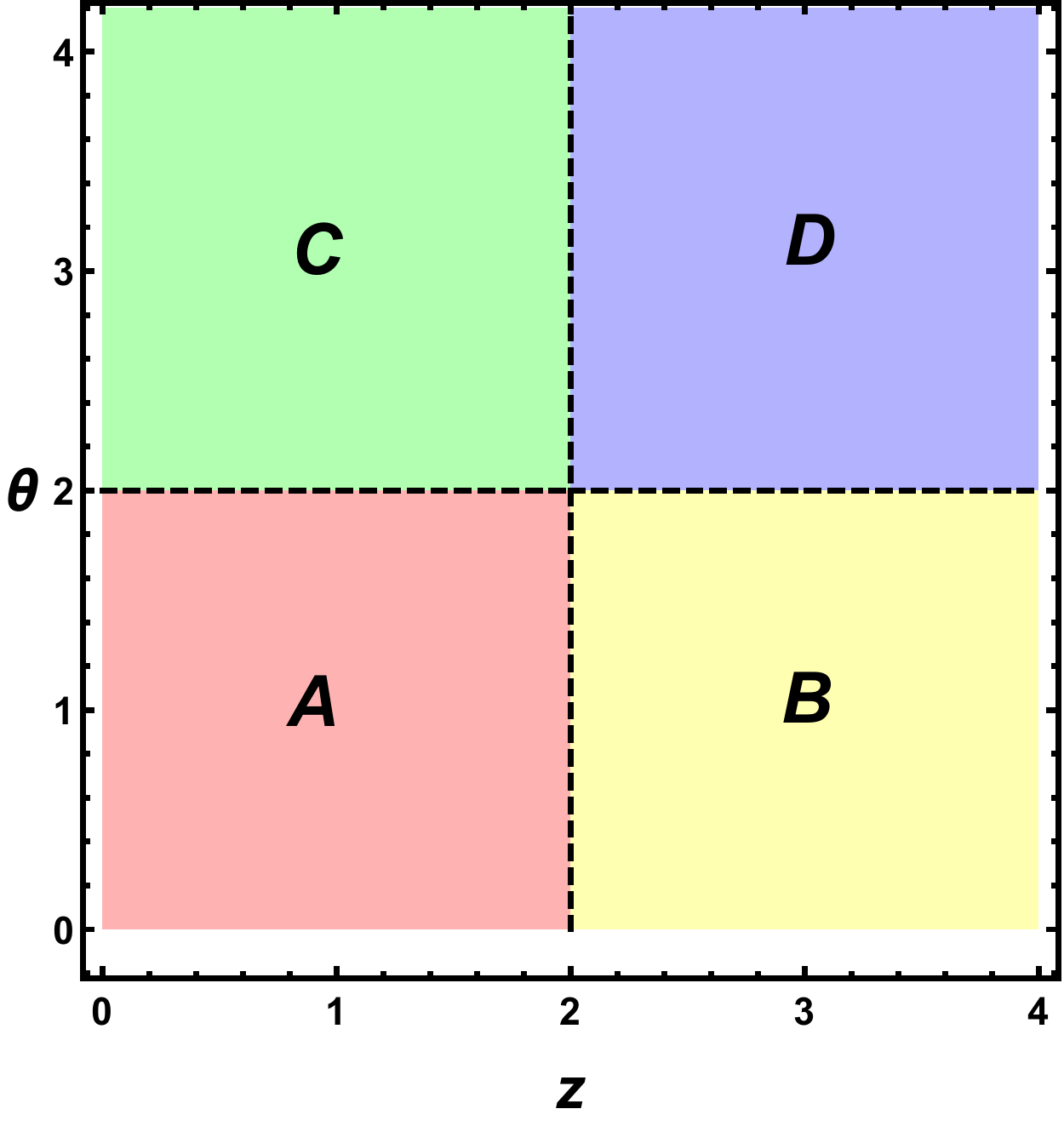} }
   \hskip 1cm
    \subfigure[Validity regime]
   {\includegraphics[width=50mm]{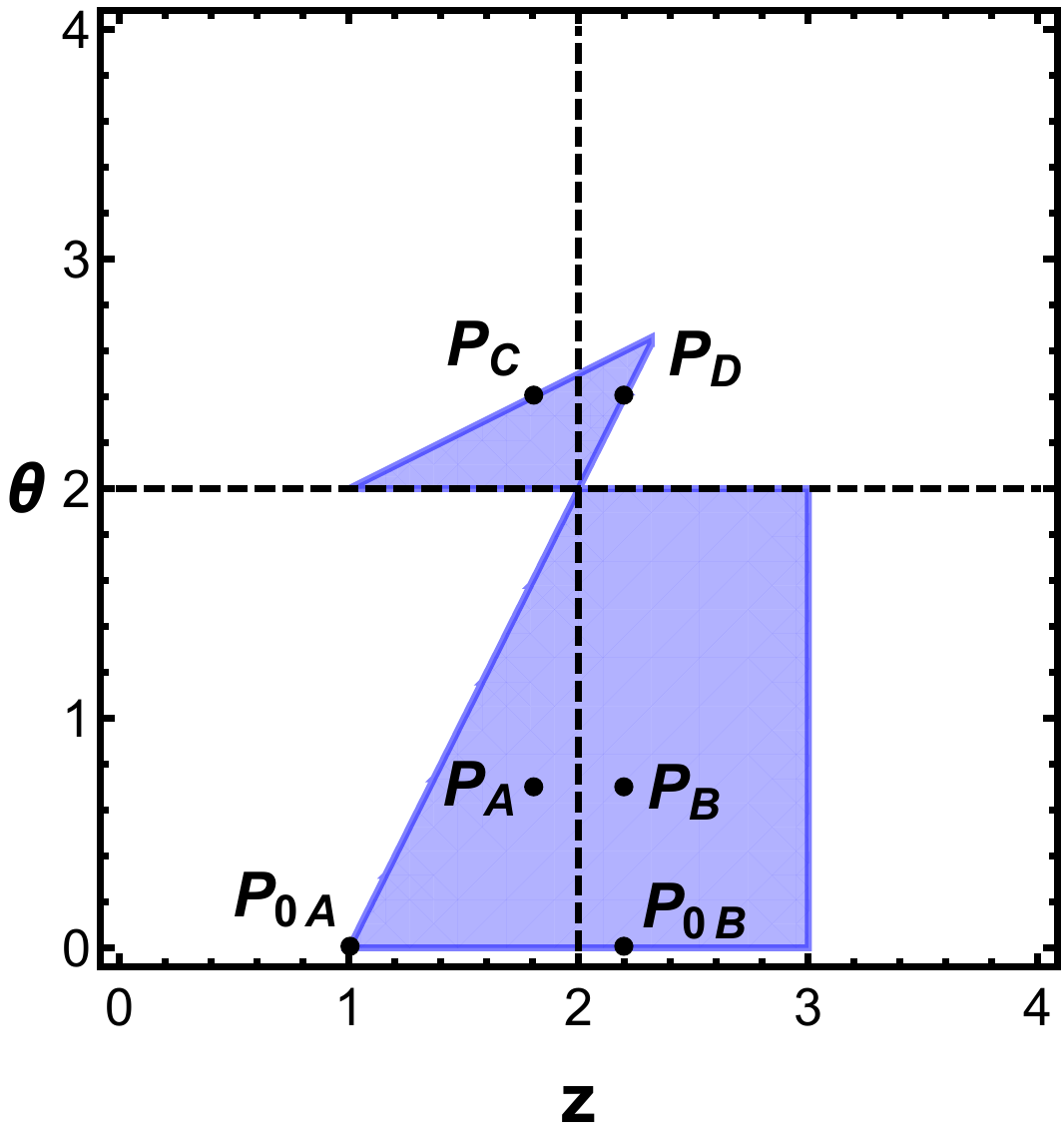} }
             \caption{ (a) 4-phases A,B,C,D due to the sign changes  at $z=2$  and $\theta=2$. 
             (b) Valid region satisfying all three conditions: (i)  $q_1$ charge reality condition, (ii) Null energy condition, (iii) Asymptotic HSV condition. Black dots indicate specific $(z,\theta)$ where we will discuss the typical behavior of transports from next section}    \label{phasetransport} 
\end{figure}

Before exploring such aspect of transports, it will be convenient to divide regions of $(\theta,z)$  as Figure \ref{phasetransport}. One can see from eq \ref{bgsol} that there's singularities at $z=2$ and $\theta=2$ . Therefore, there must be a rapid change in the transport behavior  along this boundaries. 
 We also have some restrictions on the $(z,\theta)$ which is determined by the following conditions:
\begin{itemize}
	\item  The reality condition of $q_1$ (\ref{muq}): $(2z-2)(2+z-\theta) \geq 0$
	\item Null energy condition (NEC):  ${(2-\theta)(2z-2-\theta)}\geq 0.$ See  appendix (\ref{nec}). 
	\item The condition that the background solution has hyperscaling violating geometry at the asymptotic boundary.
\end{itemize}
The regime which satisfies above three conditions is described in Figure \ref{fig:phase} (b).
In the appendices, we will examine the behaviors of transports in detail by plotting and analytic analysis and we will see that there are {\it phase transitions} at the phase boundaries of A,B,C,D {\it due to the  singularities} sitting there. 
Also there are {\it crossovers from weak localization to anti-localizations} whose boundaries are rather complicated. 
Although we list these plotting at the appendix to avoid too many figures in the main text, they are important by themselves because the plots   are predictions that can be easily compared with experiments. We will provide one such comparison in the next section.  

\section{ Application: magnetically doped surface state of  topological insulator}
\begin{figure}[h]
\centering
   \subfigure[$Q_2$ and its 3 nearby QCP's]
   {\includegraphics[width=5cm]{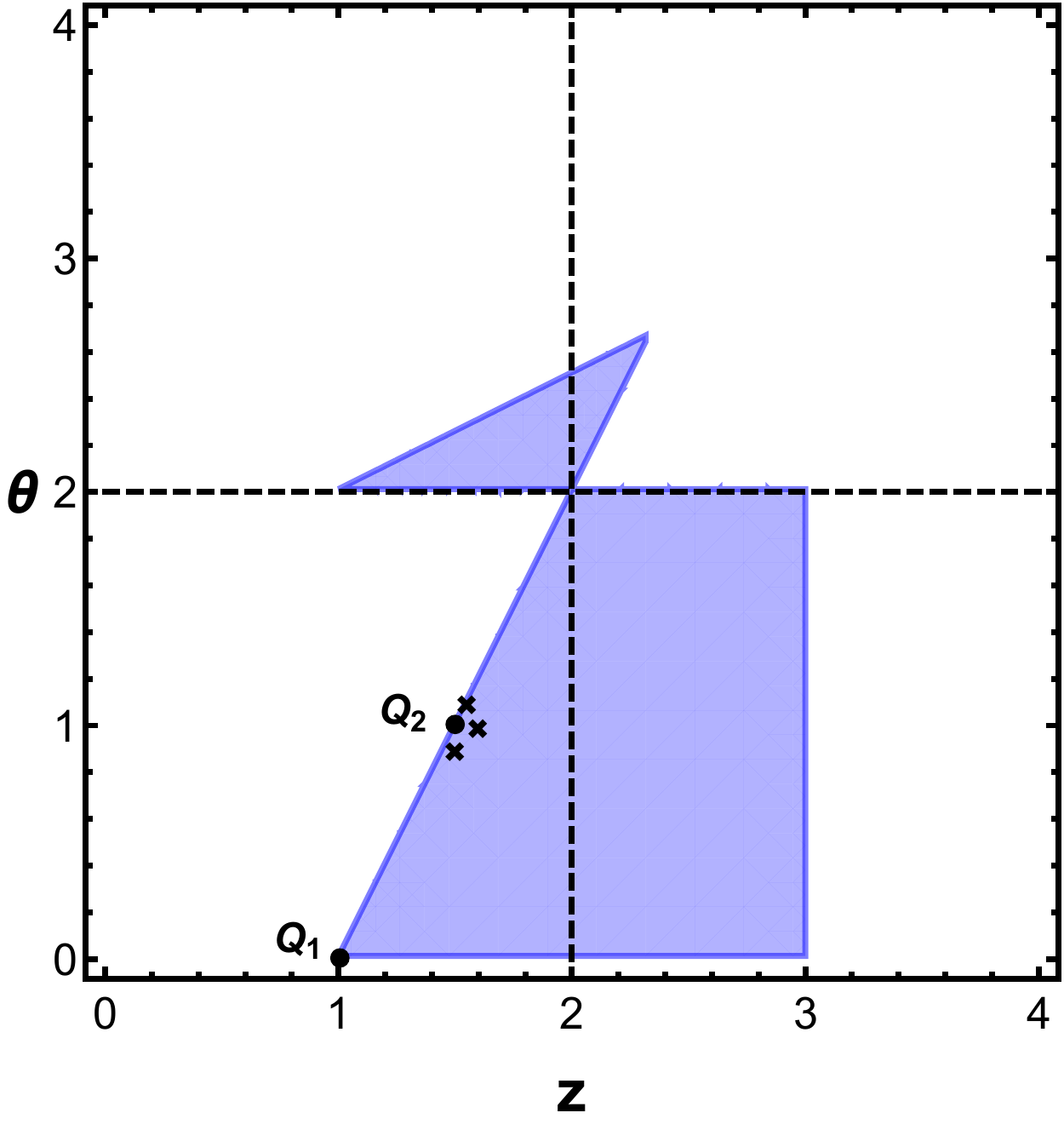}  }
    \subfigure[At $Q_2$]
   {\includegraphics[width=5.5cm]{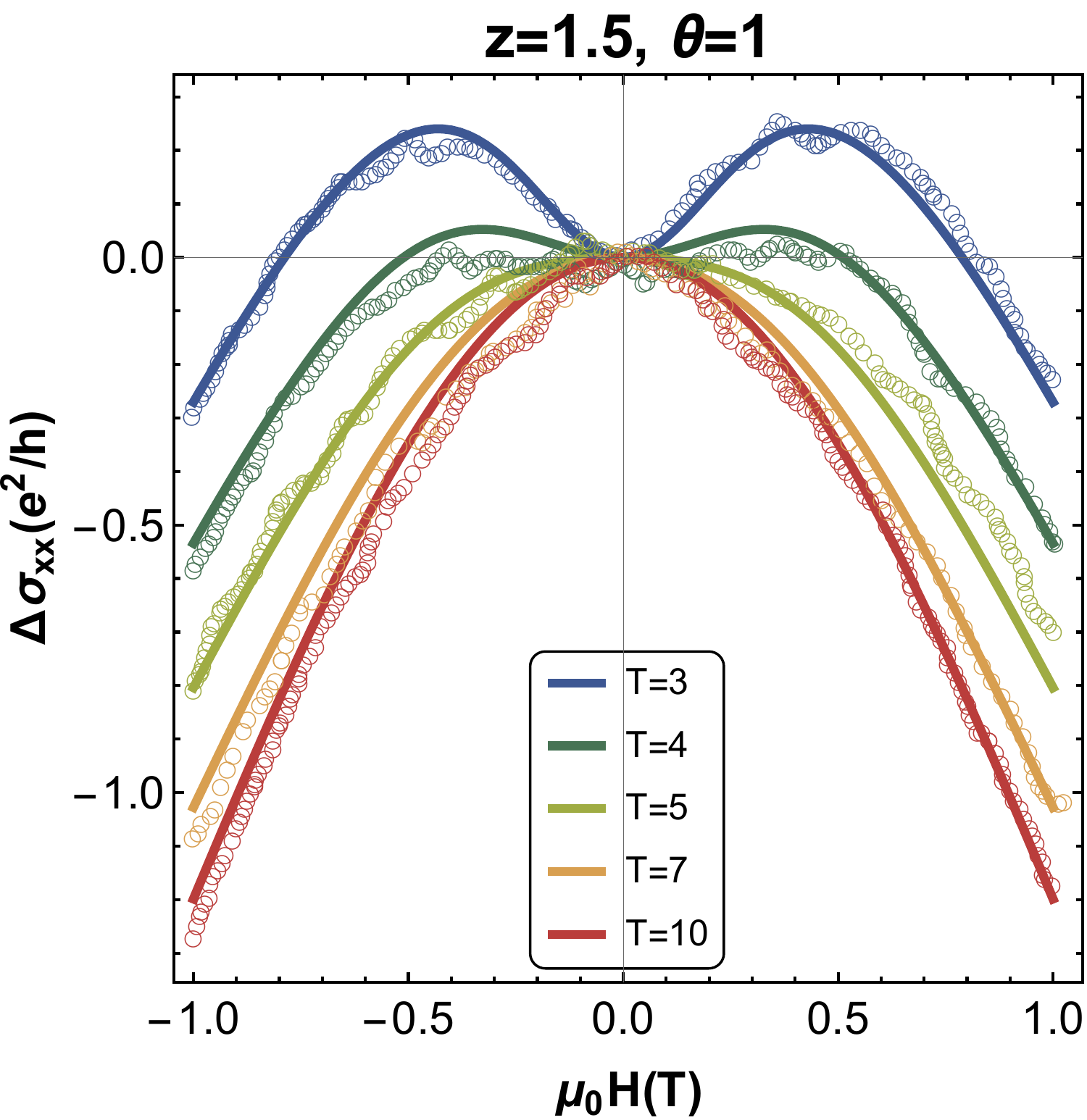}  }\\
  \subfigure[ right of $Q_2$]
   {\includegraphics[width=4.5cm]{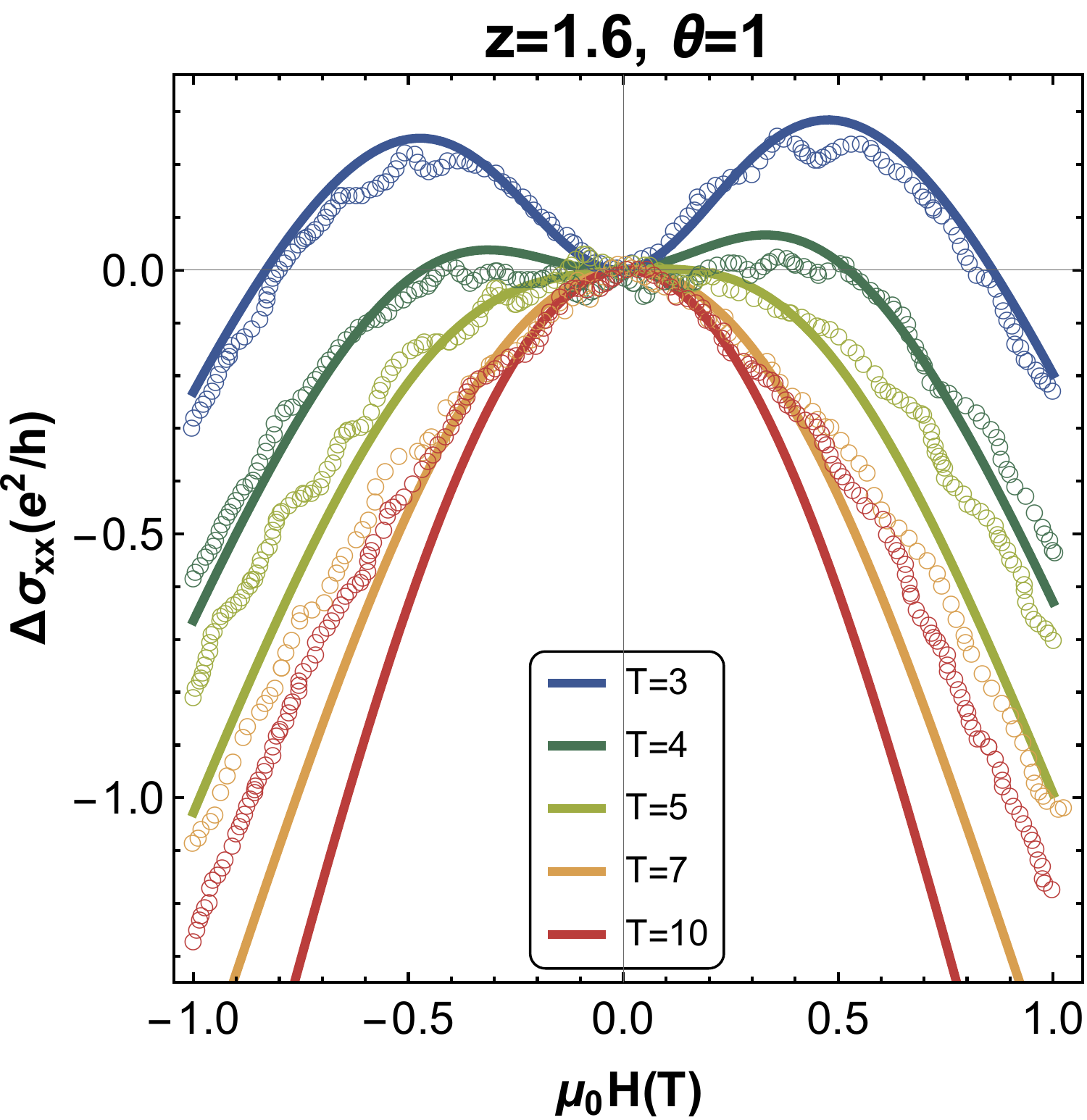}  }
 \subfigure[ below $Q_2$ ]
   {\includegraphics[width=4.5cm]{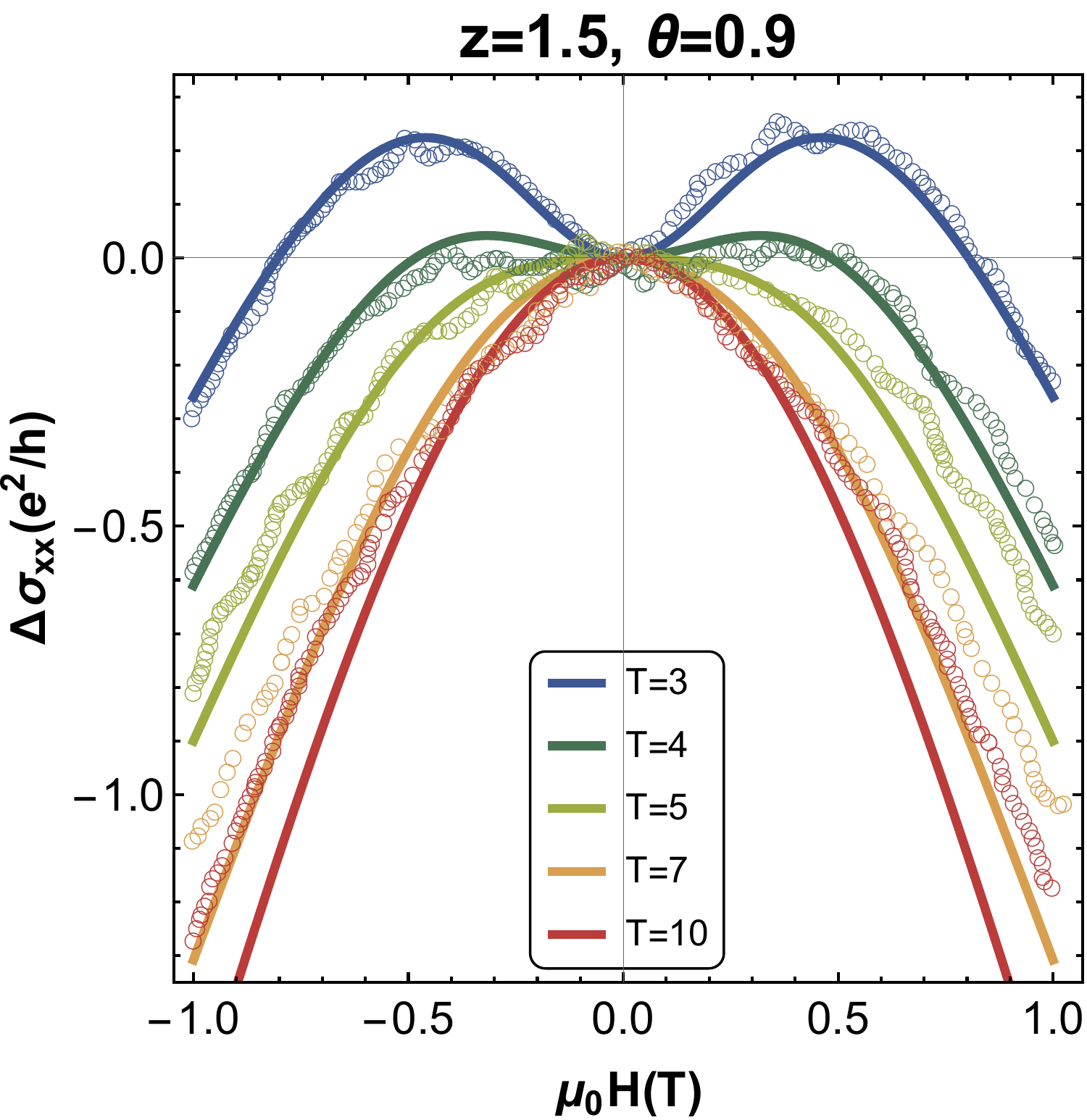}  }
   \hskip 0.5cm
 \subfigure[above $Q_2$ ]
   {\includegraphics[width=4.5cm]{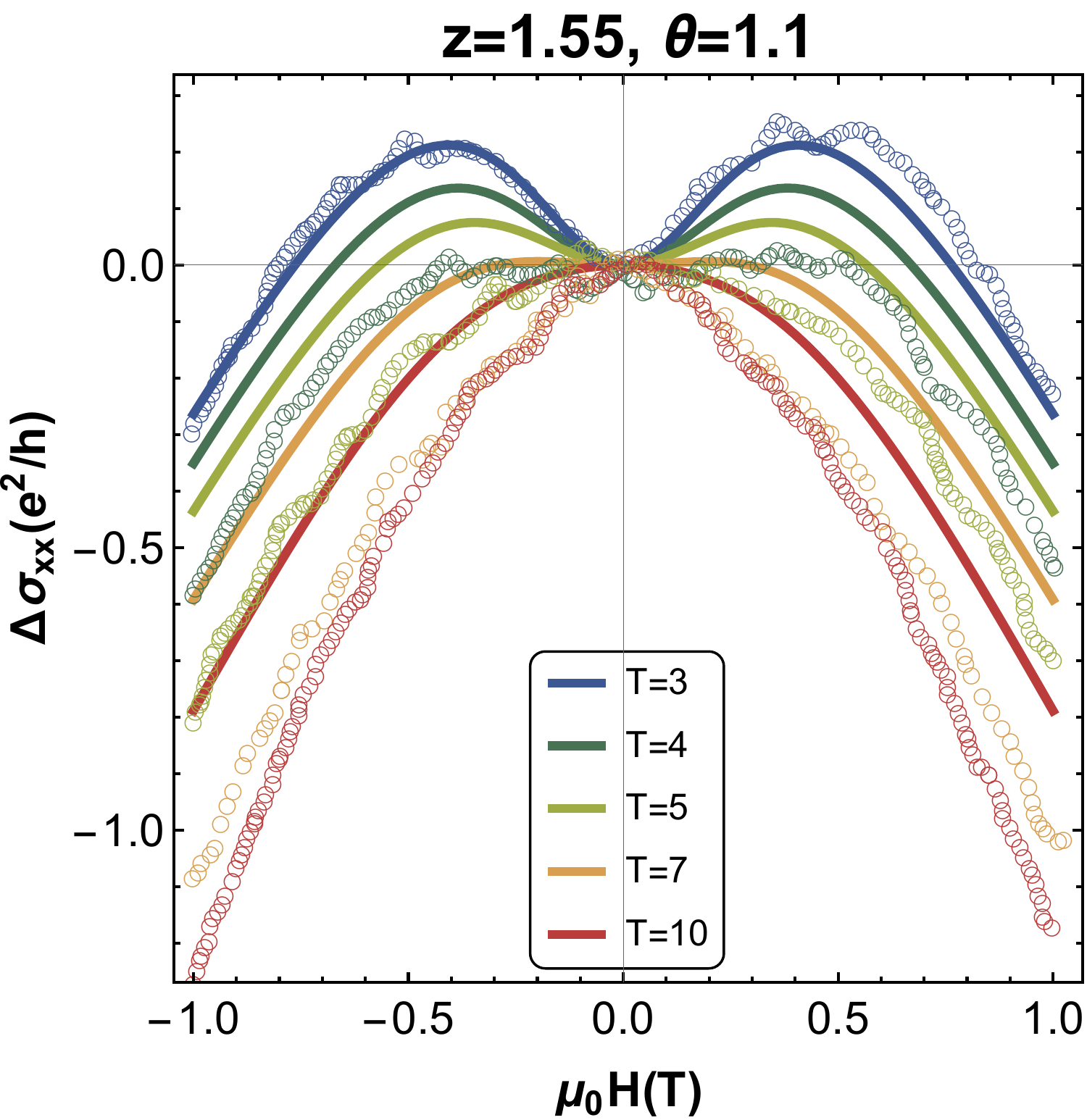}  }
    \caption{ (a) Two quantum critical point $Q_1, Q_2$: 
    $Q_1$ is for  Cr doped Bi$_2$Te$_3$ discussed in \cite{Seo:2017oyh} and $Q_2$ is for  Mn doped for Bi$_2$Se$_3$ which is the topic here.    (b,c,d,e) The uniqueness of $Q_2$, the  QCP with $(z,\theta)=(3/2, 1)$.    The data is for   Mn$_{0.04}$(Bi$_2$Se$_3$)$_{0.96}$ taken from ref. \cite{zhang2012interplay}. Solid line and circles are for  theory curve and experimental data respectively. 
  The fitting is good  at  $Q_2$ as one can see in (b).   But as we deviate from it, the fitting becomes bad   rapidly for high temperature data. See (c,d,e).   }
    \label{fig:TI}
\end{figure}

We now apply the result  to the surface state of topological insulator with gap opened by the magnetic doping\cite{zhang2012interplay},  which was one of our main motivation. 
After extensive search,  it  turns out that the best fit  comes from $(z,\theta)=(3/2, 1)$, which we call QCP $Q_2$. 
As we move away from   $(3/2, 1)$, the fitting becomes bad very rapidly as one can see in the figure \ref{fig:TI}(b,c,d).  
Therefore  there is no ambiguity in  associating  the surface state of Mn doped Bi$_3$Se$_3$ with a QCP with $(z,\theta)=(3/2,1)$. 
 In each figure of \ref{fig:TI}(b,c,d,e) we first fix the parameters such that the theory   best fit the $T=3$K data (blue line), then we use them for data at other temperatures.
 The parameters   used in figure \ref{fig:TI} are listed in table \ref{table:1}.  
\begin{table}[htp]
\caption{Used parameters in figure    \ref{fig:TI}}
\begin{center}
\begin{tabular}{|c|c|c|c|c|}
\hline
 & $(z,\theta)$  & $\beta^2 (\mu m)^2$ &  $v/(10^4m/s)$ & $q_{\chi}\gamma$ \\
 \hline\hline
  (a) &  (1.5,1)  & 2704 &  5 & 3.1 \\
   \hline
     (b) &  (1.6,1)  & 2591 &  4.3 & 2.6 \\
  \hline
     (c) &  (1.5,0.9)  & 3136 &  5 & 2.15 \\
  \hline
     (d) &  (1.5,1.1)  & 3136 &  5 & 3.3 \\
  \hline
\end{tabular}
\end{center}
\label{table:1}
\end{table}%

Here should comments on the material dependence of the 
QCP. Previously we could fit the data of Cr doped  Bi$_2$Te$_3$ with $z=1,\theta=0$ \cite{Seo:2017oyh}.
There, we also claimed that  the   data of Mn doped  Bi$_2$Se$_3$  
can also be fit with $(z,\theta)=(1, 0)$. 
However, it turns out that when we fix the parameters of the theory such that  we can fit the data at  low temperature (T=3K), higher temperature data of Bi$_2$Te$_3$ could be fit only if we allow the temperature dependence of the coupling $q_\chi$, while 
the theory can fit the data without such  tweaking if we use 
 $(z,\theta)=(3/2, 1)$. 
 
 What is the origin of  the difference between the two surface state? 
First the surface gap of  doped Bi$_2$Se$_3$ is much bigger than that of doped Bi$_2$Te$_3$.  The former has stronger Coulomb interaction between the electrons \cite{rienks2019large}  
so that  the system behaves as a strongly interacting system even at low doping.  On the other hand, doped Bi$_2$Te$_3$ is a weakly interacting   system at low doping. The surface Dirac point  of   Bi$_2$Te$_3$ is hidden in the valley of bulk band. It is   hard to see the presence of the surface gap   out of  ARPES data  while the surface gap is manifest for the doped 
 Bi$_2$Te$_3$.  See  the figure \ref{fig:111}(b and c). 
Therefore the two surface states   are  very  different   as QCP's although   they apparently look similar as surface states of TI. 

\begin{figure}[ht!]
\centering
    \subfigure[At or near QCP]
   {\includegraphics[width=4.5cm]{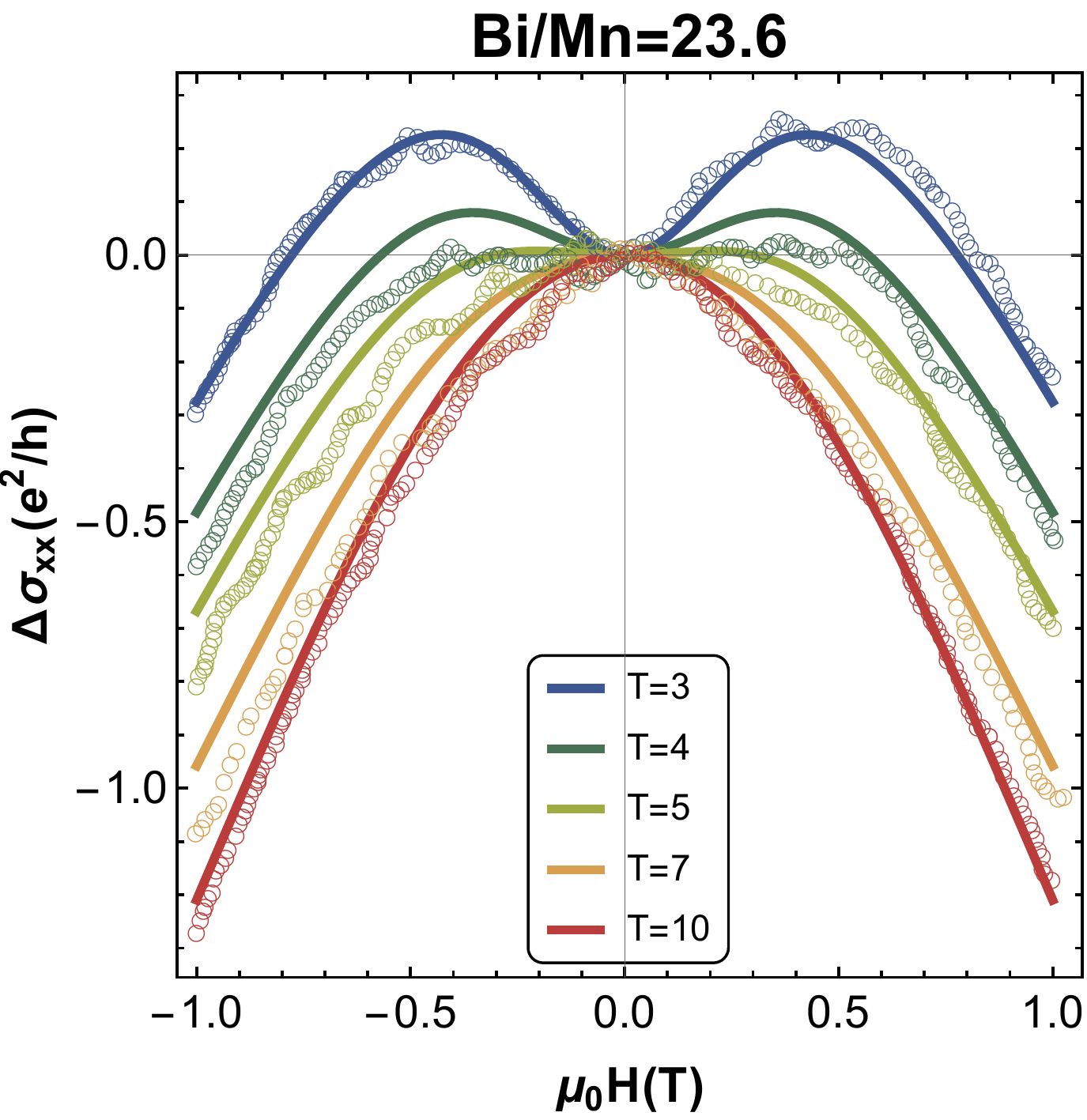}  }
  \subfigure[Far from the QCP]
   {\includegraphics[width=4.5cm]{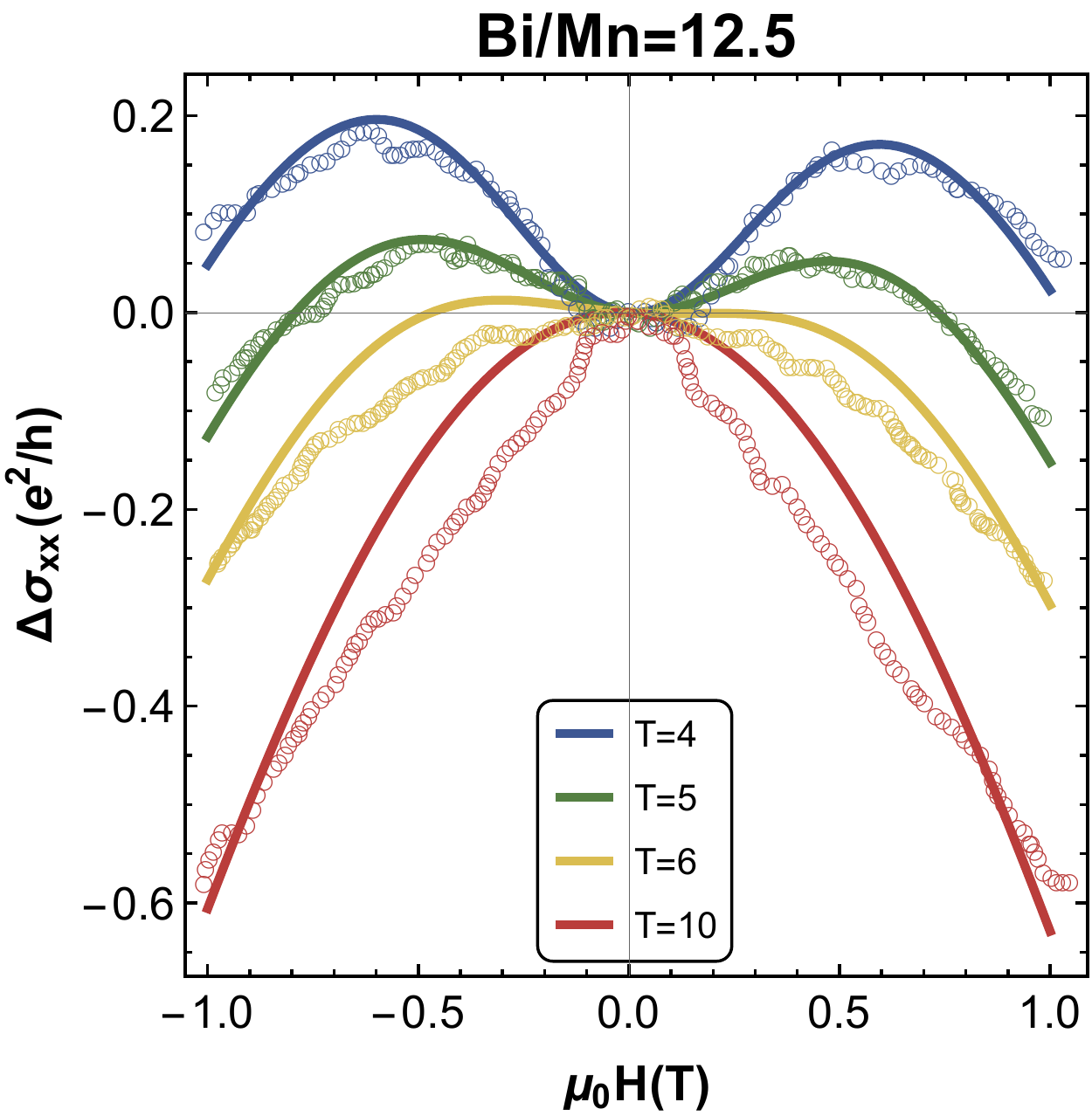}  }
 \subfigure[Farther from the QCP ]
   {\includegraphics[width=4.5cm]{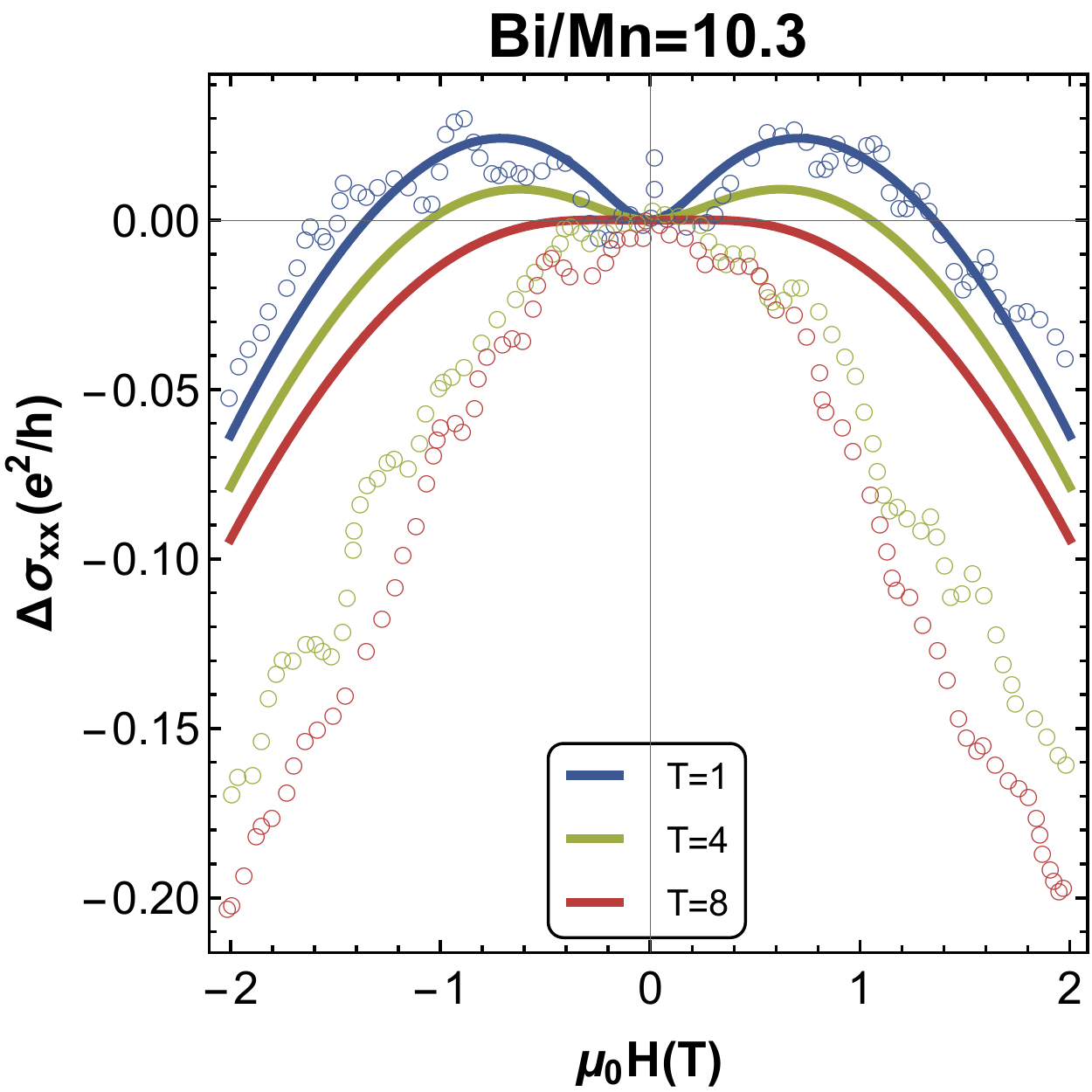}  }
     \caption{QCP  appears only at special dopings depending on the base material, where the data can be described by our theory. }
     \label{fig:Dis}
\end{figure}

 Finally we remark that quantum critical point appears only at special dopings depending on the base material.
We demonstrate this  by showing that  among three doping inverse ratio  Bi/Mn=23.6,  
12.5, 10.6 only the first data could be fit by our theory. 
See  figure \ref{fig:Dis}. 
In all three cases, we fix the theory parameter to fit the lowest temperature T=3K. We tabulated the parameters we used in table 2.  If there were a scaling behavior, the same parameter set should able to fit the all data. Only  Bi/Mn=23.6 data has such scaling property. That is only this case is at or near a QCP. 

\begin{table}[htp]
\caption{Used parameters in  figure \ref{fig:Dis}}
\begin{center}
\begin{tabular}{|c|c|c|c|c|}
\hline
 & Bi/Mn  & $\beta^2 (\mu m)^2$ &  $v/(10^4m/s)$ & $q_{\chi}\gamma$ \\
 \hline\hline
  (a) &  23.6 & 2704 &  5 & 3.1 \\
   \hline
     (b) &  12.5 & 4096 &  6.67 & 2.7 \\
  \hline
     (c) &  10.3 & 4502 &  6.67 & 0.7\\
  \hline
\end{tabular}
\end{center}
\label{table1}
\end{table}%

%
%
%
%

\section{Conclusion and discussions}
In this paper we reported a new black hole solution with  hyperscaling violation which is 
relevant  to an impurity doped quantum  materials.  
We   calculated all transport coefficients including electrical, thermo-electric and heat  conductivities. We  investigated  their properties in detail by plotting the analytic results as a function of physical parameters. 
We also investigated the   phase transitions from weak localization to weak anti-localization as the critical exponent changes. 

In the introduction we mentioned the physics of the surface state of topological insulator with magnetic doping. 
It turns out that there are  many subtle issues  in discussing the physics of QCP related to the data of topological surfaces: 
for example, the Dirac cone at the surface of the TI is due to the zero-mode soliton, which should be distinguished from a  local fermions. It is the consequence of collective motion of many electrons and it is not clear whether we can consider them as fermions satisfying the  Pauli principle. 
 We hope we can treat this issue in future work.   

One interesting question is whether we can find a strange metal properties where resistivity $\sim T$ and 
Hall angle $\Theta\sim 1/T^{2}$ for some exponent. 
It turns out that the relevant point is $(z,\theta)=(1,1)$. However it is out  of  the stability condition requested by the  null energy condition, explained in appendix C.  It is too early to abandon such regime, since we did not in-cooperate the presence of the neutralizing ionic charges into gravitational context.

Finally the most interesting project would be the matching the gravity solutions with real systems, which would request collecting all available data. We wish to come back to this issue soon.

\acknowledgments
 This  work is supported by Mid-career Researcher Program through the National Research Foundation of Korea grant No. NRF-2016R1A2B3007687.  XHG were partly supported by NSFC, China (No.11875184 $\&$ No.11805117). 
YS is  supported  by Basic Science Research Program through NRF grant No. NRF-2019R1I1A1A01057998.

\newpage
\appendix 
\noindent{\bf \Large Supplementary materials}
\section{Magneto-transports vs Quantum critical points}
In this section, we discuss the typical behaviors of magneto-transport by  taking a point from each  sector A,B,C,D of the Figure \ref{phasetransport}(a). 
We take two more points along the $\theta=0$ line.  
 Notice that the transverse quantities $\sigma_{xy}$,  $\kappa_{xy}$, and $S$ vanish when $q_{\chi}= \mu=0$. We used $\alpha=0$ and $\beta=\lambda=1.3$. 

\subsection{Classifying the magneto-conductivity   in $(z,\theta)$ plane}
\begin{figure}[ht!]
\centering
	 \subfigure[$P_{0A}$ at $q_{\chi}=0$]
   {\includegraphics[width=30mm]{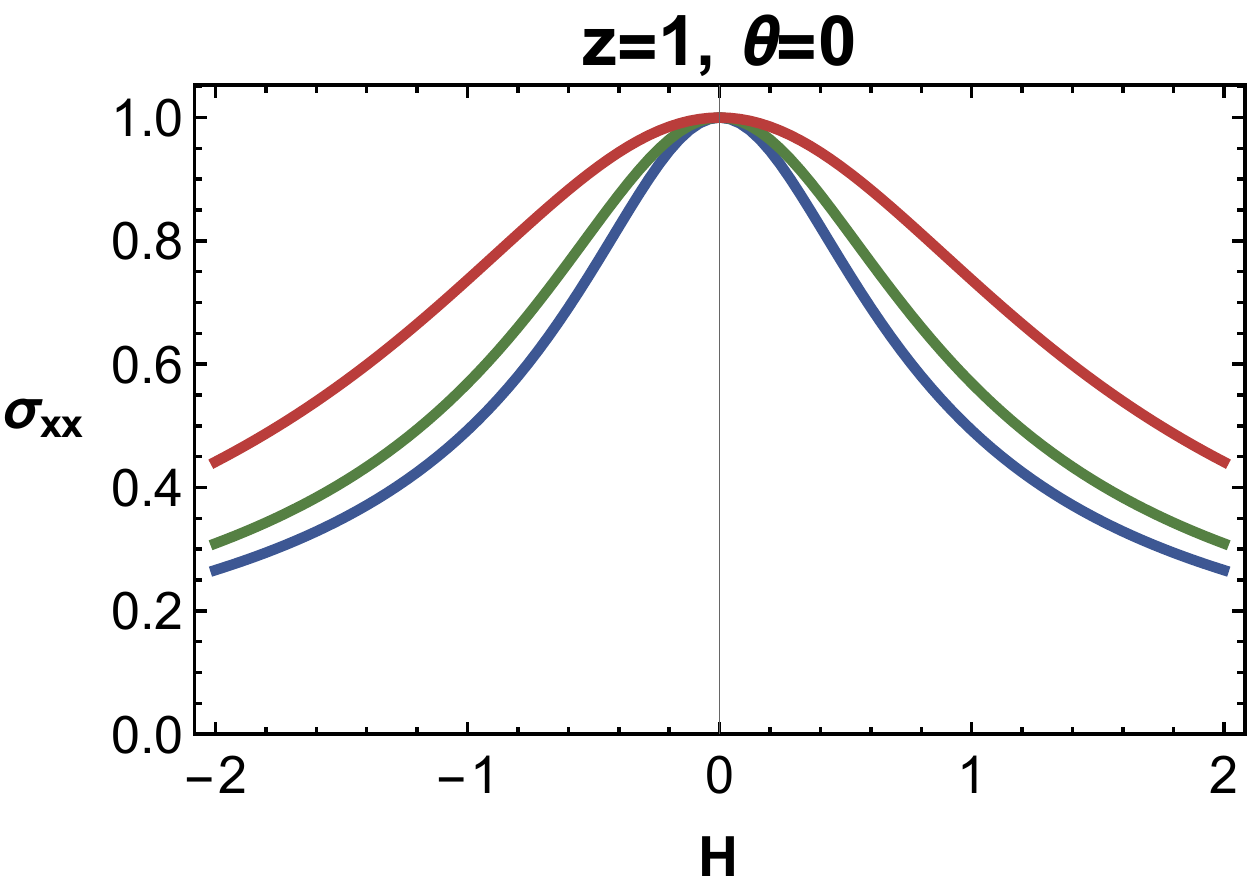} }
    \subfigure[$P_{0A}$ at  $q_{\chi}=0.7$]
   {\includegraphics[width=37mm]{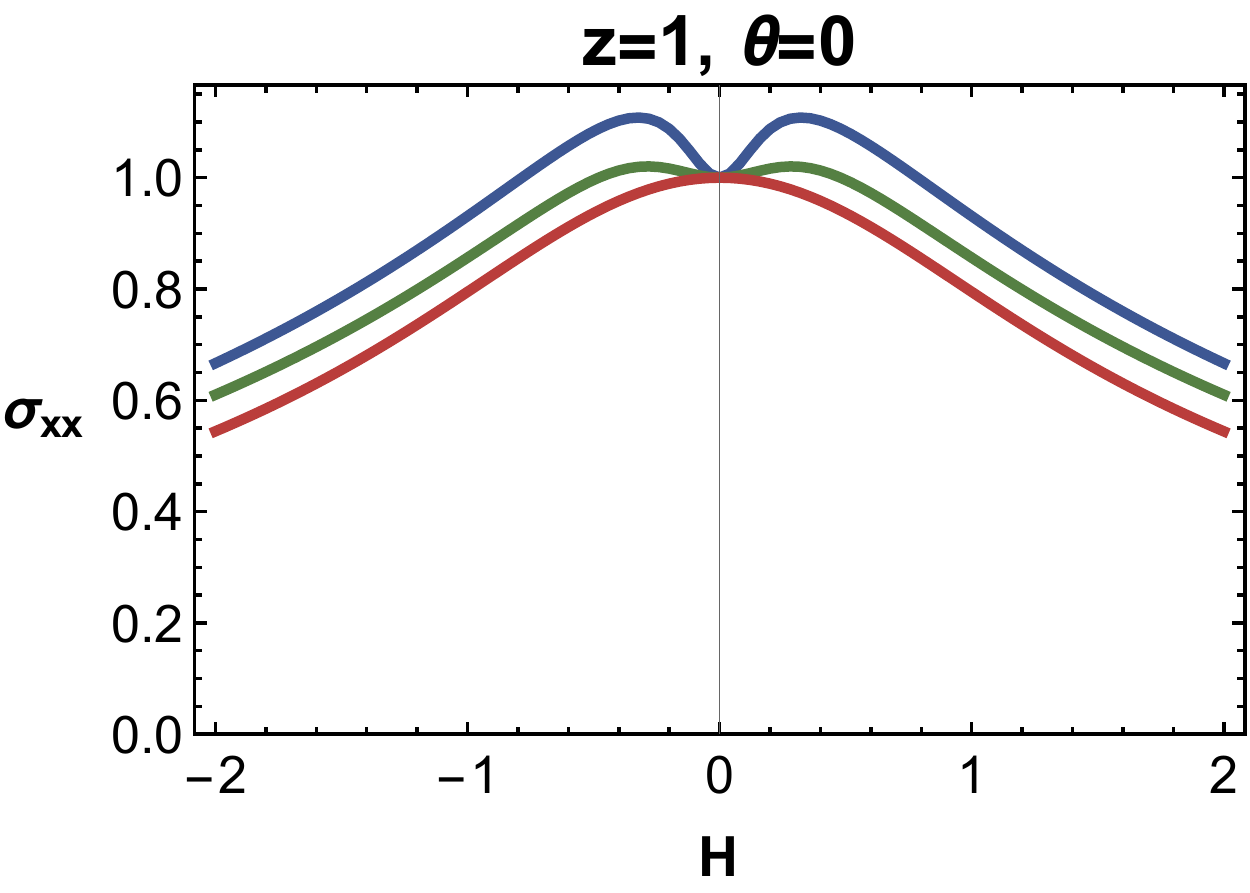} }
    \subfigure[$P_{0B}$ at $q_{\chi}=0$]
   {\includegraphics[width=30mm]{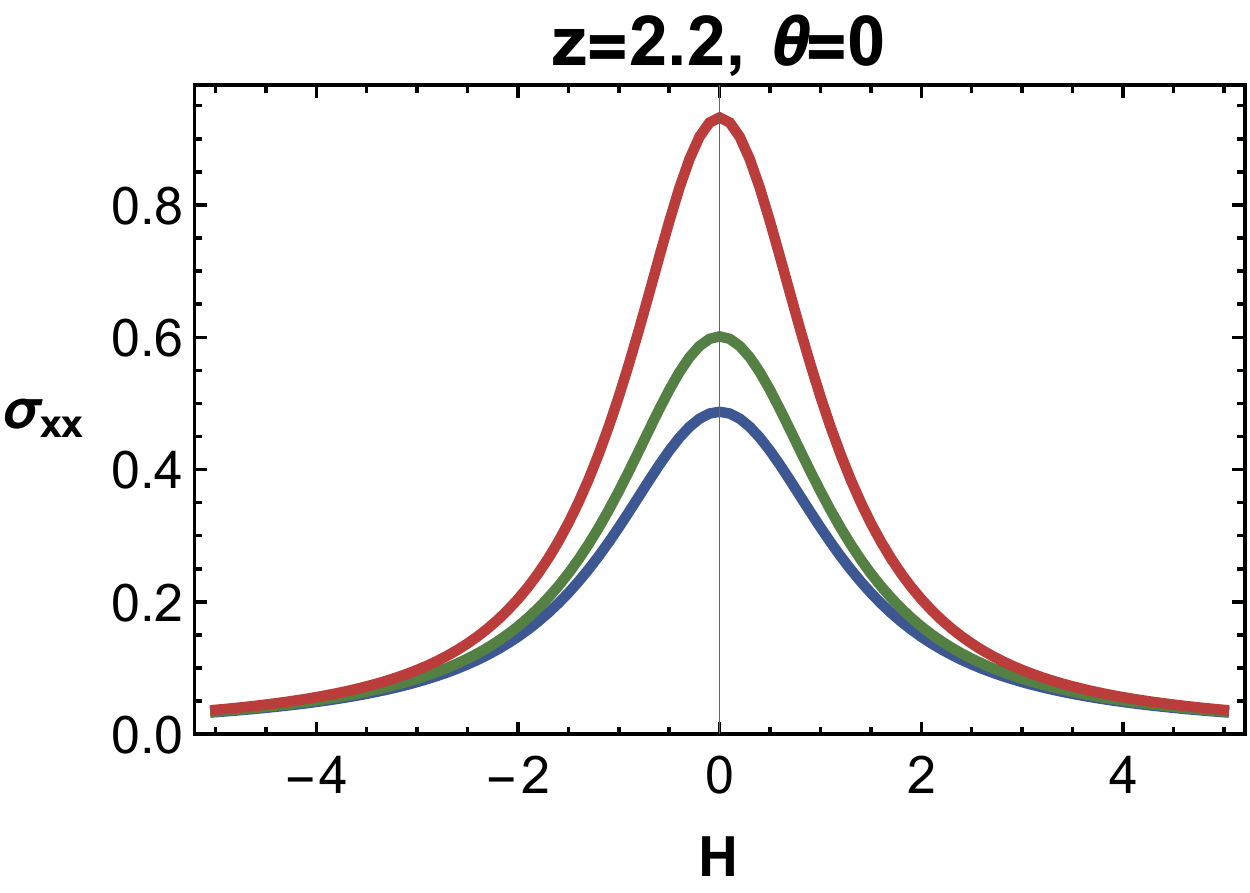} }
    \subfigure[$P_{0B}$ ]
   {\includegraphics[width=37mm]{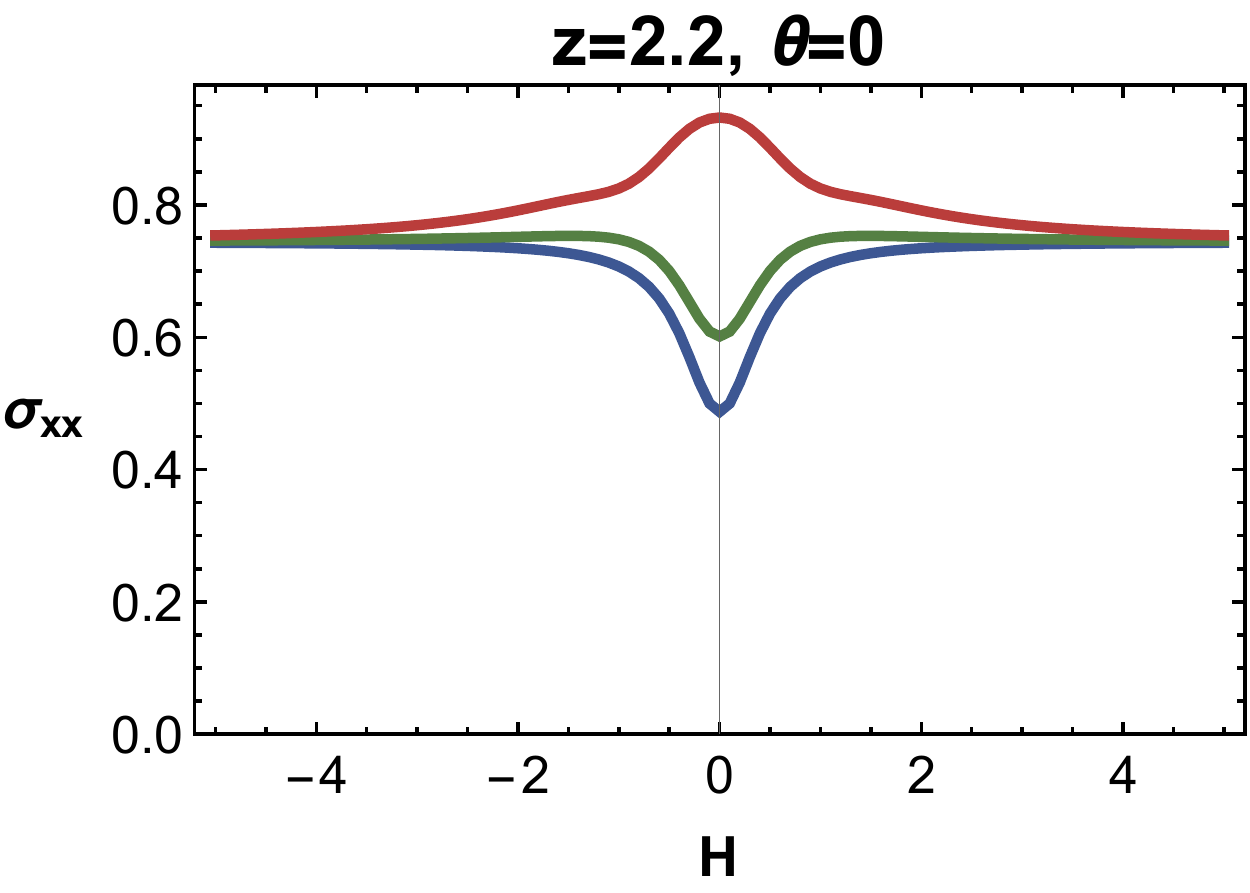} }
    \subfigure[$P_{A}$ at $q_{\chi}=0$]
   {\includegraphics[width=30mm]{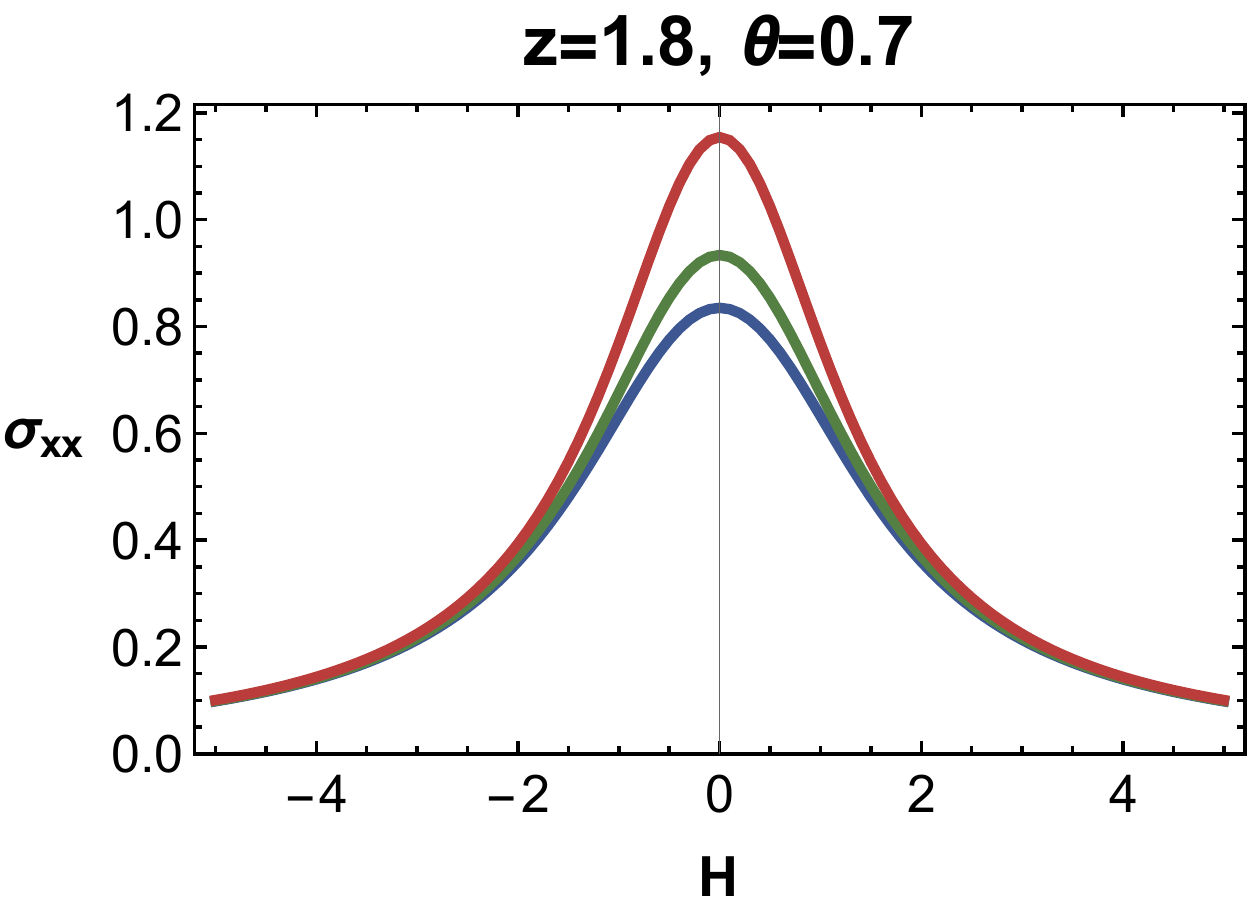} }
    \subfigure[$P_{A}$  ]
   {\includegraphics[width=37mm]{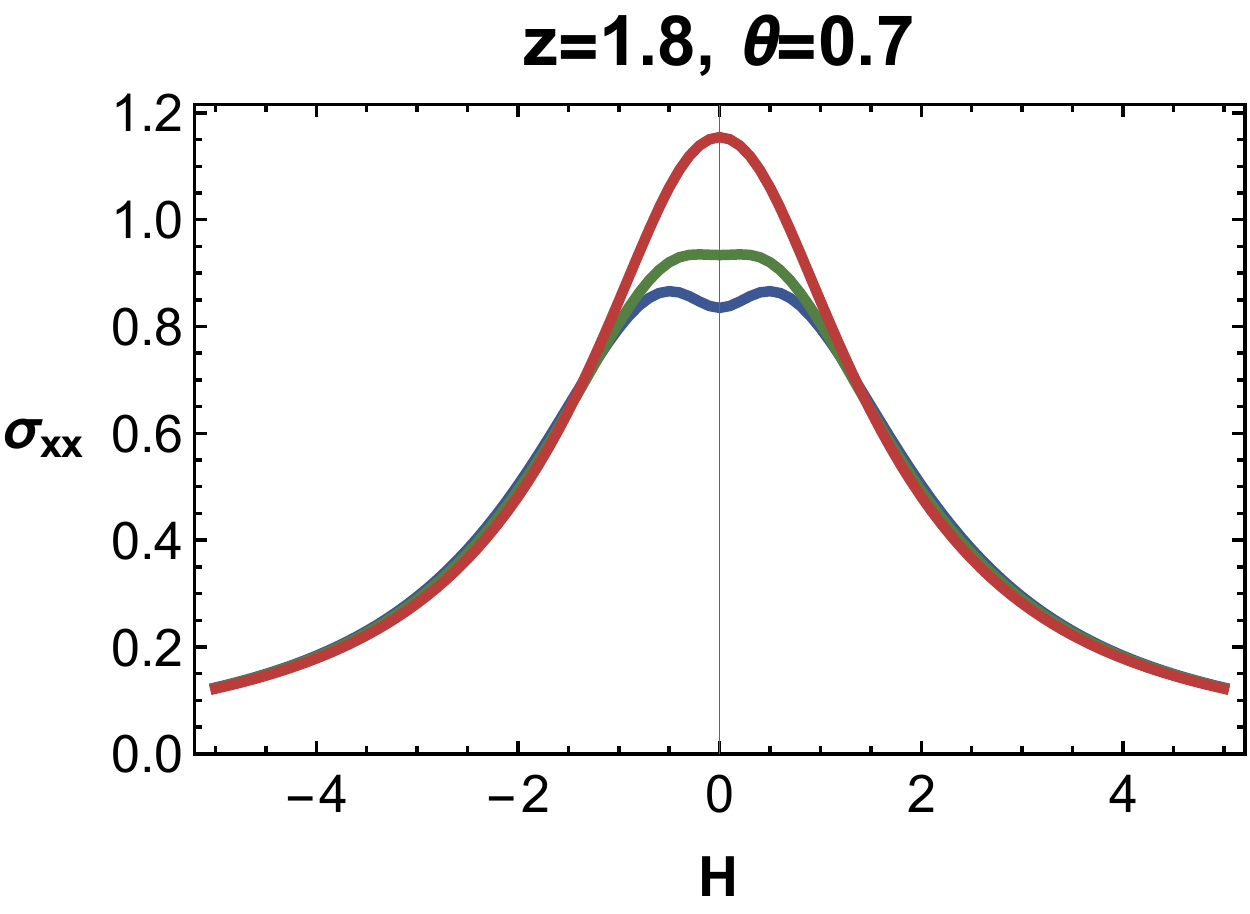} }
      \subfigure[$P_{B}$ at $q_{\chi}=0$]
   {\includegraphics[width=30mm]{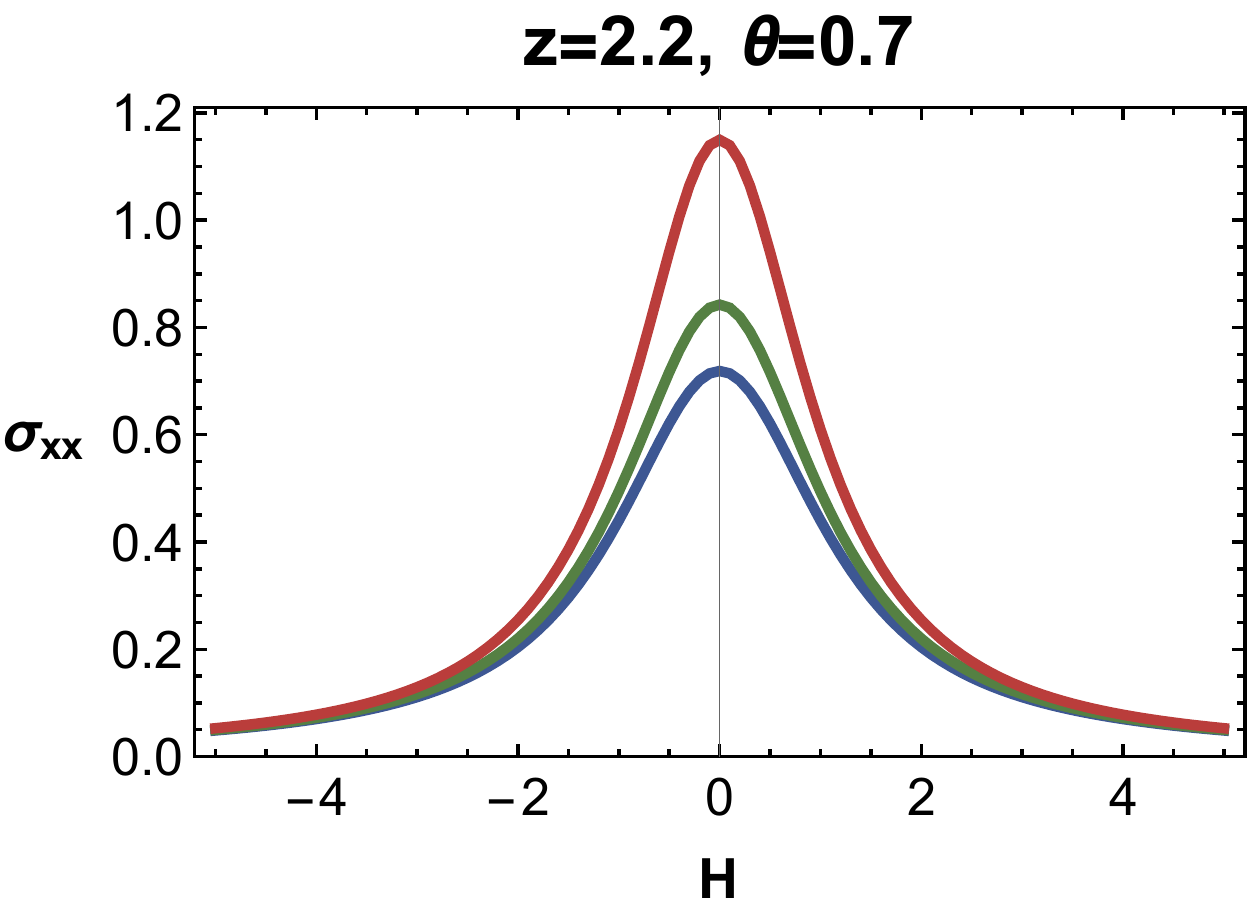} }
    \subfigure[$P_{B}$  ]
   {\includegraphics[width=37mm]{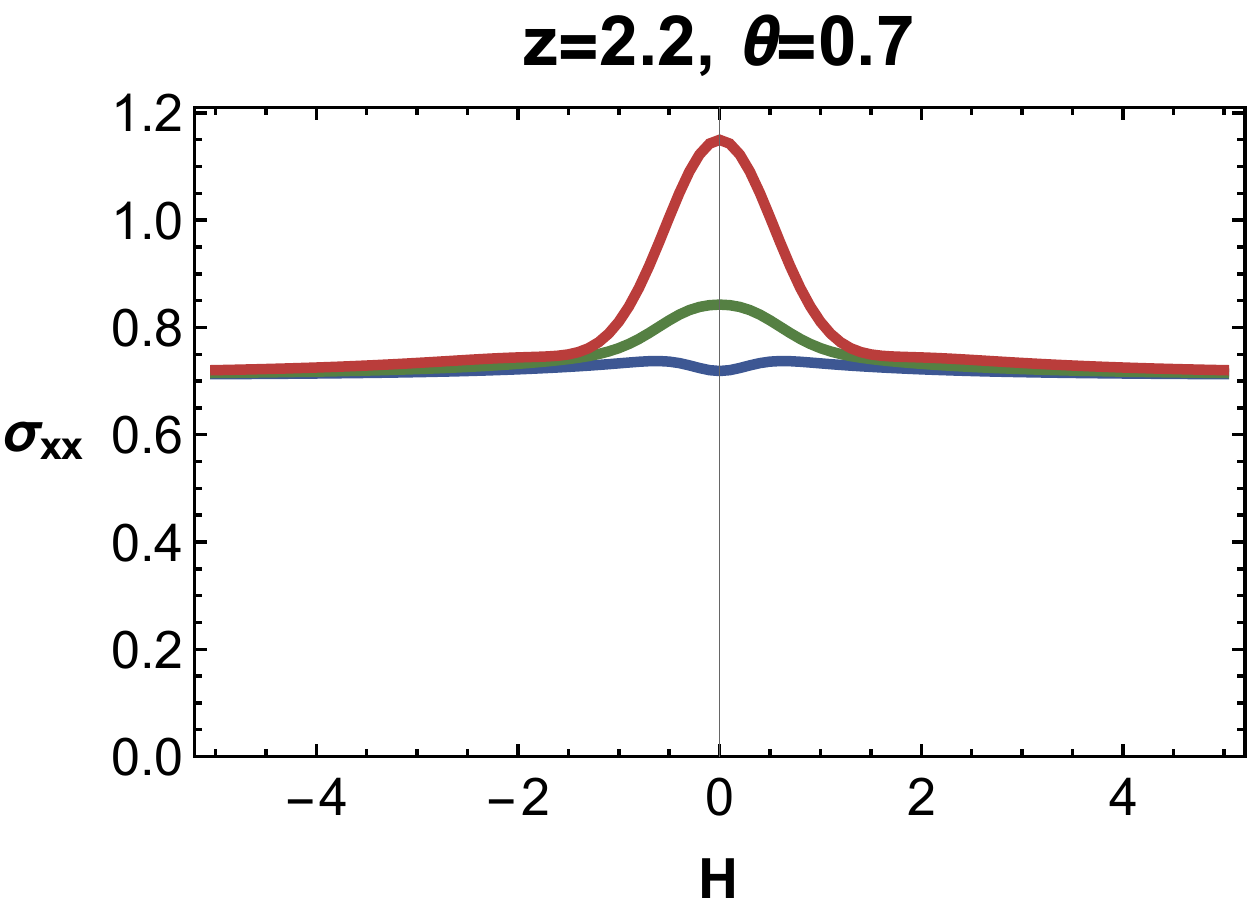} }
     \subfigure[$P_{C}$ at $q_{\chi}=0$]
   {\includegraphics[width=30mm]{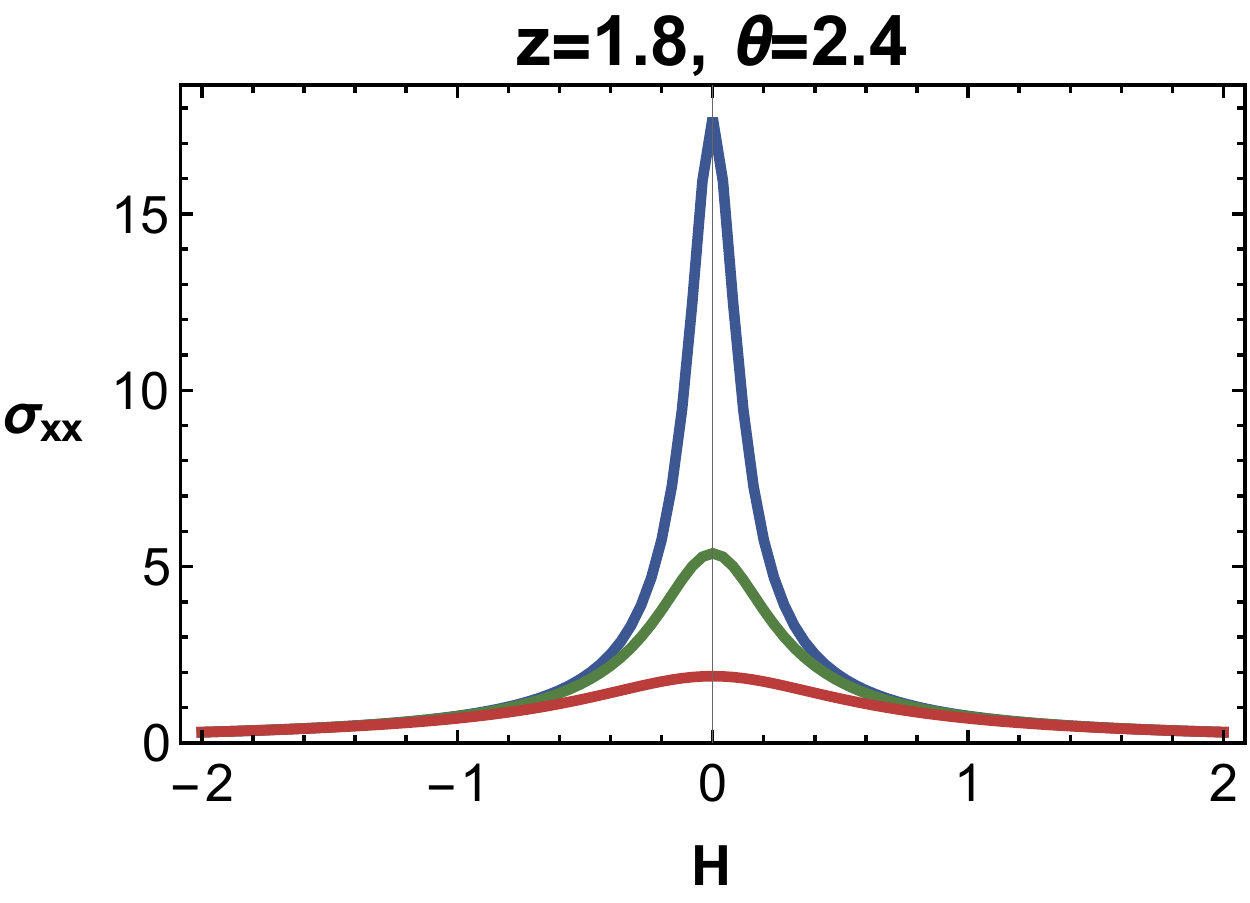} }
    \subfigure[$P_{C}$ ]
   {\includegraphics[width=37mm]{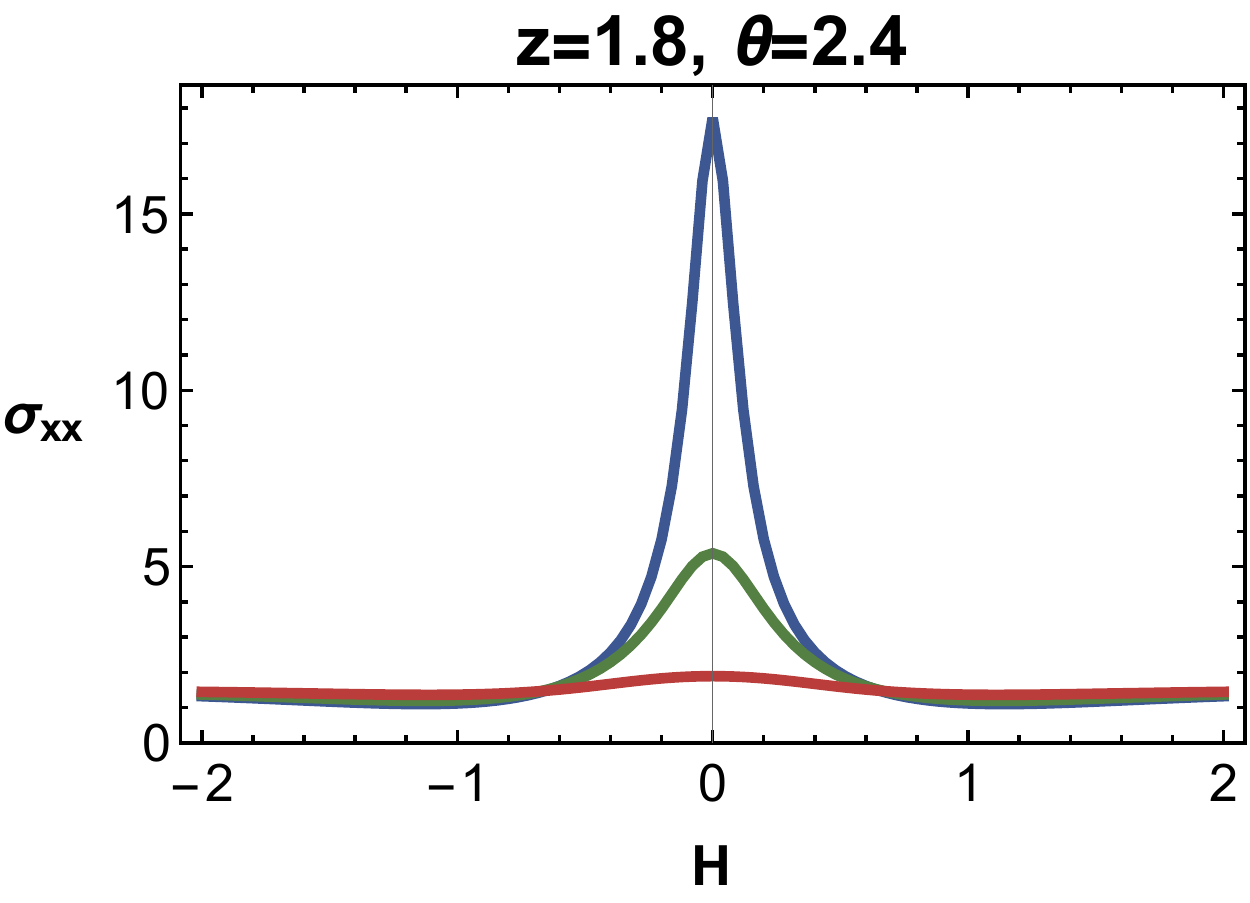} }
    \subfigure[$P_{D}$ at $q_{\chi}=0$]
   {\includegraphics[width=30mm]{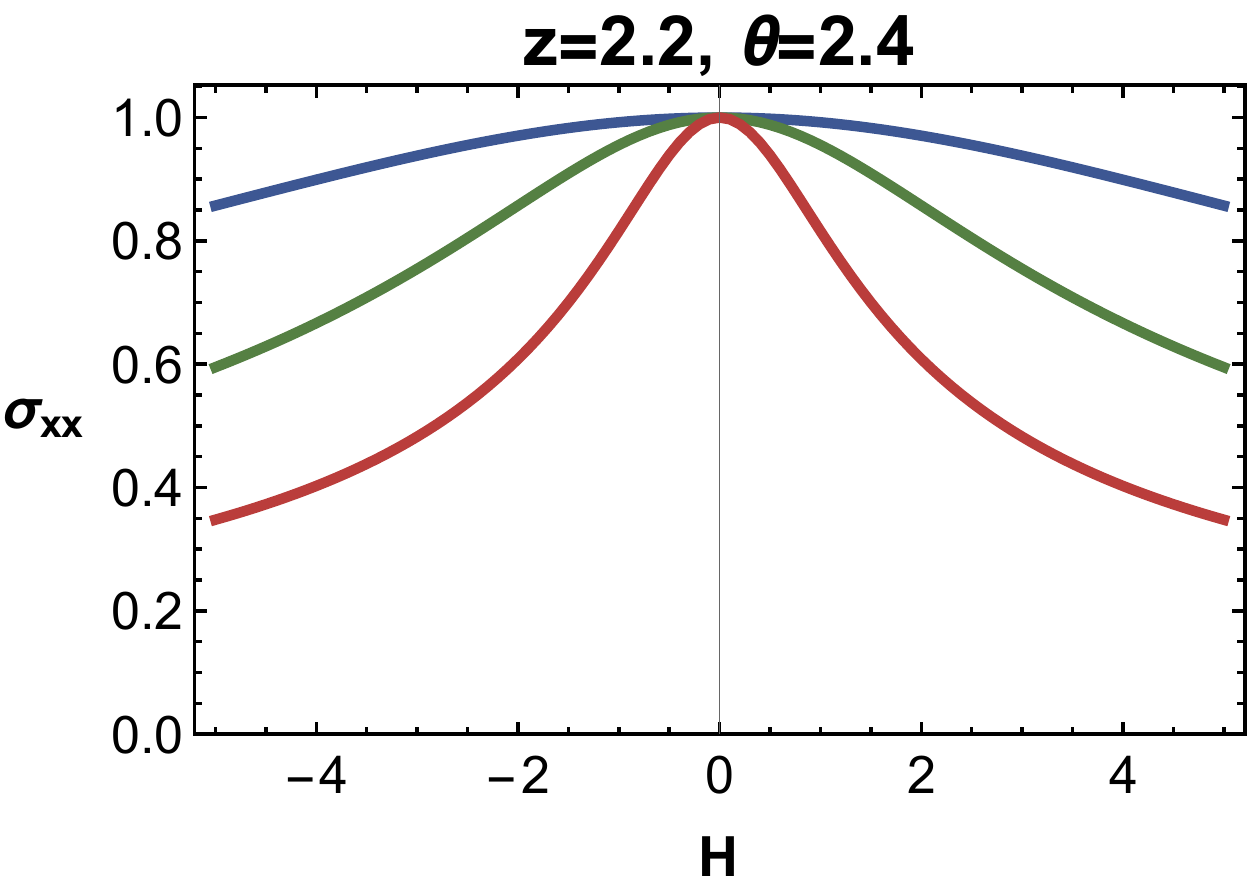} }
    \subfigure[$P_{D}$ ]
   {\includegraphics[width=37mm]{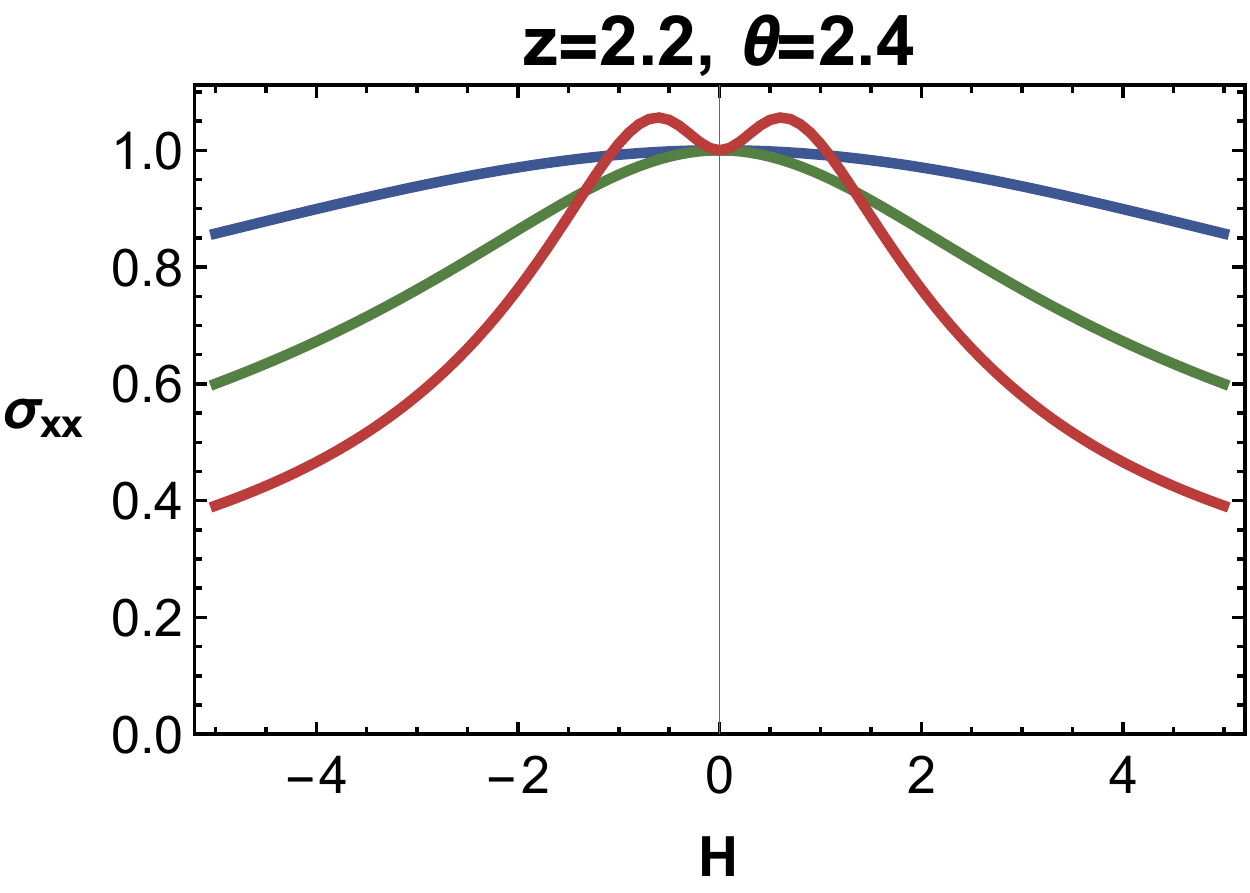} }   
                 \caption{Temperature evolution for $\sigma_{xx}(H)$ for different $(z,\theta)$. Each curves corresponds to $T=0.04,0.1,0.24$ for blue, green, and red respectively.  We used $q_{\chi}=0.7$ for all non-zero $q_{\chi}$ case.}    \label{fig:sxxH} 
\end{figure}

 Figure \ref{fig:sxxH} shows the temperature evolution of magneto-conductivity.  
 There are several interesting  features for longitudinal electric conductivity, which we will understand   by analytic calculations later : 
\begin{itemize}
 \item 
 For the regions  A and B, $\sigma^{0}_{xx}$,   the longitudinal conductivity at $H=0$,  increases as a function of temperature   while it decreases  in phase C and D. Such opposite behaviors  are  characters   for other transport coefficients also. 
 See figure \ref{fig:sxxH}. 
  \item For the regions B, C, the longitudinal electric conductivity has finite values at $H\rightarrow \infty$. It happens only when magnetic impurity  $q_{\chi}\neq0)$.
 \item At the NEC boundary in the region $A$, weak localizations appear at low temperature, while it appears at high temperature at the NEC boundary in $D$.
\end{itemize}

Although the longitudinal electric conductivity is  very complicated, we can understand a few important  aspects in   the  limit of small and large $H$.
For  $H=q_{2}=0$,  
\begin{align}\label{eq:SL0}
\sigma_{xx} |_{H=0}\equiv \sigma_{xx}^0 = r_0^{2z -\theta -2}.
\end{align}
where $r_0$ is $r_H$ at $H=q=0$. The temperature behavior of $\sigma_{xx}^0$ is different for each region in $z$-$\theta$ plane. At $H=0$,    
\begin{align}\label{eq:r0T01}
4\pi T |_{H=0}= r_0^z (2+z -\theta)+ \frac{r_0^{\theta -z} \beta^2}{\theta -2}.
\end{align}
The first term in (\ref{eq:r0T01}) is dominant for high temperature if  $\theta < 2 z$, while the second term is dominant if $\theta > 2 z$.
In the region $A$ and $B$ satisfying NEC $(\theta <2z -2)$, $\theta$ is always smaller than $ 2z$  so that   $T \sim r_H^{z}$, then    
\be\label{eq:SLA}
\sigma_{xx}^0 \sim T^{\frac{2z-\theta-2}{z}},
\ee
which is an increasing function of temperature.  This explains the temperature dependence of $\sigma_{xx}^0$ in Figure \ref{fig:sxxH} (c) - (h).

The eq. (\ref{eq:r0T01}) suggests that the regions $C$ and $D$ might be divided into two parts. When $\theta < 2 z$, the first term in (\ref{eq:r0T01}) is dominant and hence $\sigma_{xx}^0$ follows the form of (\ref{eq:SLA}) but the exponent is negative by NEC. For $\theta > 2 z $,  we have 
\begin{align}\label{eq:SLC}
\sigma_{xx}^0 \sim T^{\frac{2z-2-\theta}{\theta-z} },
\end{align}
whose  exponent is again negative when $\theta>2 z$. Therefore, in the region $C$, $D$,  $\sigma_{xx}^0$ decreases with temperature  regardless of sign of $\theta -2z$. This explains the temperature dependence of $\sigma_{xx}^0$ in Figure \ref{fig:sxxH} (i) and (j). There is the special case when $(z,~\theta)$ lies on the boundary of NEC, $2z-2-\theta =0$, where  $\sigma^{0}_{xx}=1$ is independent of $T$  by (\ref{eq:SL0}), explaining Figure \ref{fig:sxxH} (a), (b), (k) and (l). 

So far, we discussed the behavior of the longitudinal conductivity in the absence of the external magnetic field.
When the   magnetic field is finite, the relation between temperature and $r_H$ is rather  complicated  as one can see in (\ref{r0Trelation}) and the analytic expression of $r_H$ in terms of other parameters is not available. But in the small $H$ limit,   $H$ dependence of $r_H$ for  fixed  parameters $(T, \beta,~\lambda,~\theta,~z)$  can be obtained using 
\begin{align}\label{eq:dr0drHs}
\frac{\delta r_H}{\delta H}  &=- \frac{(\theta -2)\left[r_0^{4z}(2+z-2\theta)^2 +2 q_{\chi}^2 \lambda^4 r_0^{4 \theta}(z-2)(\theta-2)\right] H}{2 r_0^5 (z-2)(2+z-2\theta)^2\left[r_0^{2} z (2 z-\theta)(\theta-2)+\beta^2 r_0^{\theta}(\theta-z)\right] }+ {\cal O}(H^3) \cr
& \equiv  {\cal A}_1 H + {\cal O}(H^3),
\end{align}
where $r_0$ is $r_H(H=q=0)$.
%

\begin{figure}[ht!]
\centering
\subfigure[$q_{\chi}=0$]
   {\includegraphics[width=47mm]{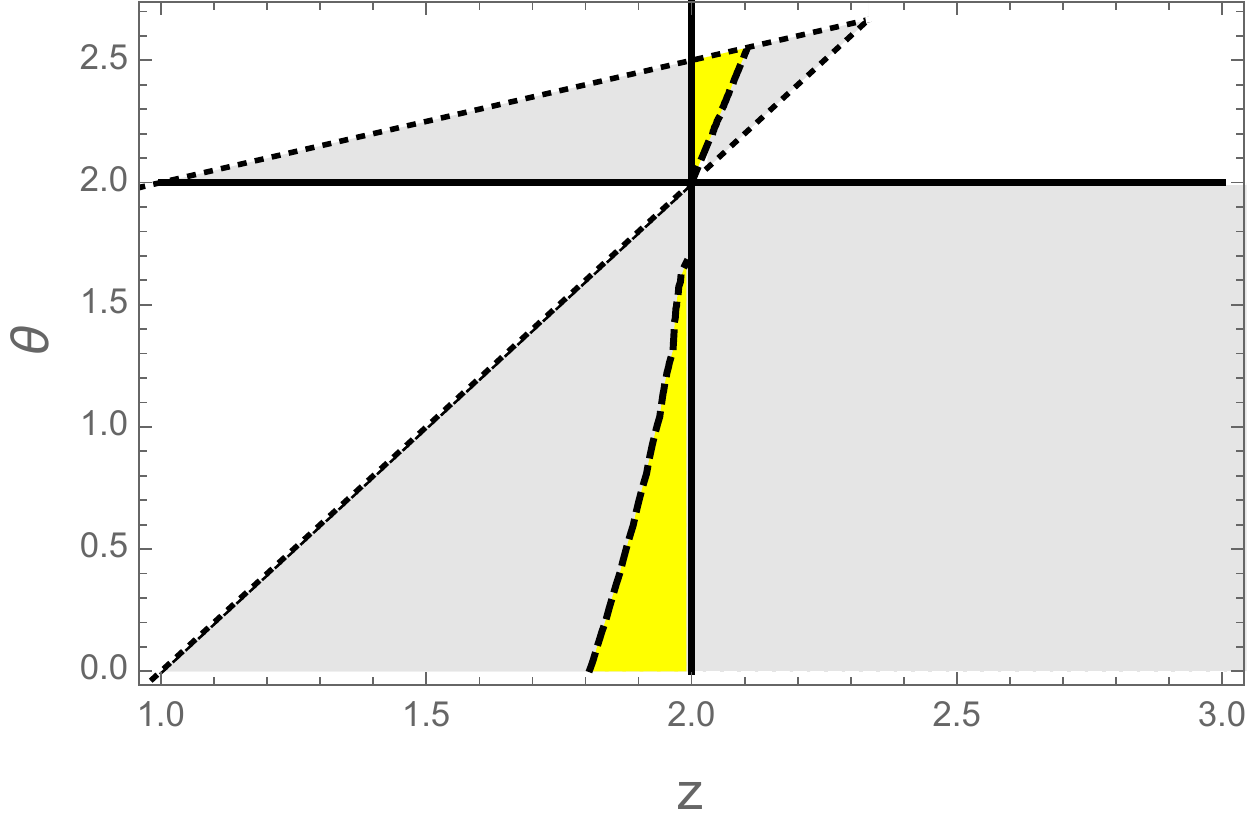} }  
\hskip.2cm
\subfigure[$q_{\chi}=0.7$]
   {\includegraphics[width=47mm]{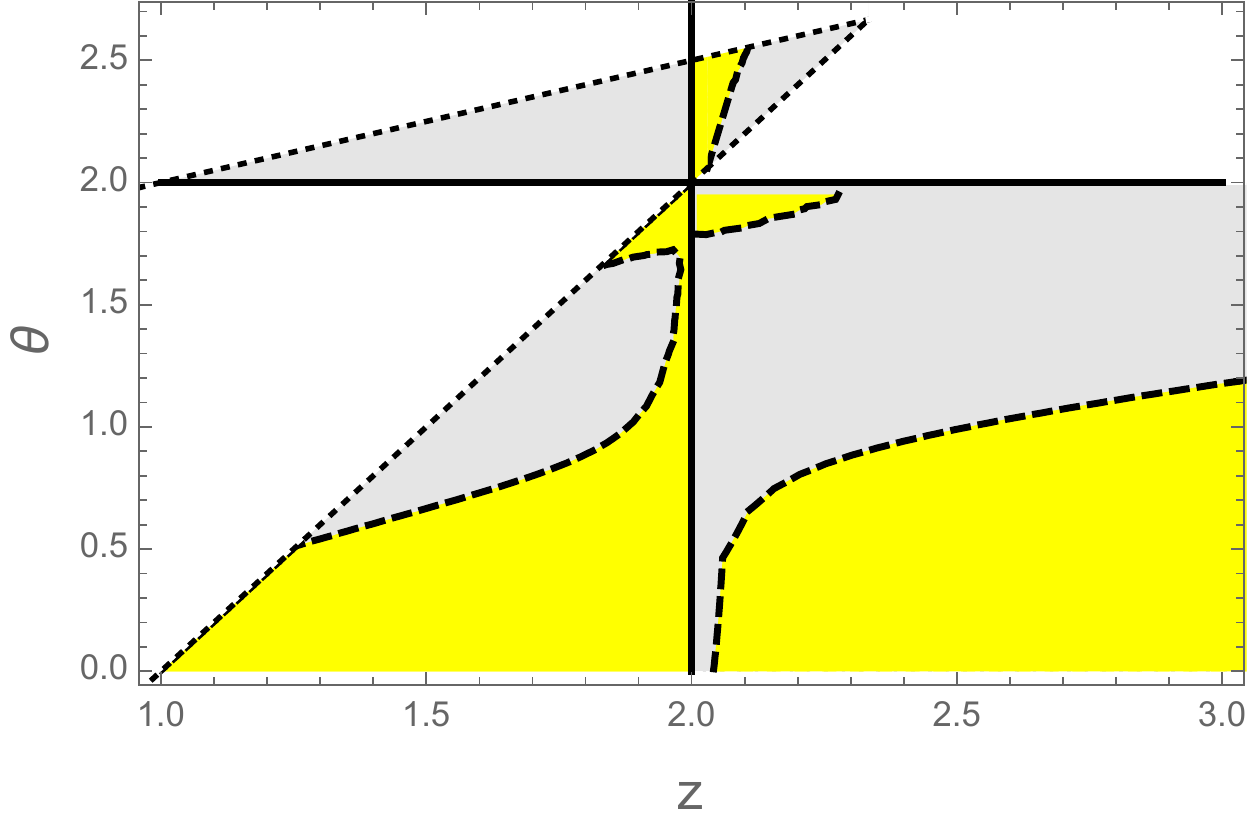} }  
\hskip.2cm   
\subfigure[$q_{\chi}=1.4$]
   {\includegraphics[width=47mm]{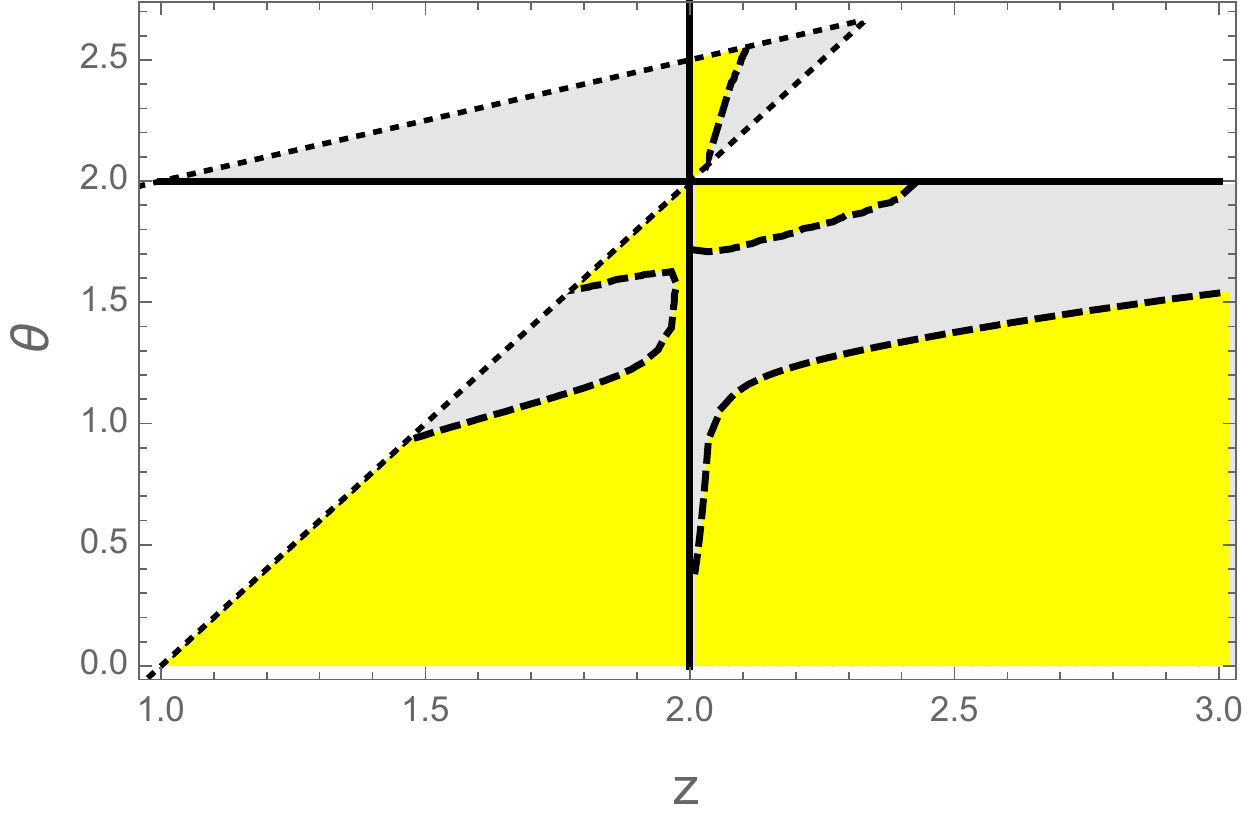} }          
           \caption{ WL (yellow) vs WAL(gray) in the validity regime.  Other  parameters are same as in  Figure \ref{fig:sxxH} and $T=0.04$. }    \label{fig:sxxSign} 
\end{figure}

Now, we have small $H$ expansion of   $\sigma_{xx}$ as
\begin{align}\label{eq:SL2}
\sigma_{xx}  = \sigma_{xx}^0 -\frac{r_0^{2z-2\theta -8}}{2} \left[ \frac{2 r_0^{4z}}{\beta^2} -{\cal A}_1 (2z-\theta -2) r_0^{\theta+5}-\frac{2 q_{\chi}^2 \lambda^4 r_0^{4\theta} (1+z-2\theta)^2}{\beta^2 (2+z-2\theta)^2}\right] H^2 +\cdots.
\end{align}
Depending on the sign of the coefficient of $H^2$, the system has weak anti-localization (WAL) for $-$ sign or weak localization (WL) for + sigin.
When $(z,~\theta)$ parameters are off the NEC boundary, the coefficient of $H^2$ in (\ref{eq:SL2}) is complicated function of $z$, $\theta$ and other parameters, but there still exists competition between  the two  terms which lead to the transition from WL 
 to WAL. Figure \ref{fig:sxxSign} is the phase diagram where yellow region denotes WL (positive sign) and gray region corresponds to WAL (negative sign). Dotted line is the boundary of the validity regime in Figure \ref{phasetransport} (b).  {  In the absence of $q_{\chi}$ most part of validity regime starts with WAL pahse. Small region near $z=2$ in $A$ and $D$ has WAL.  As $q_\chi$ increases,  the region of WL expands in the regions A, B and D while region C still remains in WAL.  }  As temperature increases, yellow regions in $A$, $B$ shrink and finally disappear, which means that the longitudinal conductivity shows weak anti-localization(WAL) behavior at high temperature. On the other hand, the yellow regions in $C$, $D$ expand and fill the whole region of $C$, $D$ at high temperature. 

For special case where $(z,~\theta)$ at the boundary of NEC ($\theta = 2z -2$), the     (\ref{eq:SL2}) has simpler form:
\begin{align}\label{eq:SL2bd}
\sigma_{xx}   
& = \sigma_{xx}^0 - \frac{r_0^{2z-4}}{\beta^2} \left[ 1- \frac{q_{\chi}^2 \lambda^4 (3z-5)^2}{(3z-6)^2} r_0^{4z -8} \right] H^2 + \cdots. 
\end{align}
In the absence of $q_{\chi}$ (\ref{eq:SL2bd}) shows weak anti-localization behavior, while it changes to weak localization behavior when $q_{\chi} \lambda^2$ is large. For the given value of $q_{\chi} \lambda^2$, temperature behavior depends on   $z$ and $\theta$.   At the NEC boundary,  the first term in (\ref{eq:r0T01}) is dominant and hence $r_H \sim T^{1/z}$. Then coefficient of $H^2$ in (\ref{eq:SL2bd}) behaves
\begin{align}
\Delta \sigma_{xx} &\sim -(1- c_{*}T^{4-\frac{8}{z}} ),
\end{align}
with a positive constant $c_{*}$, which means that weak anti-localization appears at low temperature at the  NEC boundary in A  region $z<2$ but appears  at high temperature at the NEC boundary in D $(z>2)$. See Figure \ref{fig:sxxH} (b) and (l).

Another  interesting phenomena in the region B and C is that 
  $\sigma_{xx}$ goes to constant for large magnetic field $H$ as one can see in Figure \ref{fig:sxxH} (d), (h) and (j).  
To understand  this, we first notice that large $H$ behavior of the conductivity is entirely determined by that of $r_H$, because  
eq. (\ref{transp}) says that both numerator and denominator have 
explicit $H^4$ behavior so that  in this limit the conductivity is  determined only by the 
$H$ dependence of $r_H$. 
From the expression of  eq.(\ref{r0Trelation}),  
\be
r_H=r_\infty +\frac{C}{H^2} +\cdots, \hbox{ for } H\to \infty
\ee
in the relevant regions. Namely, if $r_\infty$ is well defined, the noted behavior is explained. 
Starting with the $H$  expansion of  eq.(\ref{r0Trelation}), 
 \begin{align}
 	4\pi T= -\left(\frac{r_H^{2\theta-2-z}\Theta^2}{2(2-\theta)}+\frac{r_H^{3z-6}}{4(z-2)}\right)H^2 +\cdots,
 \end{align}
In order for the left hand side to be finite,   the coefficients of 
      $H^2$ should vanish.    Therefore 
\begin{align}
	r_H^{4z-2\theta-4}=2\left(\frac{z-2}{2-\theta}\right)\Theta^2>0
\end{align}
requesting  
\be
(z-2)(\theta-2)<0,
\ee
which is  precisely  the equation defining the  region $B$ and $C$.  
\vskip0.5cm
 
\begin{figure}[ht!]
\centering
 \subfigure[$P_{0A}$  ]
   {\includegraphics[width=45mm]{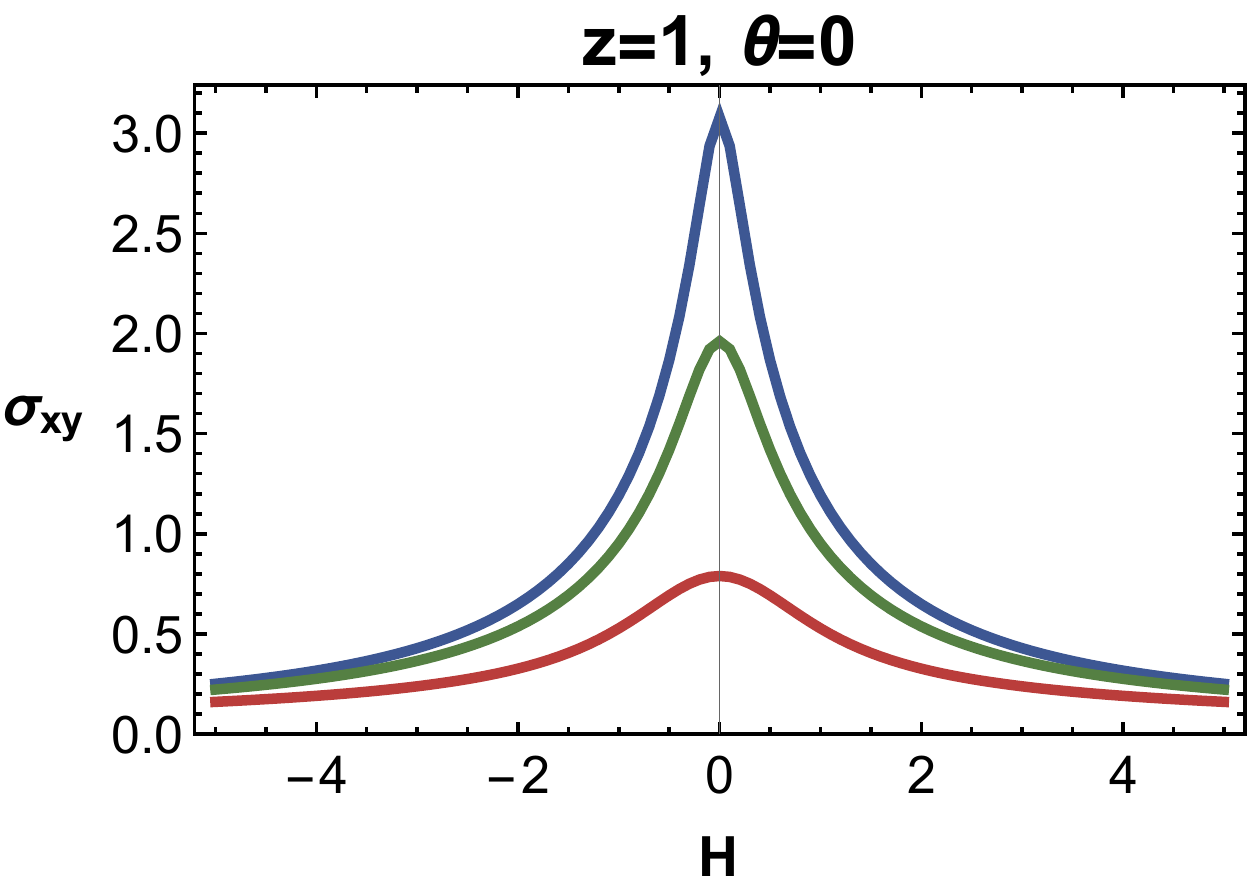} }
         \subfigure[$P_{A}$  ]
   {\includegraphics[width=45mm]{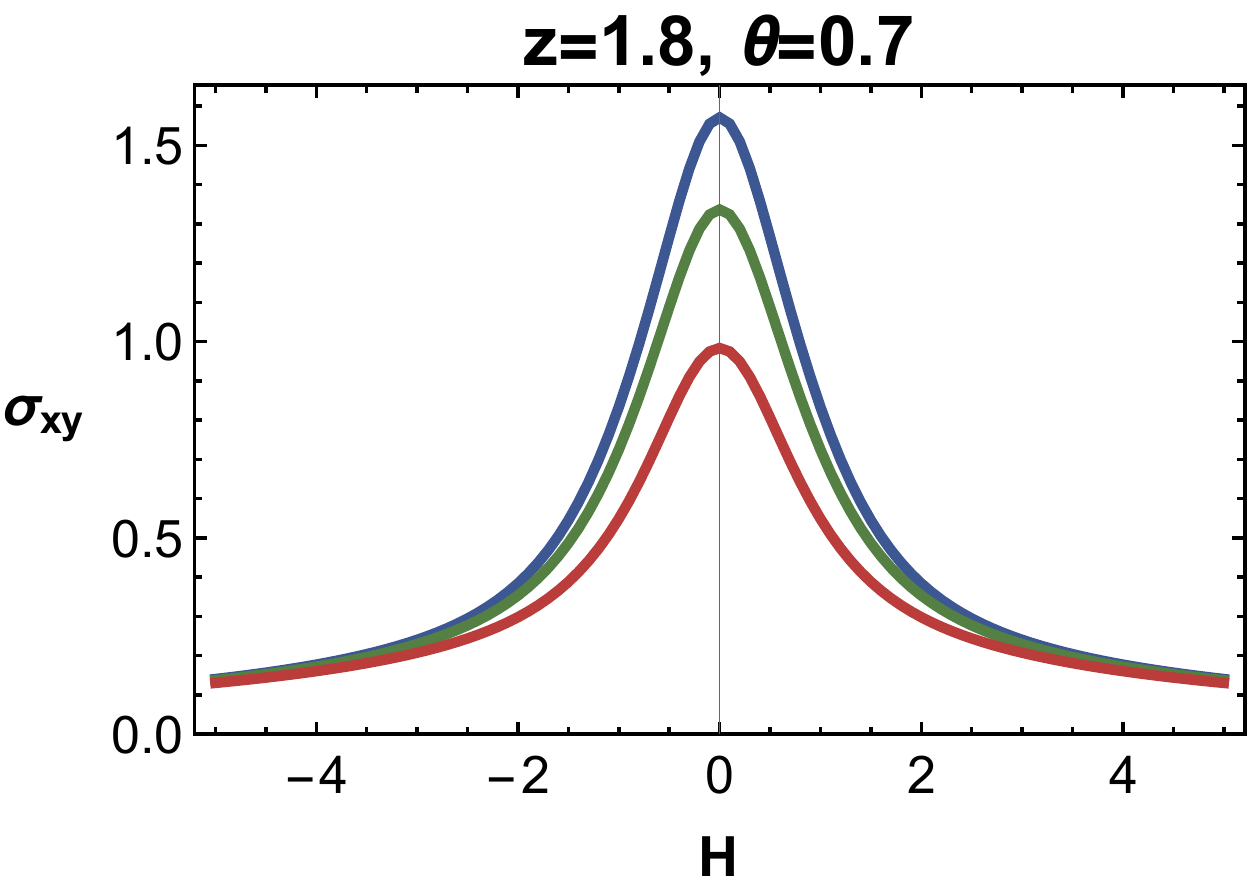} }
      \subfigure[$P_{C}$ ]
   {\includegraphics[width=45mm]{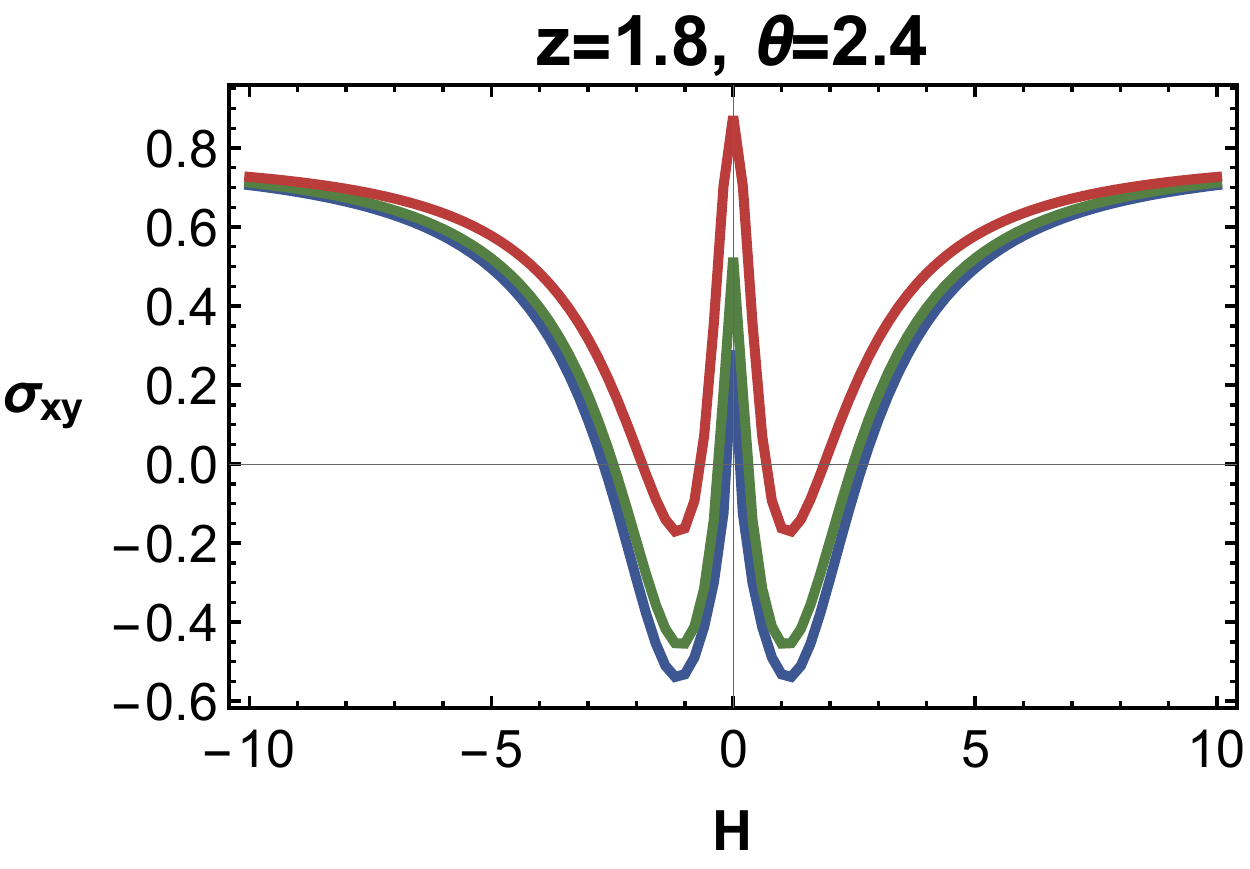} }
    \subfigure[$P_{0B}$ ]
   {\includegraphics[width=45mm]{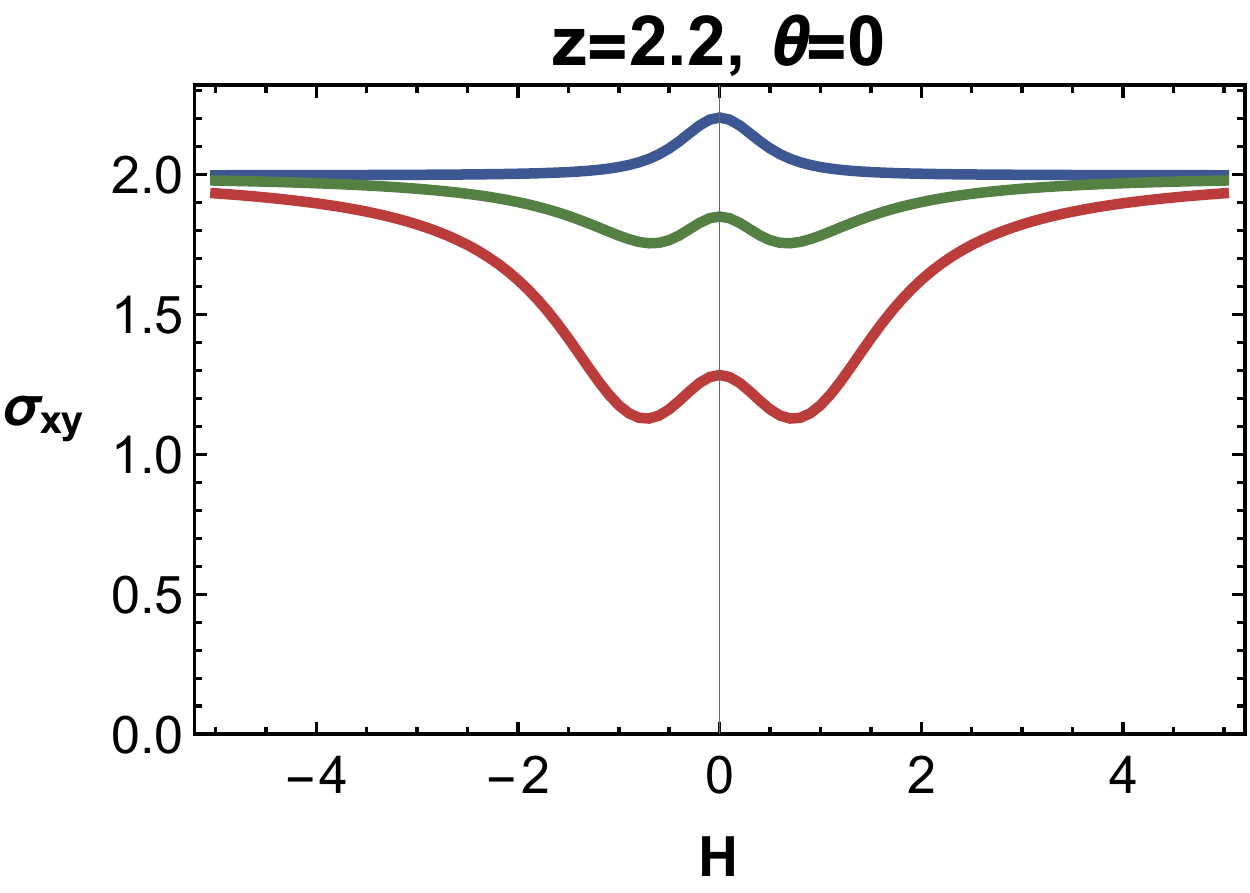} }
      \subfigure[$P_{B}$  ]
   {\includegraphics[width=45mm]{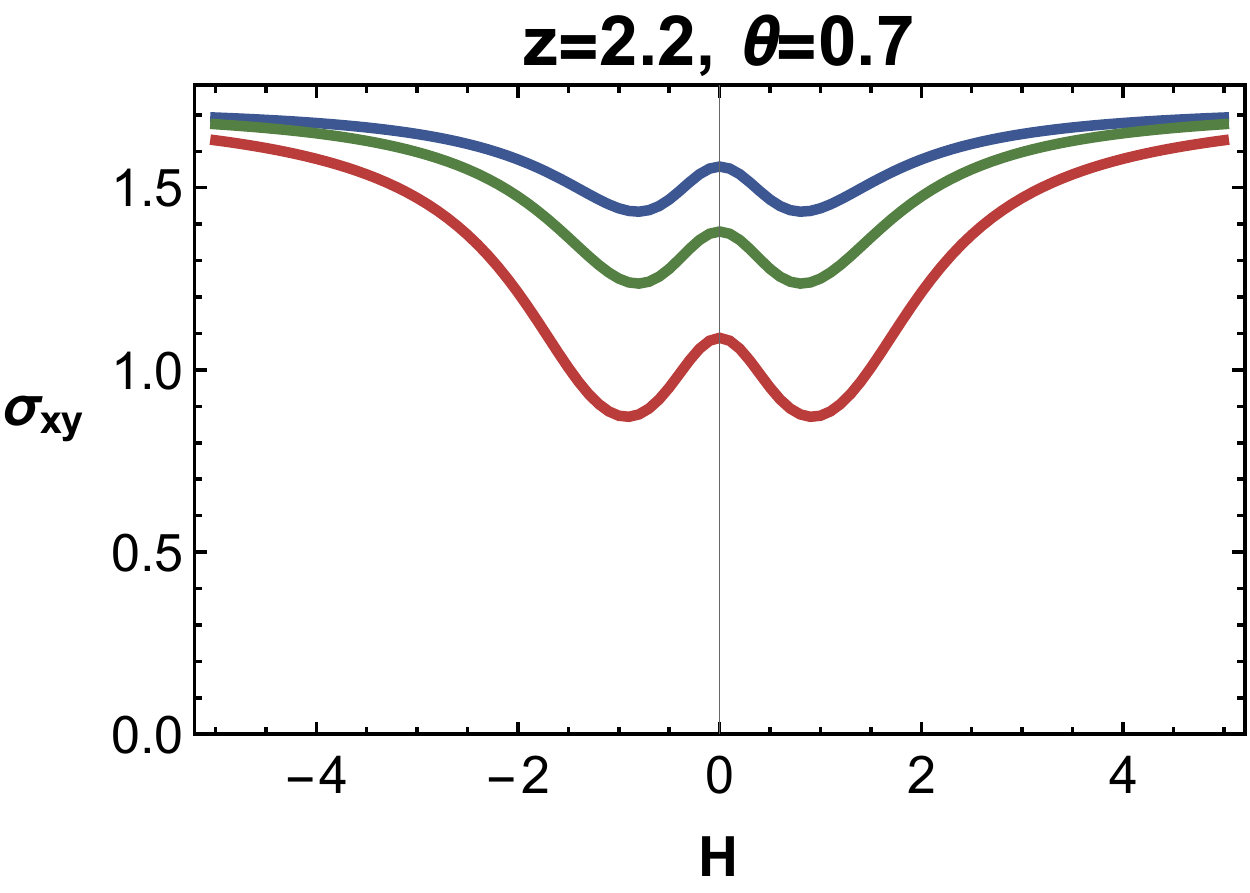} }
      \subfigure[$P_{D}$ ]
   {\includegraphics[width=45mm]{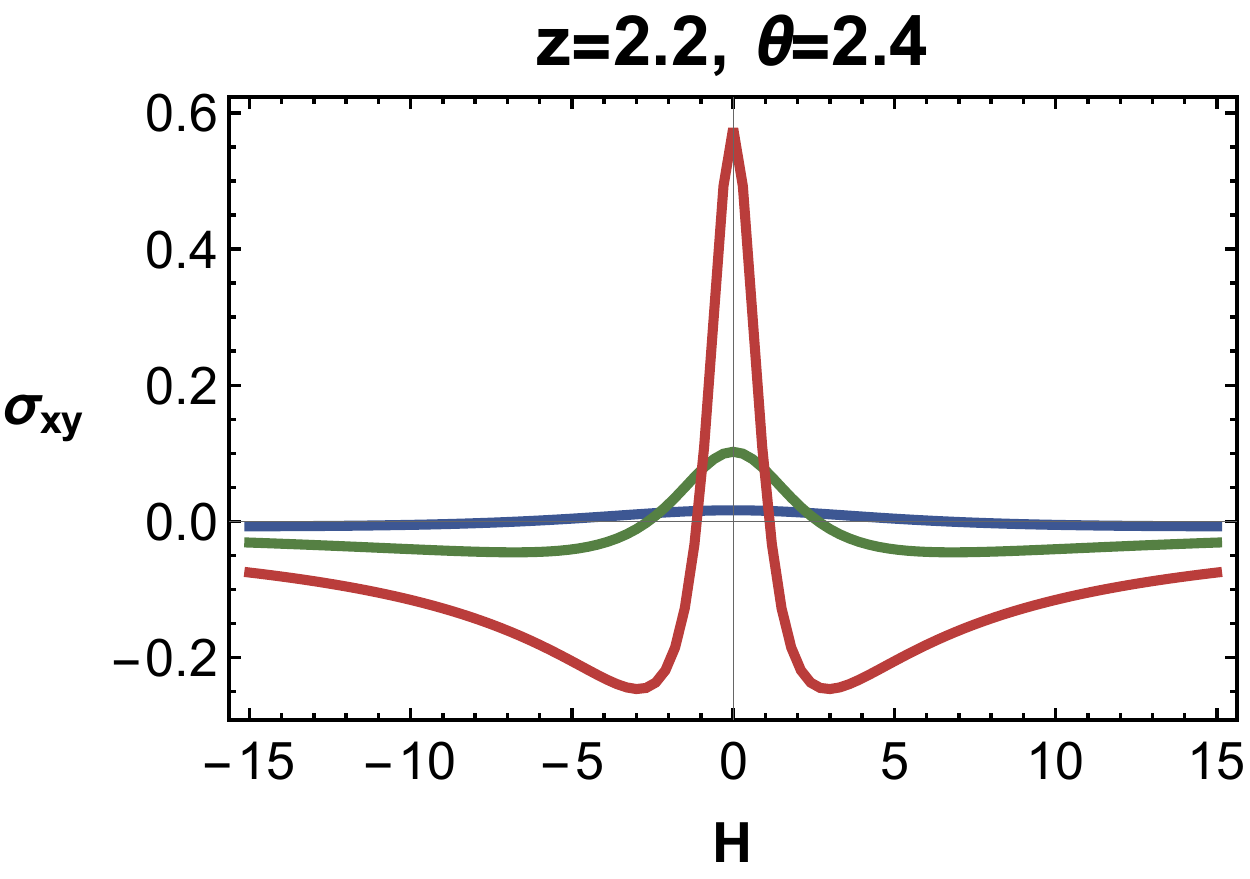} }   
                 \caption{Temperature evolution for $\sigma_{xy}(H)$ for different $(z,\theta)$.   Curves correspond to $T=0.04,0.1,0.24$ for blue, green, and red respectively. We used $q_{\chi}=0.7$.}    \label{fig:sxyH} 
\end{figure}

 Now we consider transverse magneto-conductivity. Figure \ref{fig:sxyH} shows external magnetic field dependence of  $\sigma_{xy}(H)$.
Because we are considering only zero chemical potential case,  the transverse conductivity vanishes in the absence of $q_{\chi}$, which means that the transverse conductivity is generated by the $q_{\chi}$ interaction term. This can be understood by the fact that $q_{\chi}$ term generates finite magnetization  as well as the effective charge carrier by (\ref{muq}) and therefore  transverse movement of charge carriers is also generated.

We can also expand $\sigma_{xy}$ in small magnetic field limit as previous discussion, 
\begin{align}\label{eq:Sxy}
\sigma_{xy} = \sigma_{xy}^0 - \frac{q_{\chi} \lambda^2}{2 r_0^8}\left[ \frac{4 r_0^{4z}(1+z-2\theta)}{\beta^2 (2+z-2\theta)} -(\theta-2){\cal A}_1 r_0^{\theta+5}\right] H^2 + \cdots,
\end{align}
where ${\cal A}_1$ is defined in (\ref{eq:dr0drHs}). 
$\sigma_{xy}^0$ is called `anomalous Hall conductivity' which is the transverse conductivity in the absence of the external magnetic field. It is  given by 
\begin{align}
\sigma_{xy}^0 = q_{\chi} \lambda^2 r_0^{\theta-2}.
\end{align}
Using the similar analysis as before,  we found that that $\sigma_{xy}^0$ is a decreasing function of temperature in the region $A$, $B$ but  increasing  function in the region $C$ and $D$. 

\begin{figure}[ht!]
\centering
   {\includegraphics[width=55mm]{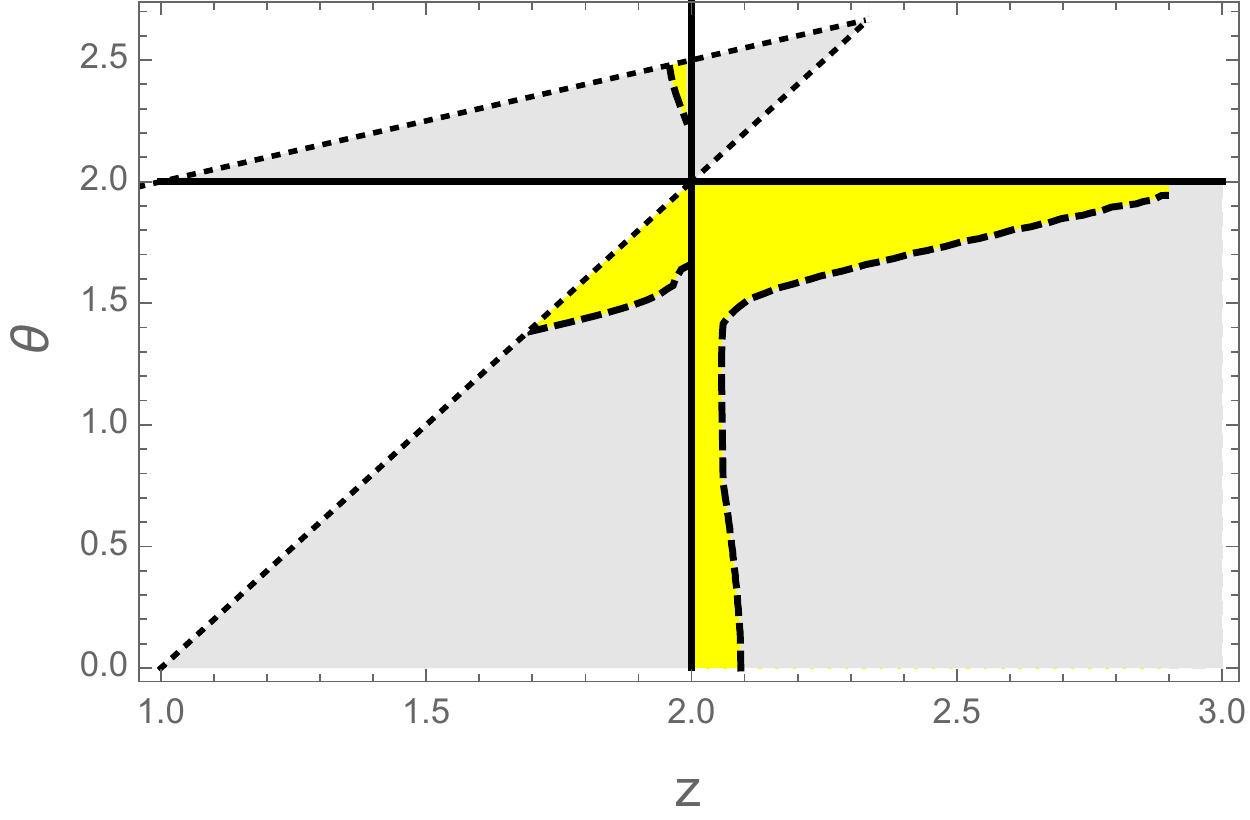} }          
           \caption{Sign of the coefficient of $H^2$ of the transverse conductivity near $H=0$. Yellow region denotes positive sign and gray region is for negative sign. Dotted line indicates validity regime. Here we use the parameters of Figure \ref{fig:sxyH} and $T=0.1$. }    \label{fig:sxySign} 
\end{figure}

The second term in (\ref{eq:Sxy}) determines curvature of the transverse conductivity near $H=0$ which is not easy to analyze. We checked numerically that the sign of the second term is always negative in all region. The result is given by the Figure \ref{fig:sxySign}. {  We find that the region of WAL and WL is insensitive of the value of $q_{\chi}$.}

\begin{figure}[ht!]
\centering
	 \subfigure[$P_{0A}$ at $q_{\chi}=0$]
   {\includegraphics[width=30mm]{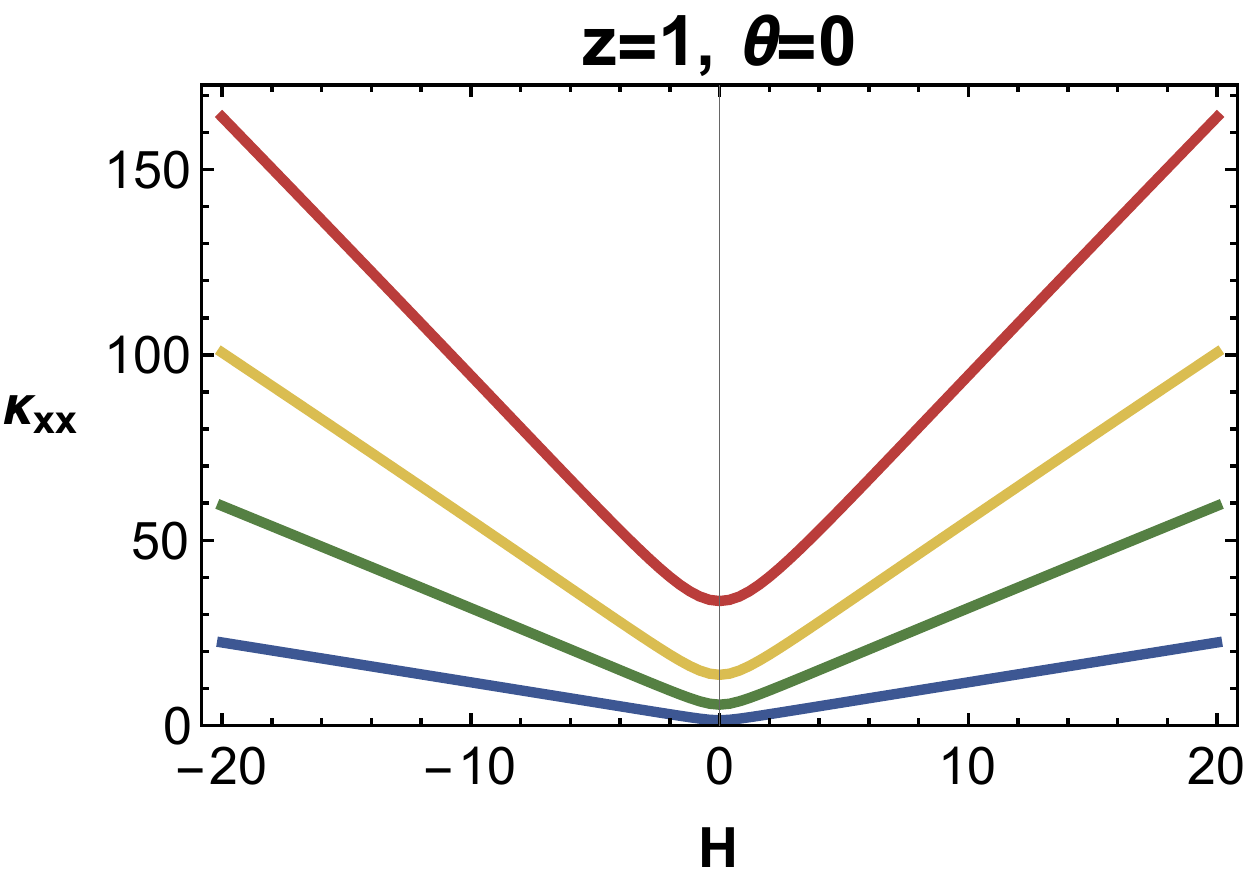} }
    \subfigure[$P_{0A}$  ]
   {\includegraphics[width=37mm]{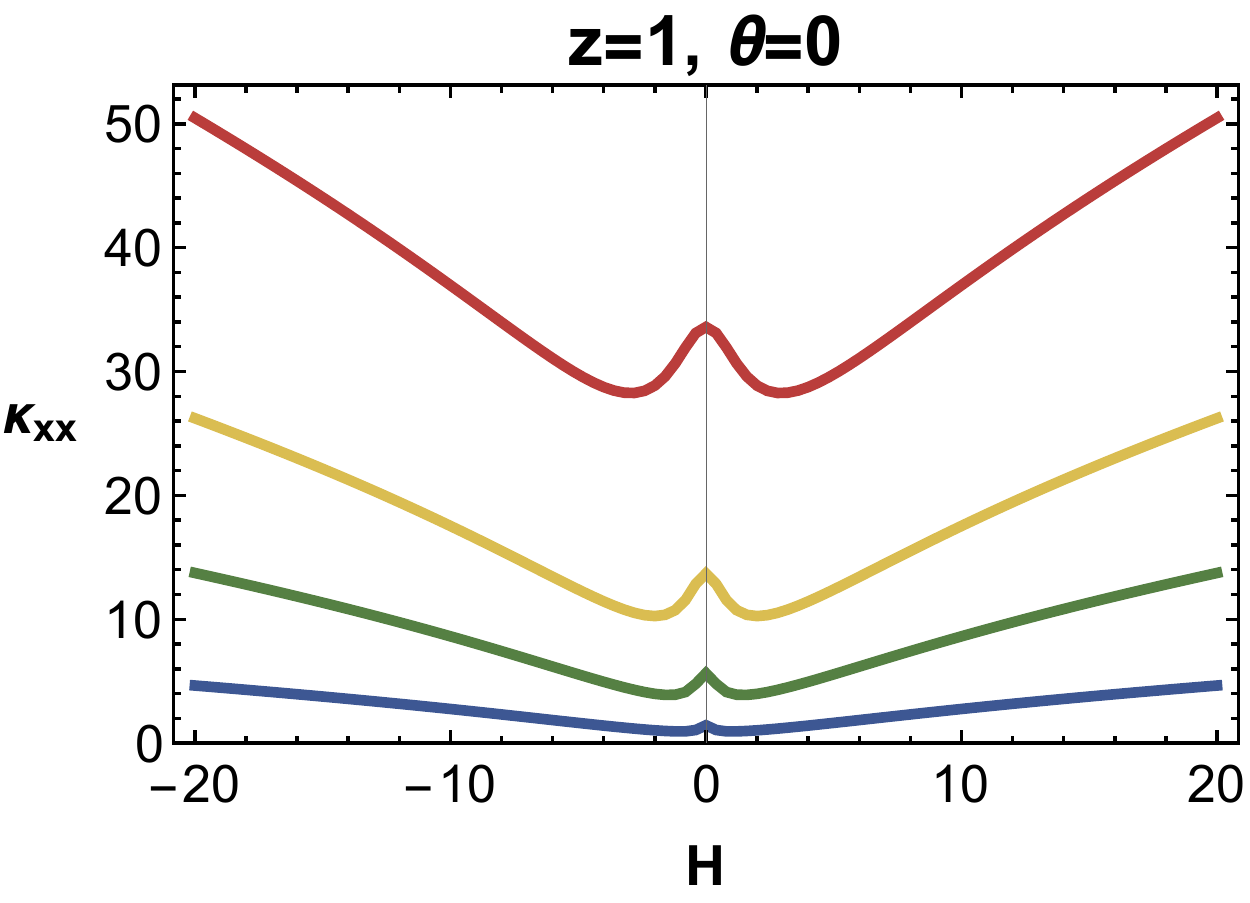} }
    \subfigure[$P_{0B}$ at $q_{\chi}=0$]
   {\includegraphics[width=30mm]{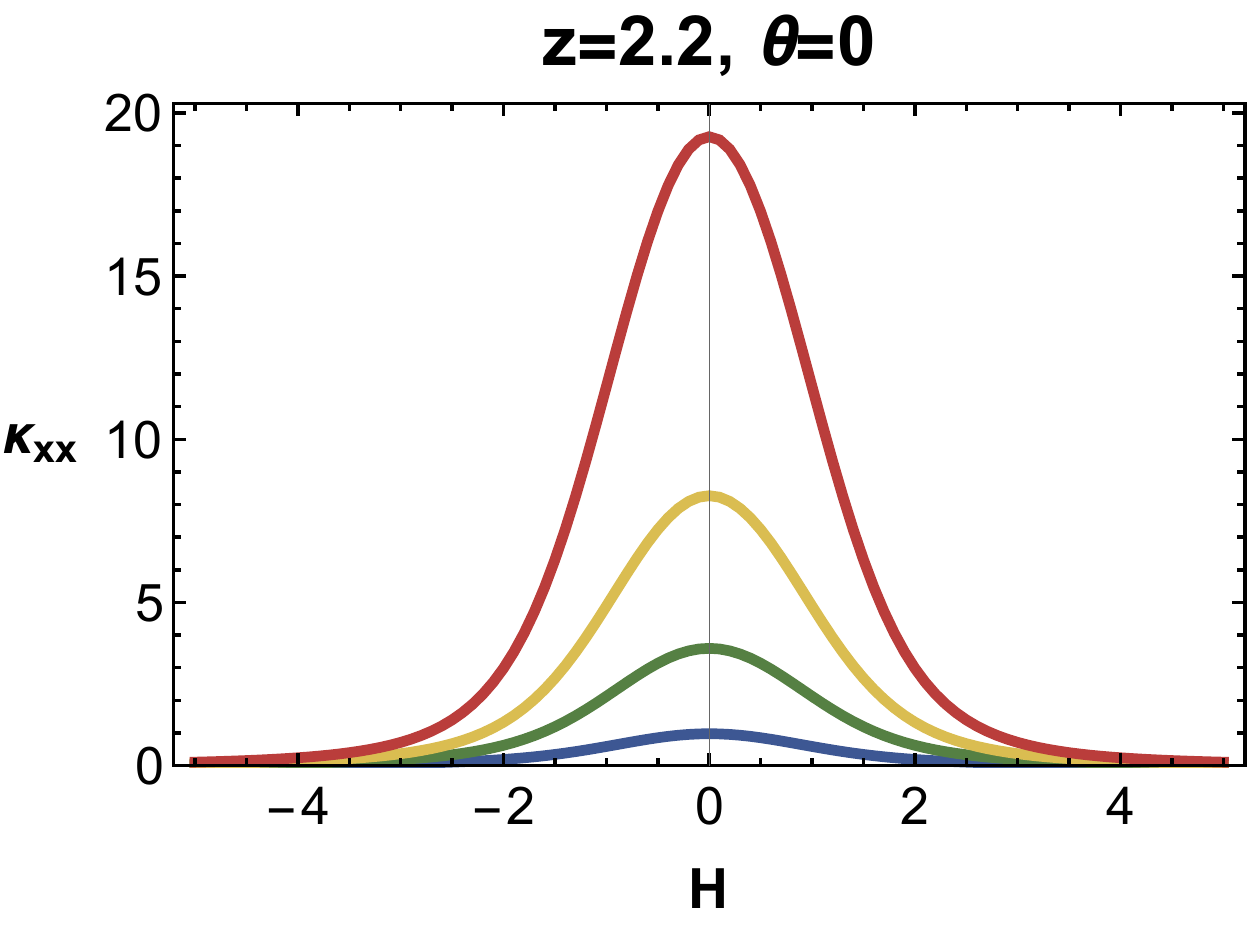} }
    \subfigure[$P_{0B}$ ]
   {\includegraphics[width=37mm]{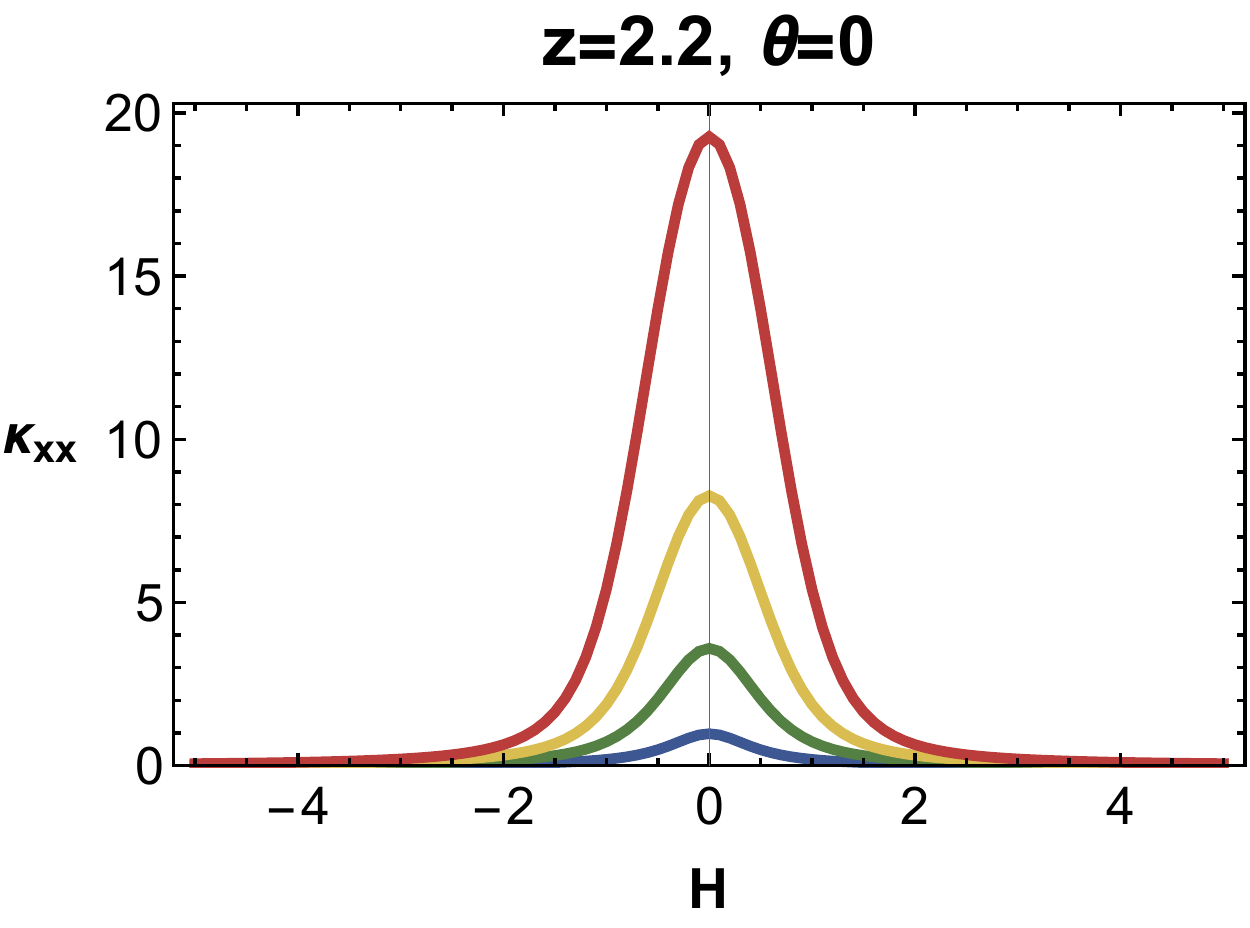} }
    \subfigure[$P_{A}$ at $q_{\chi}=0$]
   {\includegraphics[width=30mm]{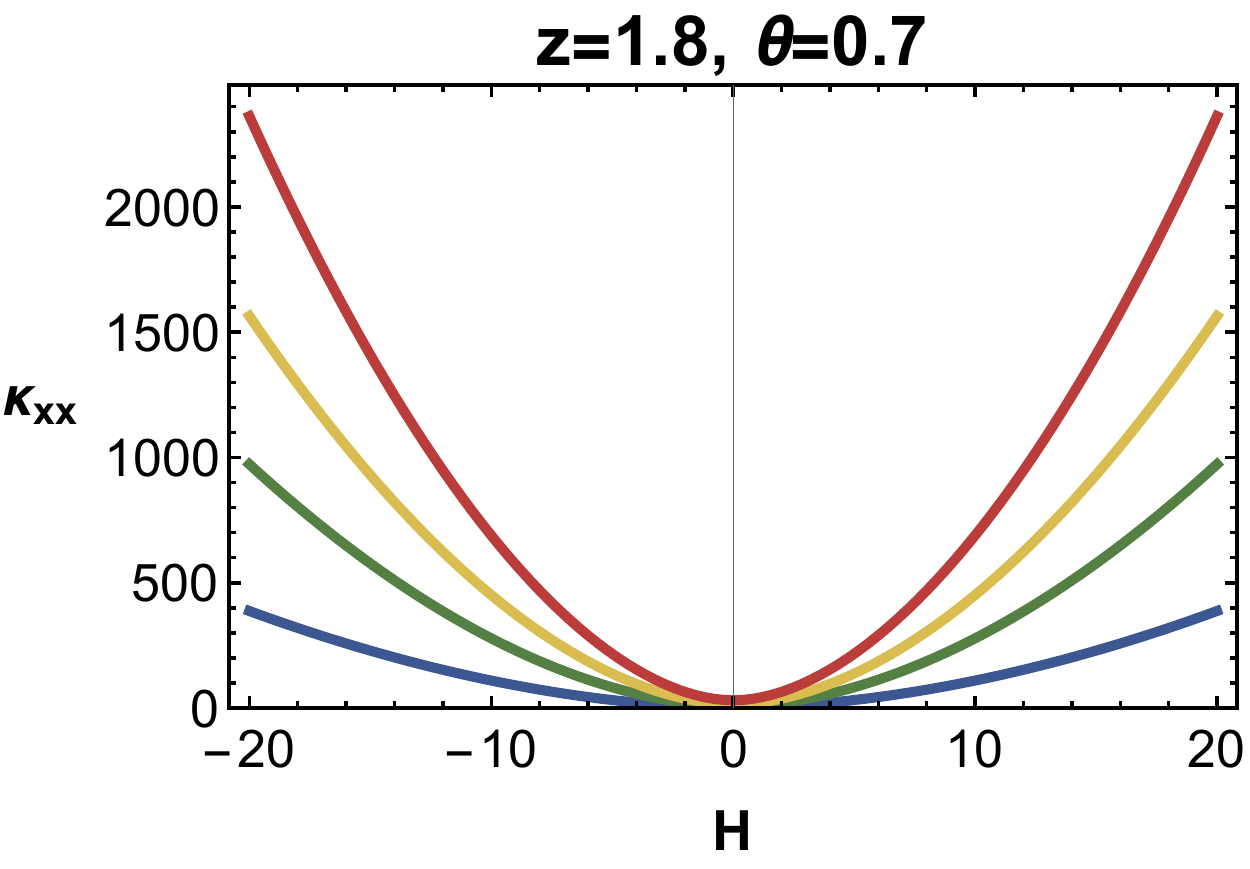} }
    \subfigure[$P_{A}$  ]
   {\includegraphics[width=37mm]{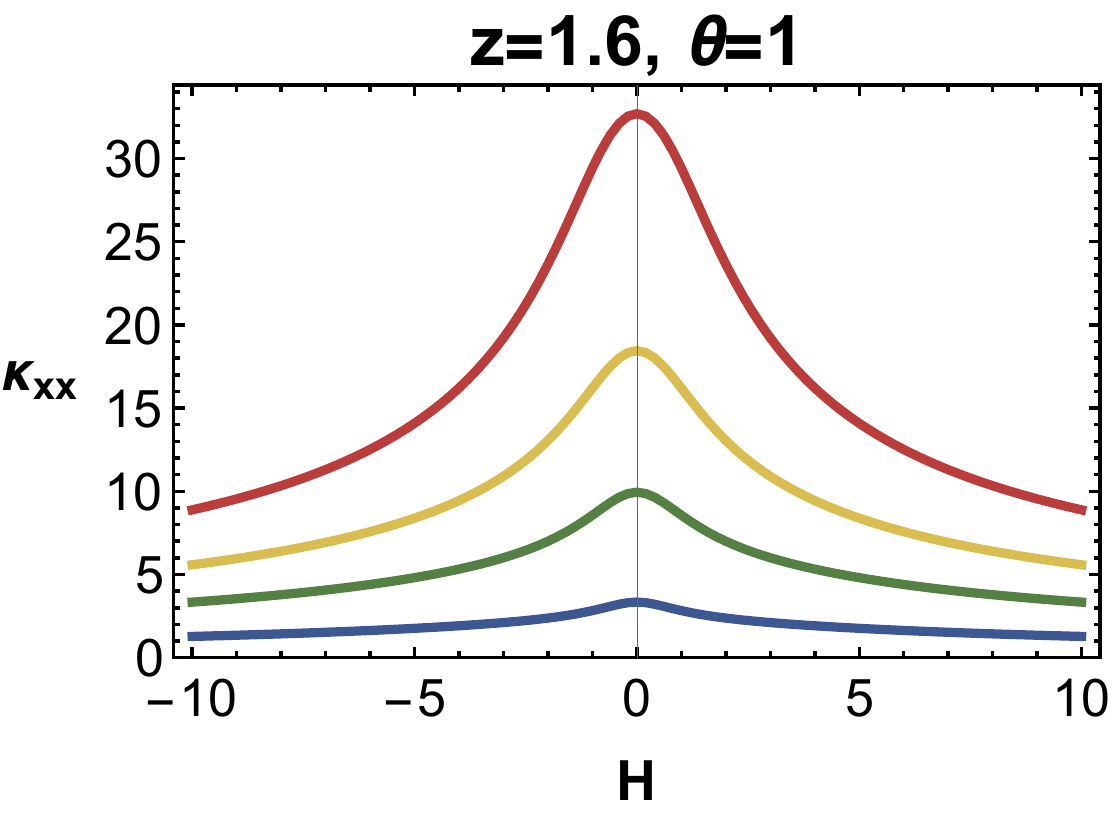} }
    \subfigure[$P_{B}$ at $q_{\chi}=0$]
   {\includegraphics[width=30mm]{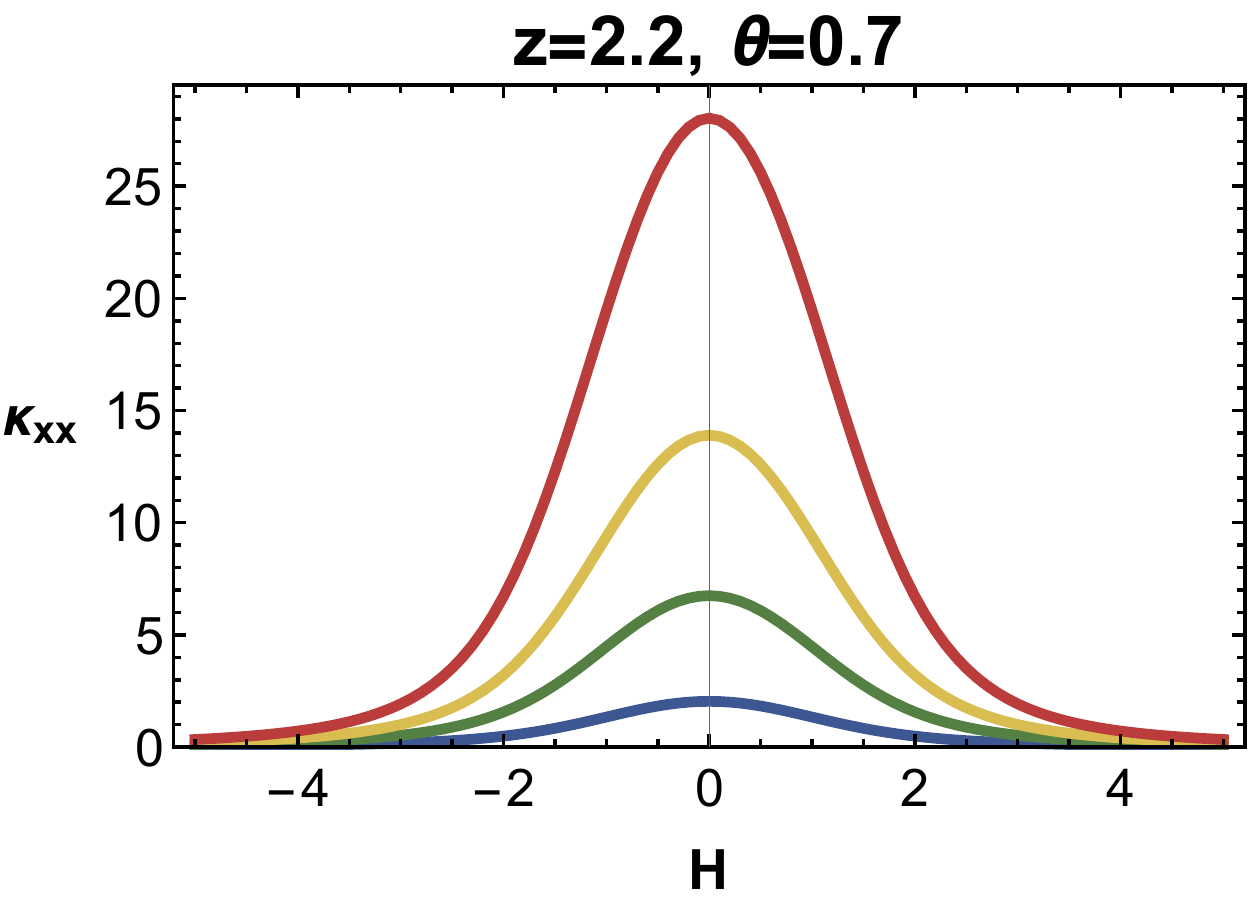} }
    \subfigure[$P_{B}$  ]
   {\includegraphics[width=37mm]{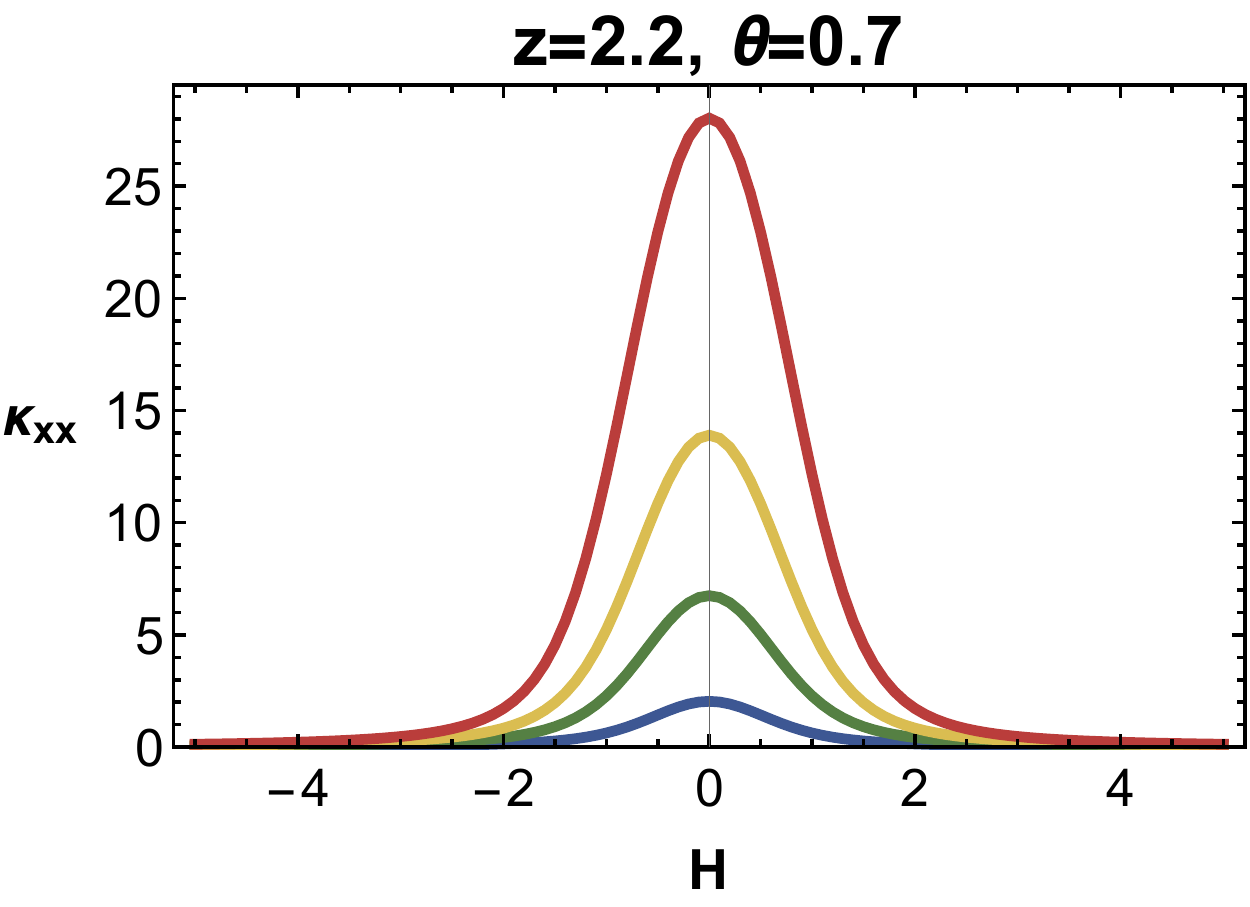} }
    \subfigure[$P_{C}$ at $q_{\chi}=0$]
   {\includegraphics[width=30mm]{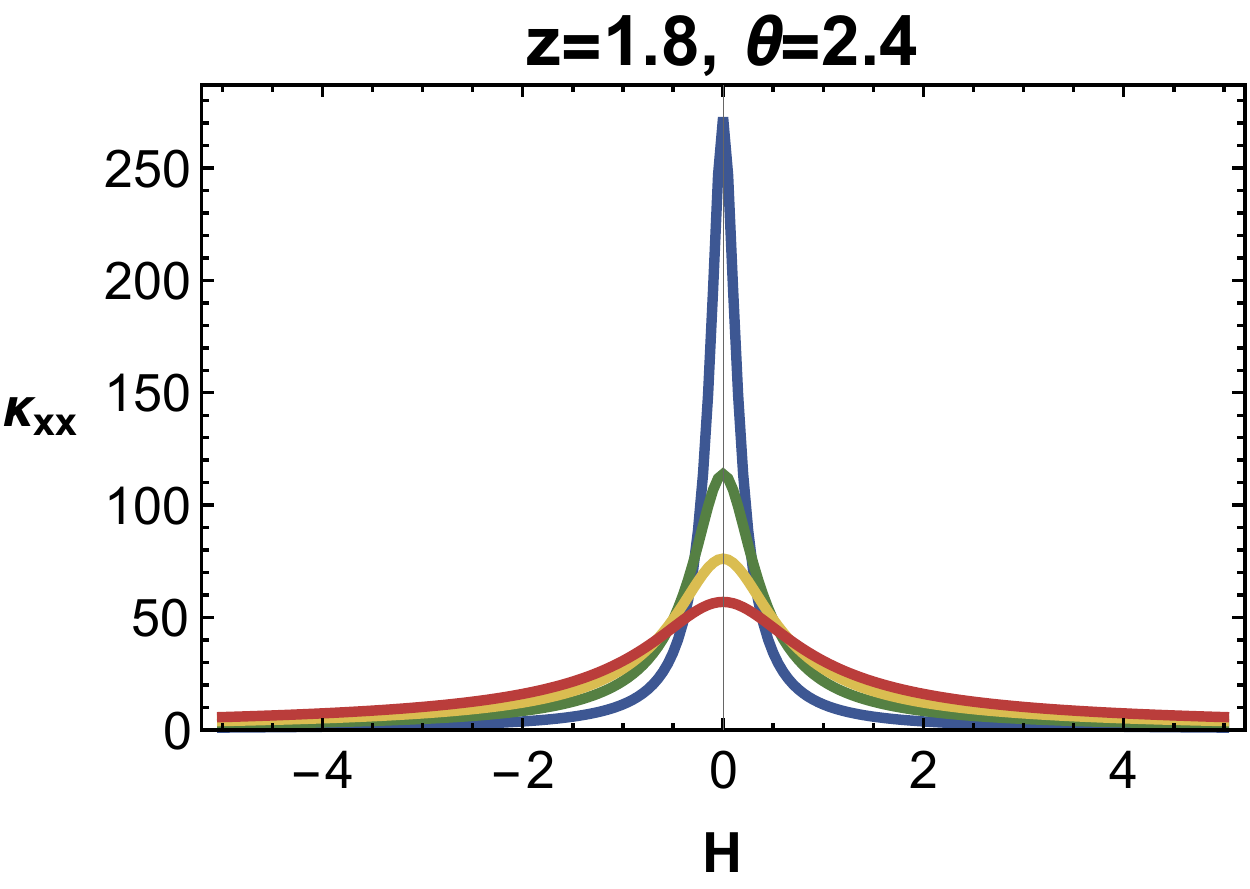} }
    \subfigure[$P_{C}$ ]
   {\includegraphics[width=37mm]{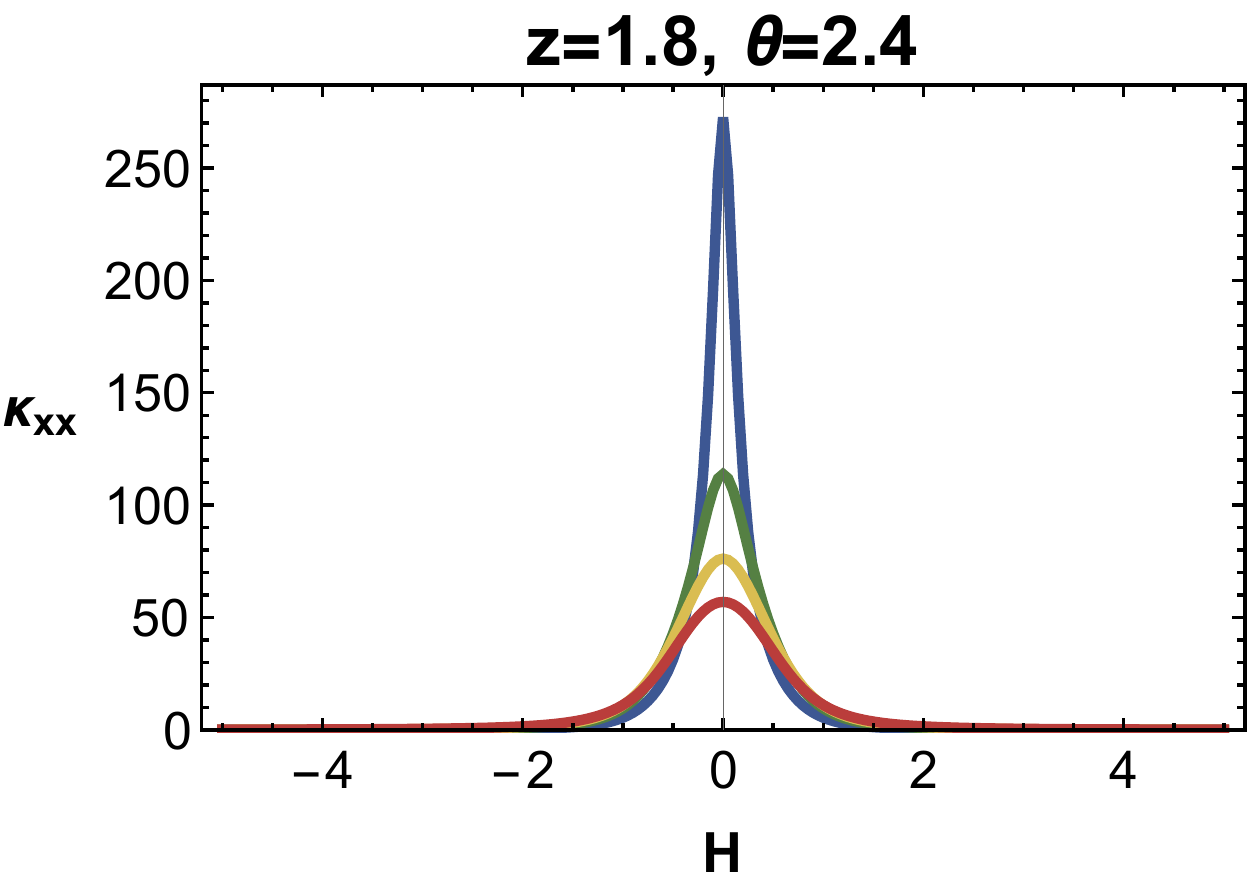} }
       \subfigure[$P_{D}$ at $q_{\chi}=0$]
   {\includegraphics[width=30mm]{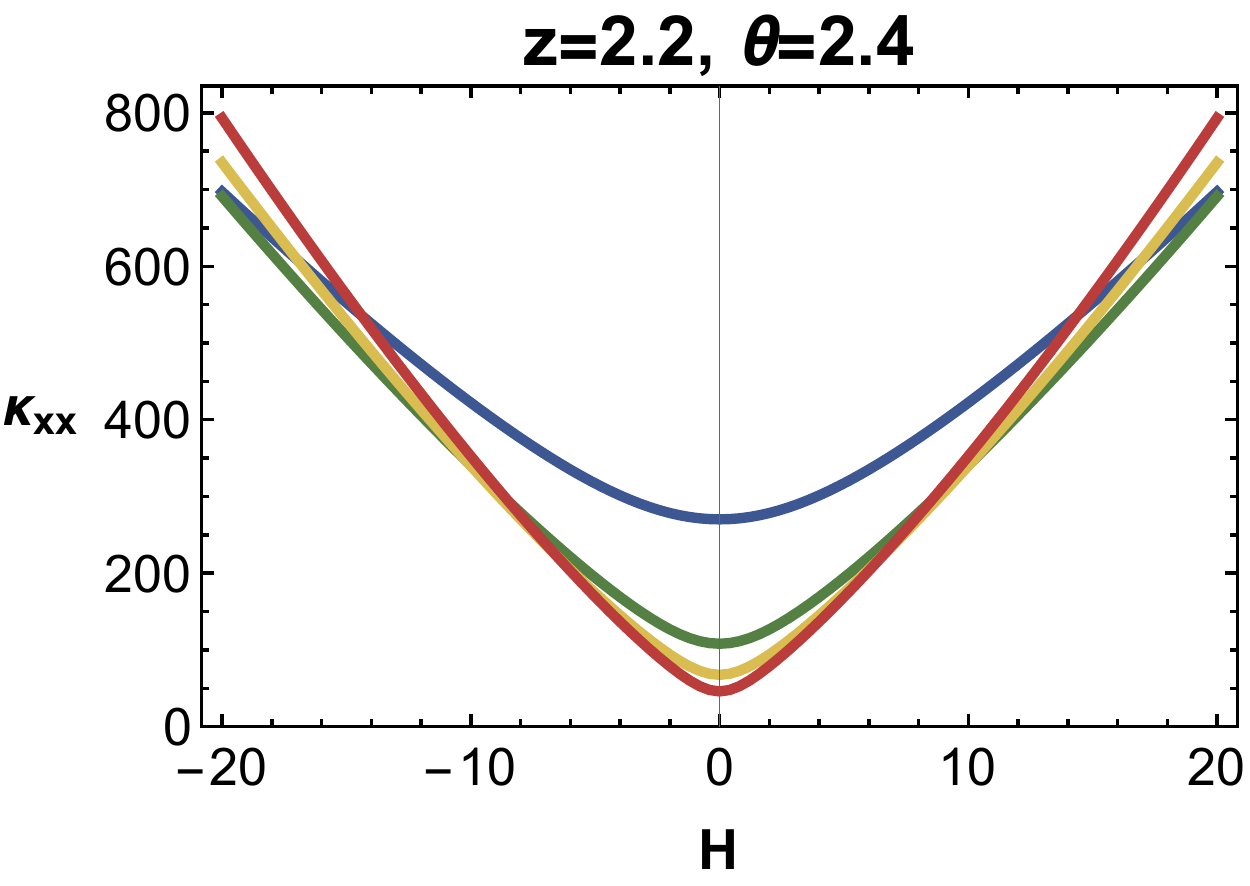} }
    \subfigure[$P_{D}$ ]
   {\includegraphics[width=37mm]{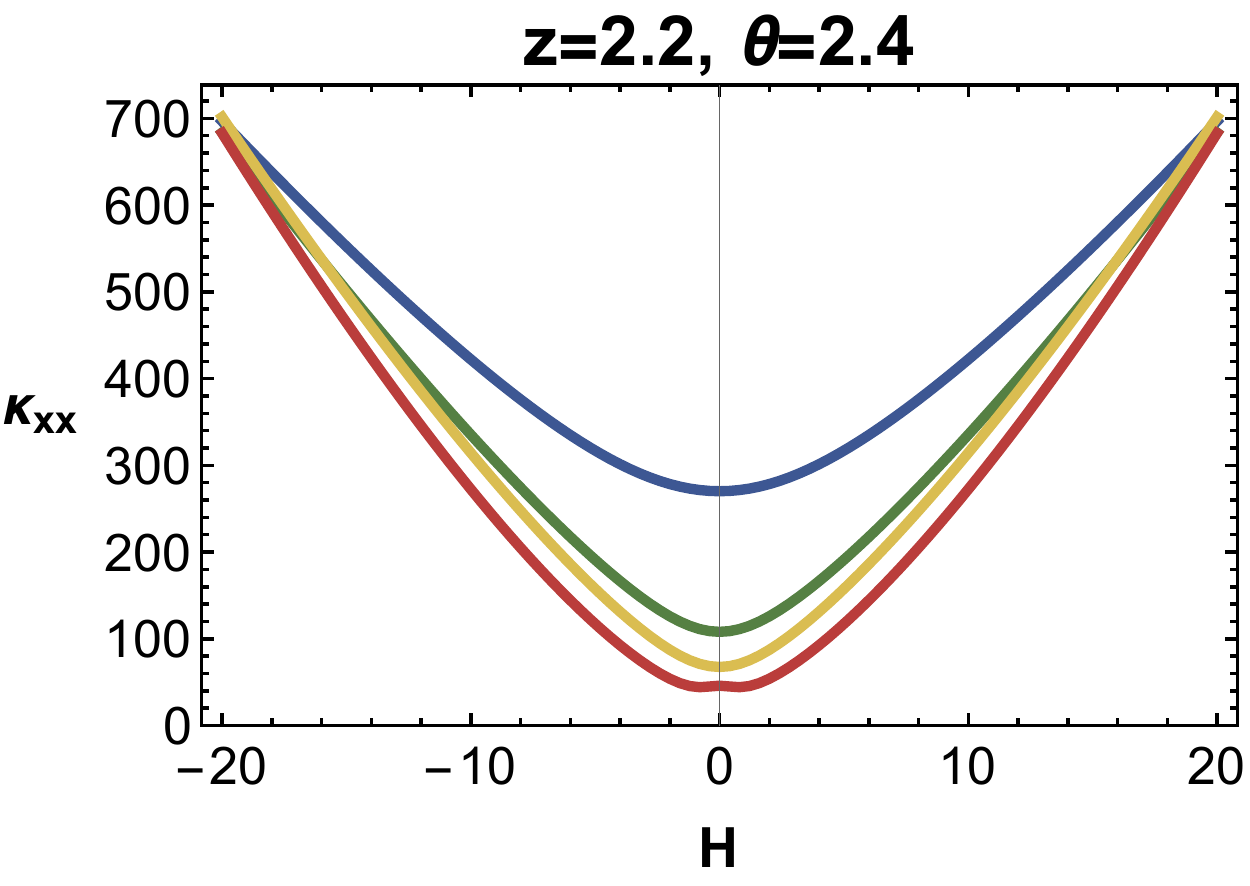} }   
                 \caption{Temperature evolution for $\kappa_{xx}(H)$ for different $(z,\theta)$. Each curve corresponds to $T=0.04,0.1,0.16, 0.24$ for blue, green, yellow and red respectively. We used $q_{\chi}=0.7$ for all non-zero $q_{\chi}$ case.}    \label{fig:kxxH} 
\end{figure}

\subsection{Classifying the magneto-thermal conductivity   in $(z,\theta)$ plane}
In this section, we will discuss magneto-thermal conductivities.    
Figure \ref{fig:kxxH} shows the external magnetic field dependence of the thermal conductivity given in (\ref{eq:kxxH}). In the small $H$ limit, the longitudinal thermal conductivity can be expanded  as
\begin{align}\label{eq:kxxH2}
\kappa_{xx} = \kappa_{xx}^0 -\frac{16\pi^2 r_0^{2z -2\theta-6} T}{\beta^4} \left[  r_0^5 \beta^2 (\theta -z) {\cal A}_1 +{\cal B}_1 \right] H^2 \cdots,
\end{align}
where ${\cal A}_1$ is defined in (\ref{eq:dr0drHs}) and 
\begin{align}
\kappa_{xx}^0 &= \frac{16 \pi^2 r_0^{2(z-\theta)}}{\beta^2} T \cr
{\cal B}_1 &=\frac{q_{\chi}^2 \lambda^4 r_0^{3\theta}}{(2+z-2\theta)^2} \left[(1+z-2\theta)(2+z-2\theta) +\frac{1}{1+q_{\chi}^2 \lambda^4 r_0^{-4z +4\theta}}\right].
\end{align}

The temperature dependence of $\kappa_{xx}^0$ can be understood using the same analysis we used for the electric conductivity. In the region $A$ and $B$, horizon radius   $\sim T^{1/z}$ and $z>\theta$. Therefore, $\kappa_{xx}^0$ increases at high temperature. One can easily check that the exponent becomes negative in the region $C$ and $D$ and hence $\kappa_{xx}^0$ is suppressed at high temperature.

\begin{figure}[ht!]
\centering
\subfigure[$q_{\chi}=0$]
   {\includegraphics[width=45mm]{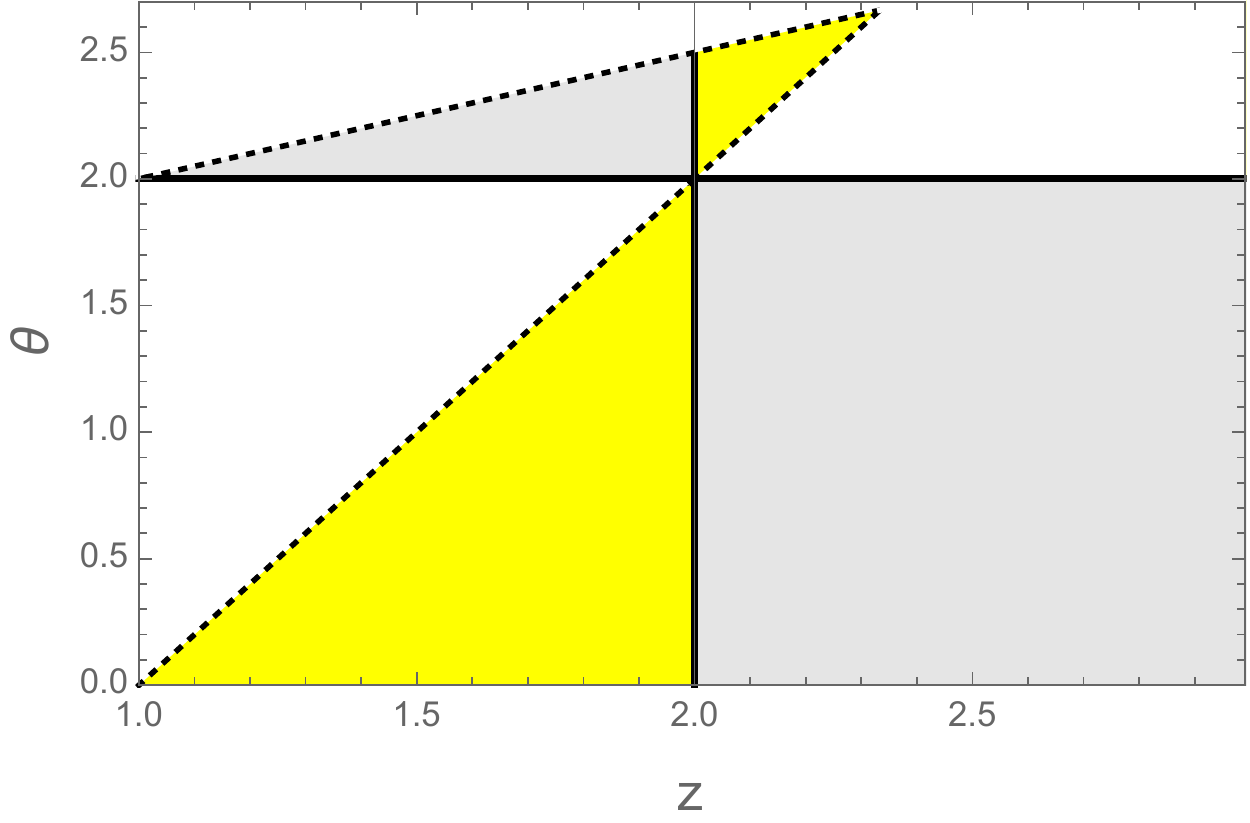} }    
\hskip.2cm
\subfigure[$q_{\chi}=0.7$]
   {\includegraphics[width=45mm]{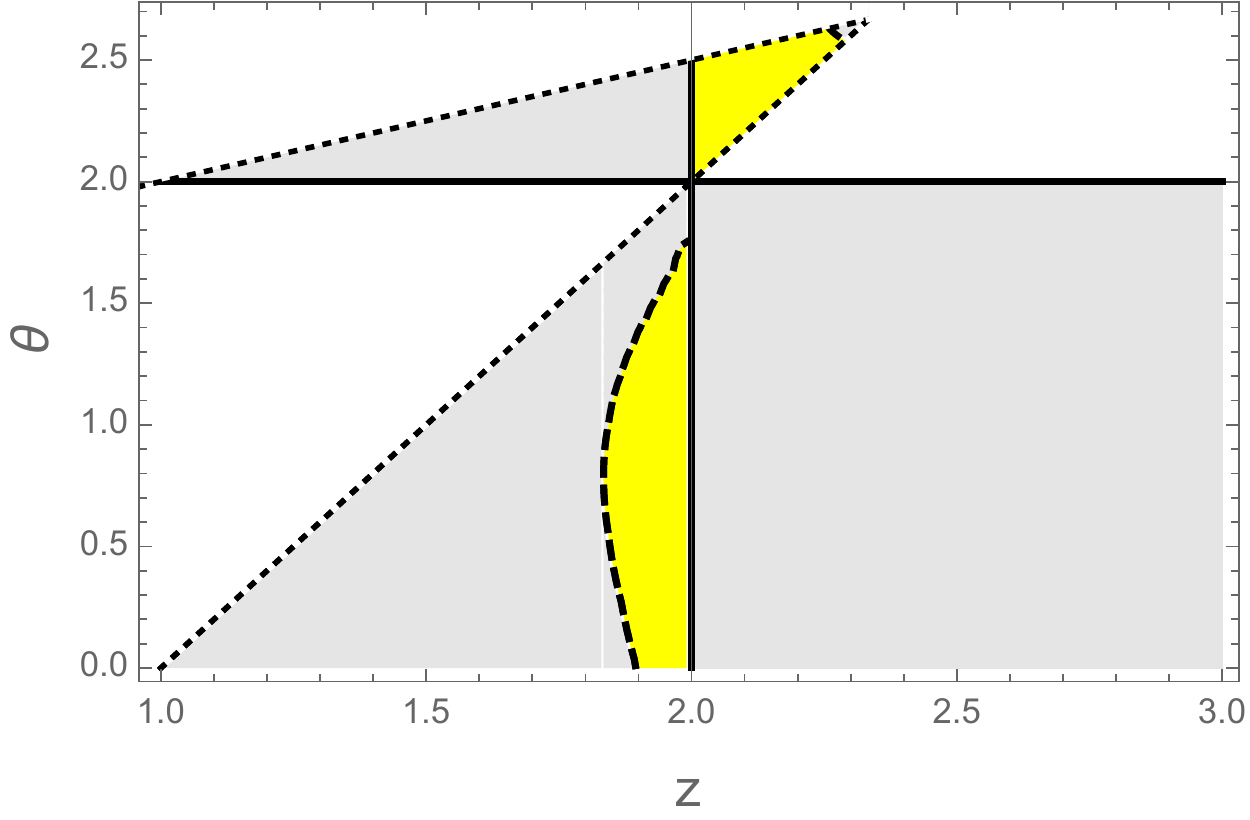} }    
\hskip.2cm
\subfigure[$q_{\chi}=1.4$]
   {\includegraphics[width=45mm]{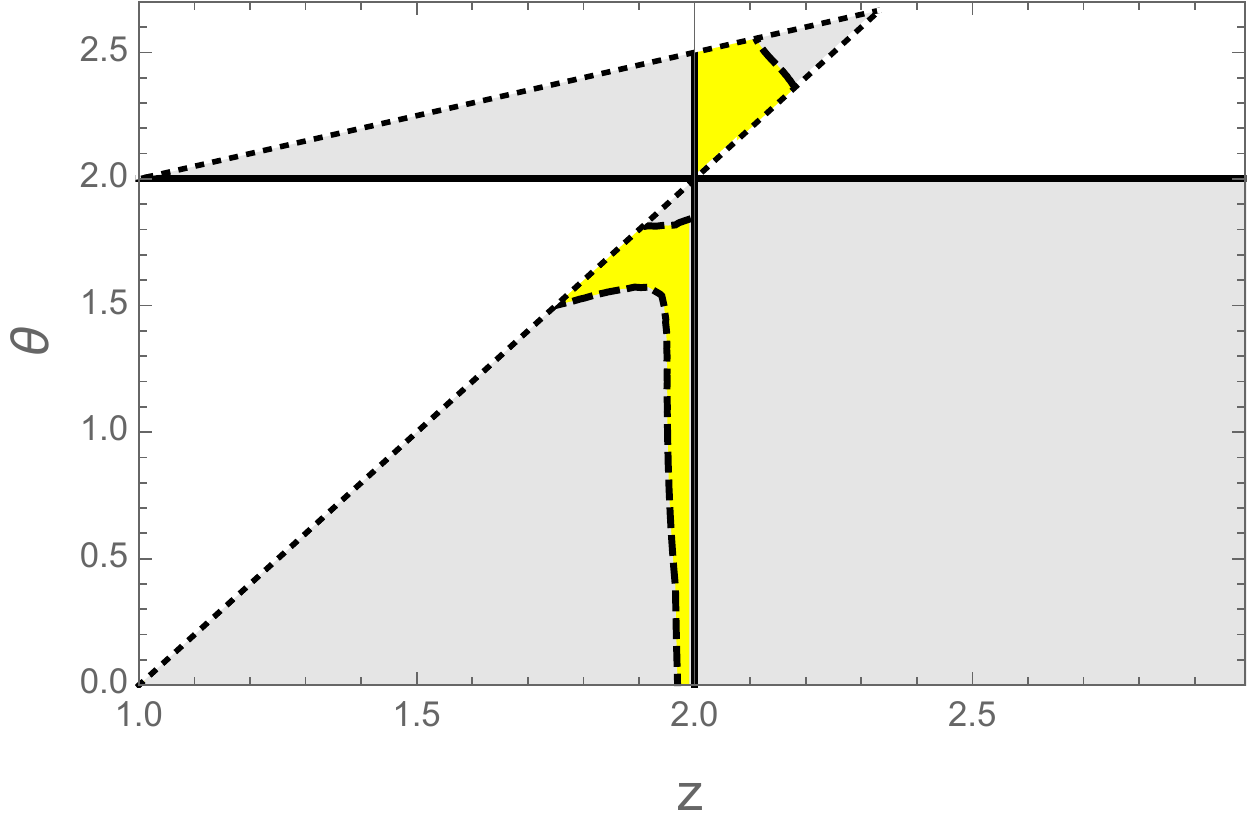} }          
           \caption{Sign of the coefficient of $H^2$ of the thermal conductivity near $H=0$. Yellow region denotes positive sign(WL) and gray region is for negative sign(WAL). Dotted line is for NEC. Here we use the same parameter of Figure \ref{fig:kxxH} and $T=0.16$. }    \label{fig:kxxSign} 
\end{figure}

\begin{figure}[ht!]
\centering
 \subfigure[  WL ]
   {\includegraphics[width=45mm]{kxxz18th07.pdf} }
     \subfigure[ Near phase boundary]
   {\includegraphics[width=45mm]{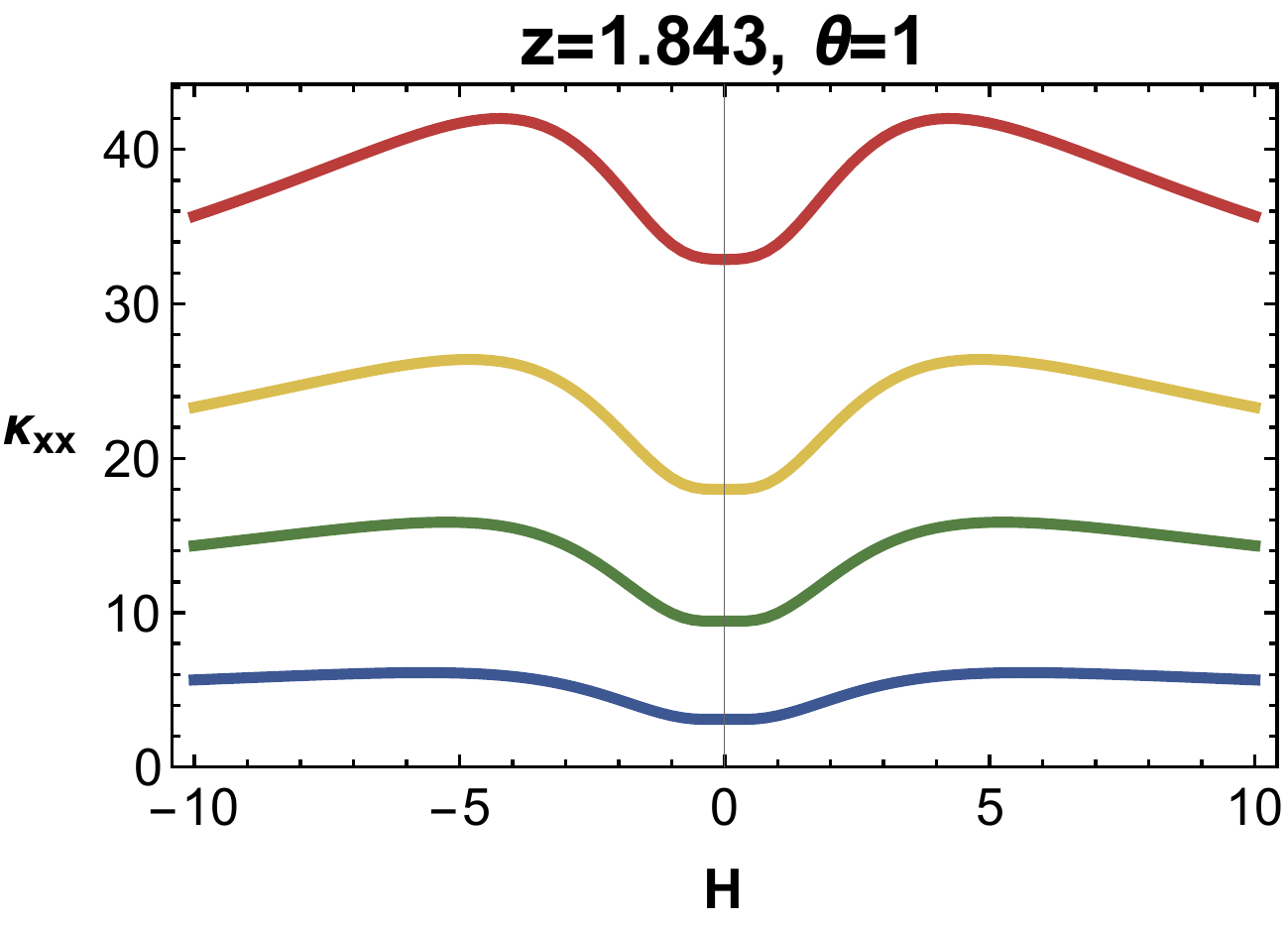} }
     \subfigure[WAL ]
   {\includegraphics[width=45mm]{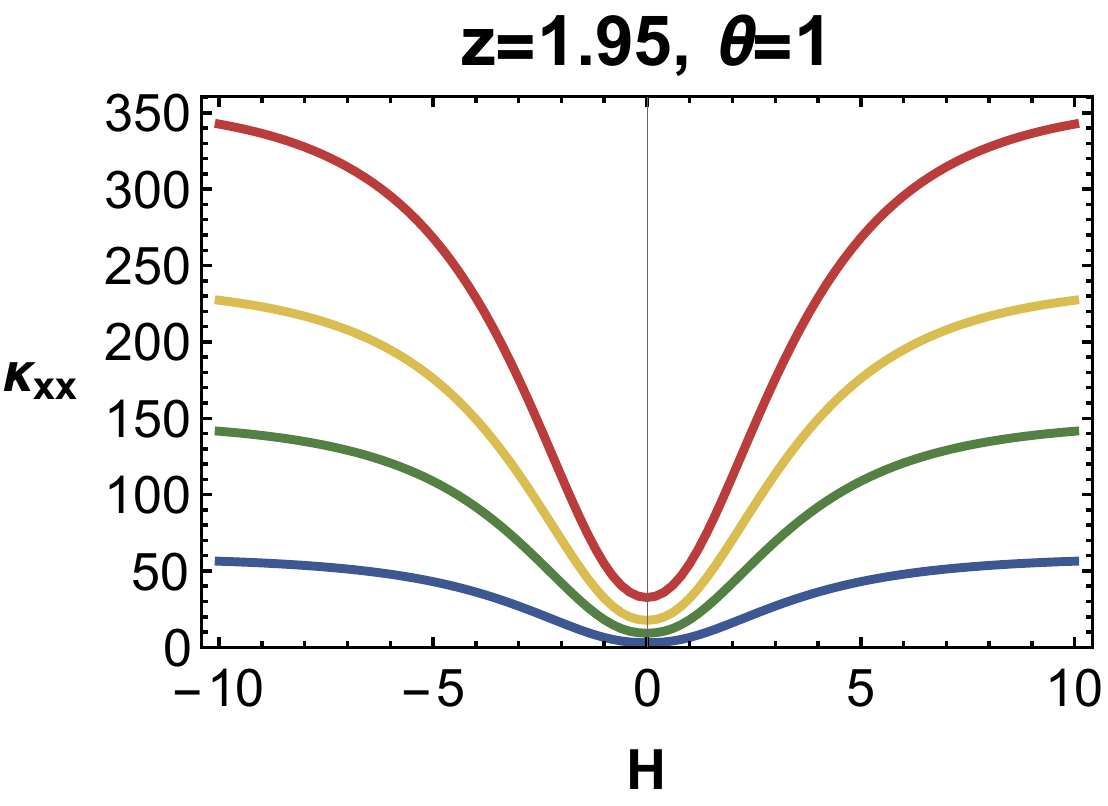} }
                 \caption{Phase transition  from WL  to WAL behavior   of $\kappa_{xx}(H)$ as $z$ increases along $\theta=1$ line. The transition occur near  $z=1.843$.   Curves corresponds to $T=0.04,0.1,0.16, 0.24$ for blue, green, yellow, and red respectively. We used $q_{\chi}=0.7$.}    \label{fig:kxxph} 
\end{figure}

The coefficient of $H^2$ in (\ref{eq:kxxH2}) determines a shape of the thermal conductivity near $H=0$. However, the coefficient is too complicated function for analytic approaches. The numerical result for the sign of the coefficient of $H^2$ is drawn in Figure \ref{fig:kxxSign}, 
where yellow region denotes   the positive sign  corresponding  to the weak localization. Gray region corresponds to the negative sign and  weak anti-localization. In region $B$ and $C$, the coefficient is always negative but in region $A$ and $D$, there are domains of positive  sign.  In the absence of $q_{\chi}$. the region $A$ and $D$ are filled with WL as shown in the Figure \ref{fig:kxxH}. As we increase $q_{\chi}$, the region of WAL start to appear and increases. But the shape is complicated. The sign of the coefficient is positive near $z=2$ and turns to negative at other region.
See Figure \ref{fig:kxxSign}. 
Figure \ref{fig:kxxph} demonstrates the phase transition 
from WL to WAL due to the sign flip of $H^{2}$ coefficient.

\begin{figure}[ht!]
\centering
 \subfigure[$P_{0A}$  ]
   {\includegraphics[width=45mm]{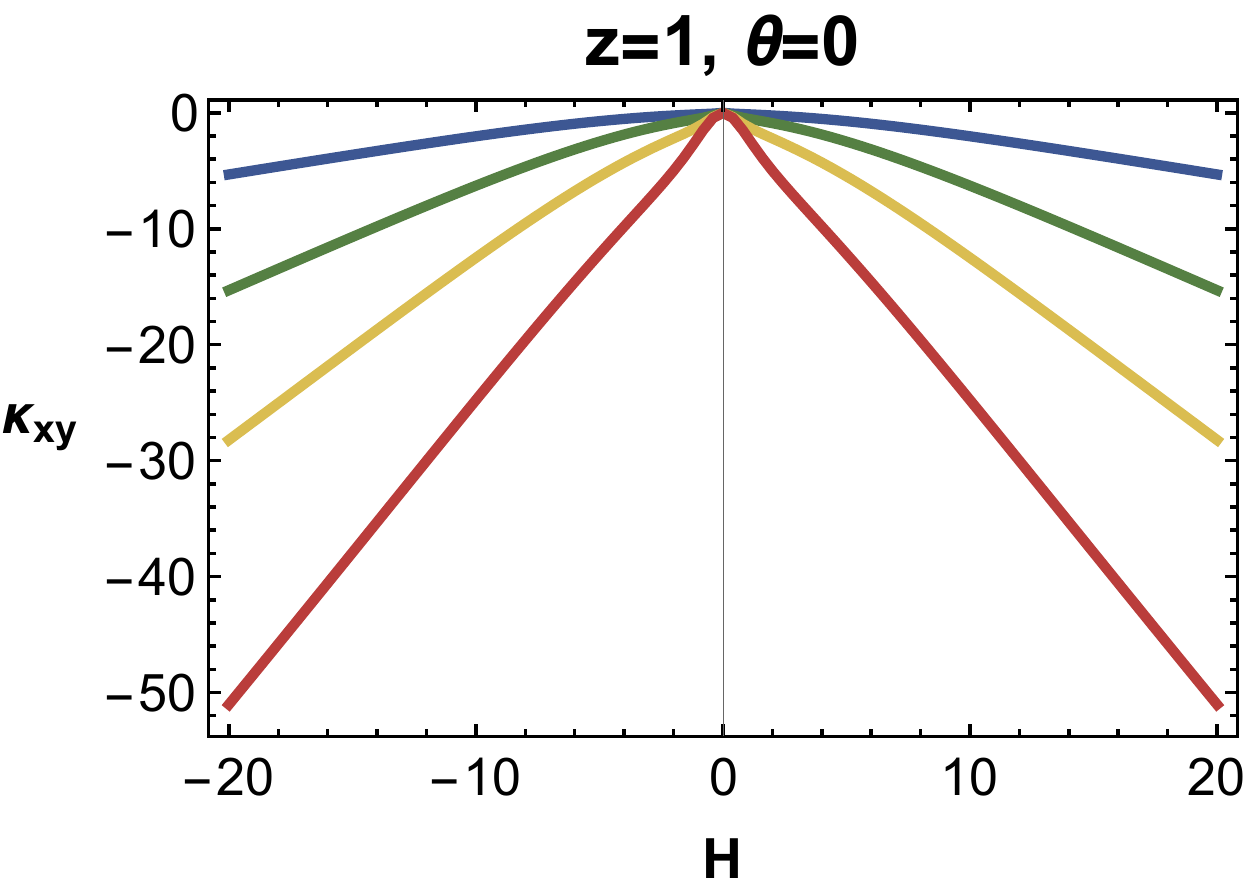} }
         \subfigure[$P_{A}$  ]
   {\includegraphics[width=45mm]{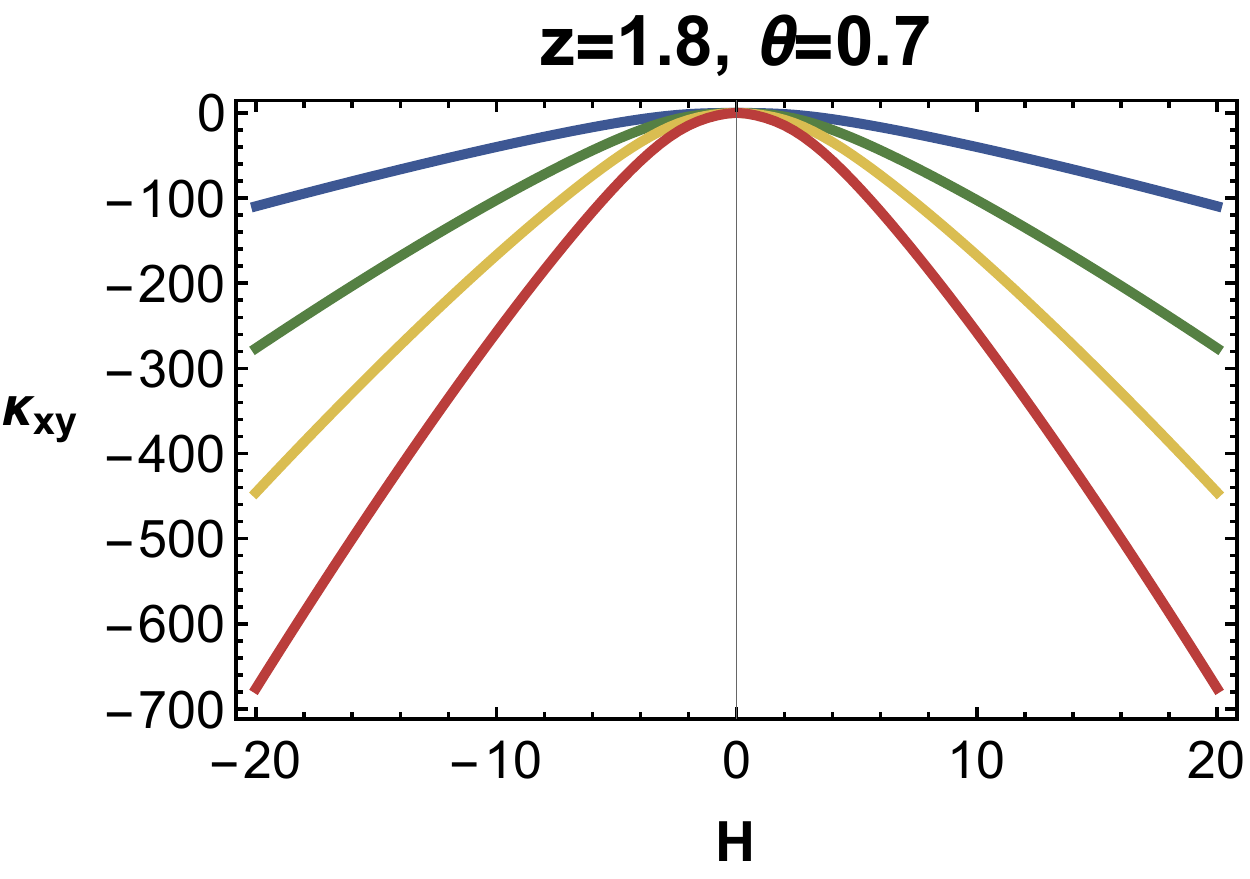} }
      \subfigure[$P_{C}$ ]
   {\includegraphics[width=45mm]{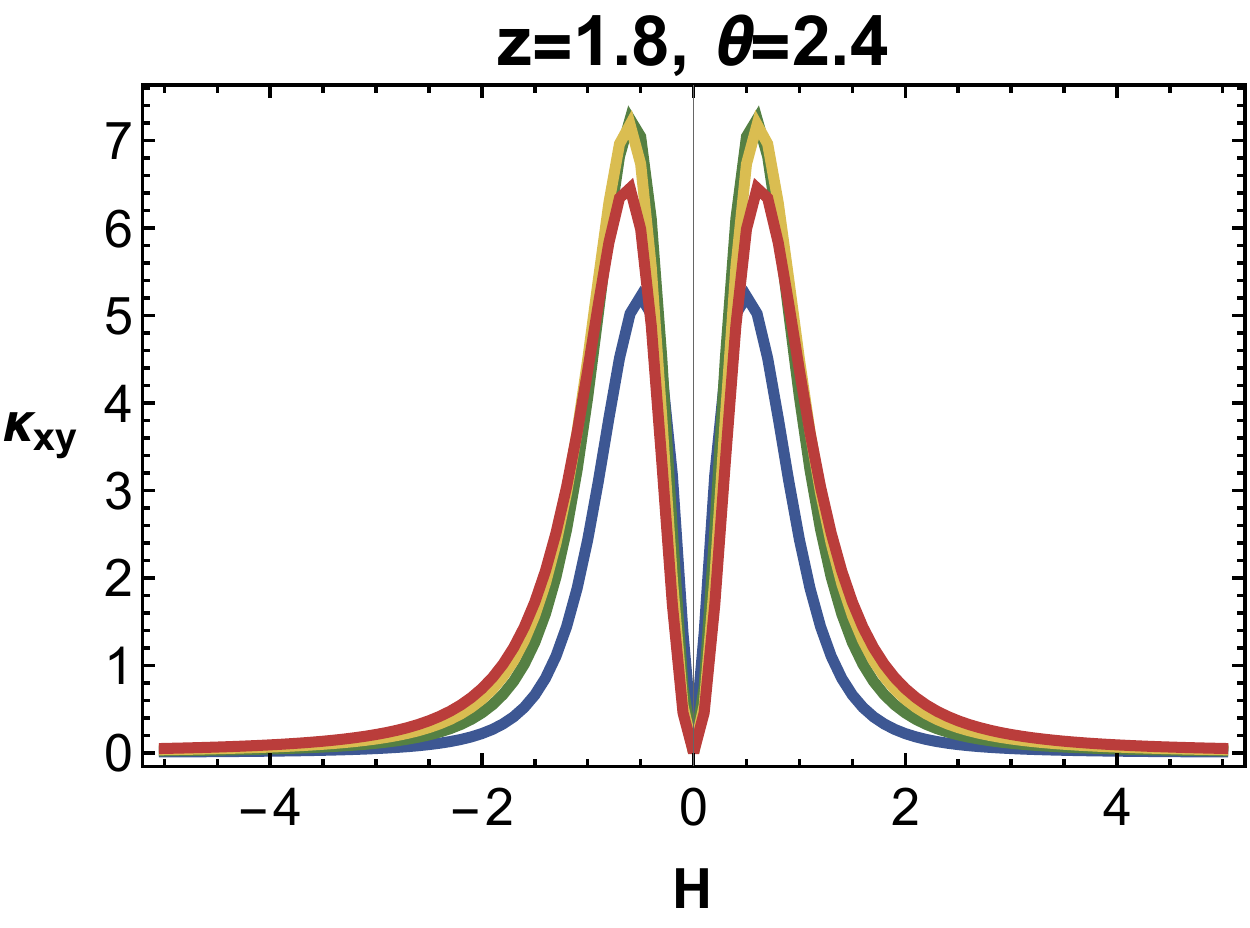} }
    \subfigure[$P_{0B}$ ]
   {\includegraphics[width=45mm]{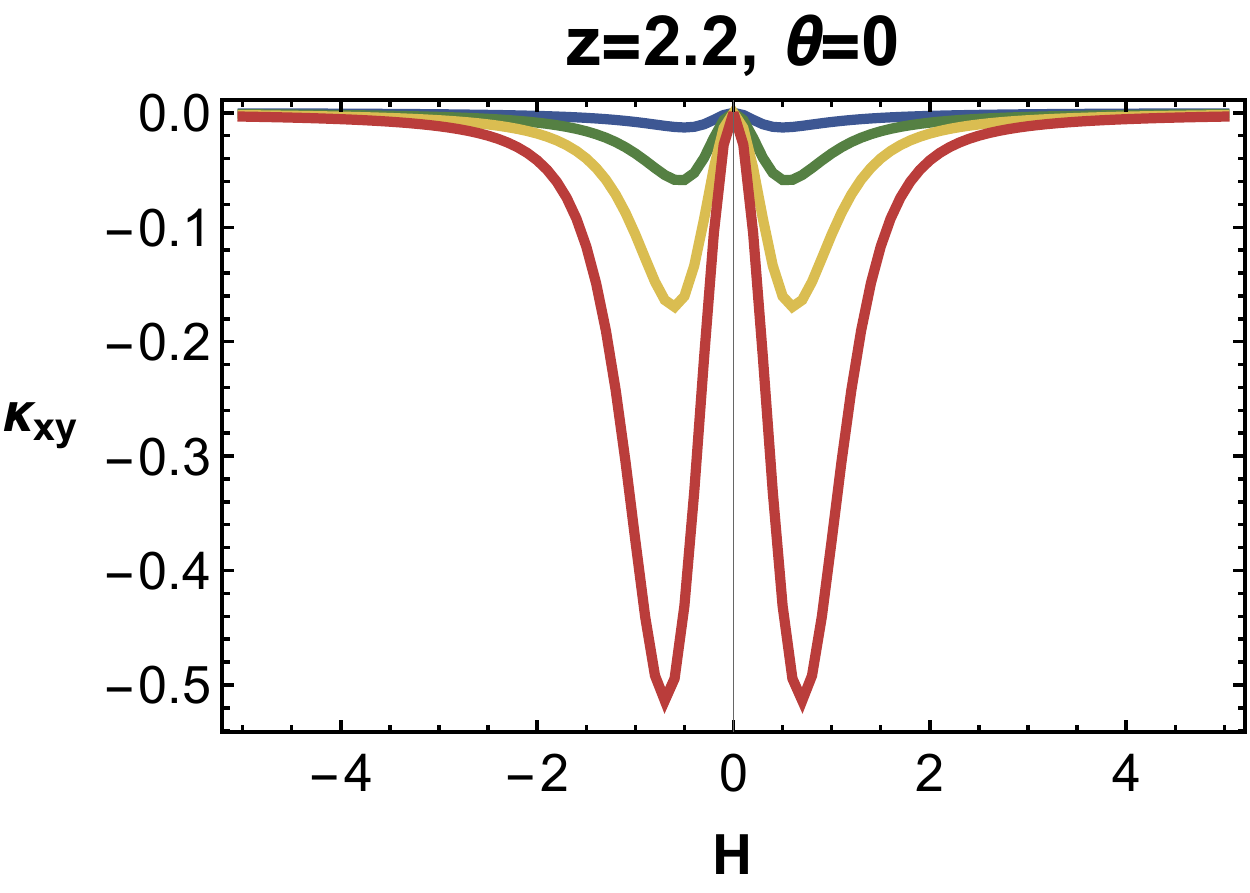} }
      \subfigure[$P_{B}$  ]
   {\includegraphics[width=45mm]{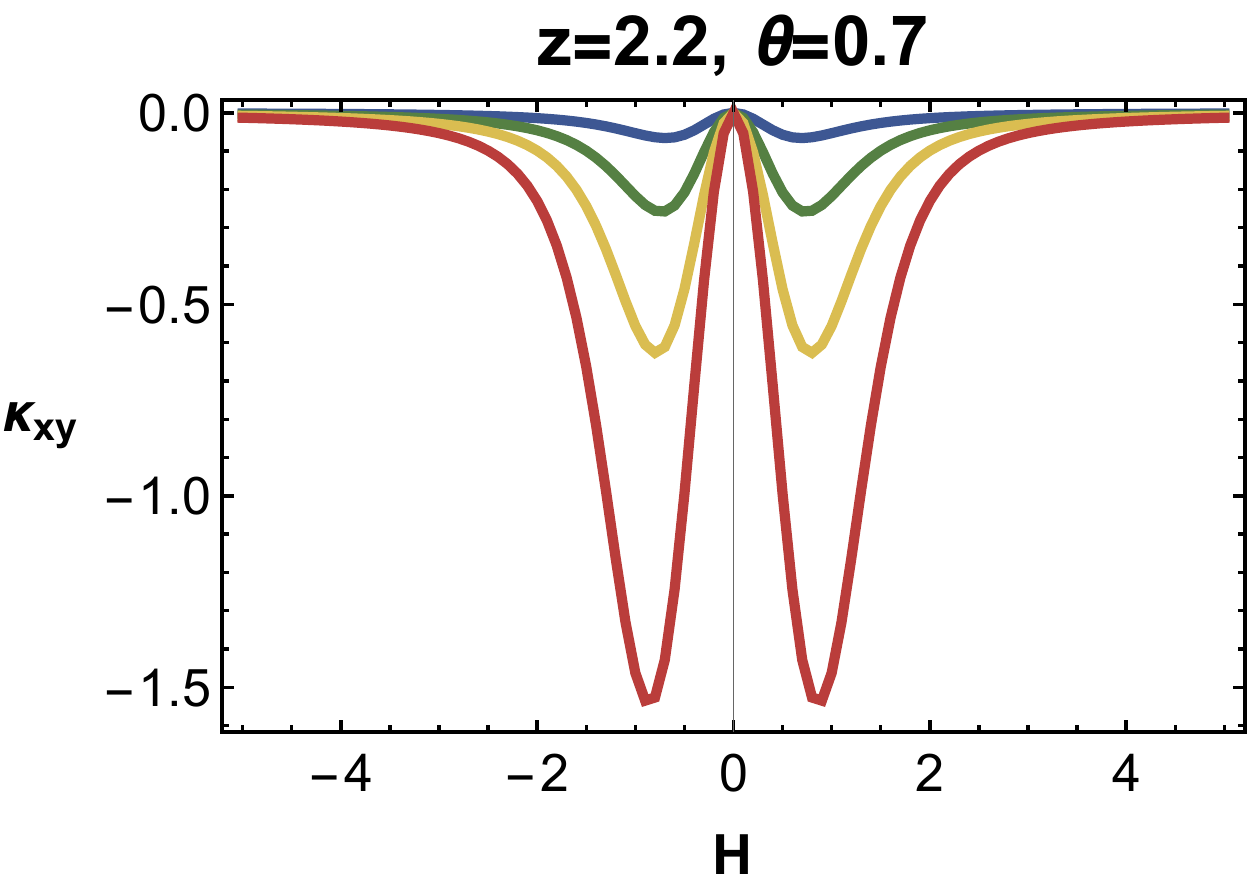} }
      \subfigure[$P_{D}$ ]
   {\includegraphics[width=45mm]{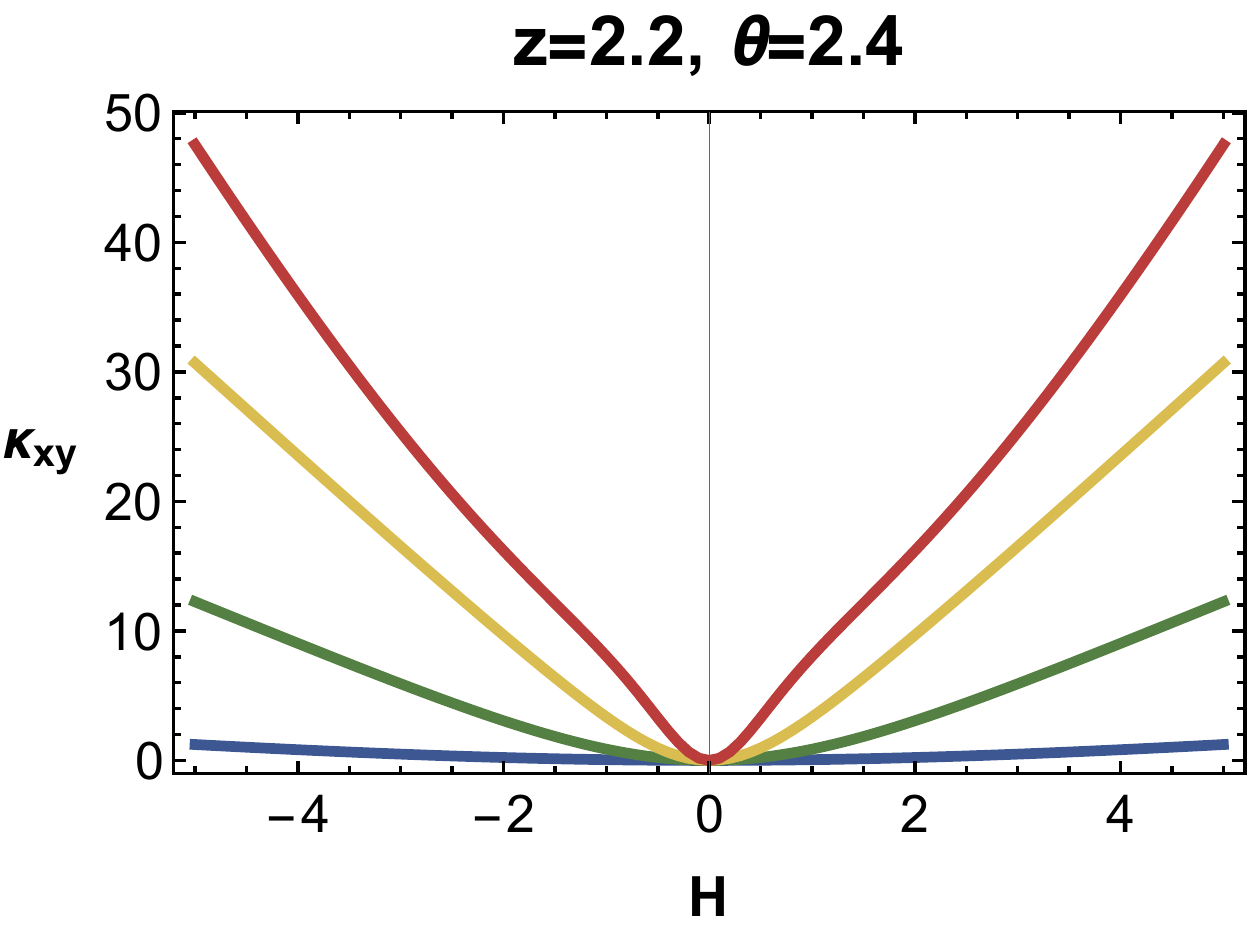} }   
                 \caption{Temperature evolution for $\kappa_{xy}(H)$ for different $(z,\theta)$. Each curves corresponds to $T=0.04,0.1,0.16, 0.24$ for blue, green, yellow, and red respectively. We used $q_{\chi}=0.7$.}    \label{fig:kxyH} 
\end{figure}

\begin{figure}[ht!]
\centering
  \subfigure[$\kappa_{xy}(q_{\chi}=0.7)$]
   {\includegraphics[width=50mm]{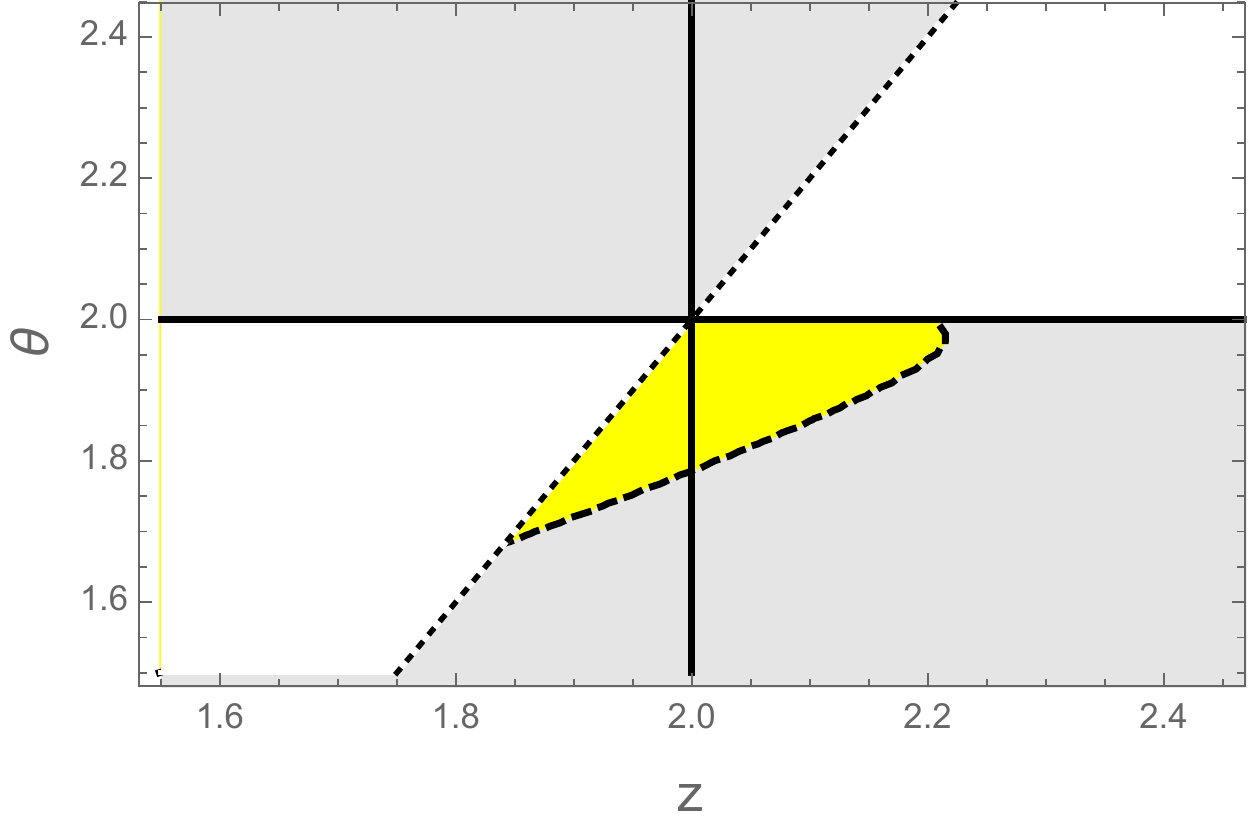} }    
   \hskip1cm
    \subfigure[$\kappa_{xy}(q_{\chi}=1.4)$]
   {\includegraphics[width=50mm]{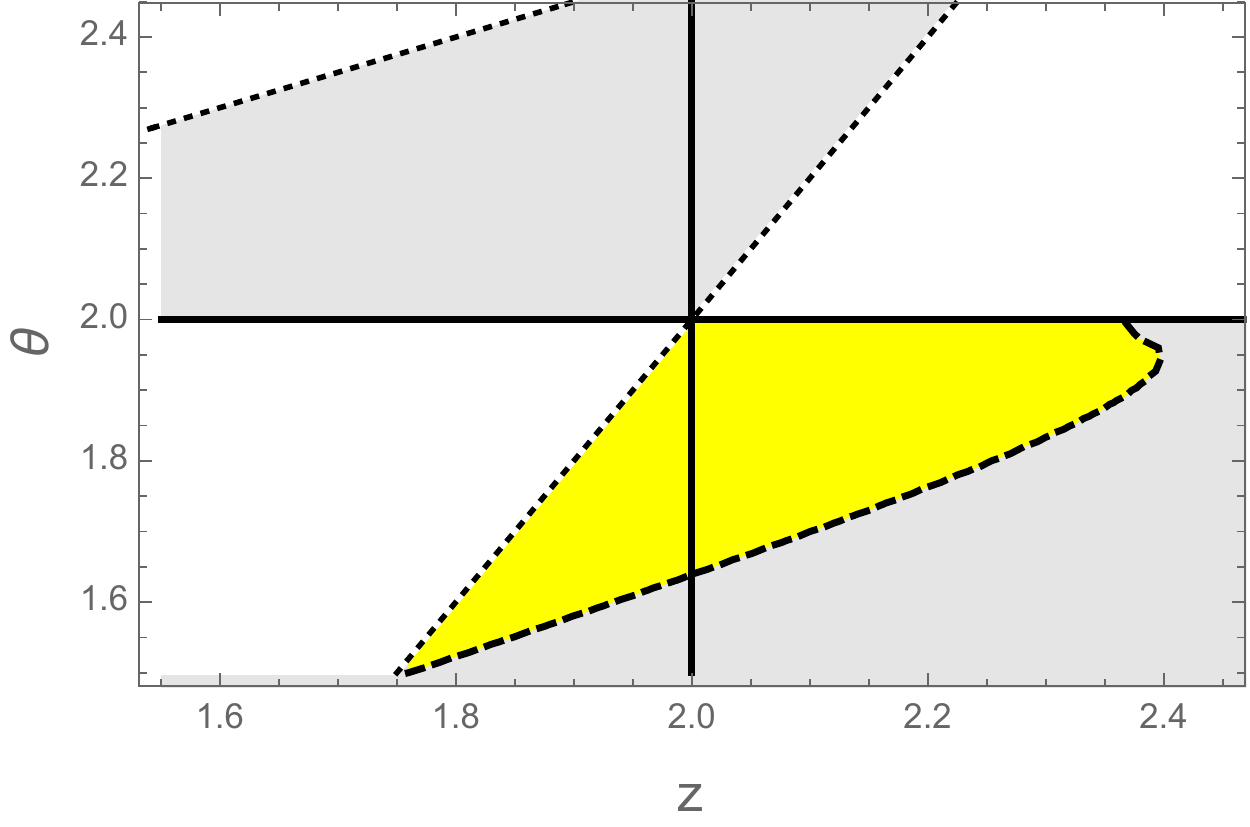} }    
       \subfigure[$N(q_{\chi}=0.7)$]
   {\includegraphics[width=50mm]{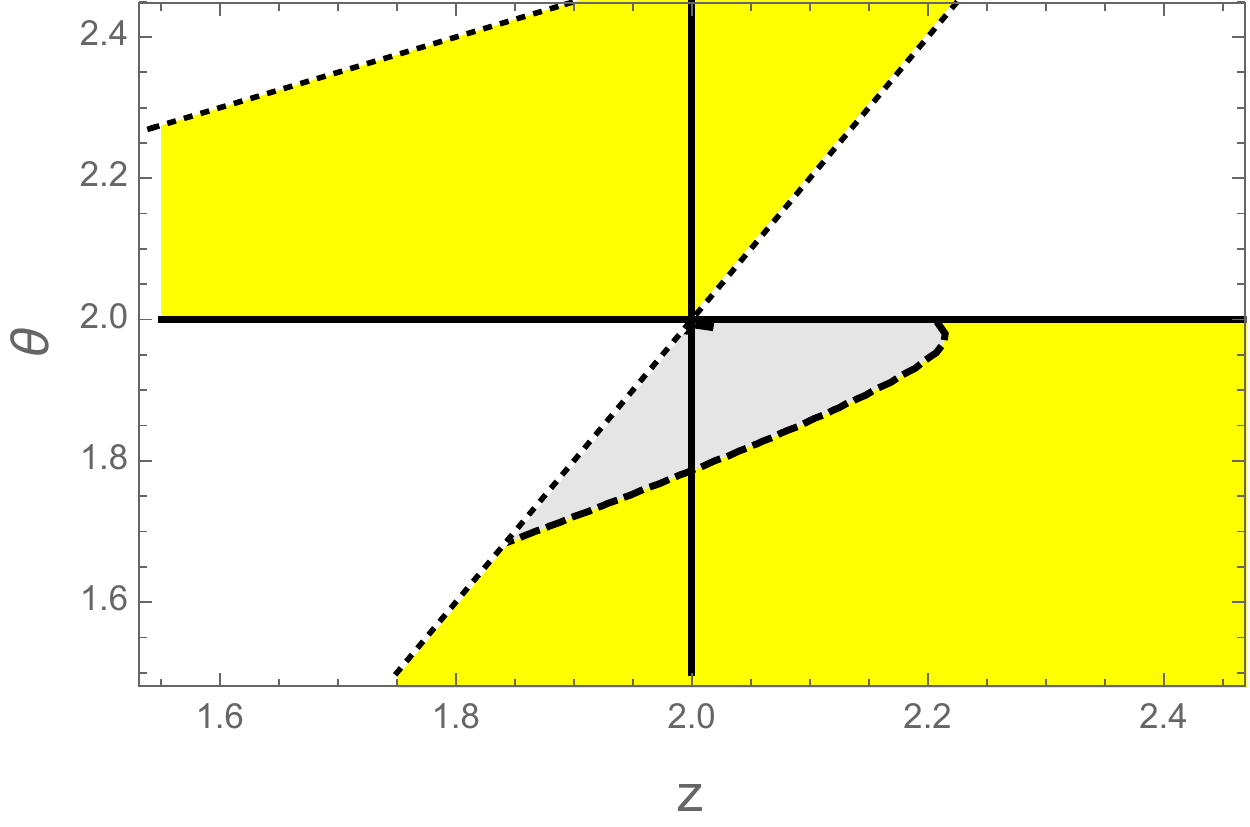} }  
\hskip1cm
     \subfigure[$N(q_{\chi}=1.4)$]
   {\includegraphics[width=50mm]{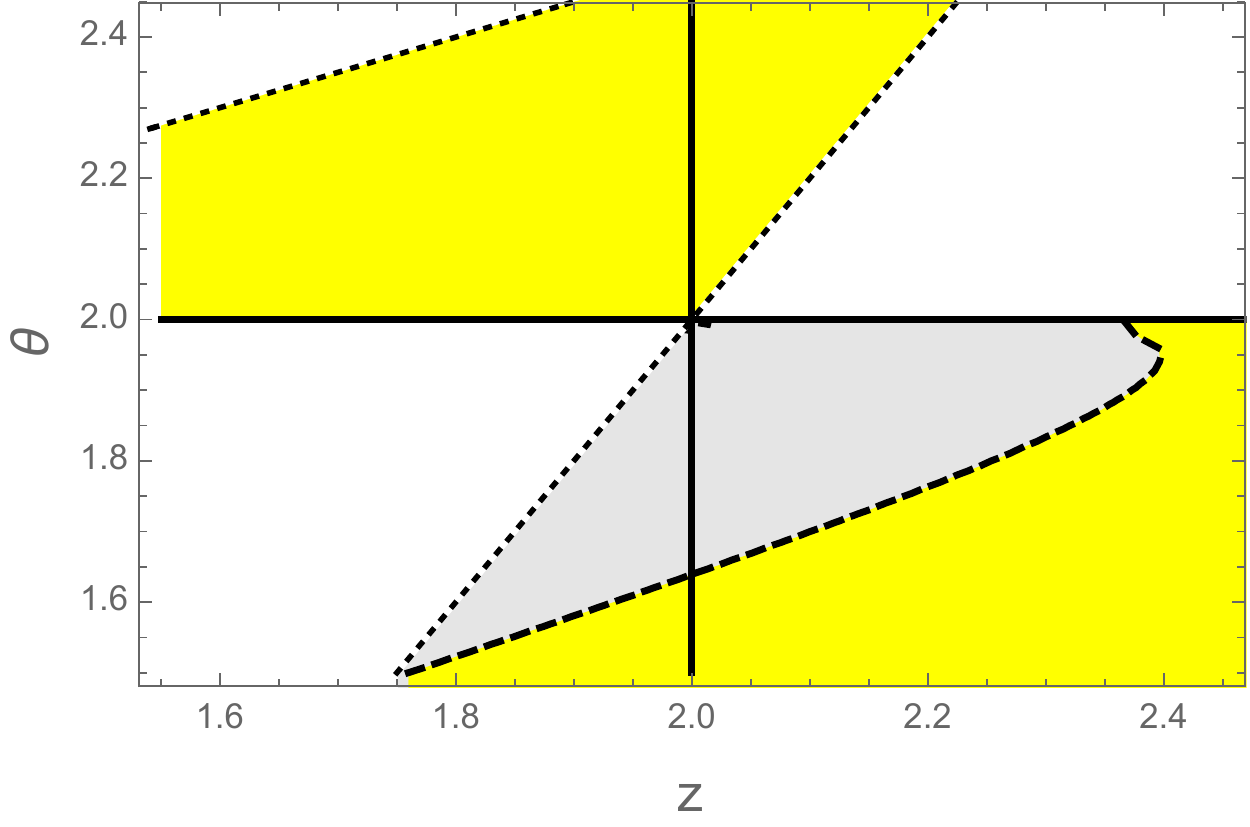} }          
           \caption{(a,b) WL vs WAL in  thermal conductivity.  (c,d) WL vs WAL in Nernst signal. 
           They are determined by sign of the coefficient of $H^2$  in case of $\kappa_{xy}$ or  $H$ in case of N. Yellow region denotes positive sign(WL) and gray region is for negative sign(WAL). Dotted line is for NEC. Here we use the same parameter of figures and  $T=0.24$. }    \label{fig:SignKN} 
\end{figure}

The transverse thermal conductivity can be expanded for small $H$ as
\begin{align}\label{eq:kxyH2}
\kappa_{xy} = -\frac{16 \pi^2 q_{\chi}\lambda^2 r_0^{4z -\theta-6}}{\beta^4}\, \left[\frac{ (2+z-2\theta)r_0^{4z} + q_{\chi}^2 \lambda^4 (1+z-2\theta) r_0^{4\theta}}{ (2+z-2\theta)^2(r_0^{4 z} +q_{\chi}^2 \lambda^4 r_0^{4\theta})} \right] H^2 + \cdots.
\end{align}
When $q_{\chi} =0$, the sign of coefficient is only depends on the sign of $-(2+z-2\theta)$ because all other terms are positive definite. This sign  is negative in region $A$, $B$ and positive in $C$, $D$. The full magnetic field dependence of the transverse thermal conductivity is drawn in Figure \ref{fig:kxyH}.
When we turn on $q_{\chi}$, the second term can change the value of $(z,~\theta)$ where the sign of the coefficient flipped. Numerical calculation indicates that the sign of $H^2$ changes near $(z,~\theta)=(2,2)$ in the region $A$, $B$.  This region increase as $q_{\chi}$ increases, see Figure \ref{fig:SignKN} (a).  

\begin{figure}[ht!]
\centering
	 \subfigure[$P_{0A}$ at $q_{\chi}=0$]
   {\includegraphics[width=30mm]{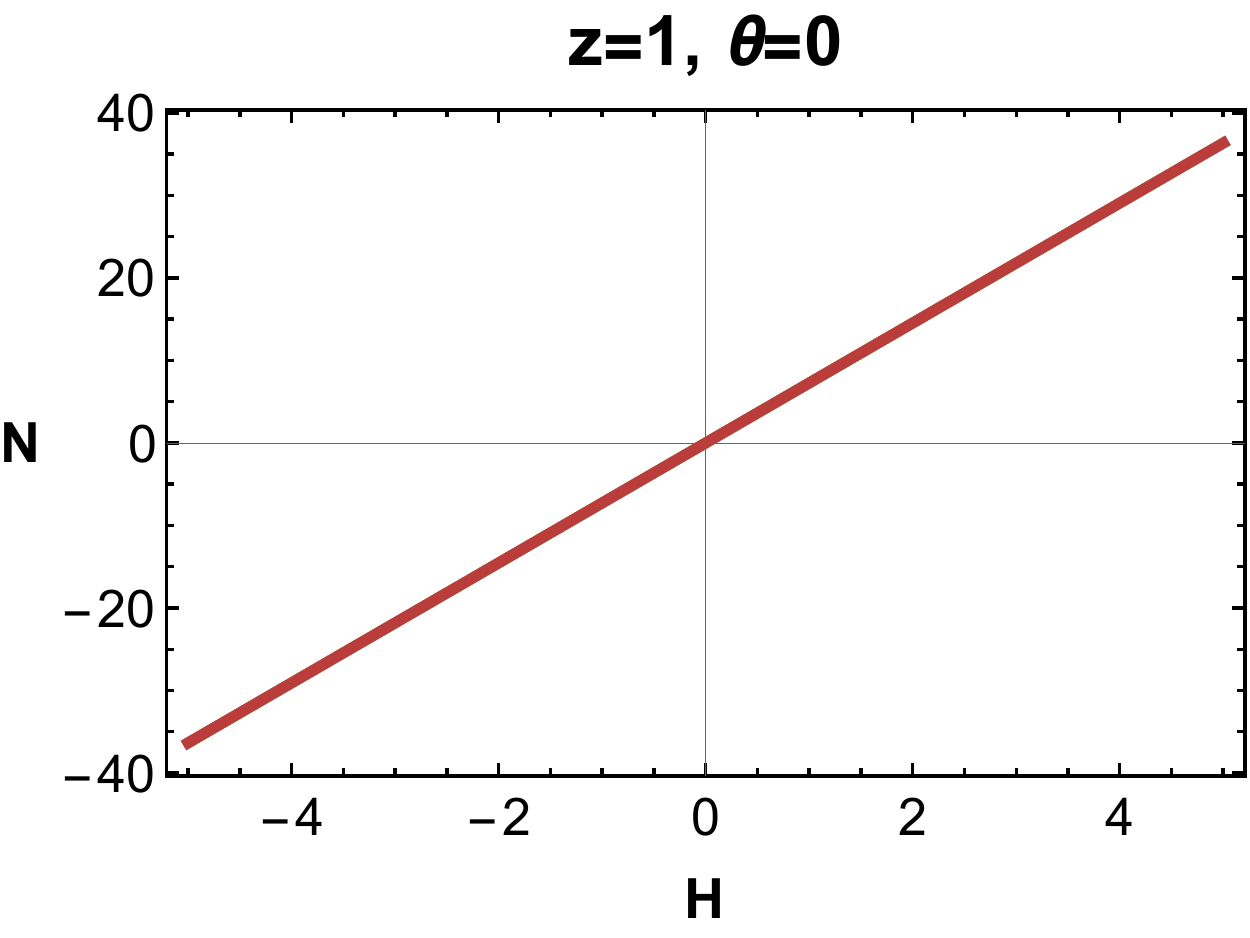} }
    \subfigure[$P_{0A}$  ]
   {\includegraphics[width=37mm]{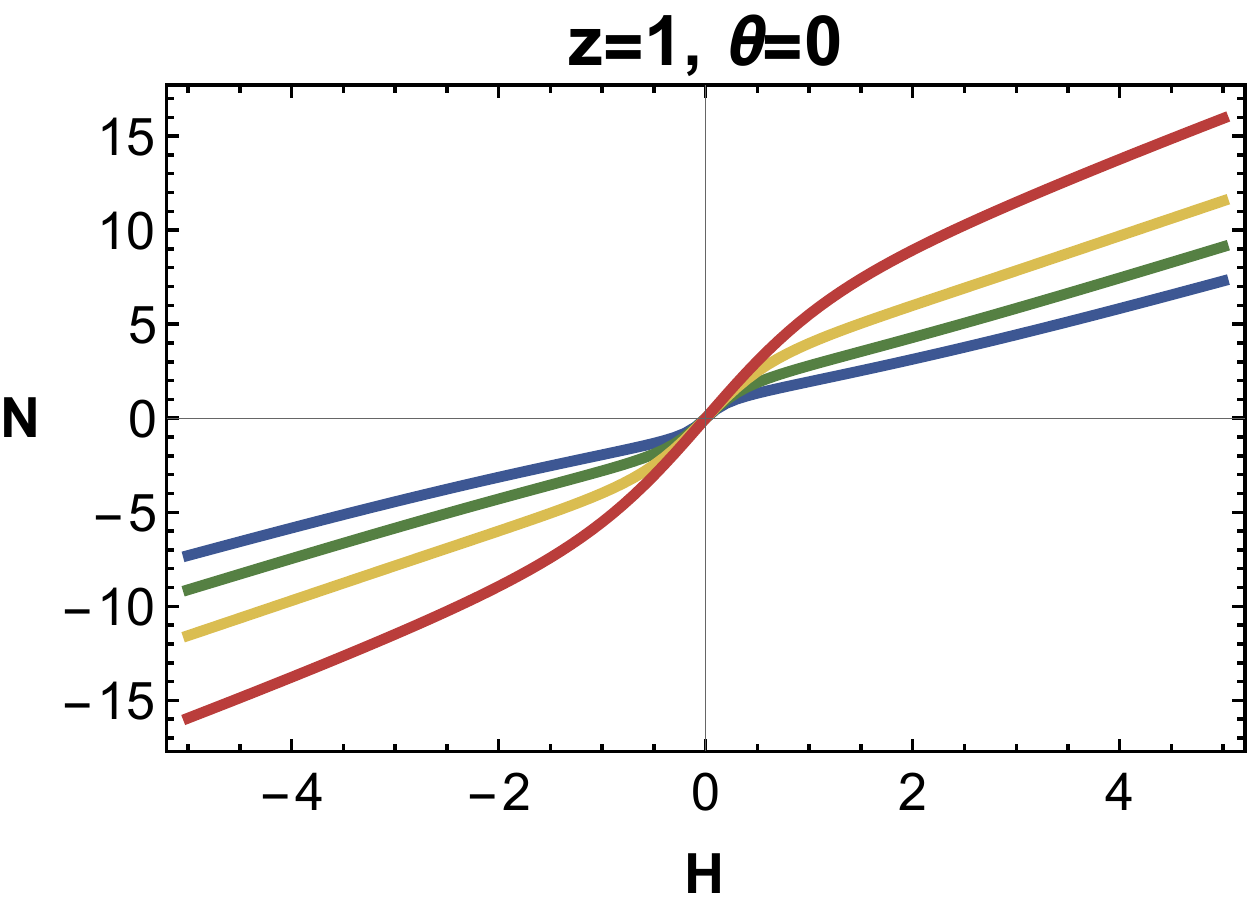} }
    \subfigure[$P_{0B}$ at $q_{\chi}=0$]
   {\includegraphics[width=30mm]{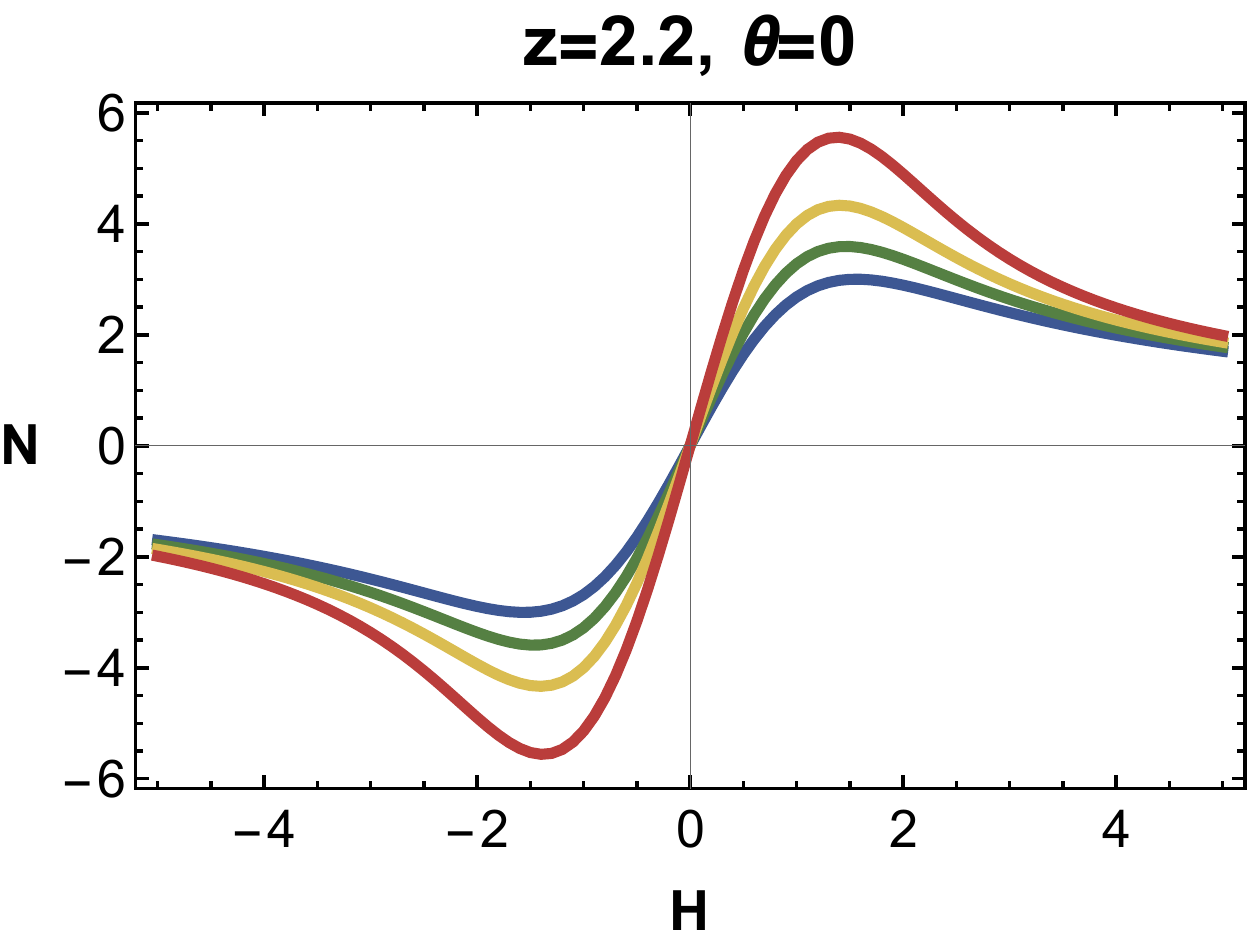} }
    \subfigure[$P_{0B}$ ]
   {\includegraphics[width=37mm]{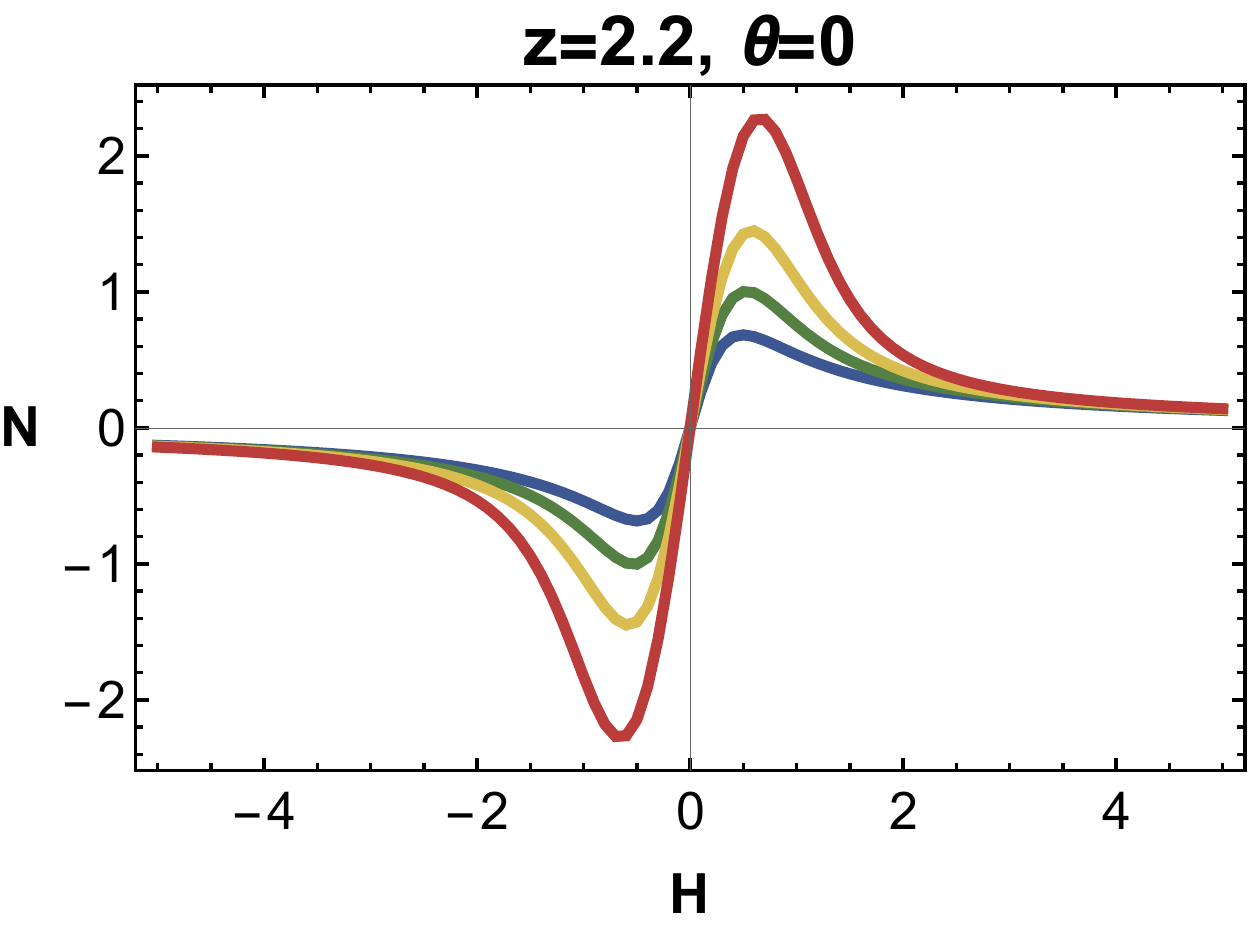} }
    \subfigure[$P_{A}$ at $q_{\chi}=0$]
   {\includegraphics[width=30mm]{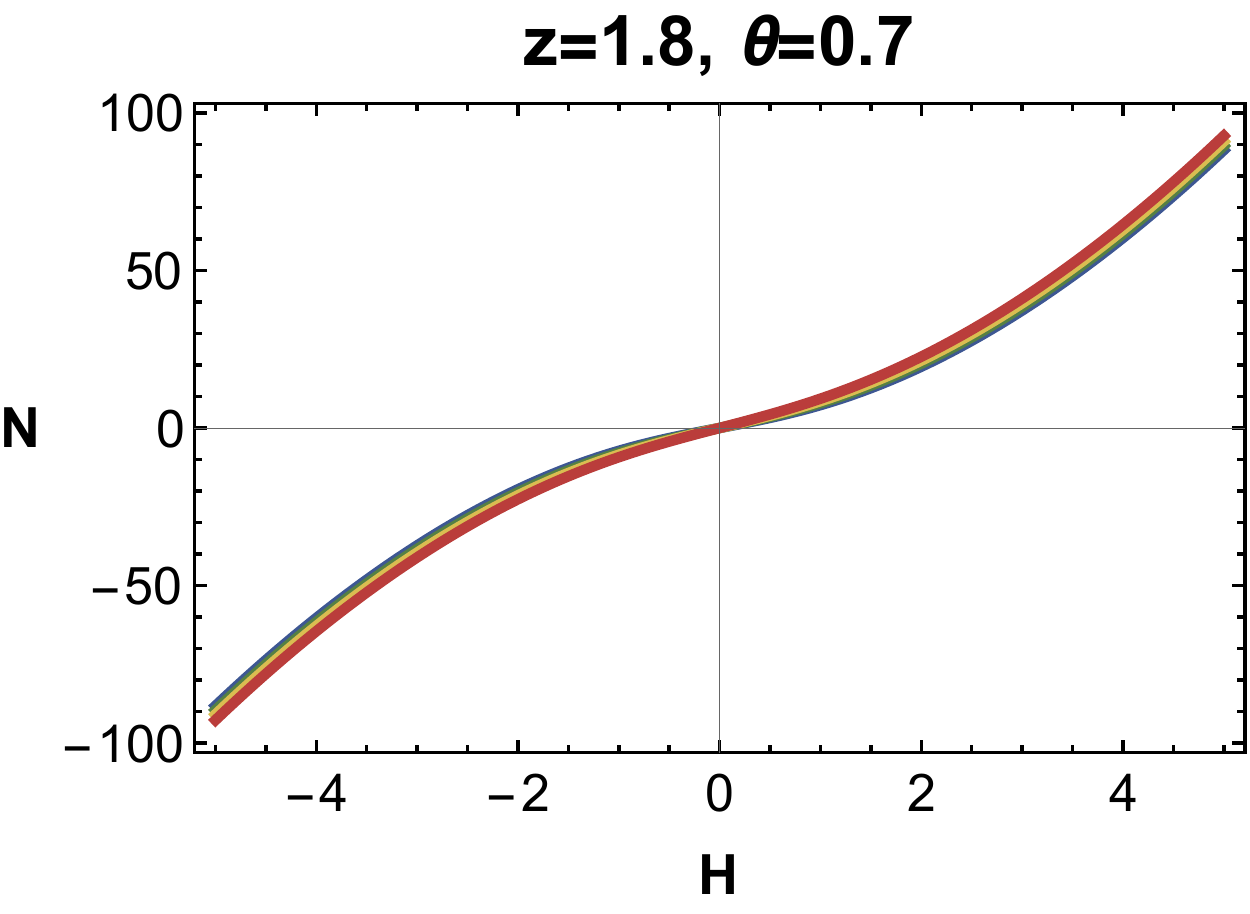} }
    \subfigure[$P_{A}$  ]
   {\includegraphics[width=37mm]{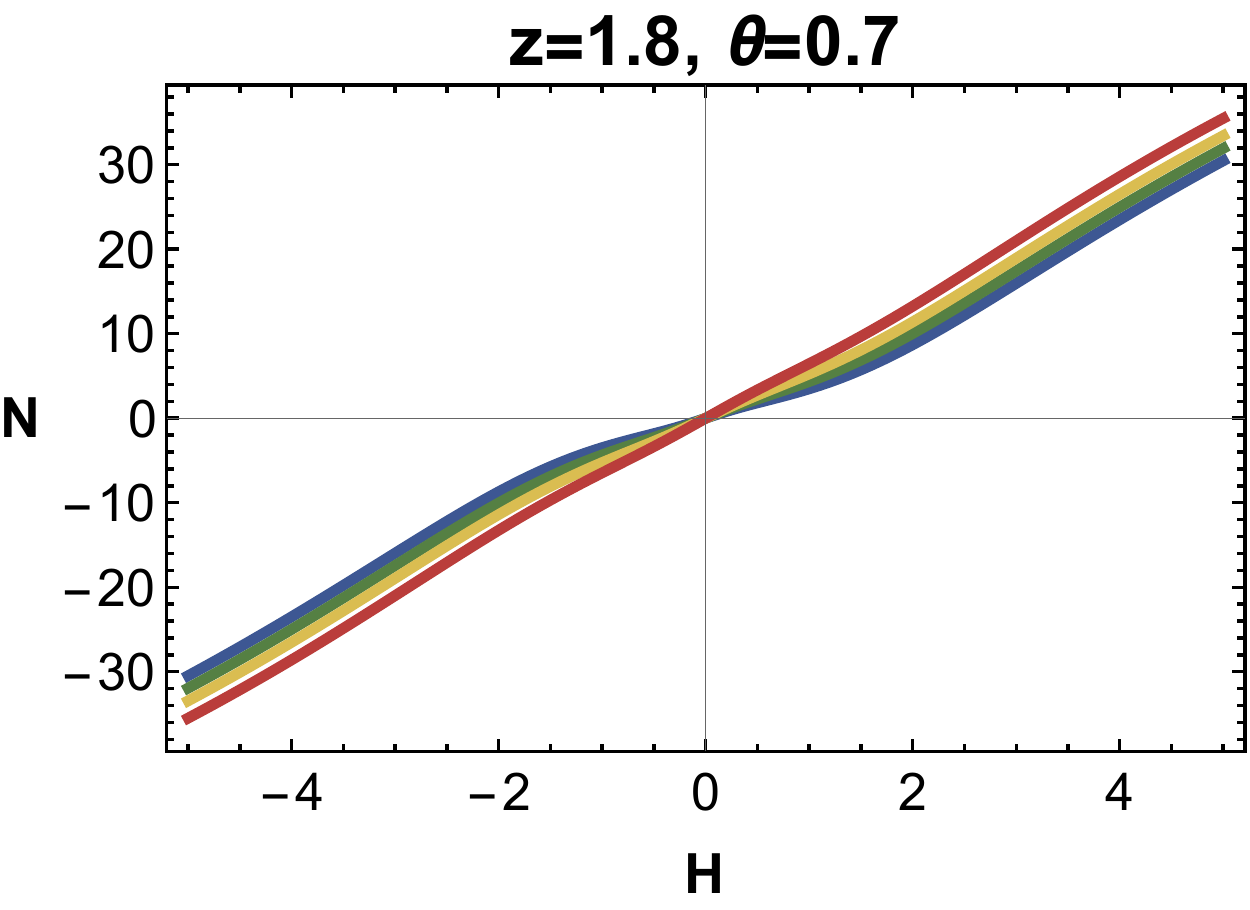} }
      \subfigure[$P_{B}$ at $q_{\chi}=0$]
   {\includegraphics[width=30mm]{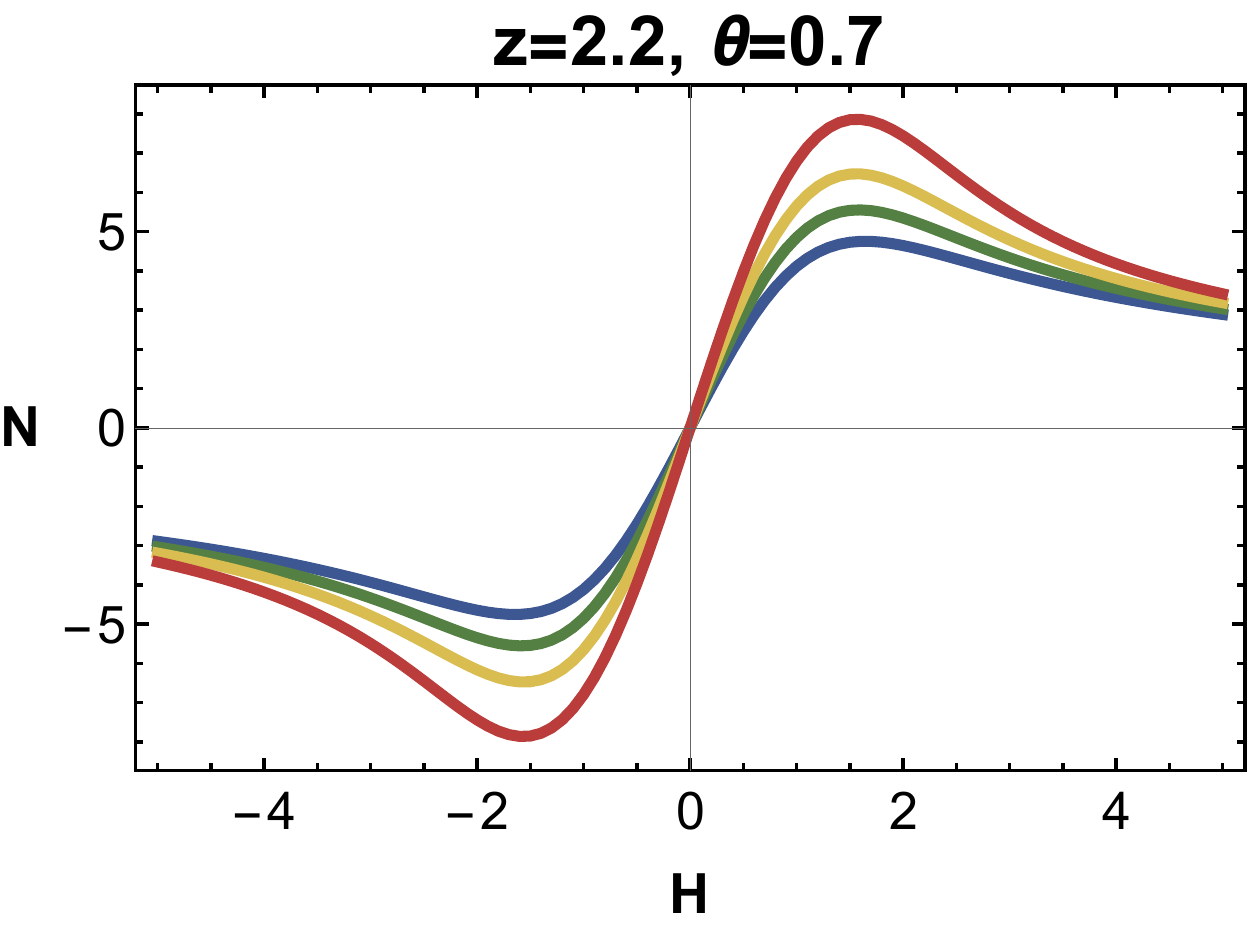} }
    \subfigure[$P_{B}$  ]
   {\includegraphics[width=37mm]{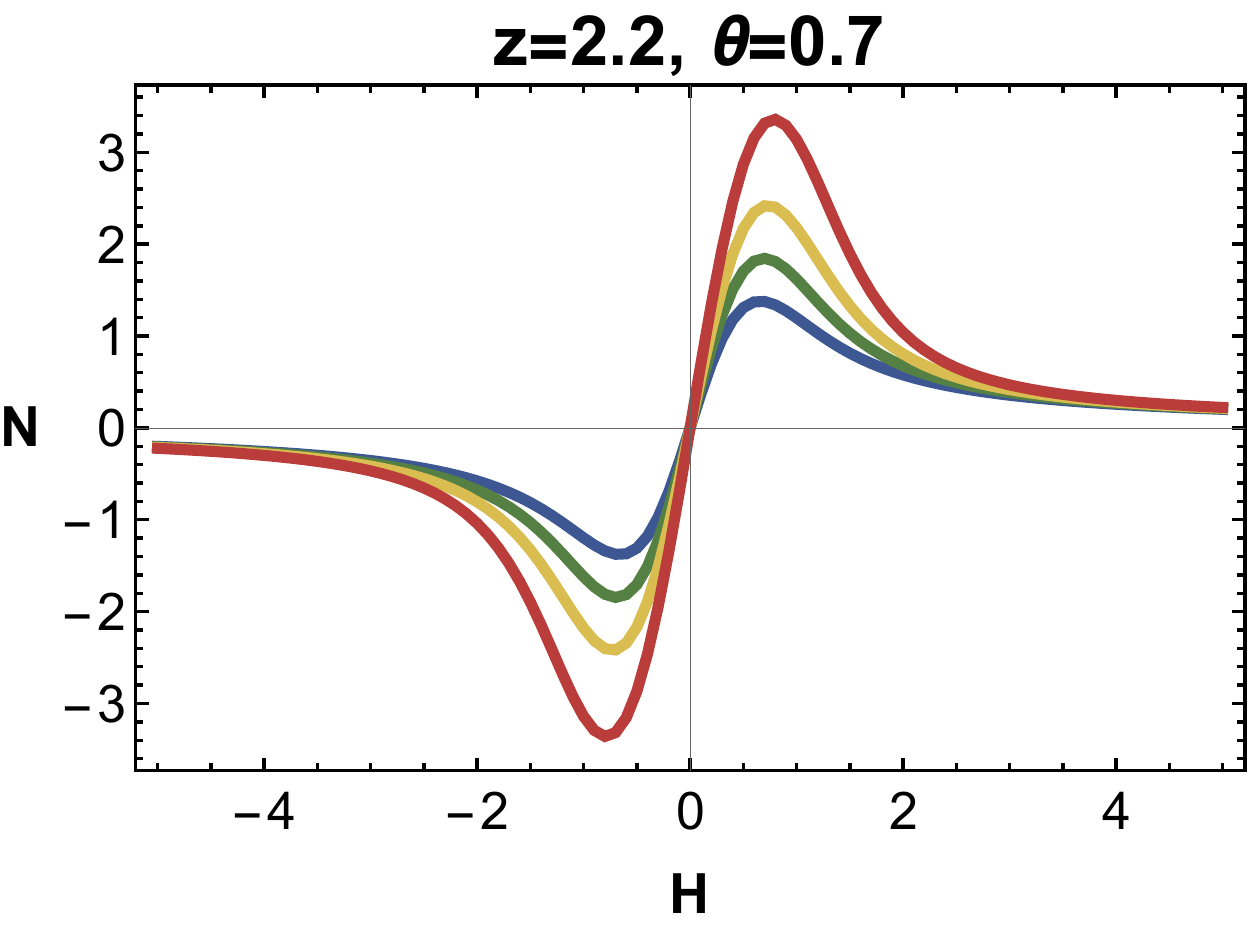} }
    \subfigure[$P_{C}$ at $q_{\chi}=0$]
   {\includegraphics[width=30mm]{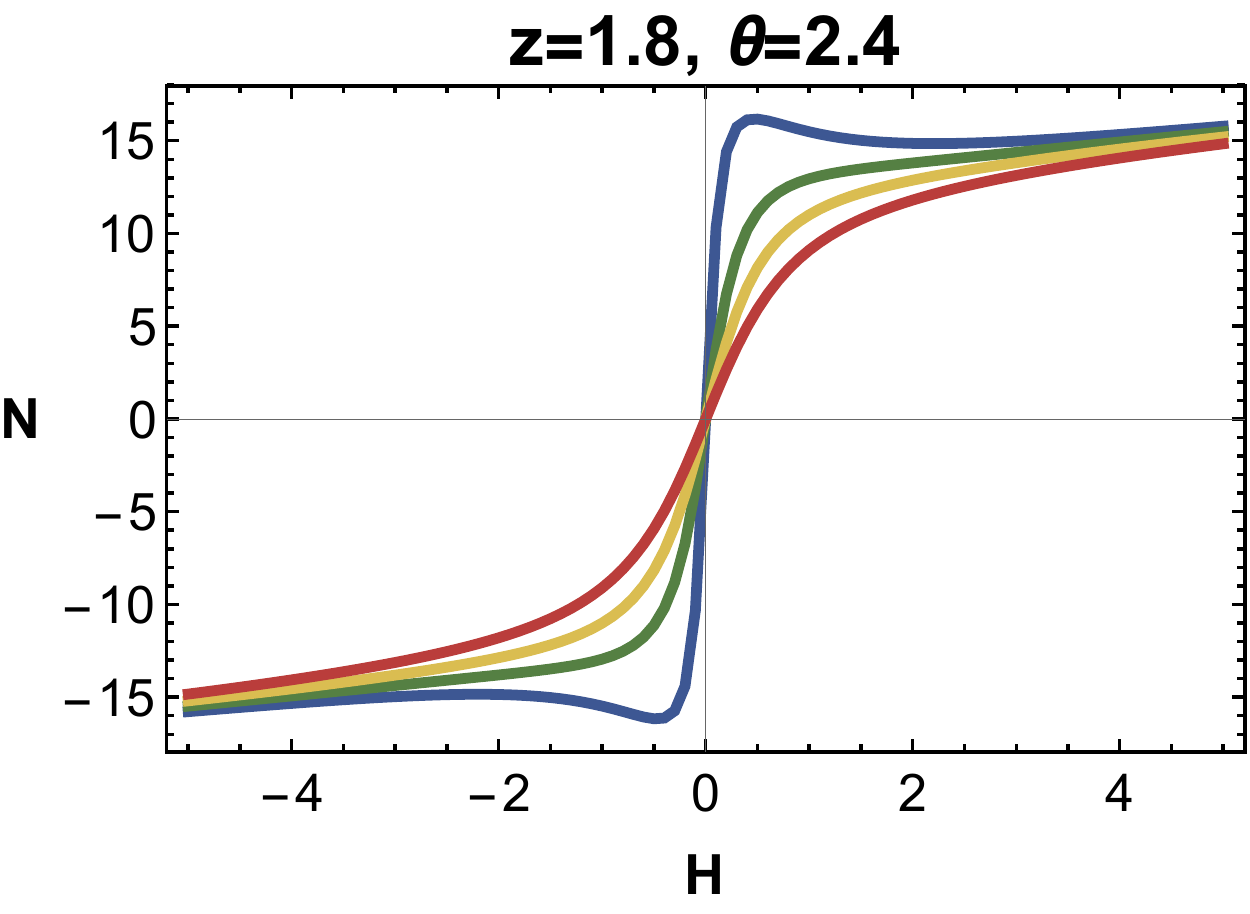} }
    \subfigure[$P_{C}$ ]
   {\includegraphics[width=37mm]{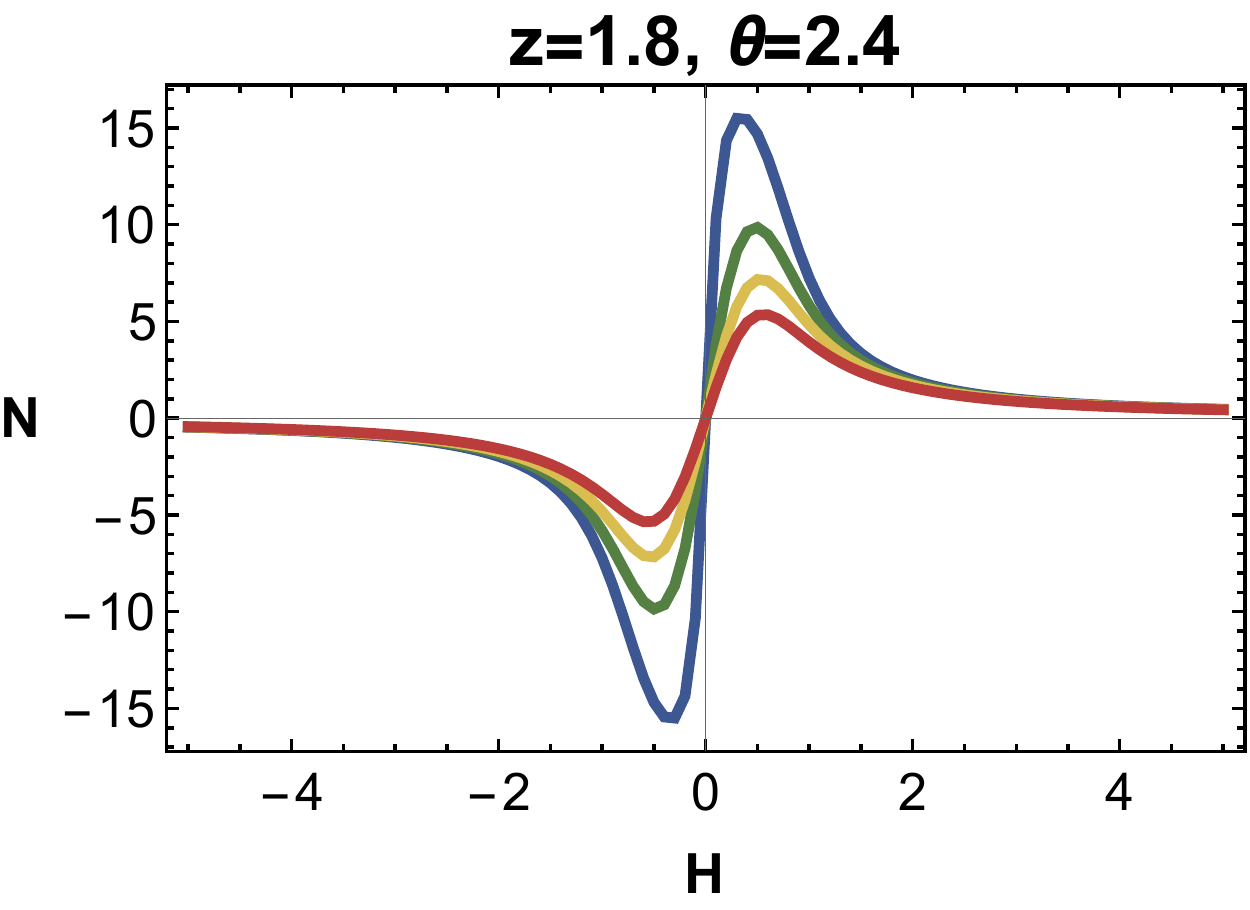} }
     \subfigure[$P_{D}$ at $q_{\chi}=0$]
   {\includegraphics[width=30mm]{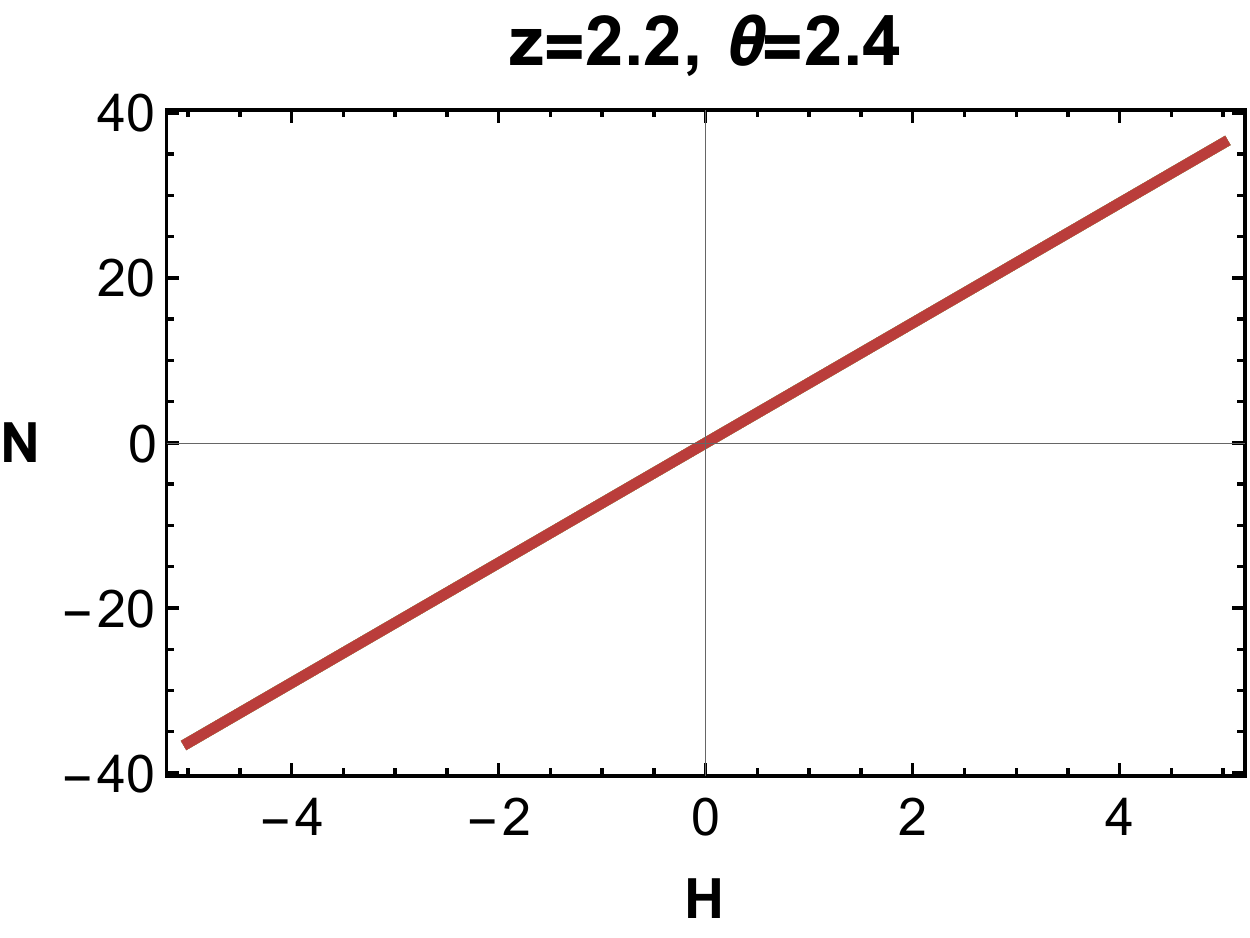} }
    \subfigure[$P_{D}$ ]
   {\includegraphics[width=37mm]{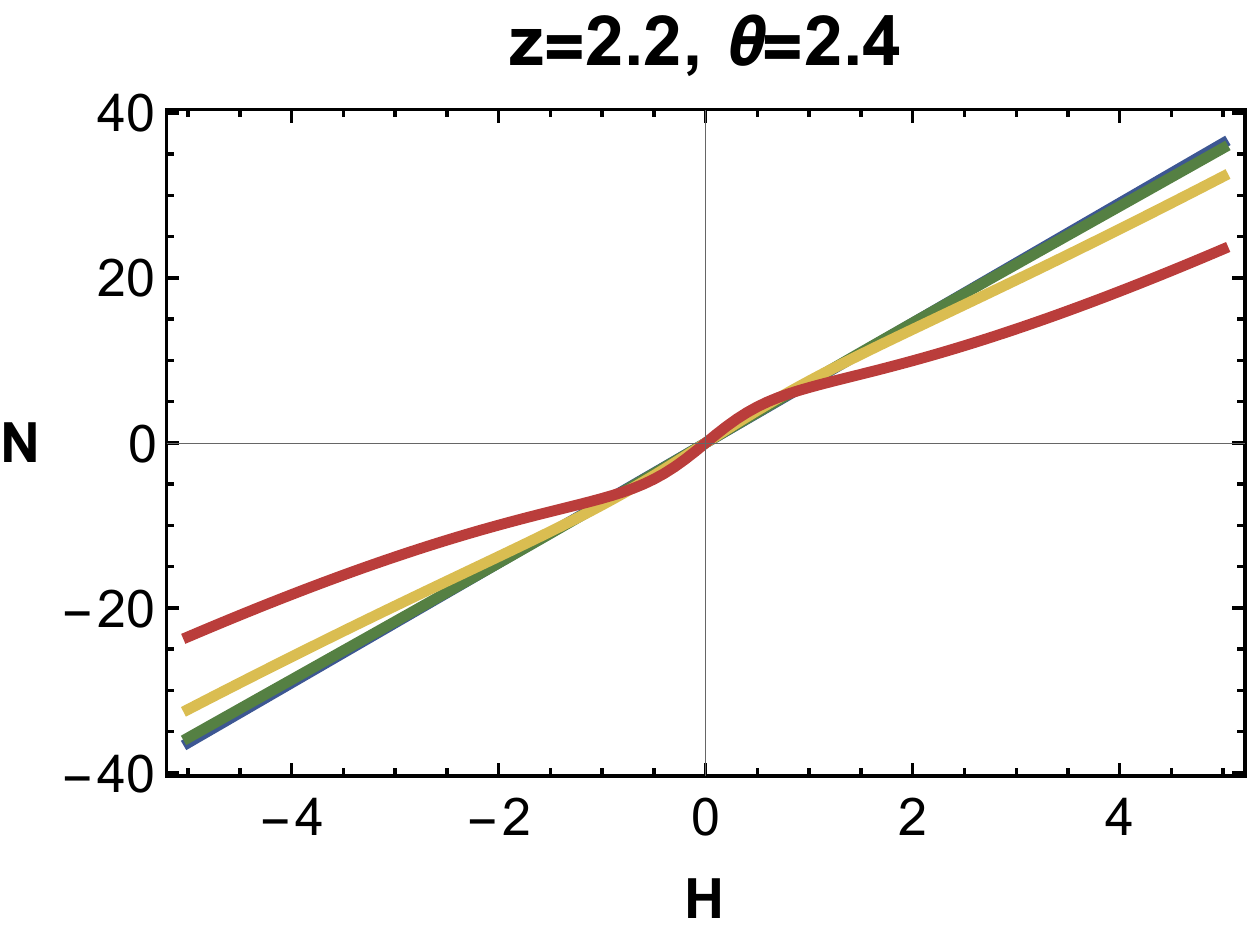} }   
                 \caption{Temperature evolution for $N(H)$ for for each region of $(z,\theta)$. Curves corresponds to $T=0.04,0.1,0.16, 0.24$ for blue, green, yellow, and red respectively. We used $q_{\chi}=0.7$ for all non-zero $q_{\chi}$ case. }    \label{fig:NH} 
\end{figure}

\begin{figure}[ht!]
\centering
 \subfigure[$P_{0A}$   ]
   {\includegraphics[width=37mm]{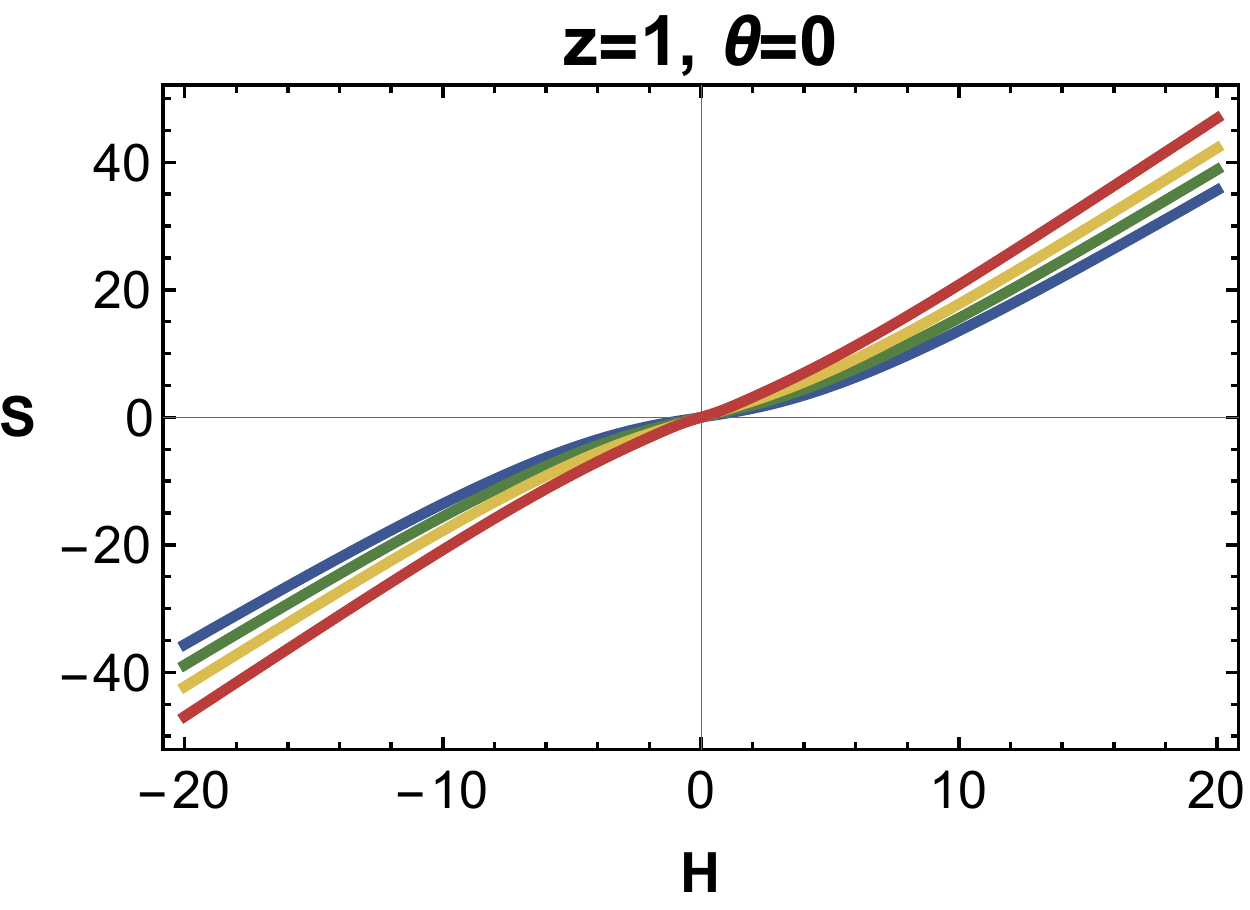} }
     \subfigure[$P_{A}$  ]
   {\includegraphics[width=37mm]{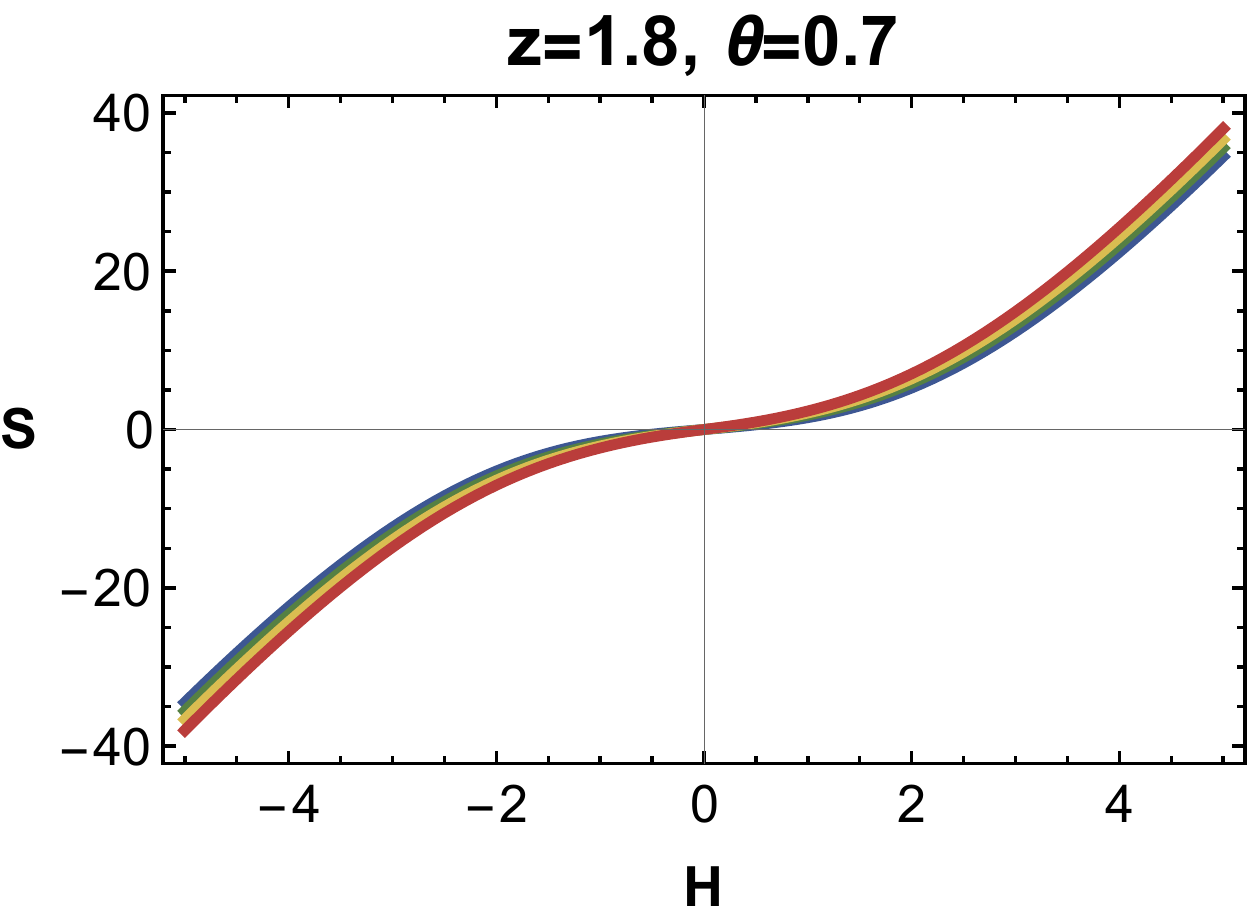} }
      \subfigure[$P_{C}$ ]
   {\includegraphics[width=37mm]{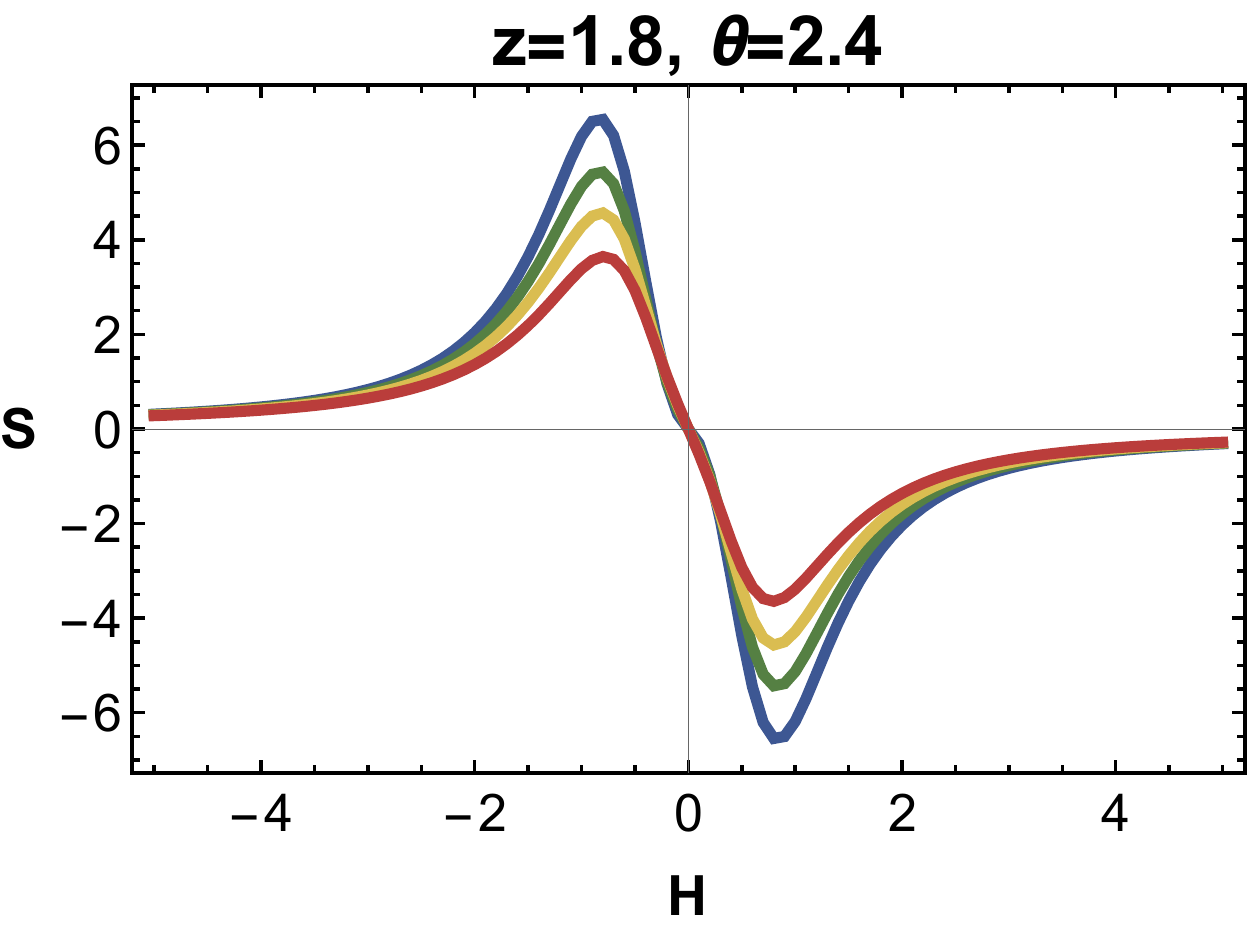} }
    \subfigure[$P_{0B}$ ]
   {\includegraphics[width=37mm]{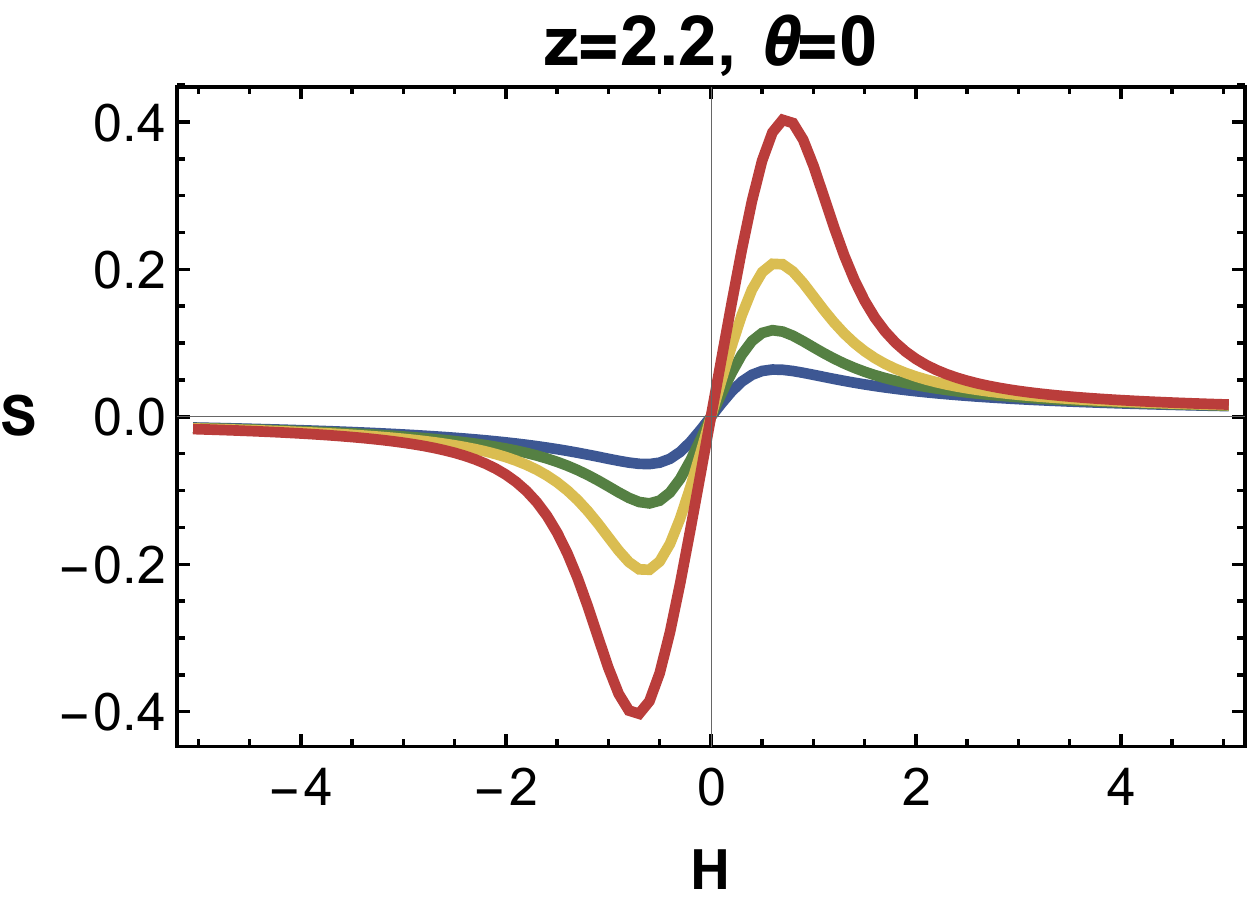} }
      \subfigure[$P_{B}$  ]
   {\includegraphics[width=37mm]{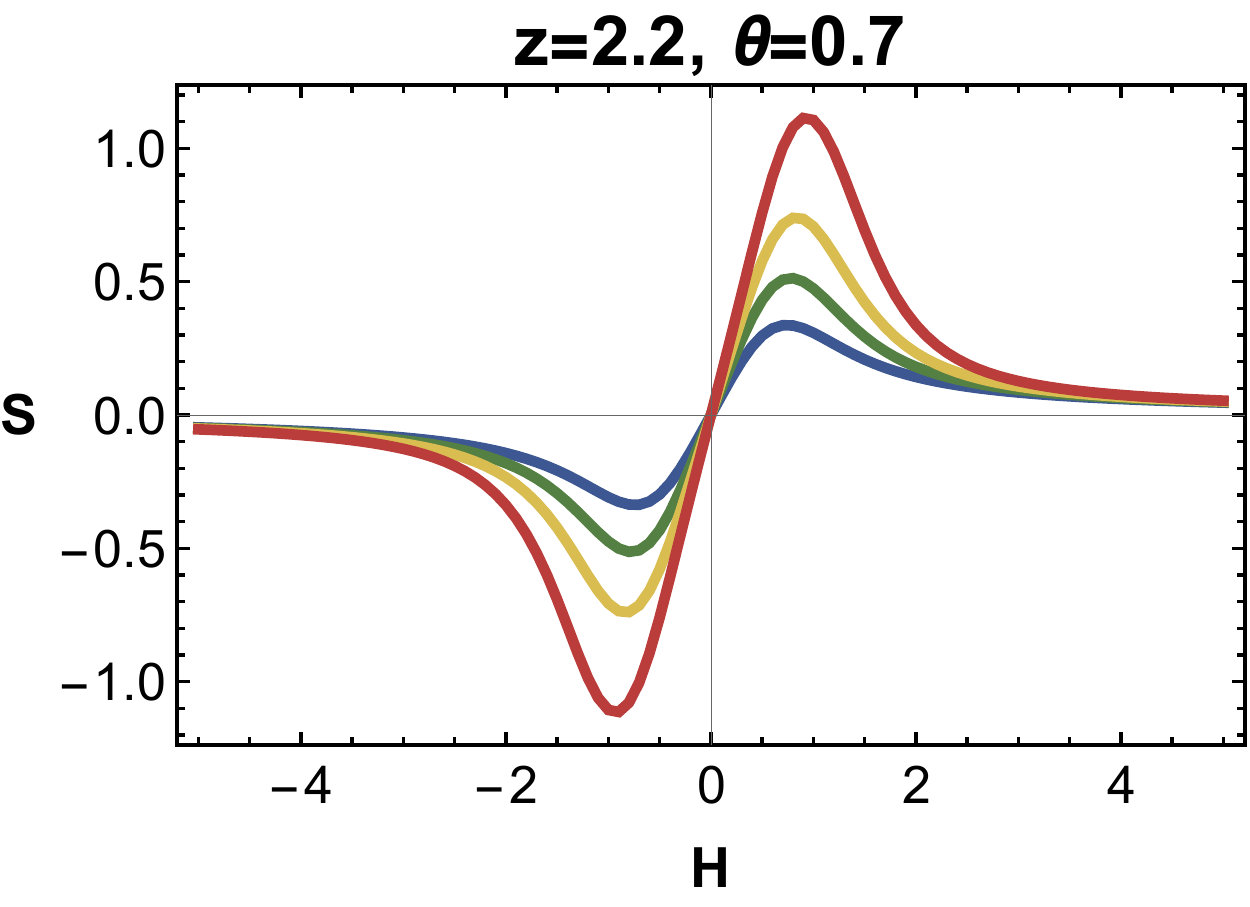} }
      \subfigure[$P_{D}$ ]
   {\includegraphics[width=37mm]{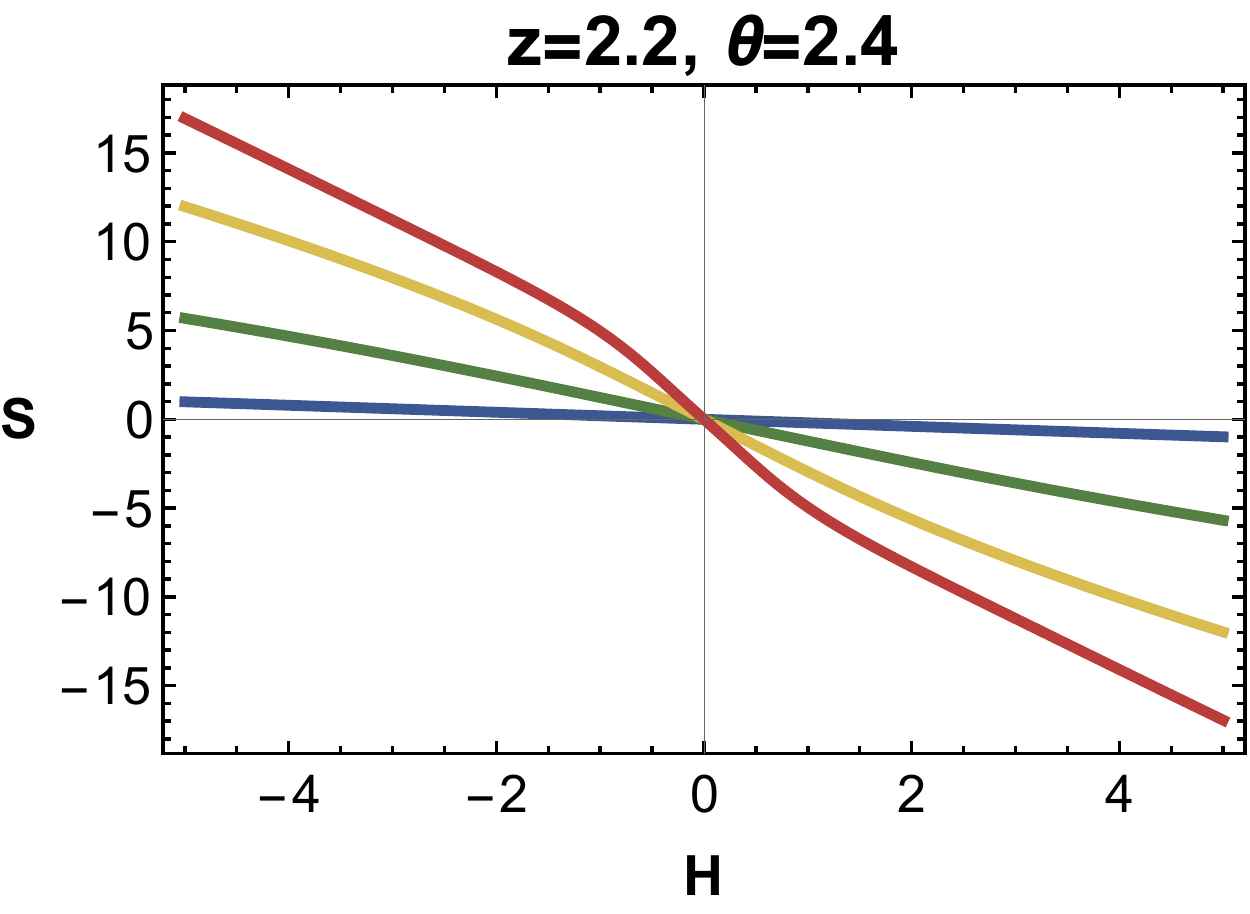} }   
                 \caption{Temperature evolution for $S(H)$ for each region of $(z,\theta)$. Each curves corresponds to $T=0.04,0.1,0.16, 0.24$ for blue, green, yellow, and red respectively.  In all figure we used  $q_{\chi}=0.7$.}     \label{fig:SH} 
\end{figure}

\vskip0.1cm
Figure \ref{fig:NH} and Figure \ref{fig:SH} show the external magnetic field dependence of the Seebeck coefficient and Nernst signal which are related to the thermoelectric conductivity $\alpha_{ij}$. These quantities have different behavior from the other transport coefficients.  They are odd function of the magnetic field. The small field expansion of the Seebeck coefficient and Nernst signal become
\begin{align}
N &\sim \frac{4\pi r_0^{2z-\theta -2}}{\beta^2} \left[ \frac{(2+z-2\theta)r_0^{4z}+q_{\chi}^2\lambda^4 (1+z-2\theta)r_0^{4\theta} }{(2+z-2\theta)(r_0^{4z}+q_{\chi}^2\lambda^4 r_0^{4\theta})}\right] H +\cdots , \cr
S &\sim \frac{4\pi}{\beta^2} \, \frac{q_{\chi} \lambda^2 r_0^{4z+\theta -2}}{(2+z-2\theta)(r_0^{4z} + q_{\chi}^2 \lambda^4 r_0^{4\theta})} H + \cdots,
\end{align}
which are enough to explain the linearity of $N$ and $S$ 
near $H=0$ in the figures \ref{fig:NH} and \ref{fig:SH}. In the absence of $q_\chi$, the slop of Nernst signal is alway positive, but it can be negative depending on the value of $q_{\chi}$. The coefficient of $H$ is same as the coefficient of $H^2$ of the transverse thermal conductivity (\ref{eq:kxyH2}) with opposite sign. Therefore, the region where the sign changes should be the same as $\kappa_{xy}$. See Figure \ref{fig:SignKN} (b). 

The sign of the slope of Seebeck coefficient  only depends on $(2+z-2\theta)$ in the denominator. Together with NEC, the sign is positive in the region $A$, $B$ and negative in $C$, $D$.

In the figure \ref{fig:NH}, we draw  
the temperature evolution for $N(H)$ for different $(z,\theta)$ which are denoted as big  dots  in the figure \ref{phasetransport}(b). 
In the figure \ref{fig:SH}, we draw  
the temperature evolution for $S(H)$ for different $(z,\theta)$ which are denoted as big  dots  in the figure \ref{phasetransport}(b).

\section{  Density dependence of Transports}
In this section, we discuss  the $q_{2}$, density,  dependence of transport coefficients. For the ease of the analysis 
we consider only only zero   external magnetic field cases. 
 Notice that $\rho_{yx}$, $\kappa_{xy}$, and $N$ vanish when $q_{\chi}=0$ and $H=0$. 

\begin{figure}[ht!]
\centering
 \subfigure[ ]
   {\includegraphics[width=45mm]{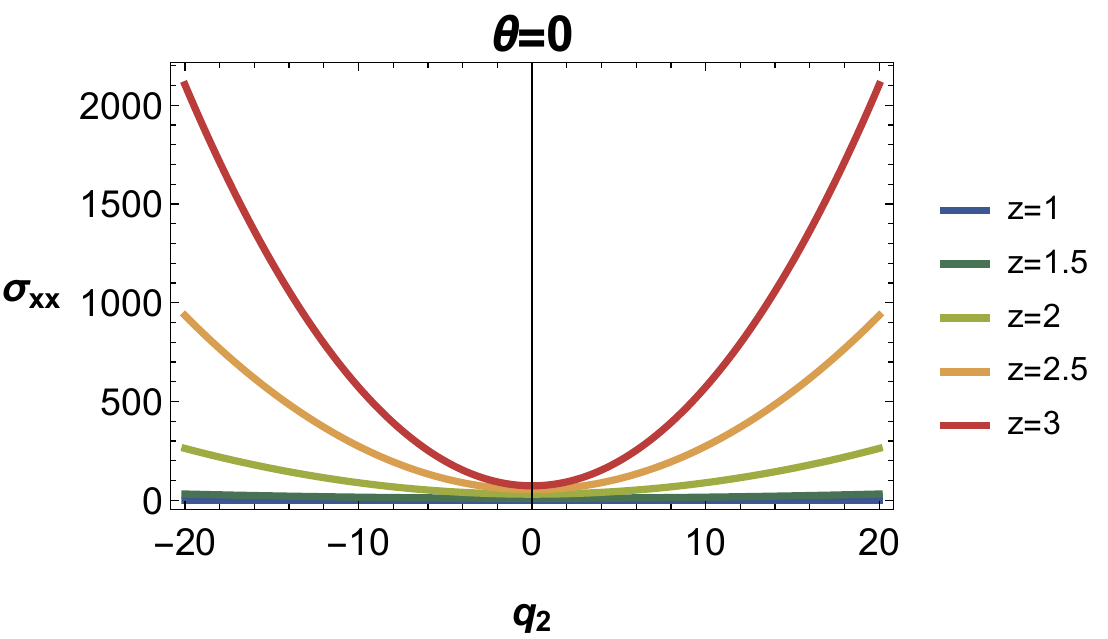} }
     \subfigure[]
   {\includegraphics[width=45mm]{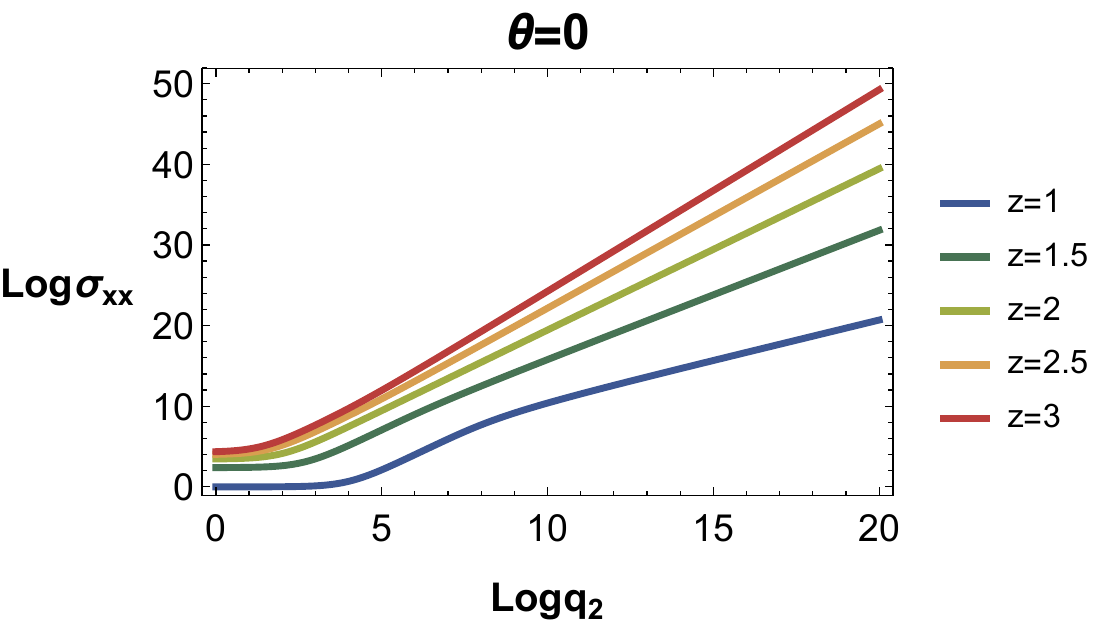} }
     \subfigure[ ]
   {\includegraphics[width=45mm]{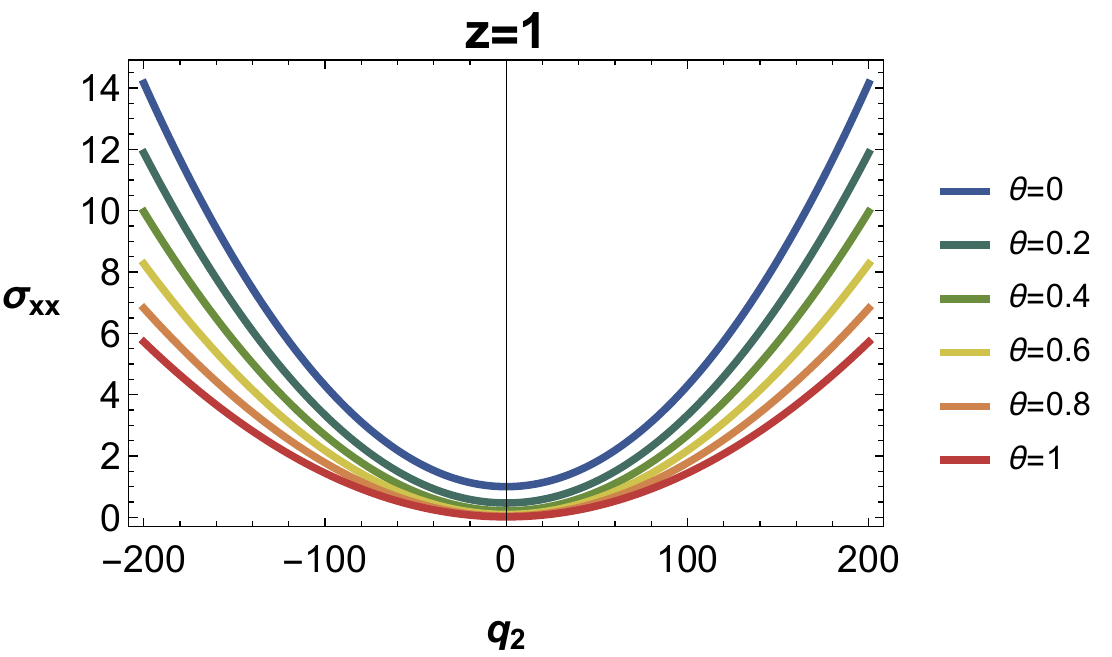} }
      \subfigure[]
   {\includegraphics[width=45mm]{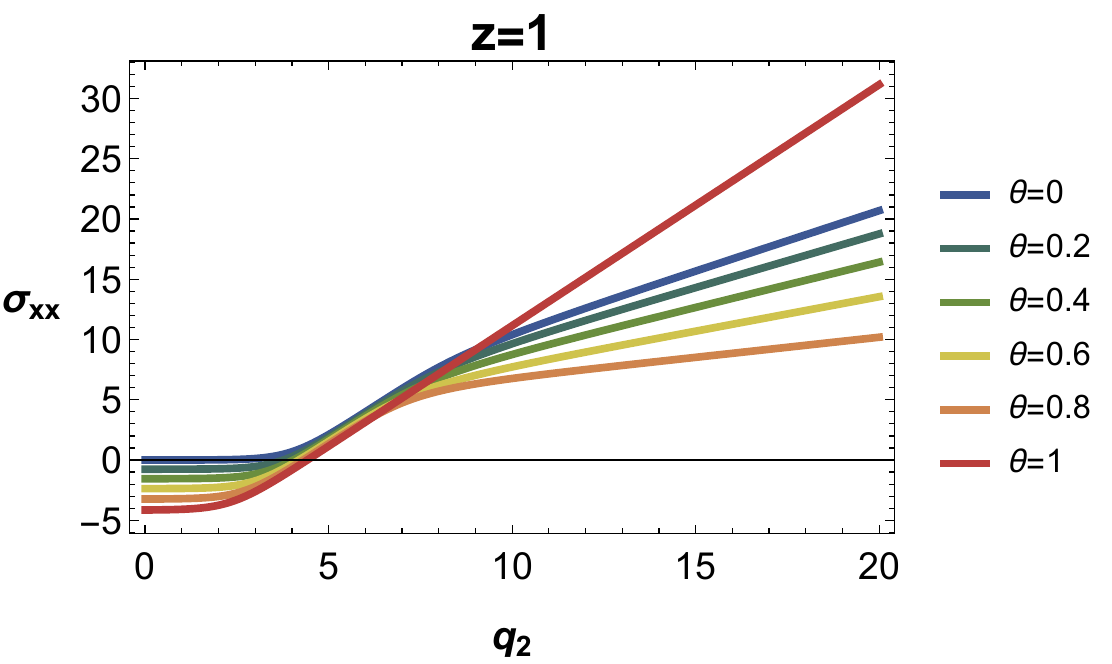} }
      \subfigure[ ]
   {\includegraphics[width=50mm]{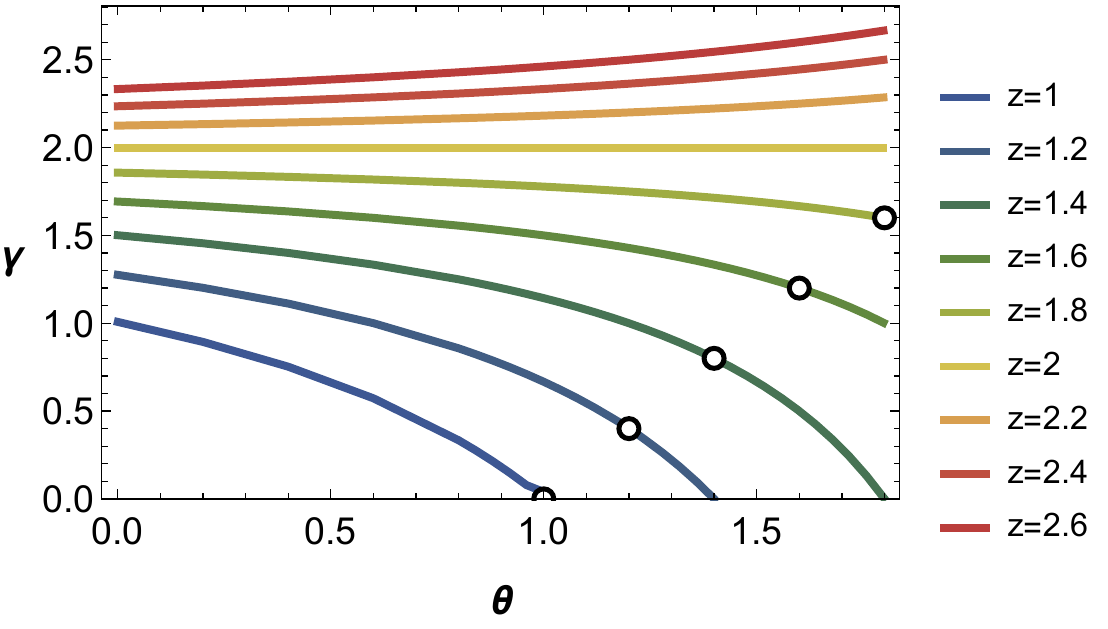} }
      \subfigure[]
   {\includegraphics[width=50mm]{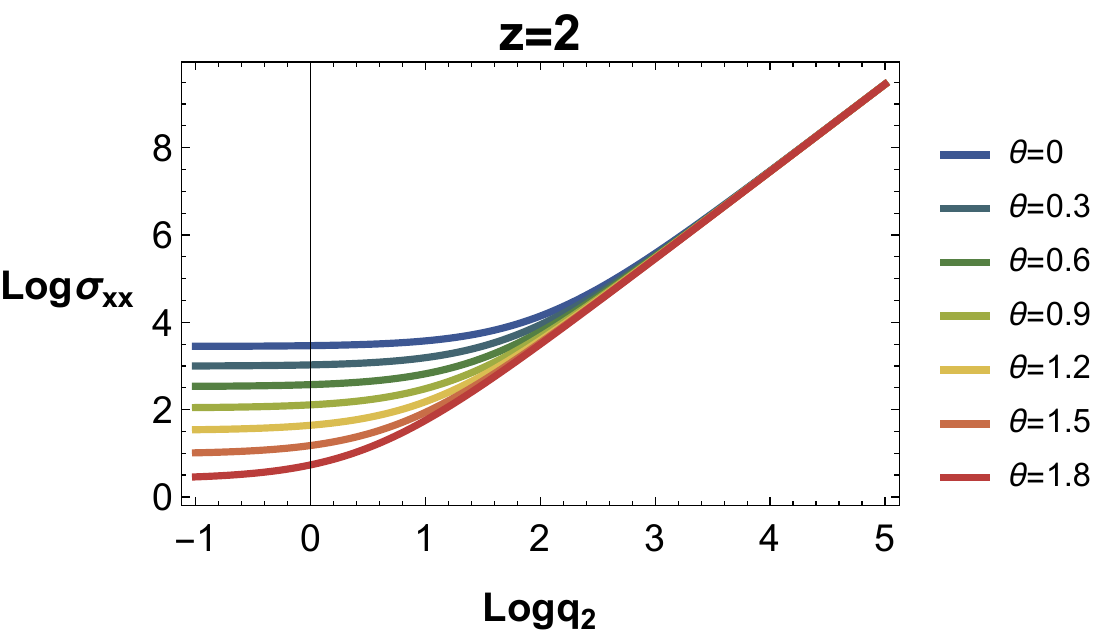} }   
                 \caption{Scaling behavior of $\sigma_{xx}(q_2)$ at large $q_2$ and $T=10$. (a) $z$-evolution of $\sigma_{xx}(q_2)$ at $\theta=0$.  (b) log-log plot of (a), (c) $\theta$-evolution of $\sigma_{xx}(q_2)$ at $z=1$. (d) log-log plot of (c),  (e) The exponent $\gamma$ of $\sigma_{xx}(q_2)|_{q_2\rightarrow \infty} \sim q_2^{\gamma}$ for various $(z,\theta)$. Whenever $z=\theta$, the value of $\gamma$ is discontinuous, which is indicated by empty circles in the figure. At all the empty circle points, $\gamma=2$ which is discontinuous value from its neighboring point. Here, we analyzed this only in Phase A and B. (f) When $z=2$, $\gamma=2$ independent of $\theta$, which is indicated by yellow line in  (e)  }    \label{fig:scbehavior} 
\end{figure}

In the regions A and B, the longitudinal electric conductivity has a scaling behavior for large density, namely,  
$\gamma=\partial (\log\sigma_{xx})/\partial (\log q_2)$ is a constant at $q_2\rightarrow\infty$.  
To see this, we consider 
the temperature  at $H=0$: 
\begin{align}
	4\pi T =r_H^z(2+z-\theta)+\frac{q_2^2r_H^{-z-2+2\theta}(z-\theta)^2}{2(\theta-2)}+ \frac{r_H^{\theta -z} \beta^2}{\theta -2}.
\end{align}
In regions A and B satisfying NEC, $\theta<2z$ so that we can neglect the third term for large $q_2$. 
For the regions A and B where $\theta<2$,  we have 
$r_H\sim q_2^{\frac{1}{z-\theta+1}}$. Rewriting the longitudinal conductivity from eq (\ref{transp}) with this horizon behavior for large density,
\begin{align}
	\sigma_{xx}=r_H^{2z-4}\left(r_H^{2-\theta}+\frac{q_2^2}{\beta^2}\right)\sim q_2^{\gamma} \qquad \text{with} \qquad \gamma=\frac{2(2z-\theta-1)}{z-\theta+1} .     
\label{slargeq}
\end{align}
Notice that the first term in (\ref{slargeq}) is subleading so the   behavior is governed by the explicit $q^2$ dependent term.
Fig. \ref{fig:scbehavior} demonstrate the scaling behavior of $\sigma_{xx}(q_2)$. Notice that when $z=\theta$, $r_0$ has no dependence on $q_2$  which means $\gamma$ is fixed as $2$. 

 \begin{figure}[ht!]
\centering 
	 \subfigure[$P_{0A}$ at $q_{\chi}=0$]
   {\includegraphics[width=30mm]{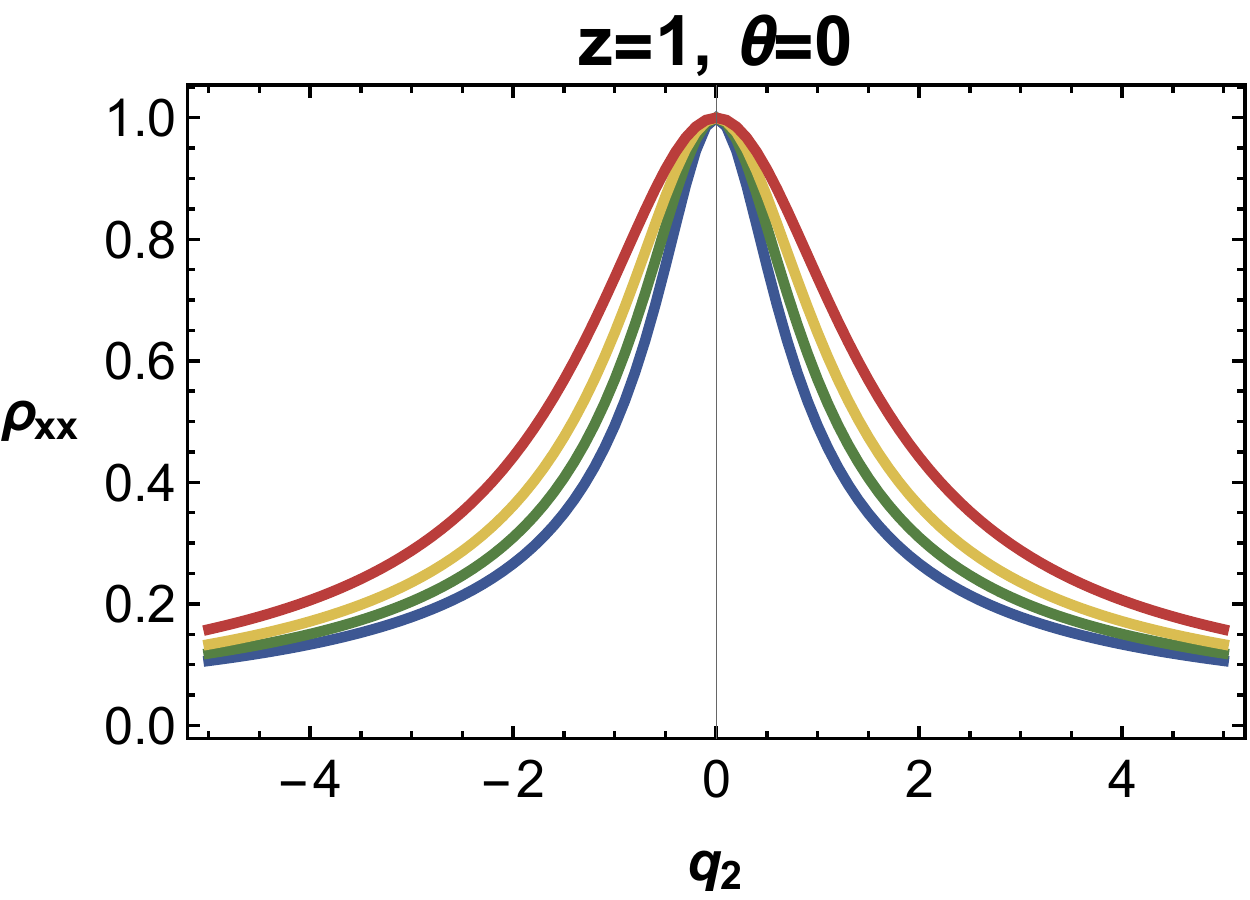} }
    \subfigure[$P_{0A}$  at $q_{\chi}=5$ ]
   {\includegraphics[width=37mm]{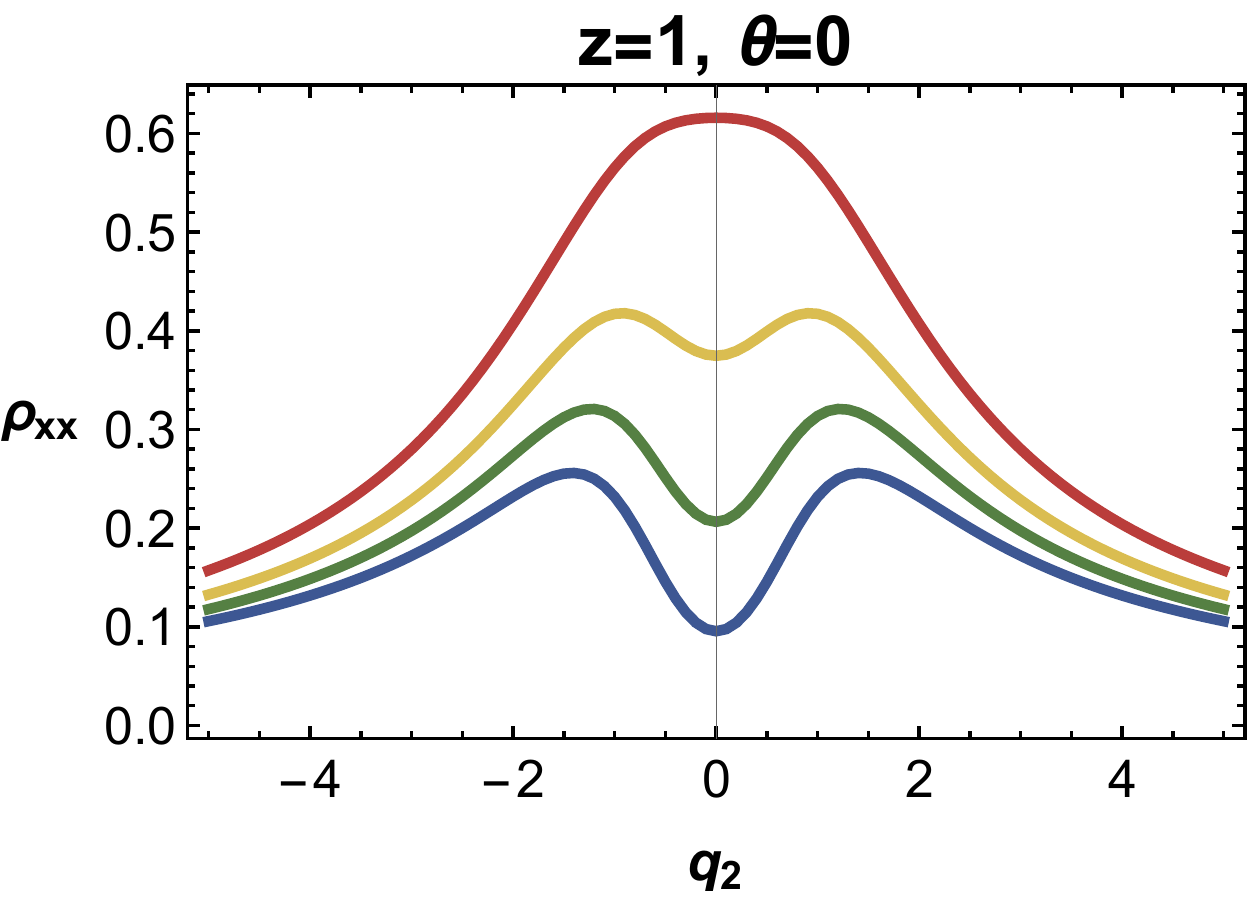} }
    \subfigure[$P_{0B}$ at $q_{\chi}=0$]
   {\includegraphics[width=30mm]{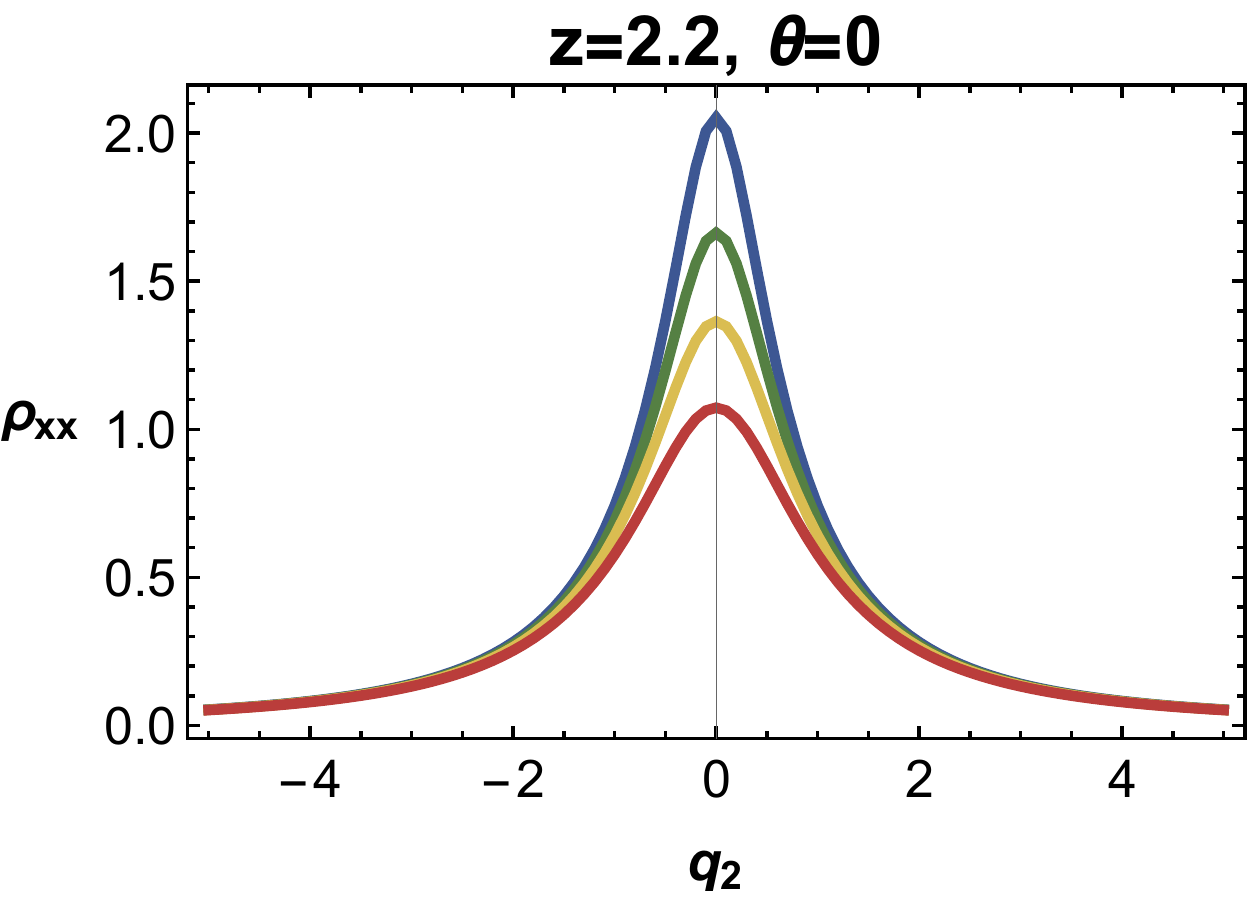} }
    \subfigure[$P_{0B}$ ]
   {\includegraphics[width=37mm]{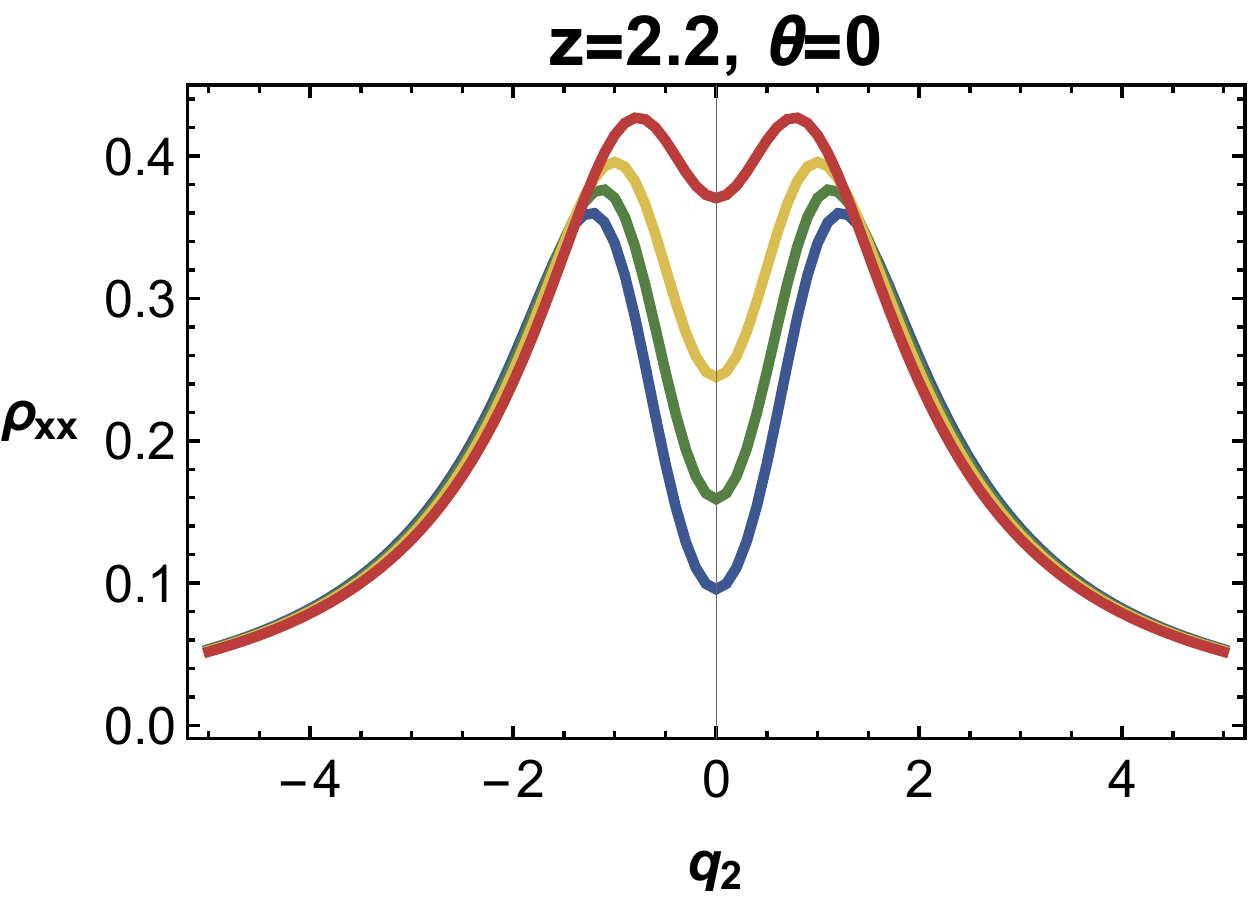} }
    \subfigure[$P_{A}$ at $q_{\chi}=0$]
   {\includegraphics[width=30mm]{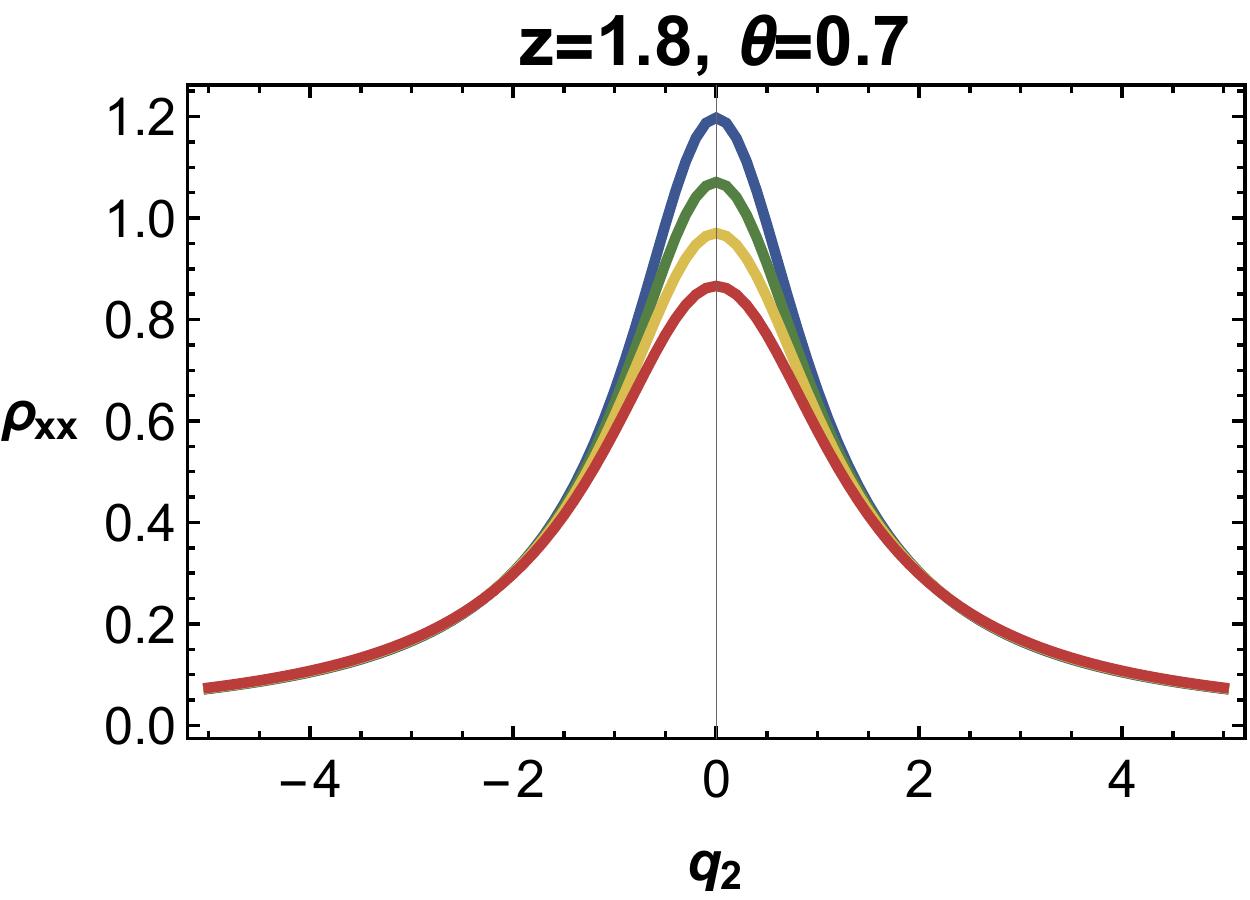} }
    \subfigure[$P_{A}$  ]
   {\includegraphics[width=37mm]{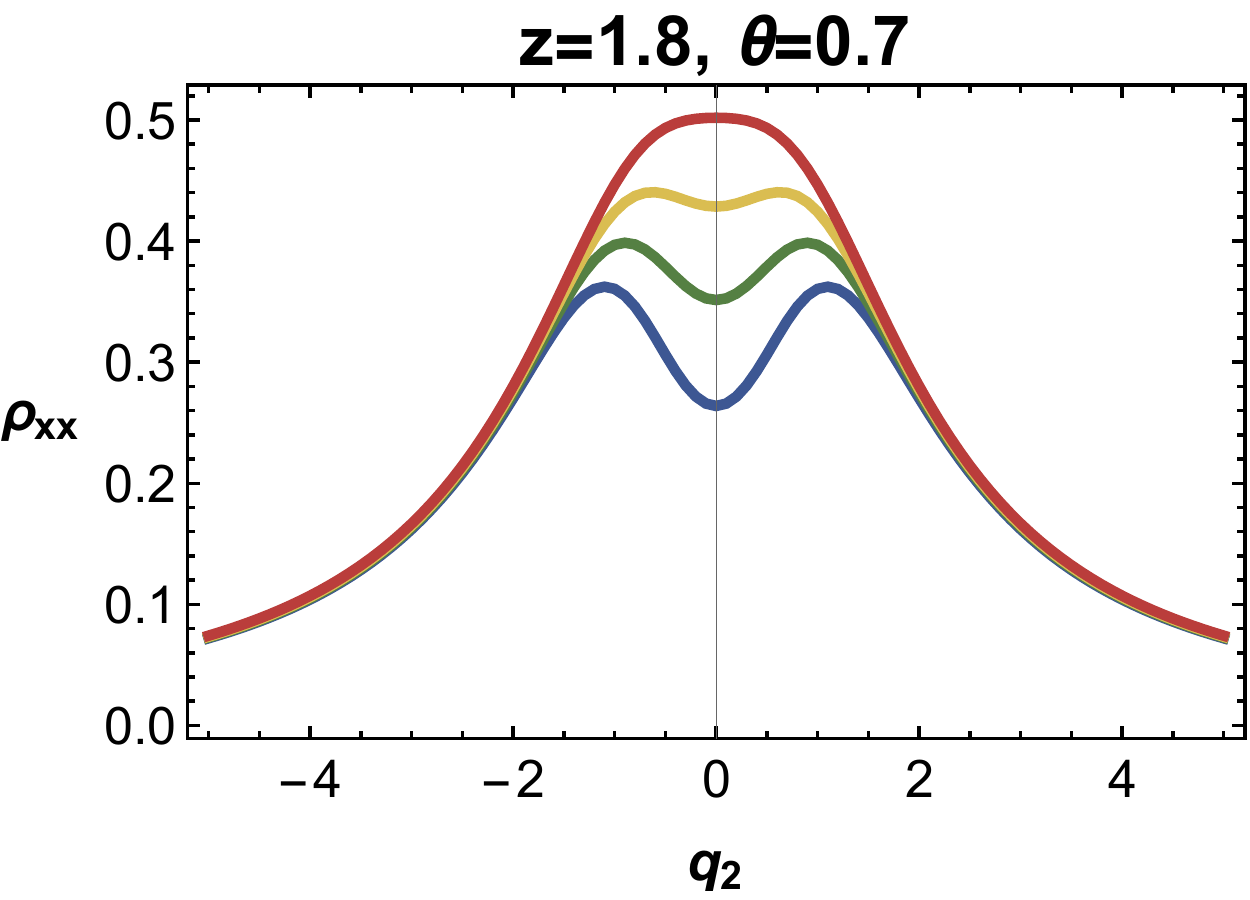} }
   \subfigure[$P_{B}$ at $q_{\chi}=0$]
   {\includegraphics[width=30mm]{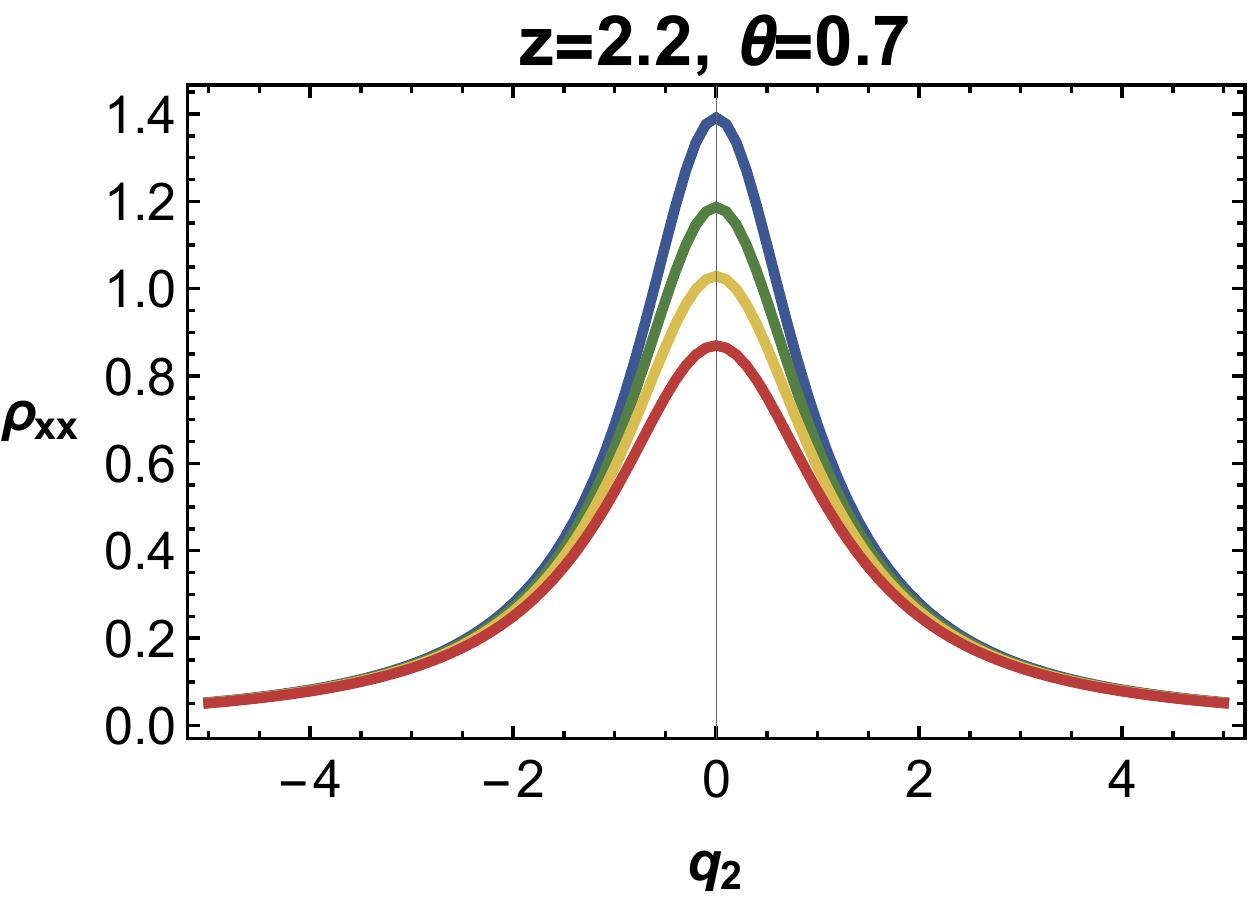} }
    \subfigure[$P_{B}$  ]
   {\includegraphics[width=37mm]{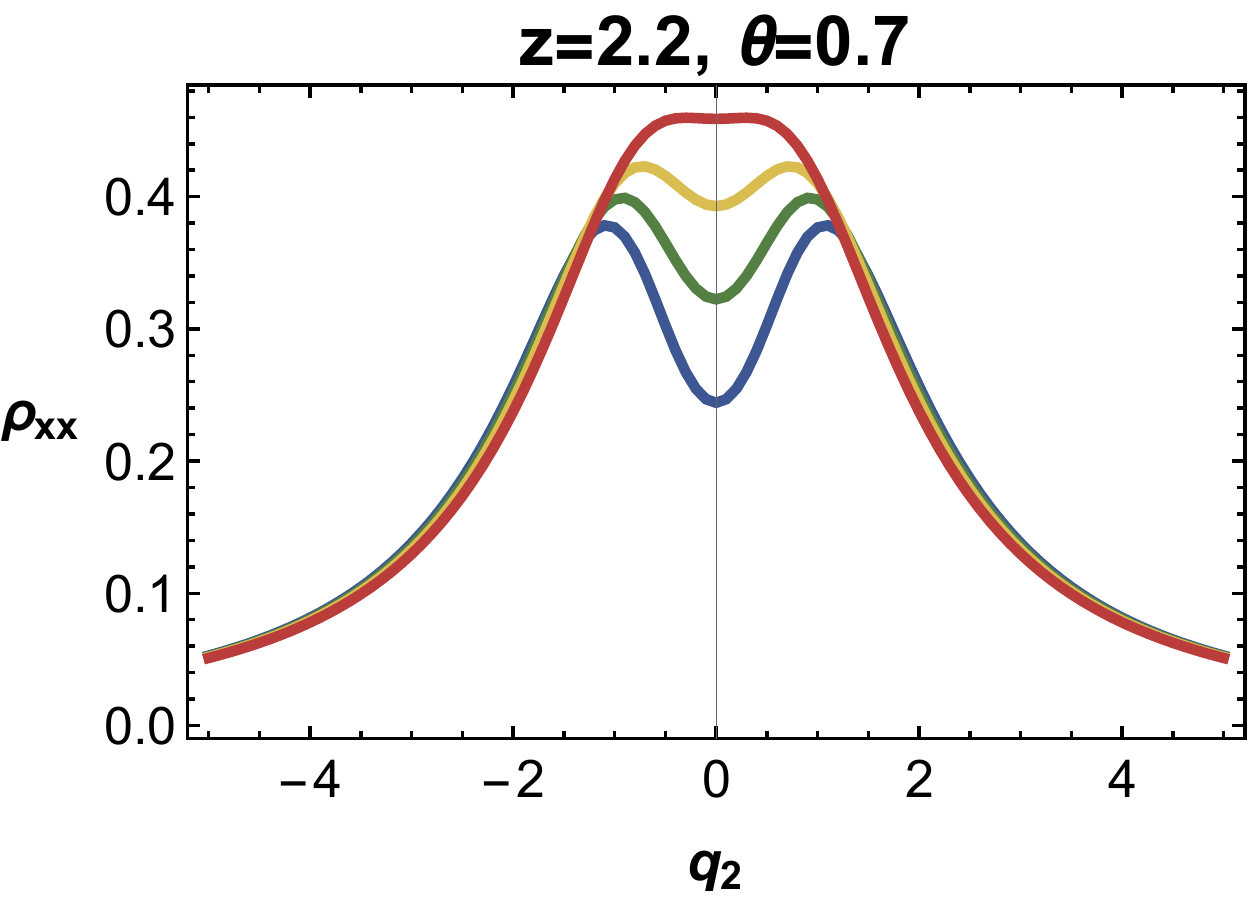} }
    \subfigure[$P_{C}$ at $q_{\chi}=0$]
   {\includegraphics[width=30mm]{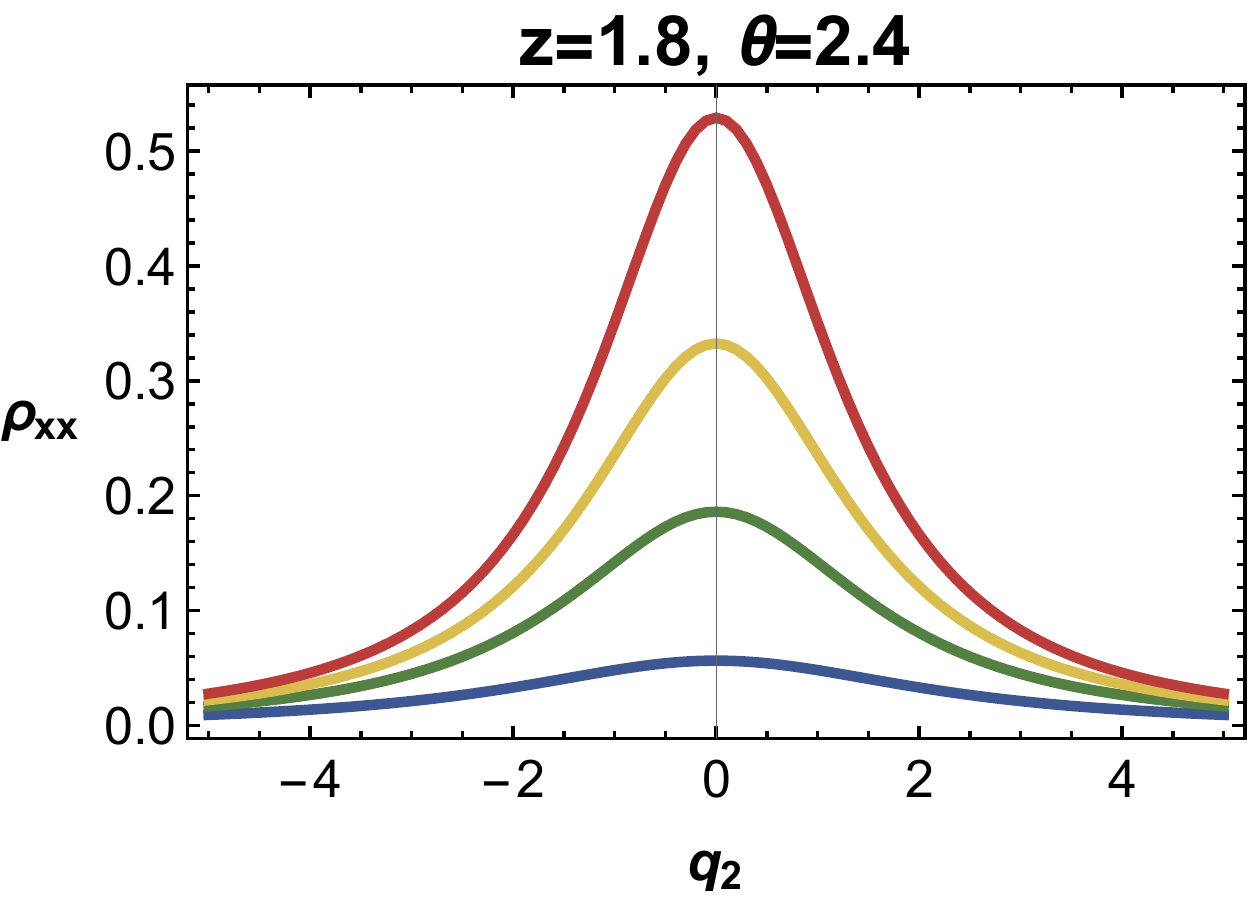} }
    \subfigure[$P_{C}$ ]
   {\includegraphics[width=37mm]{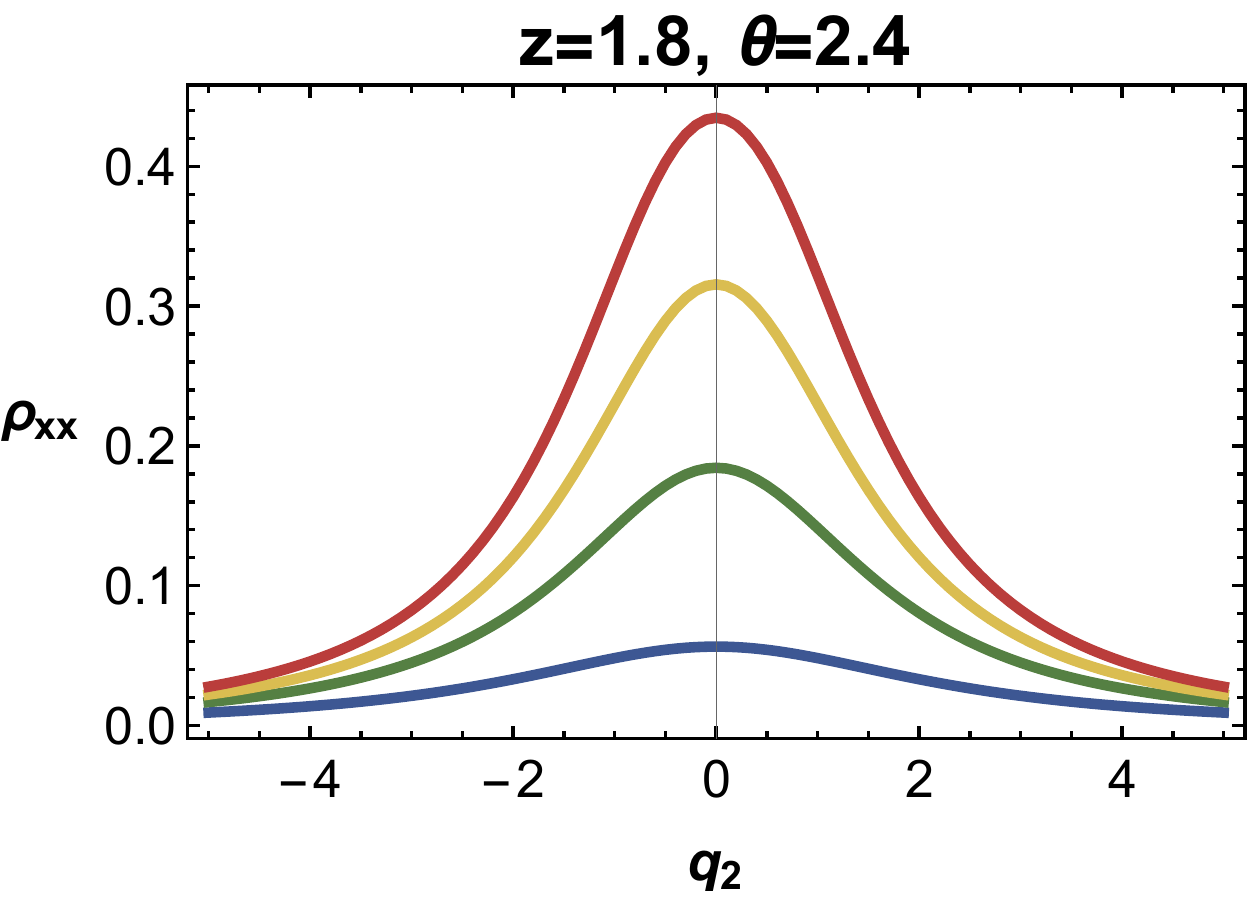} }
        \subfigure[$P_{D}$ at $q_{\chi}=0$]
   {\includegraphics[width=30mm]{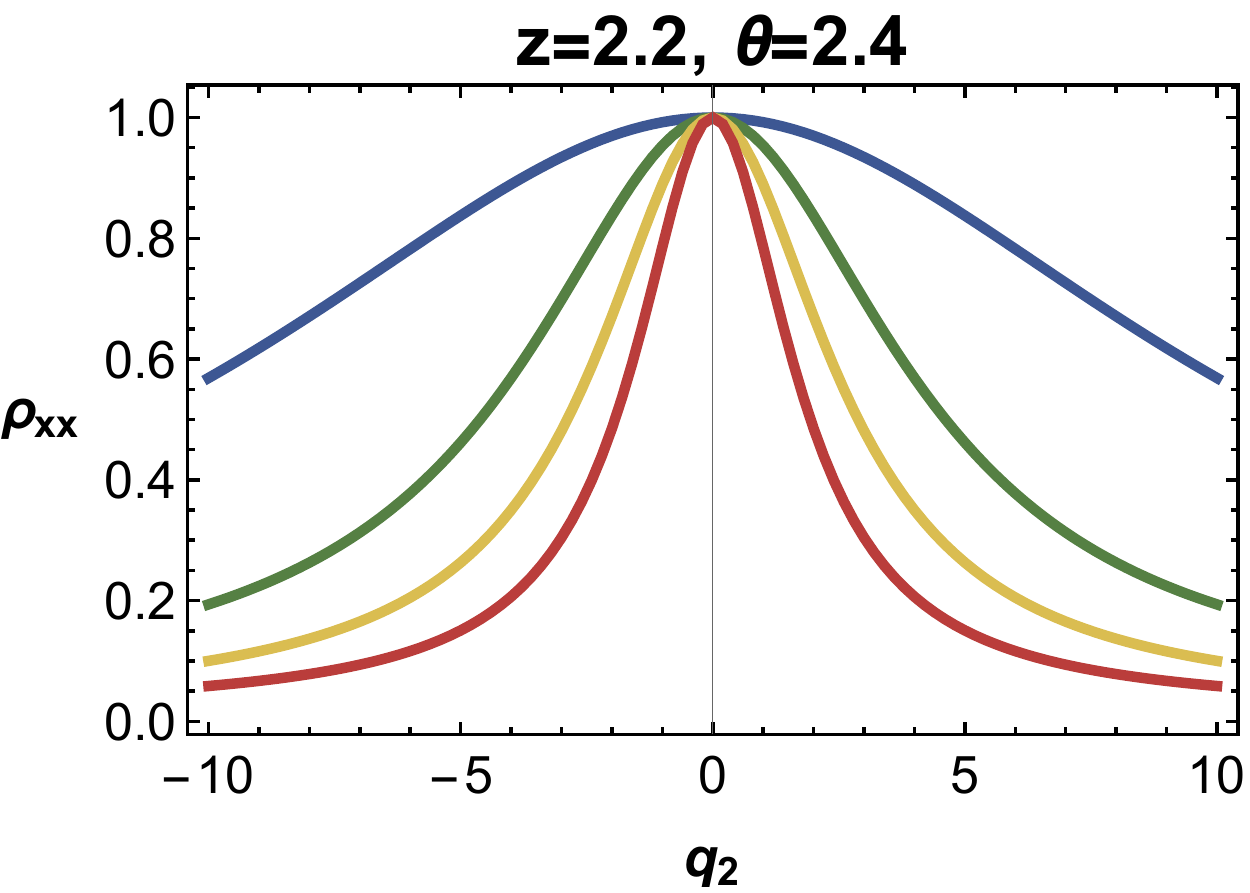} }
    \subfigure[$P_{D}$ ]
   {\includegraphics[width=37mm]{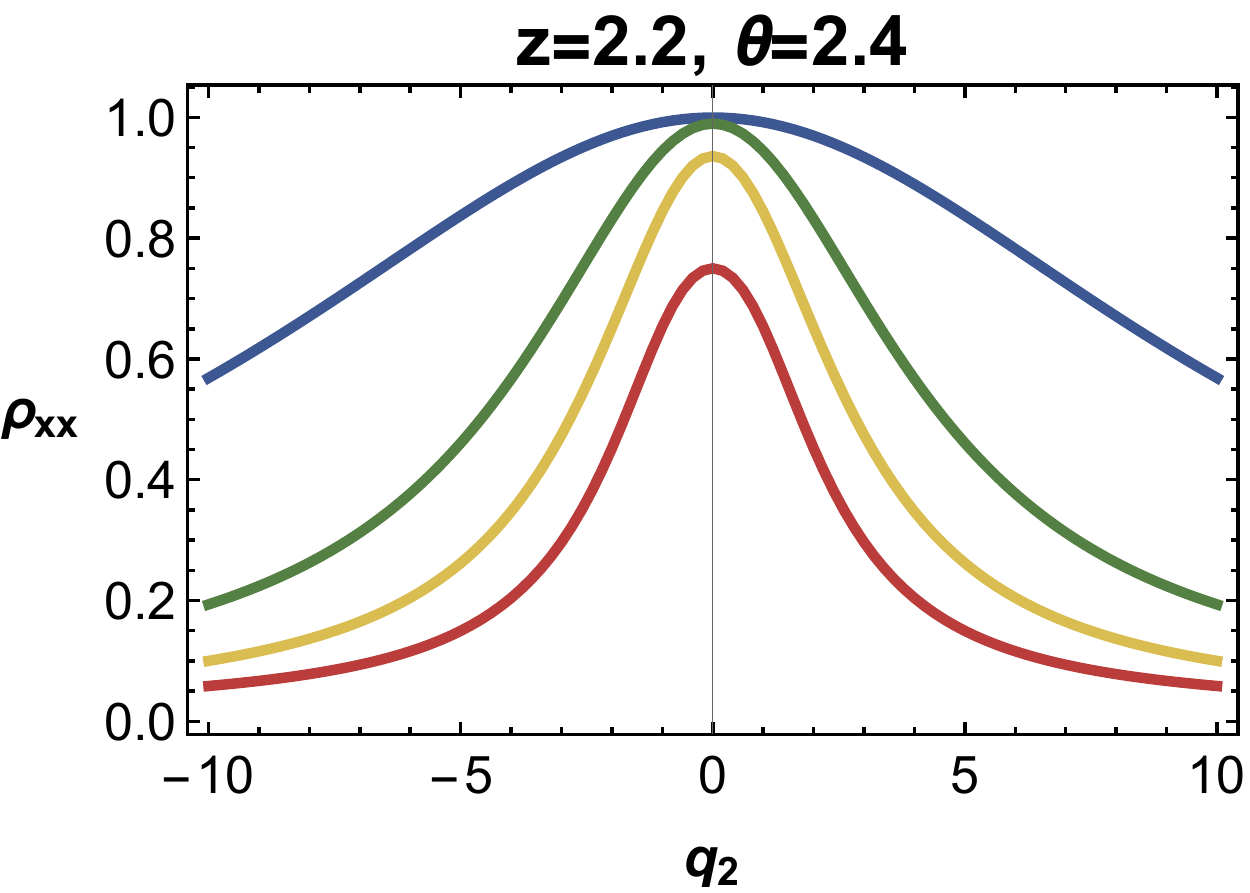} }   
                 \caption{Temperature evolution for $\rho_{xx}(q_2)$ for various $(z,\theta)$.   Curves correspond to $T=0.04,0.1,0.16, 0.24$ for blue, green, yellow, and red respectively. In all figure, $q_{\chi}=5$ unless it is specified.}   \label{fig:rxx} 
\end{figure}
\begin{figure}
\centering
\subfigure[$q_{\chi}=0$]
   {\includegraphics[width=45mm]{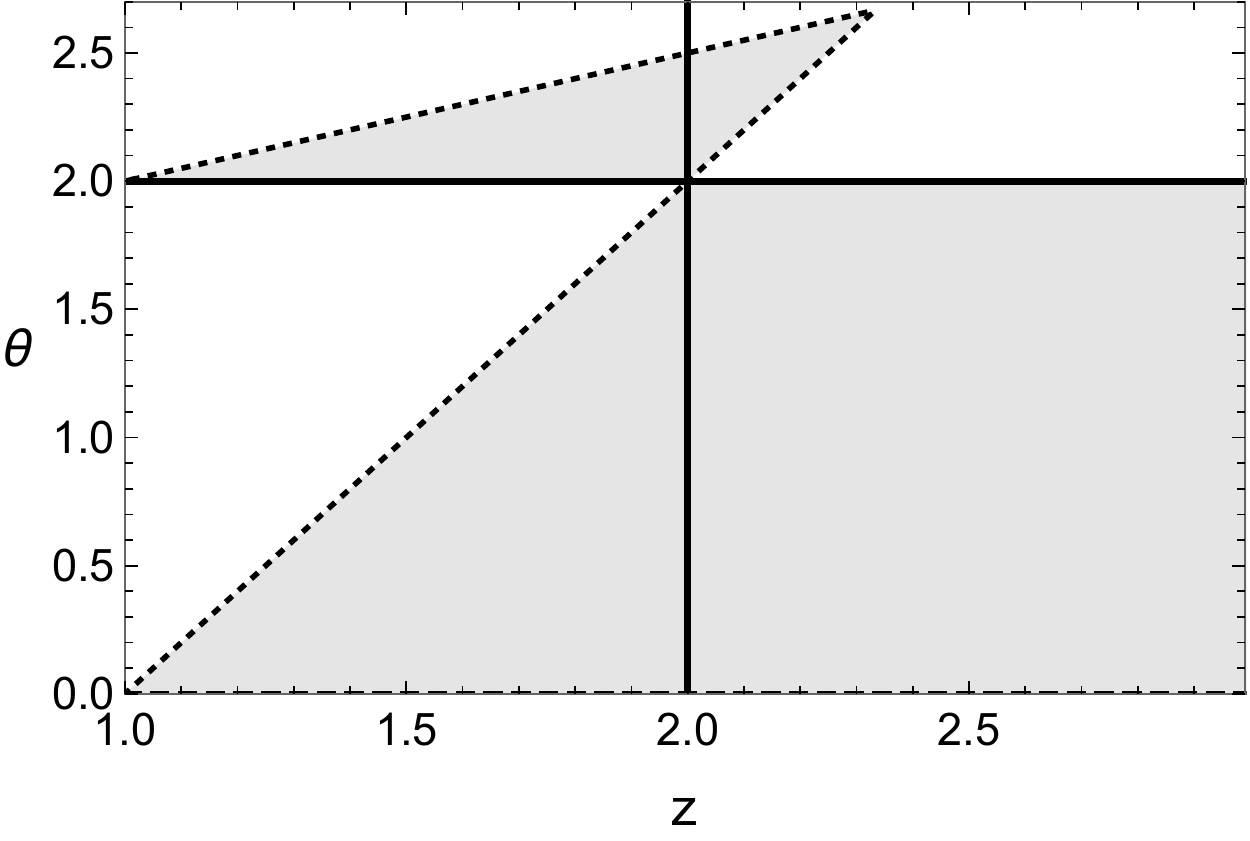} }    
\hskip.2cm
\subfigure[$q_{\chi}=0.7$]
   {\includegraphics[width=45mm]{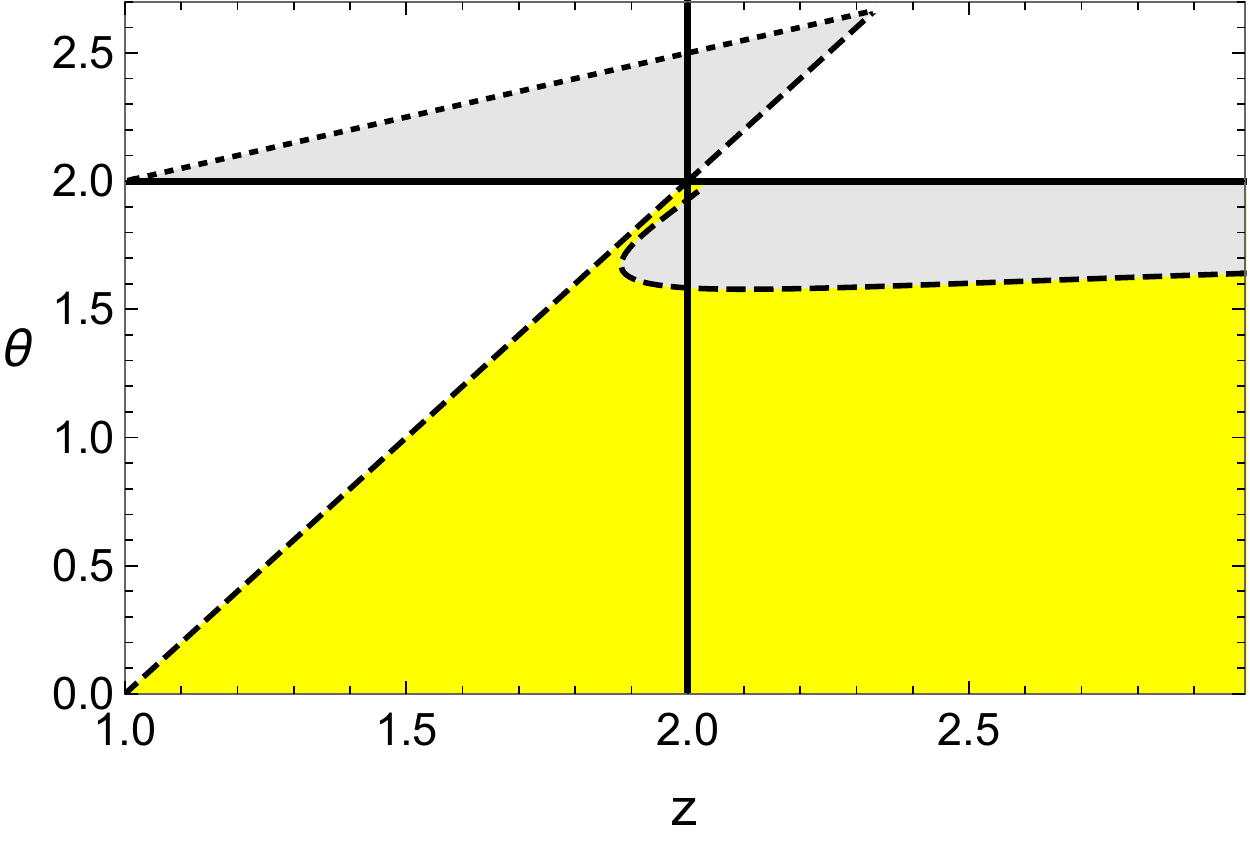} }    
\hskip.2cm
\subfigure[$q_{\chi}=1.4$]
   {\includegraphics[width=45mm]{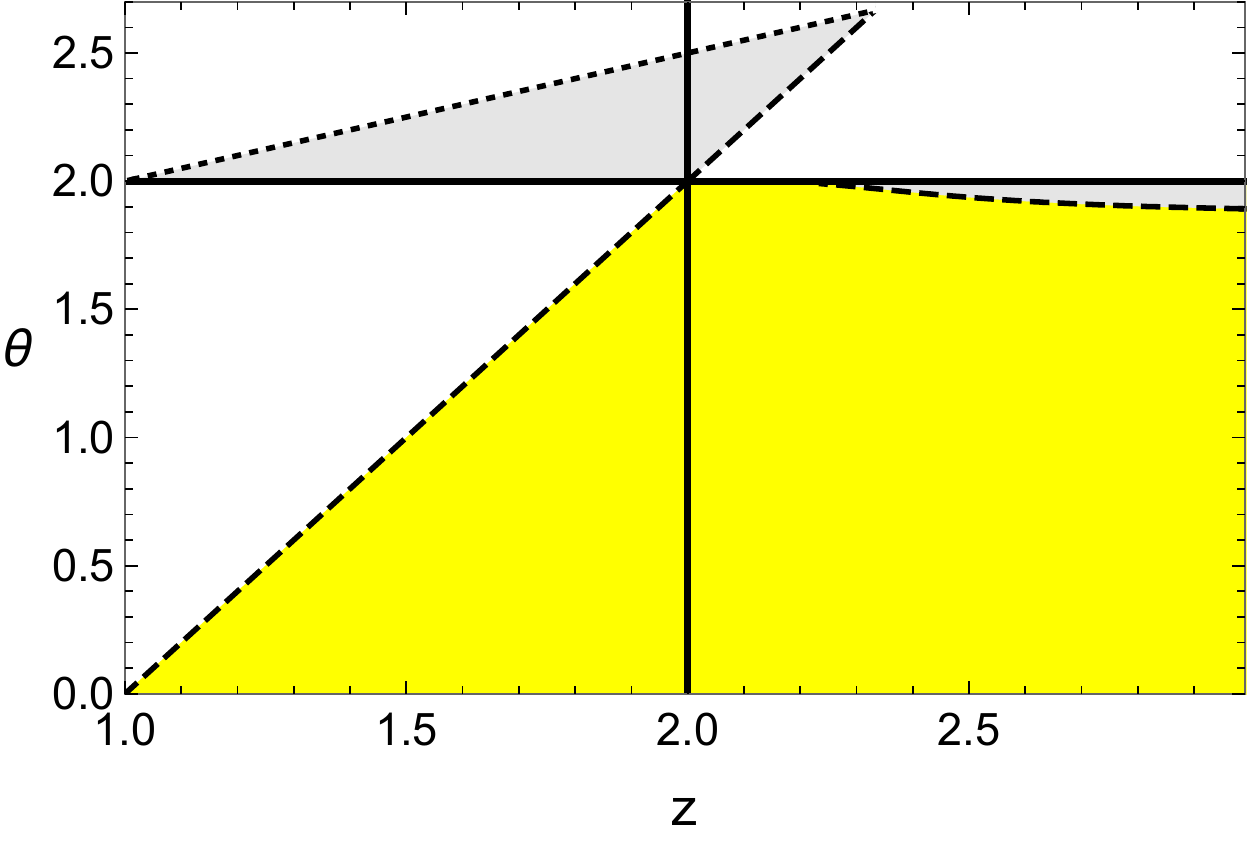} }        
   	\caption{WL vs WAL in the longitudinal resistivity. They are determined by the   sign of the coefficient of $q_2^2$  near $q_2=0$. Yellow region denotes positive sign(WL) and gray region is for negative sign(WAL). Dotted line is NEC. Here we use the same parameters as Figure \ref{fig:rxx} and $T=0.04$.}\label{fig:rxxq}
\end{figure}

 Now we consider the density dependence of the longitudinal resistivity. Figure \ref{fig:rxx} shows the density dependence of $\rho_{xx}(q_2)$.   When $q_2=0$, the longitudinal resistivity is given by
 \begin{align}
 	\rho_{xx}|_{q_2=0}\equiv\rho_{xx}^0=\frac{r_0^{2z+\theta+2}}{r_0^{4z}+q_{\chi}^2\lambda^4r_0^{4\theta}}
 \end{align}
 where $r_0$ is $r_H(q_2=0)$. Different from electric conductivity, the resistivity does not have the scaling behavior in $T$, due to the $q_{\chi}^2$ term in denomenators. 
 
  For the finite density case, the relation between $r_H$ and $T$  is still complicated so we will take similar way to previous section. In the small $q_2$ limits, $q_2$ dependece of $r_H$ for fixed parameters $(T, \beta,\lambda,\theta,z)$ can be obtained from
 \begin{align}\label{rhq2}
 	\frac{\delta r_H}{\delta q_2}\bigg |_{q_2 \ll 1}&\sim -\frac{(z-\theta)^2r_0^{2\theta-1}}{z(z-\theta+2)(\theta-2)r_0^{2z}+(\theta-z)\beta^2r_0^{\theta}}q_2+\mathcal{O}(q_2^4)\nonumber\\	&\equiv \mathcal{A}_2q_2+\mathcal{O}(q_2^4) 	
 \end{align}
 Now, we have small $q_2$ expansion of $\rho_{xx}$ as
 \begin{align}\label{rxxexp}
 	\rho_{xx}|_{q_2\ll 1}&=\rho_{xx}^0-\frac{1}{\mathcal{D}^2}\bigg(\frac{2r_0^{2(3z+\theta)}-2q_{\chi}^2\lambda^4r_0^{2(z+3\theta)}}{\beta^2}\nonumber\\
 	&-\frac{\mathcal{A}_2}{2}r_0^{2z+\theta+1}\left(r_0^{4z}(2-2z+\theta)+q_{\chi}^2\lambda^4r_0^{4\theta}(2+2z-3\theta)\right)\bigg)q_2^2+\cdots
 \end{align}
 where $\mathcal{D}=r_0^{4z}+q_{\chi}^2\lambda^4r_0^{4\theta}$. The coefficient of $q_2^2$ in (\ref{rxxexp}) determines a shape of the resistivity near $q_2=0$. Roughly speaking, $q_{\chi}$ can flip the sign of curvature flipped in  $(z,\theta)$ dependent way, which is difficult to analyze. We calculated the   the sign of the second term numerically and the result is in  Figure \ref{fig:rxxq}. 
 When $q_{\chi}=0$, the sign of coefficients of $q_2^2$ is always negative. In regions A,B, $q_{\chi}$ interaction suppress the resitivity near $q_2=0$, but the effect of $q_{\chi}$ diminishes as $\theta$ increases. In region C,D, the effect of $q_{\chi}$ terms become negligible due to the large $\theta$.
\begin{figure}[ht!]
\centering
 \subfigure[$P_{0A}$  ]
   {\includegraphics[width=45mm]{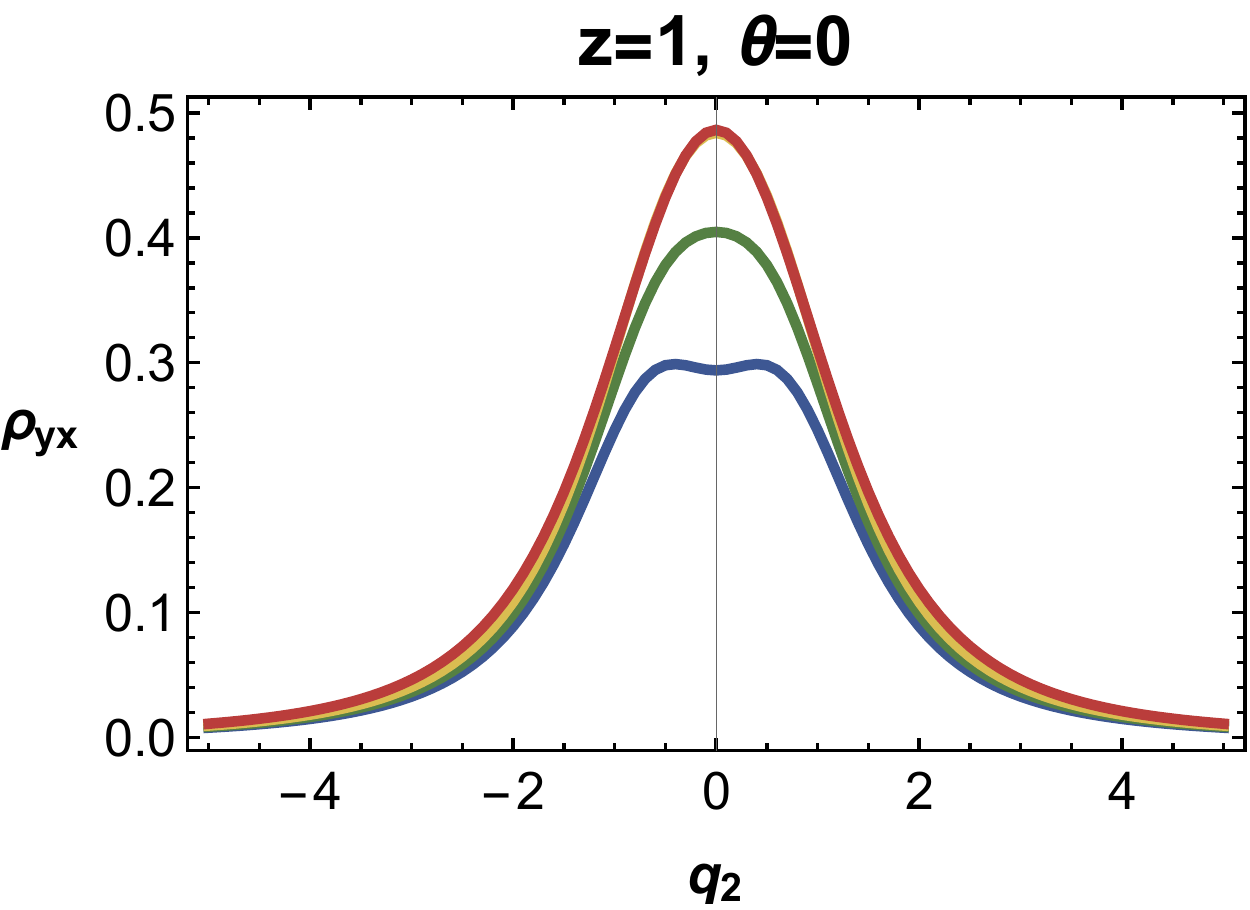} }
     \subfigure[$P_{A}$  ]
   {\includegraphics[width=45mm]{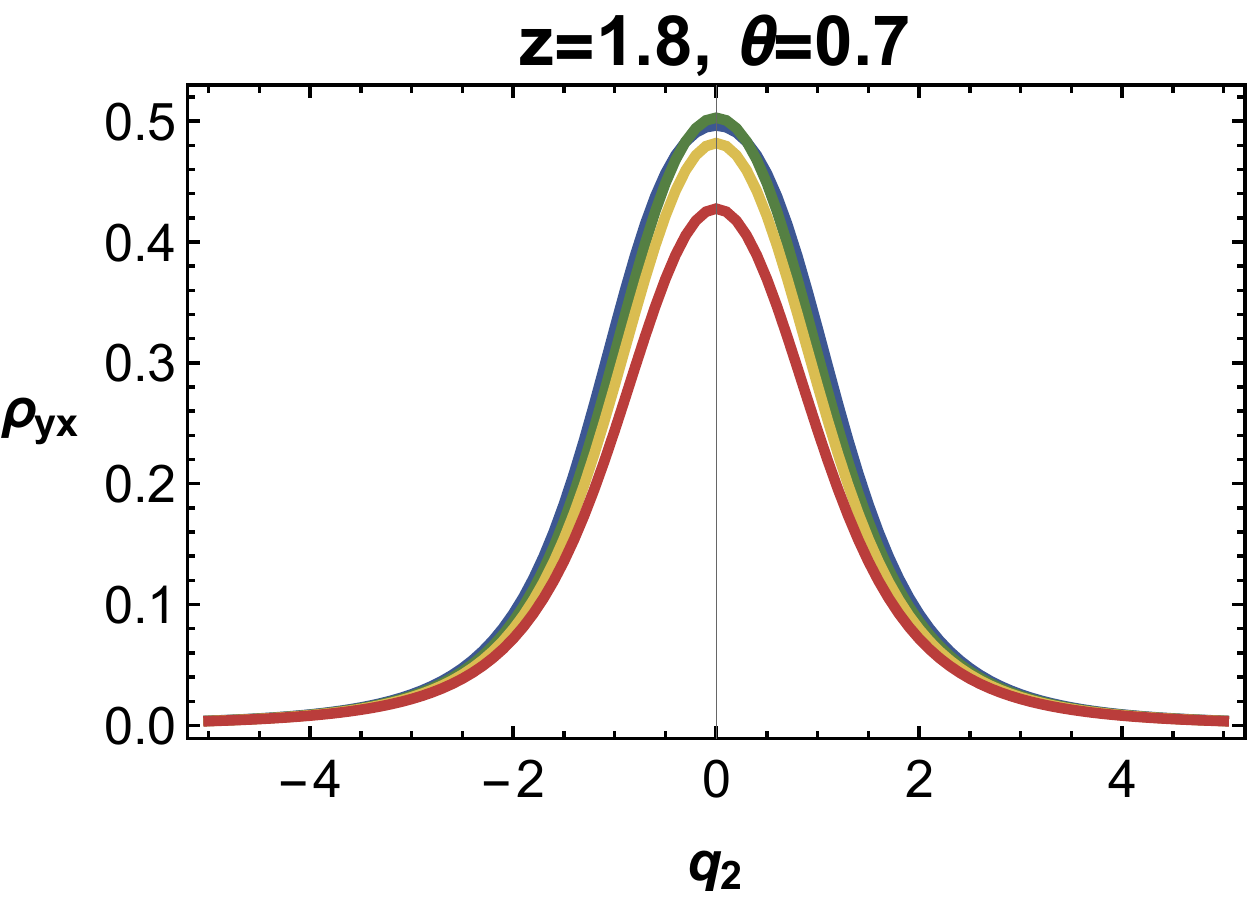} }
      \subfigure[$P_{C}$ ]
   {\includegraphics[width=45mm]{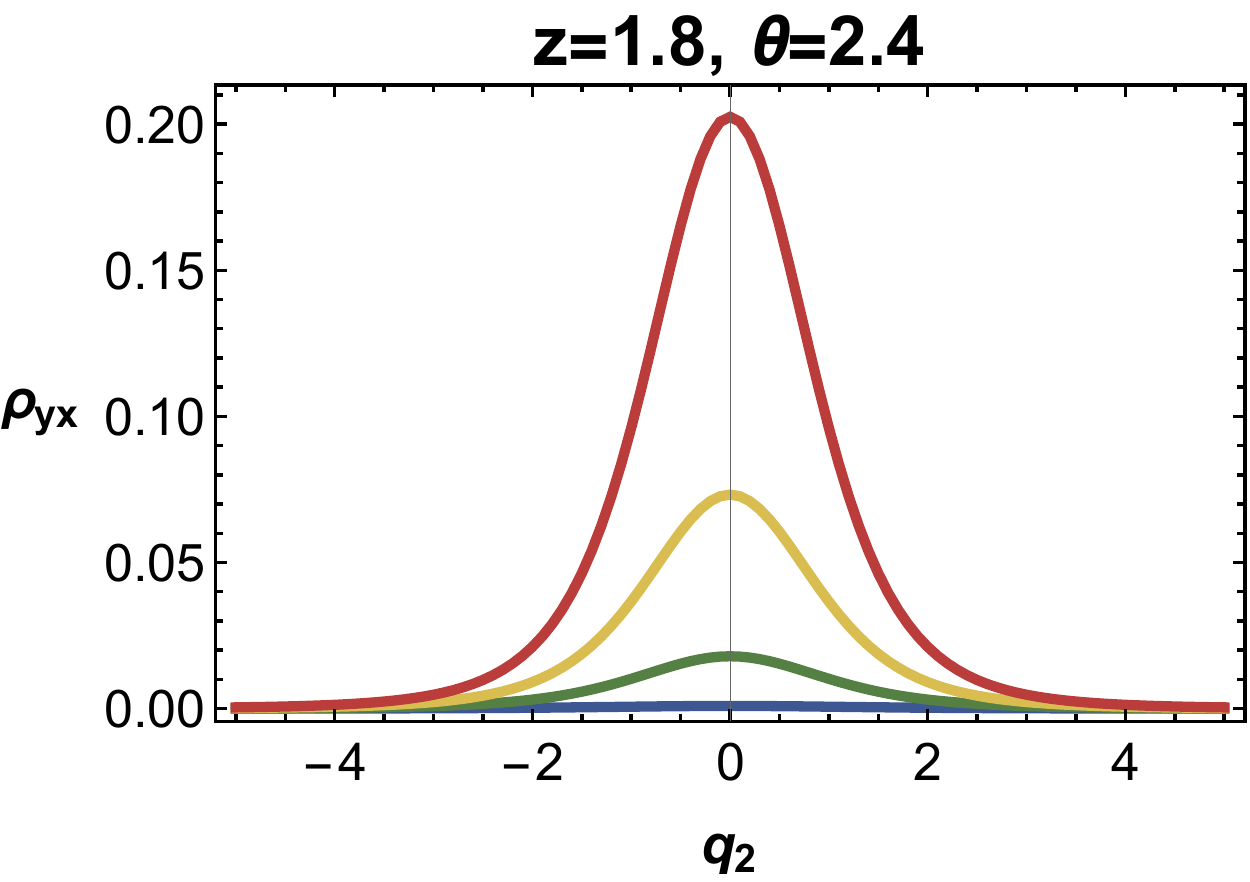} }
     \subfigure[$P_{0B}$ ]
   {\includegraphics[width=45mm]{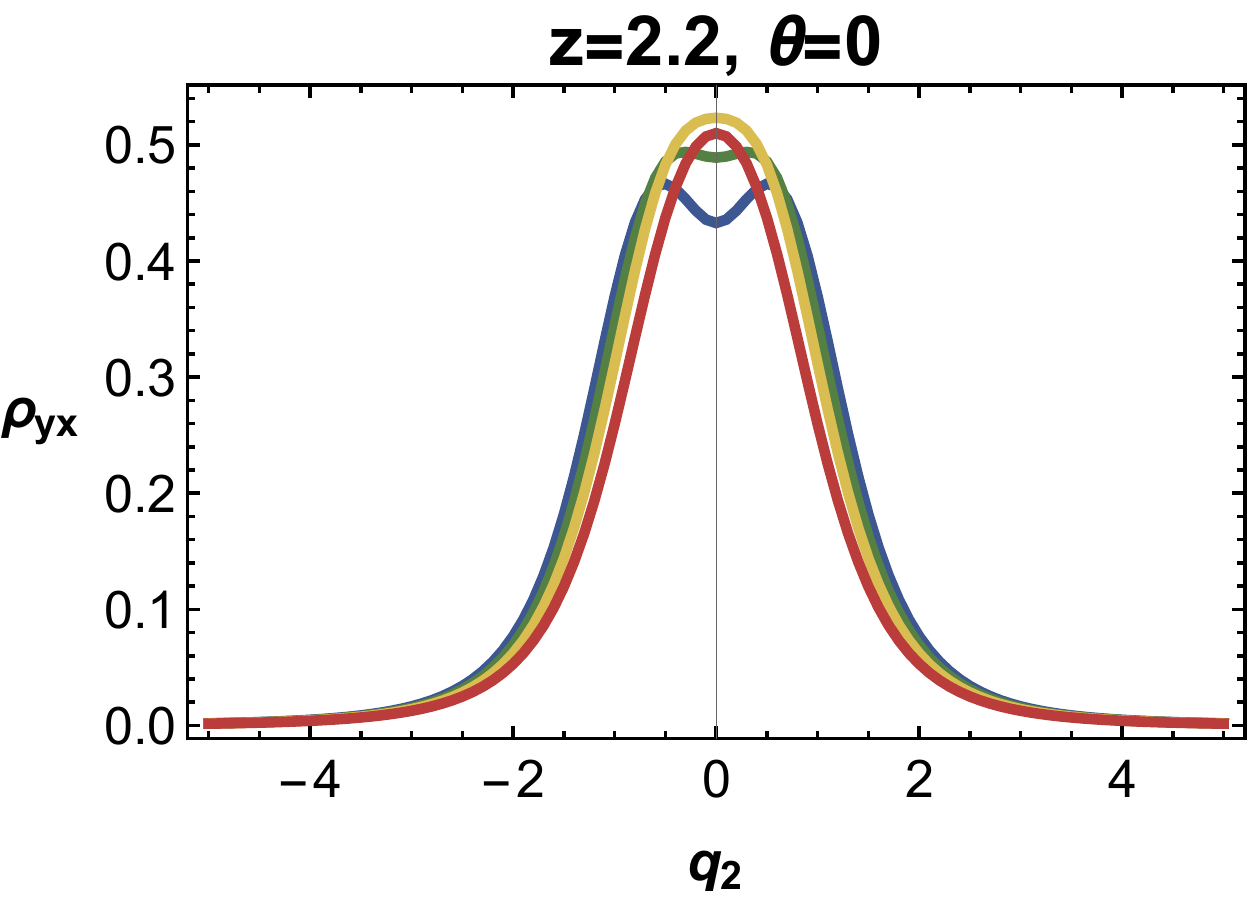} }
      \subfigure[$P_{B}$  ]
   {\includegraphics[width=45mm]{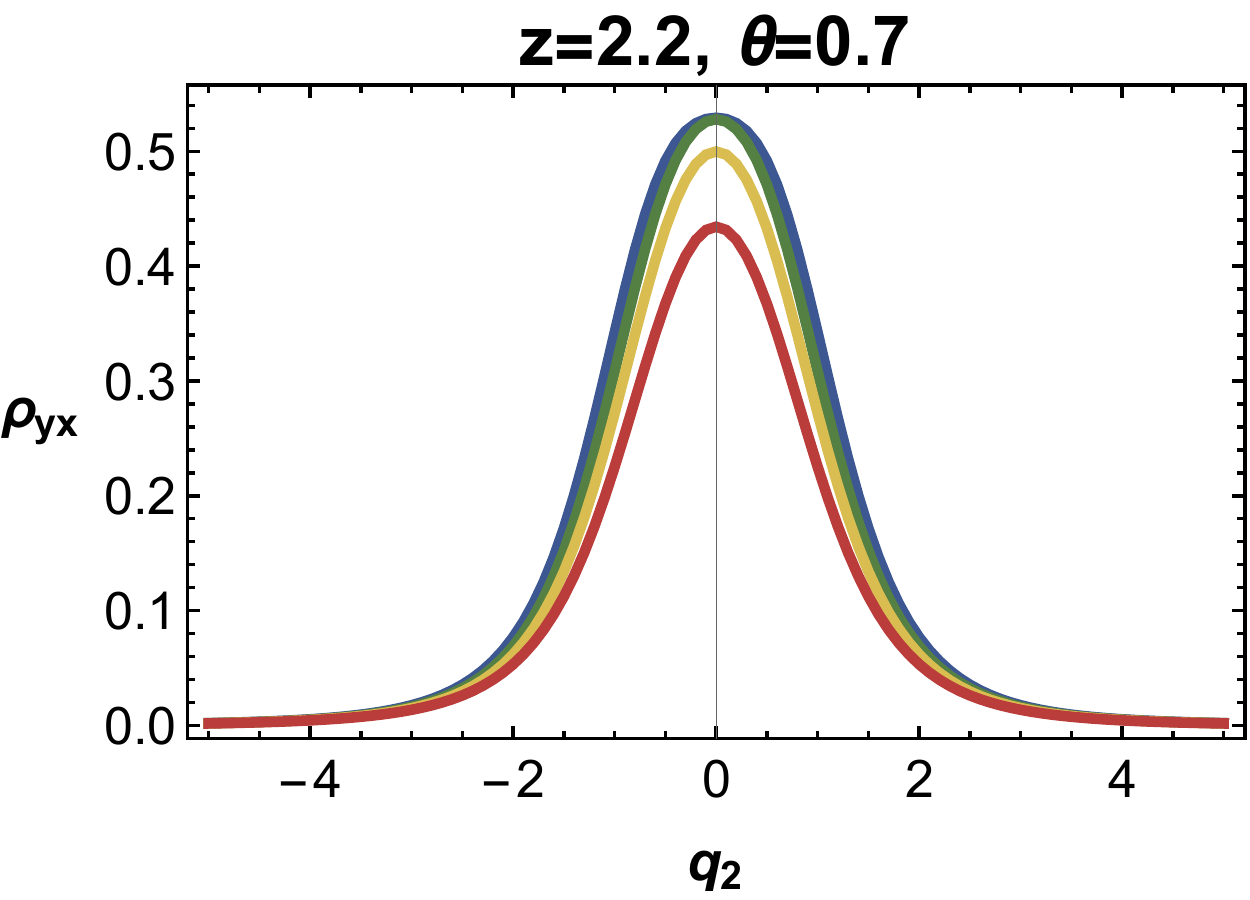} }
      \subfigure[$P_{D}$ ]
   {\includegraphics[width=45mm]{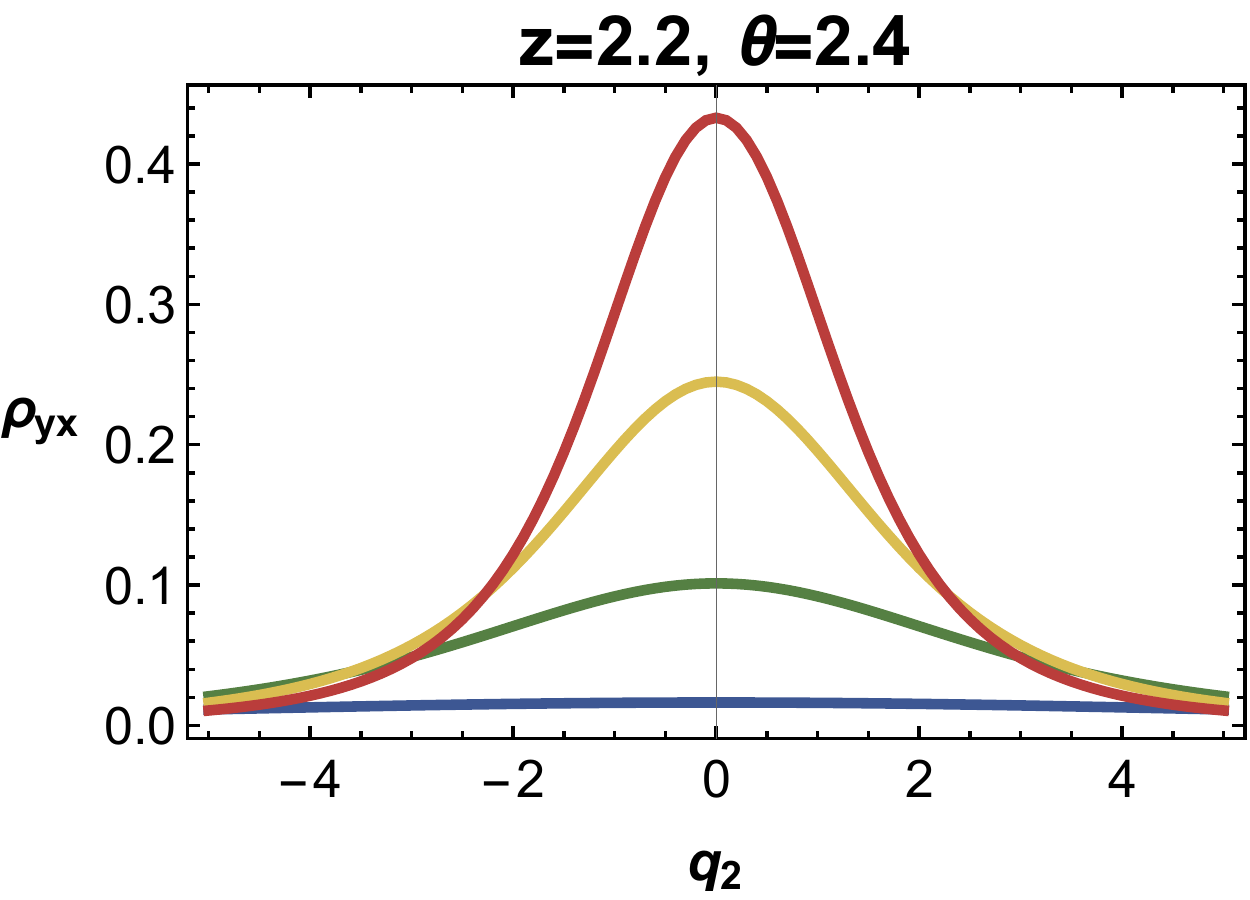} }   
                 \caption{Temperature evolution for $\rho_{yx}(q_2)$ for different $(z,\theta)$. Each curves corresponds to $T=0.04,0.1,0.16, 0.24$ for blue, green, yellow, and red respectively. In all figure, $q_{\chi}=5$.}    \label{fig:ryx} 
\end{figure}

 Next, we consider the density dependence of the Hall conductivity. Figure \ref{fig:ryx} shows the density dependence of $\rho_{yx}(q_2)$.    When $q_2=0$, the longitudinal resistivity is given by 
  \begin{align}
 	\rho_{yx}^0 \equiv \rho_{yx}|_{q_2=0}=\frac{q_{\chi}\lambda^4r_0^{2+3\theta}}{r_0^{4z}+q_{\chi}^2\lambda^4r_0^{4\theta}},
 \end{align}
 where $r_0$ is $r_H(q_2=0)$. Also, we can expand $\rho_{yx}$ in small density limit as previous discussion,
 \begin{align}\label{ryxexp}
 	\rho_{yx}(q_2)&=\rho_{yx}^0-\frac{1}{2\mathcal{D}^2}\bigg(\frac{4q_{\chi}\lambda^2r_0^{4(z+\theta)}}{\beta^2}\nonumber\\
 	& +q_{\chi}\lambda^2r_0^{3\theta+1}\mathcal{A}_2\left(r_0^{4z}(4z-3\theta-2)+q_{\chi}^2\lambda^4 r_0^{4\theta}(\theta-2)\right)   \bigg)q_2^2+\cdots, 
 \end{align}
 where $\mathcal{D}=(r_0^{4z}+q_{\chi}^2\lambda^4r_0^{4\theta})$. 
 \begin{figure}
 \centering
	\subfigure[$q_{\chi}=0.7$]
   {\includegraphics[width=45mm]{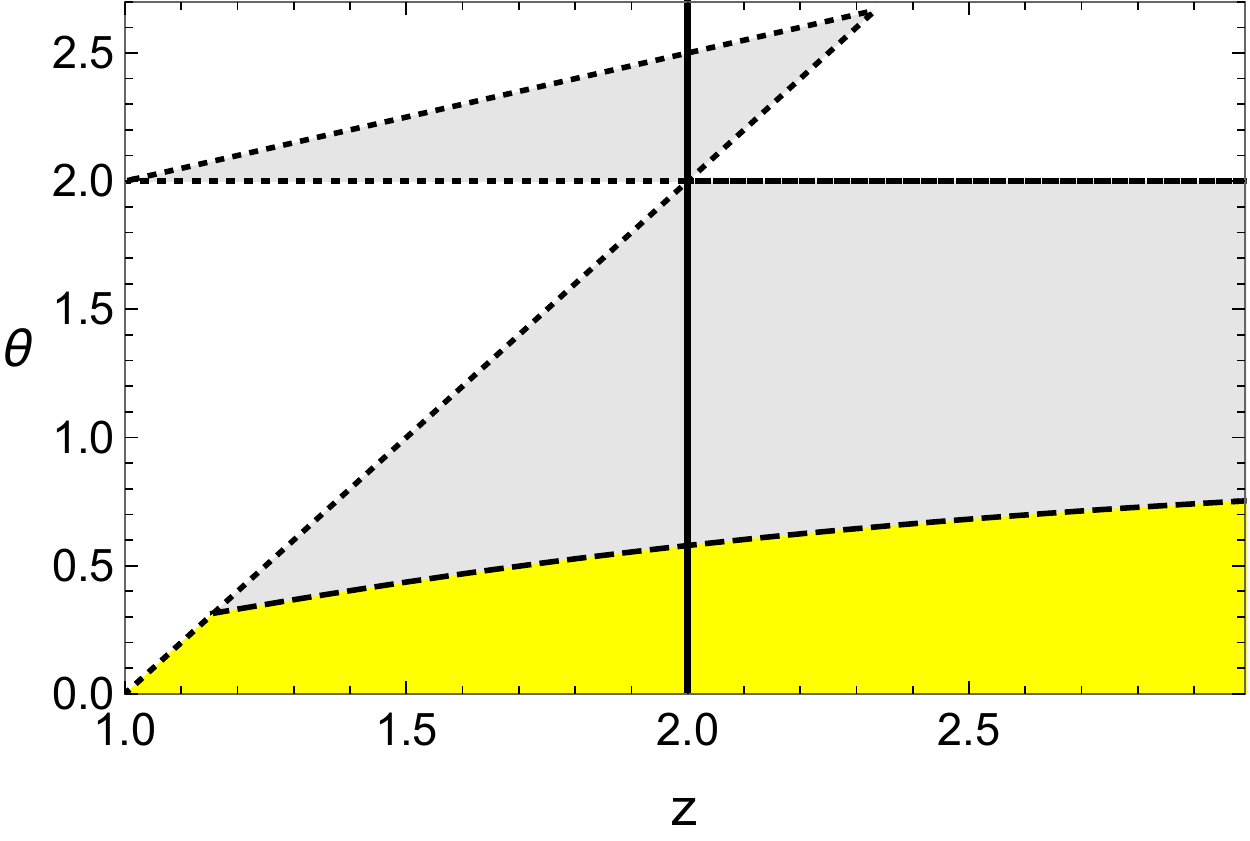} }    
\hskip.2cm
\subfigure[$q_{\chi}=1.4$]
   {\includegraphics[width=45mm]{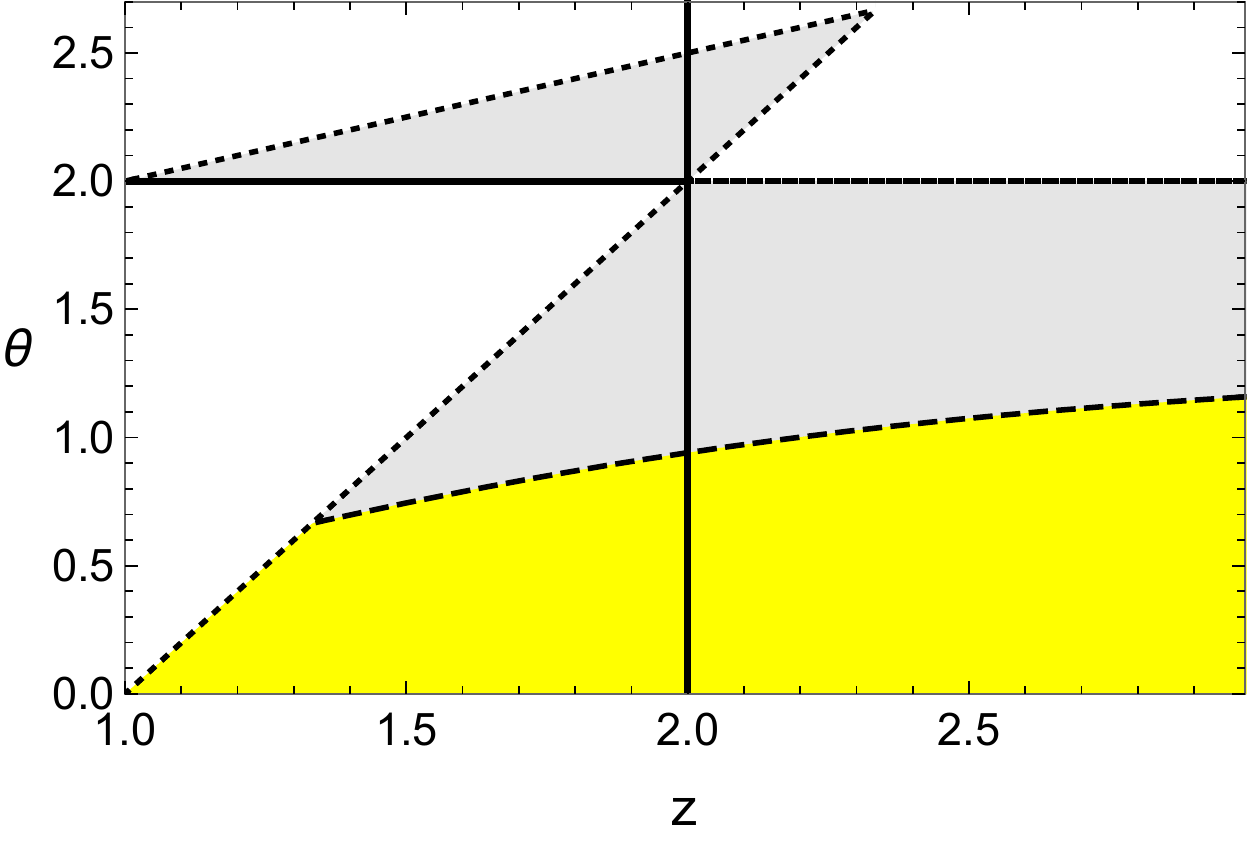} }   	
   \caption{Sign of the coefficient of $q_2^2$ of the transverse resistivity near $q_2=0$. Yellow region denotes positive sign(WL) and gray region is for negative sign(WAL). Dotted line is NEC. Here we use the same parameters as Figure \ref{fig:rxx} and $T=0.04$.}
	\label{fig:ryxq}
\end{figure}

Roughly speaking, $q_{\chi}$ can flip  the sign of curvature  $(z,\theta)$ dependent way, which is difficult to analyze analytically. We investigated it numerically and the result is in Figure \ref{fig:ryxq}. 
 In regions A, B, $q_{\chi}$-term suppress the resistivity near $q_2=0$, but the effect of $q_{\chi}$ is reduced  as $\theta$ increases. In region C,D, the effect of $q_{\chi}$ terms becomes negligible due to the large $\theta$.
 
 Now let's turn to  the density dependence of the   thermal conductivity. We can perform the same analysis as the resistivity: 
   \begin{align}\label{kappaexp}
   	\kappa_{xx}(q_2)=\kappa_{xx}^0-\frac{16\pi^2Tr_0^{2z-2\theta-1}}{\beta^2\mathcal{D}}\left(\frac{r_0^{4z+\theta+1}}{\beta^2}-(z-\theta)\mathcal{A}_2\mathcal{D}\right)q_2^2+\cdots,
   \end{align}
  $\mathcal{A}_2$ terms in (\ref{kappaexp}) determines the curvature of the $\kappa_{xx}$ near the $q_2=0$.  
The sign of the coefficients is shown as Figure \ref{fig:kxxq}. 
  \begin{figure}[ht!]
\centering
	 \subfigure[$P_{0A}$ at $q_{\chi}=0$]
   {\includegraphics[width=30mm]{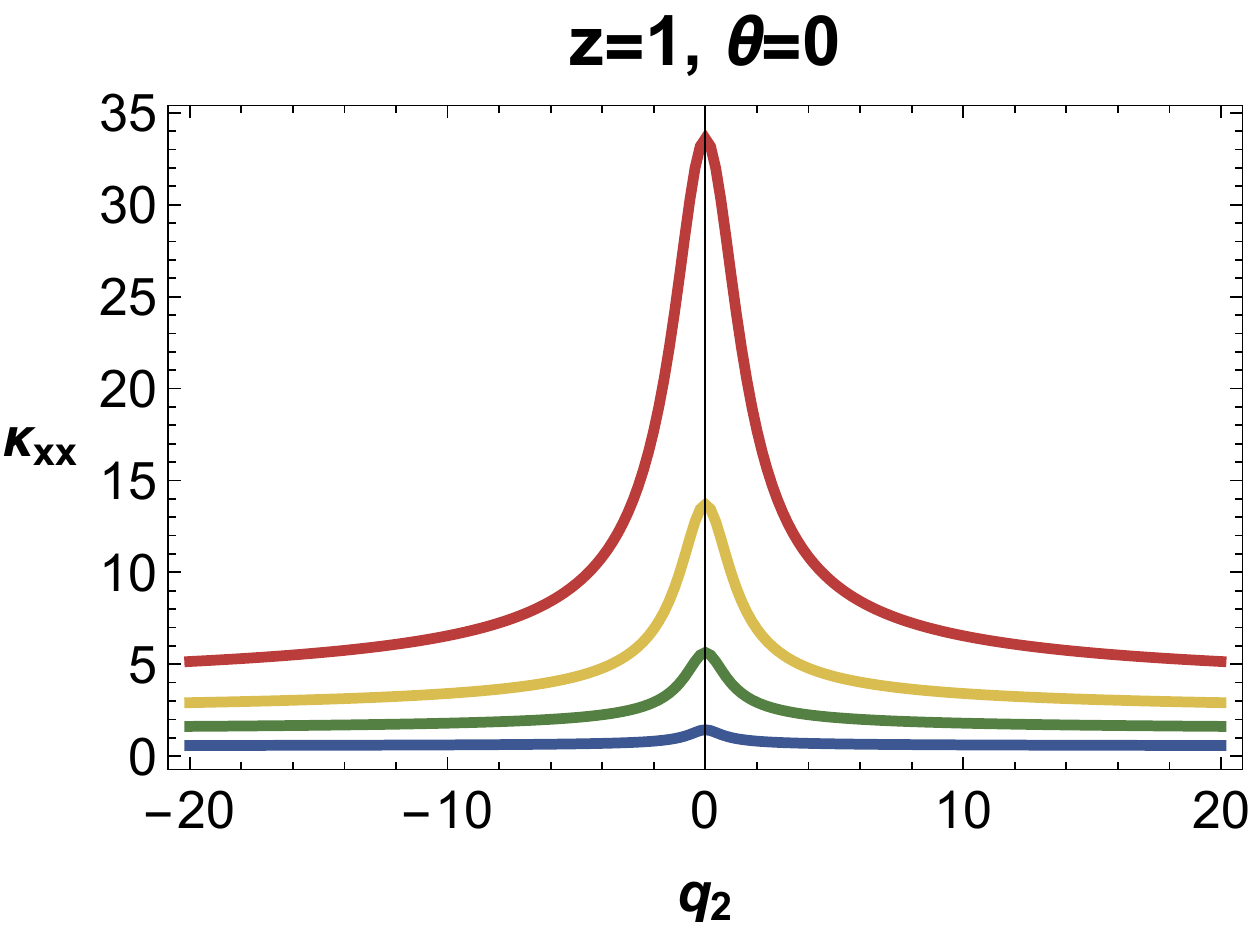} }
    \subfigure[$P_{0A}$    ]
   {\includegraphics[width=37mm]{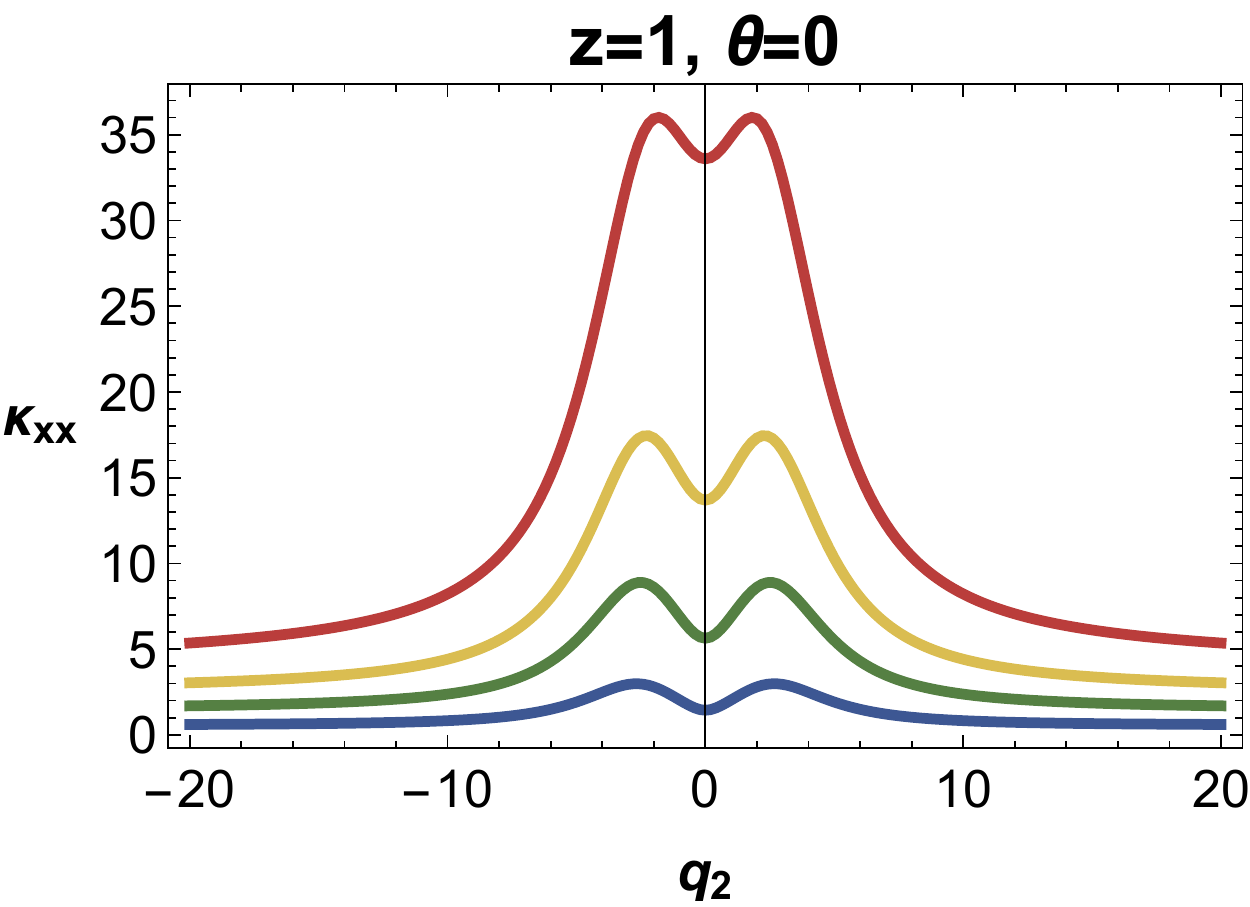} }
      \subfigure[$P_{0B}$ at $q_{\chi}=0$]
   {\includegraphics[width=30mm]{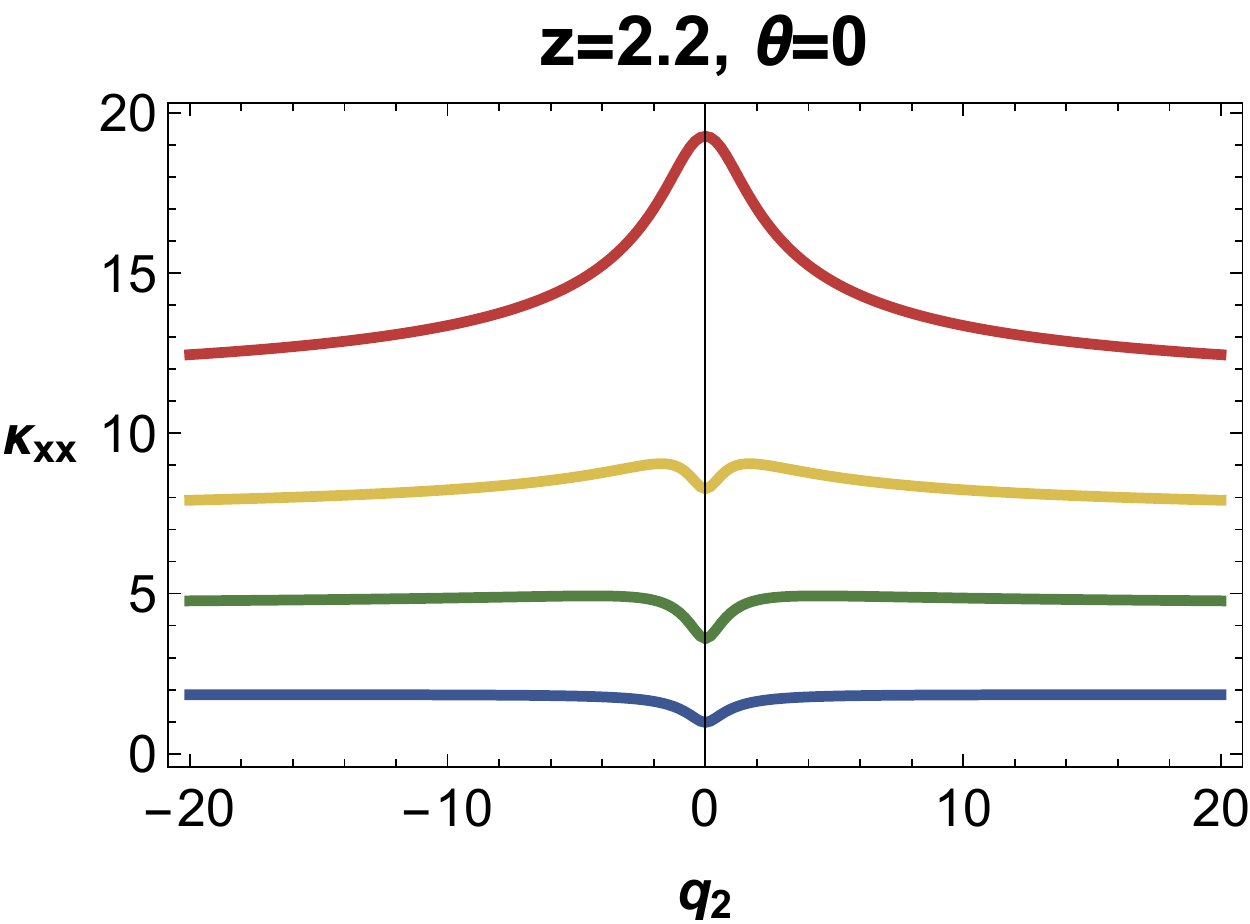} }
    \subfigure[$P_{0B}$   ]
   {\includegraphics[width=37mm]{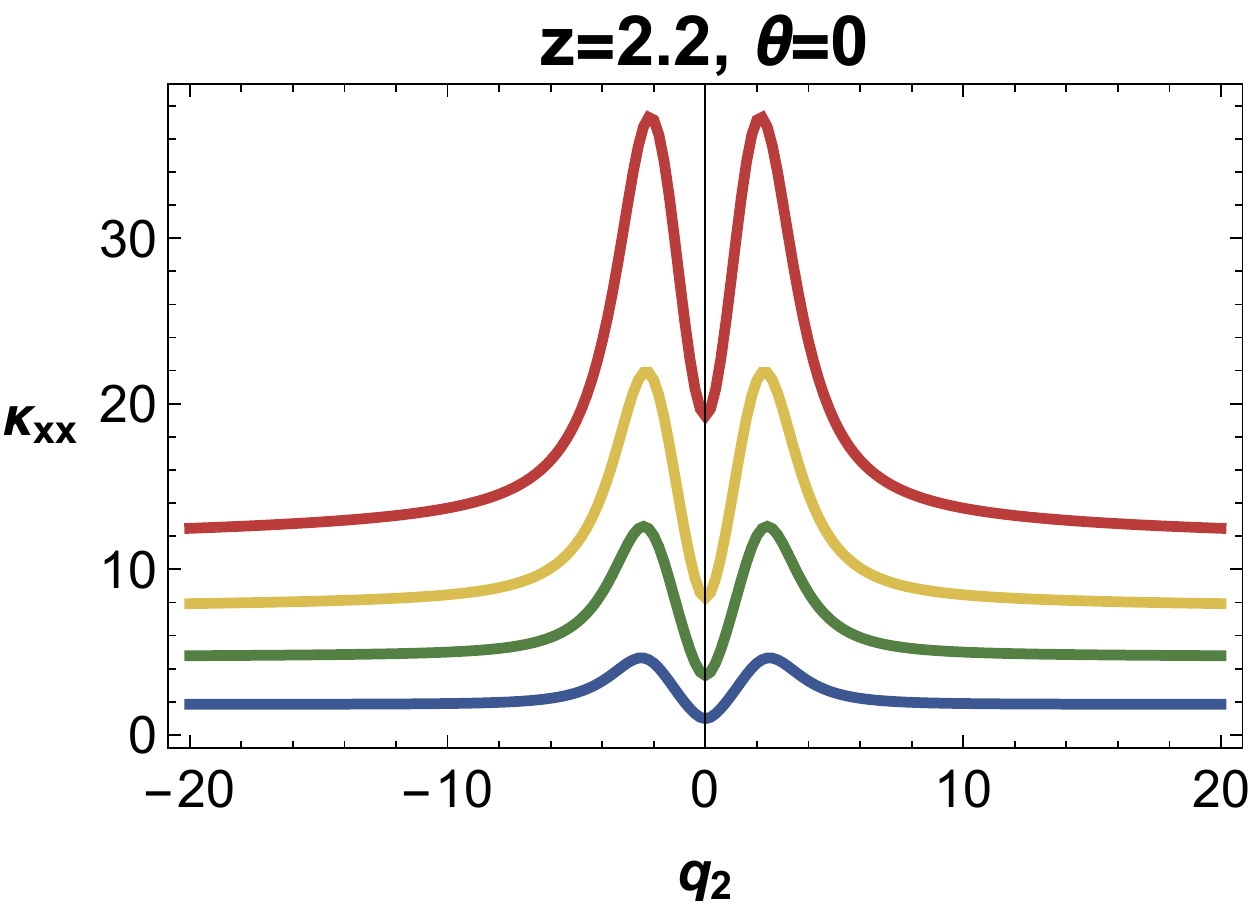} }
     \subfigure[$P_{A}$ at $q_{\chi}=0$]
   {\includegraphics[width=30mm]{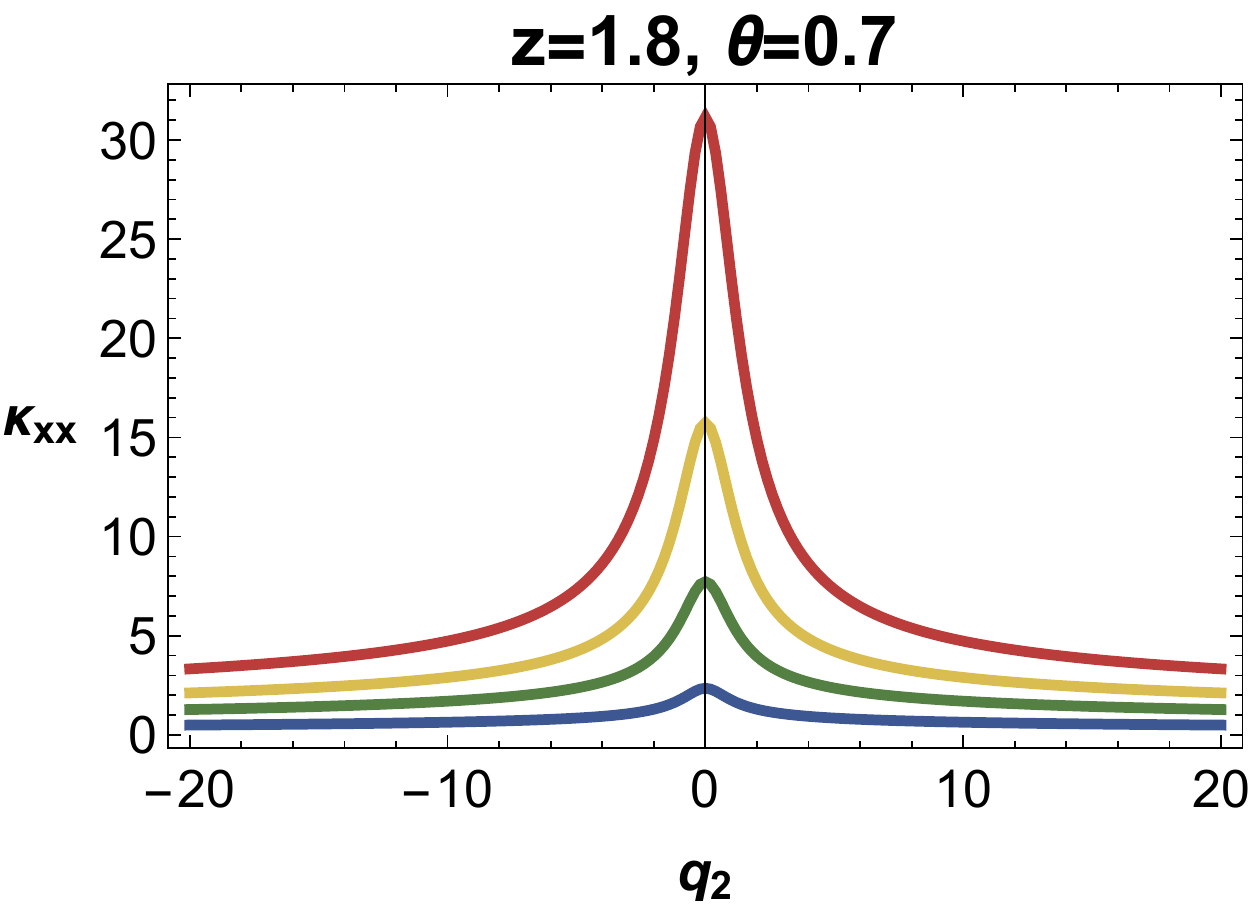} }
    \subfigure[$P_{A}$    ]
   {\includegraphics[width=37mm]{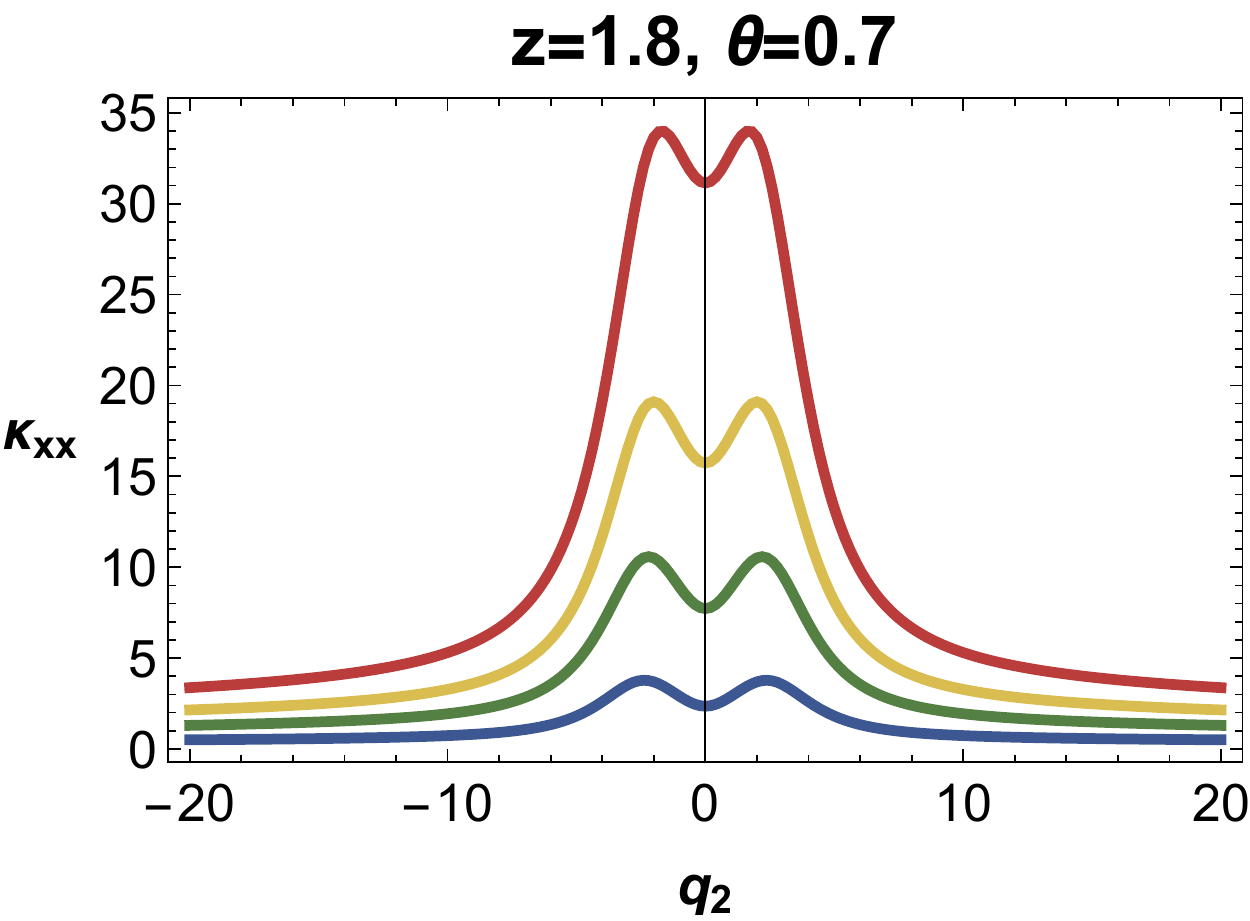} }
  \subfigure[$P_{B}$ at $q_{\chi}=0$]
   {\includegraphics[width=30mm]{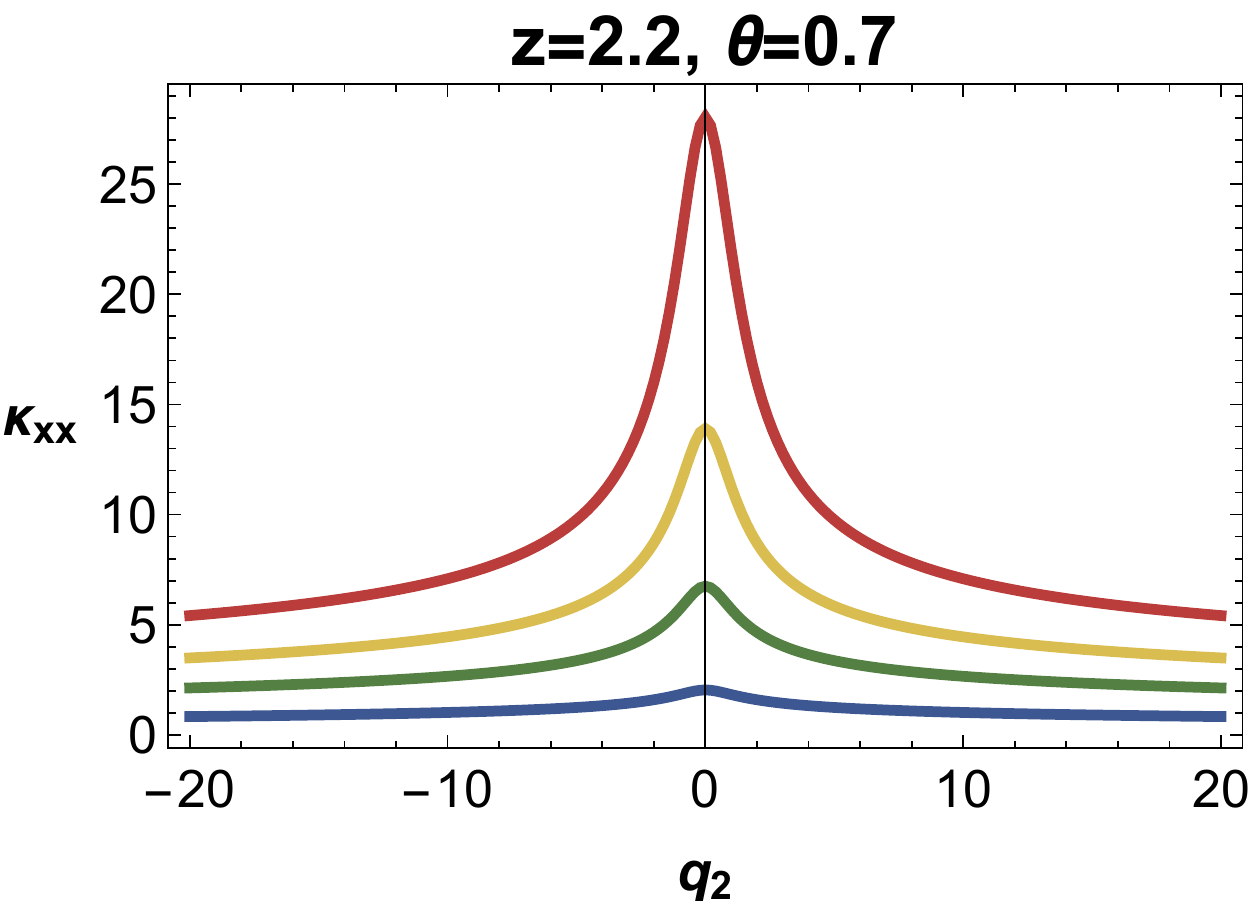} }
    \subfigure[$P_{B}$    ]
   {\includegraphics[width=37mm]{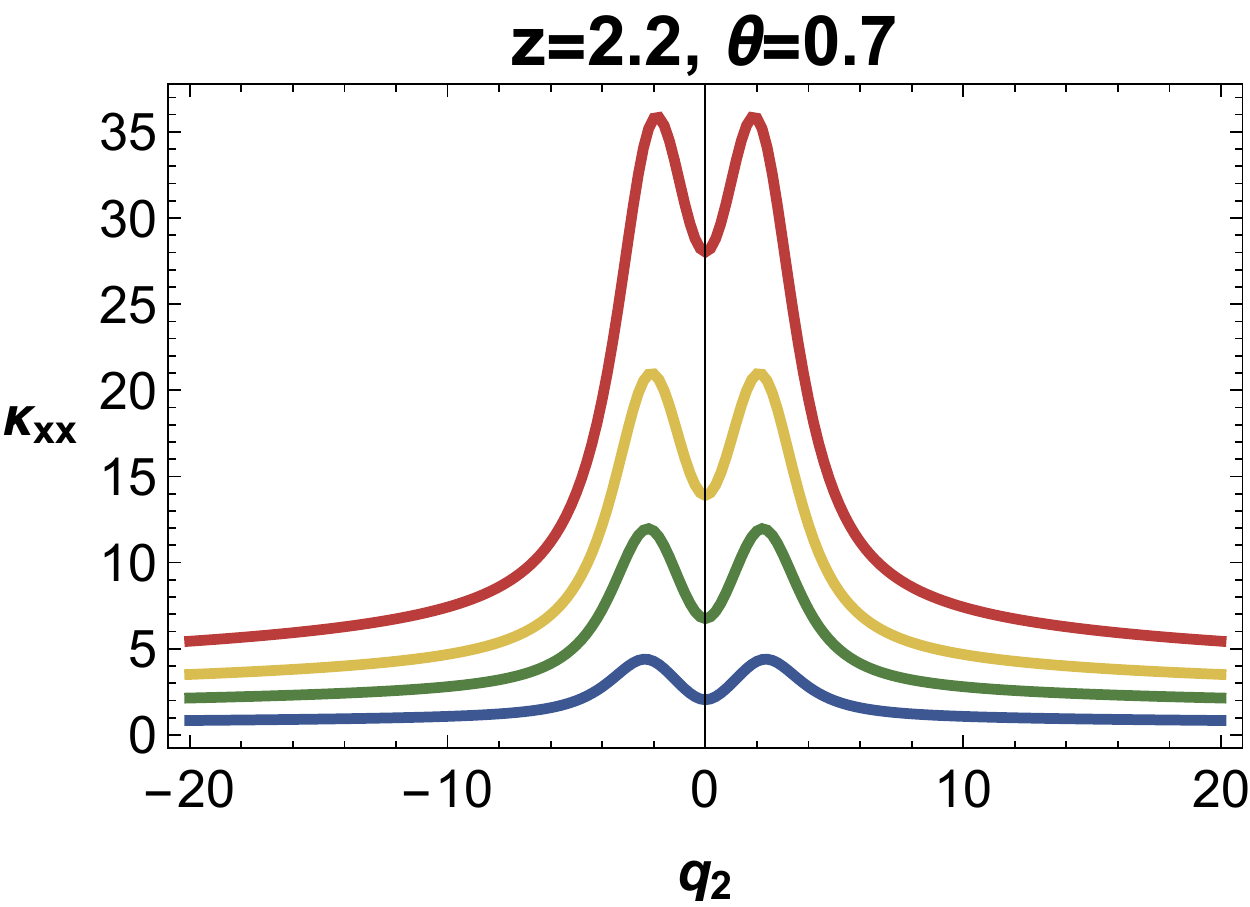} }
    \subfigure[$P_{C}$ at $q_{\chi}=0$]
   {\includegraphics[width=30mm]{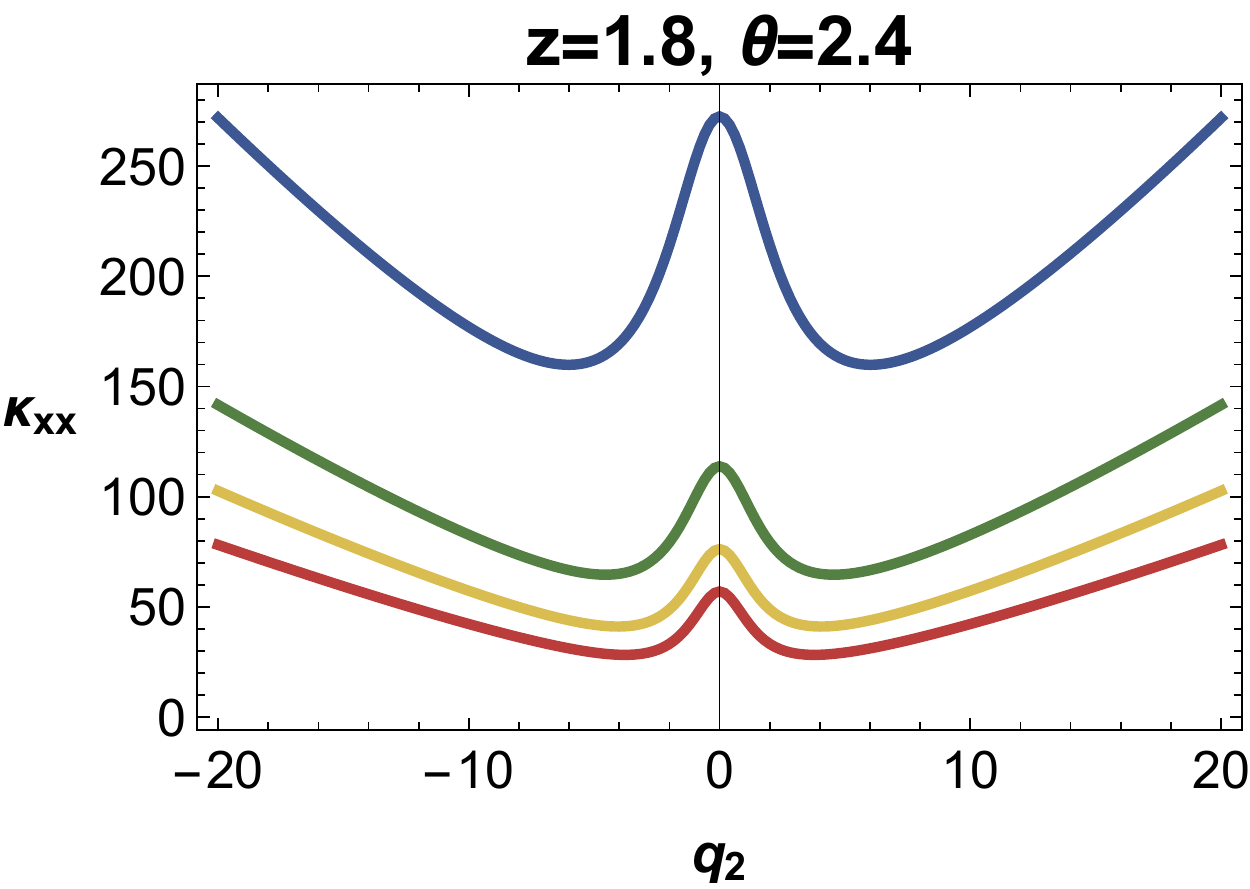} }
    \subfigure[$P_{C}$   ]
   {\includegraphics[width=37mm]{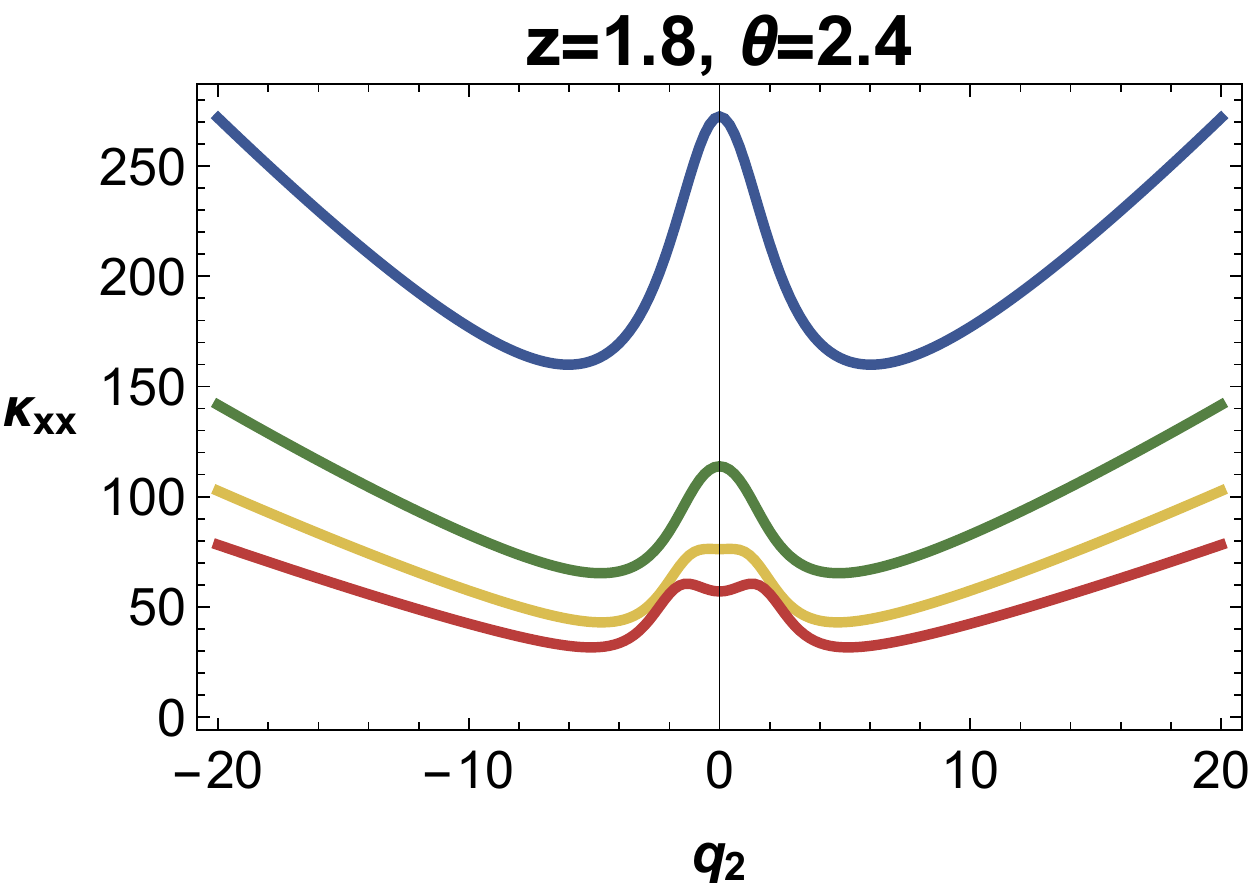} }
      \subfigure[$P_{D}$ at $q_{\chi}=0$]
   {\includegraphics[width=30mm]{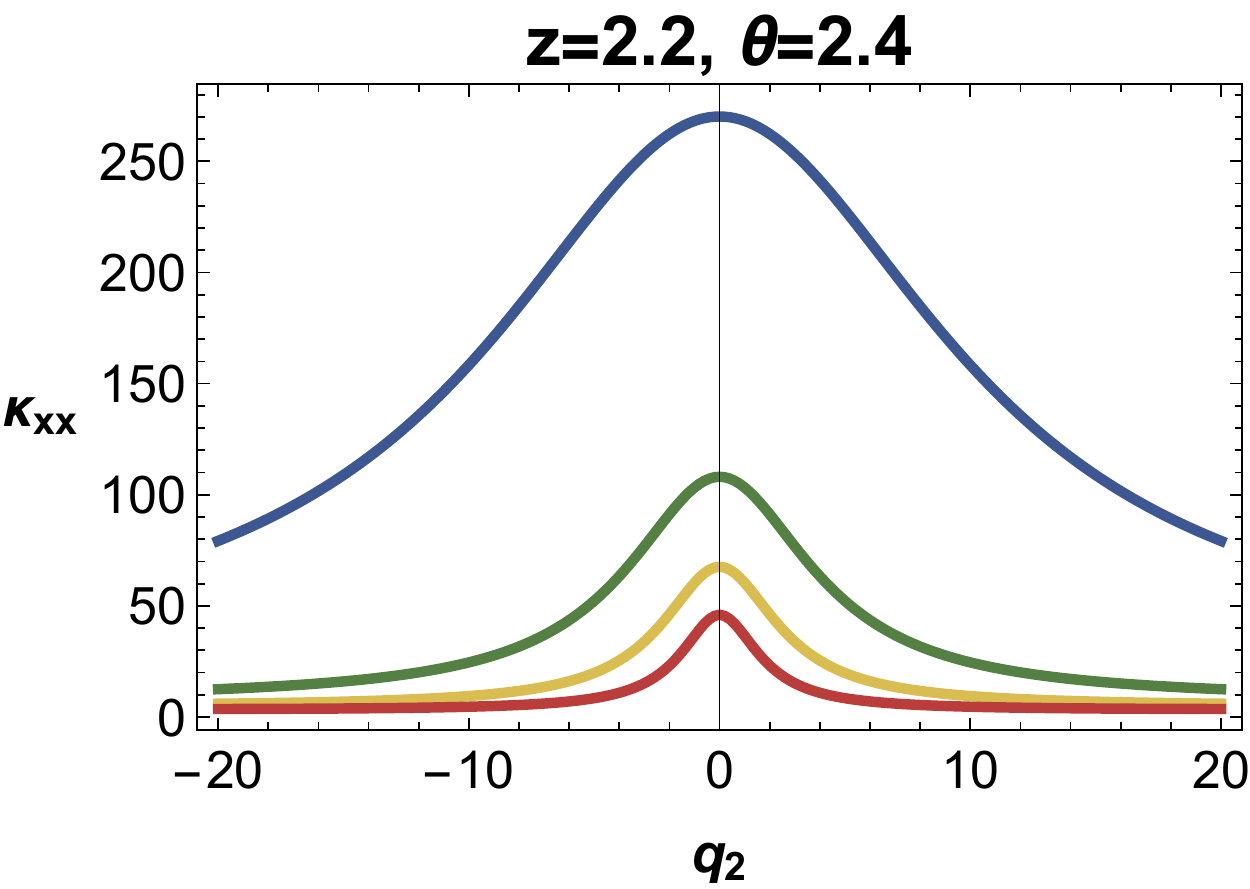} }
    \subfigure[$P_{D}$   ]
   {\includegraphics[width=37mm]{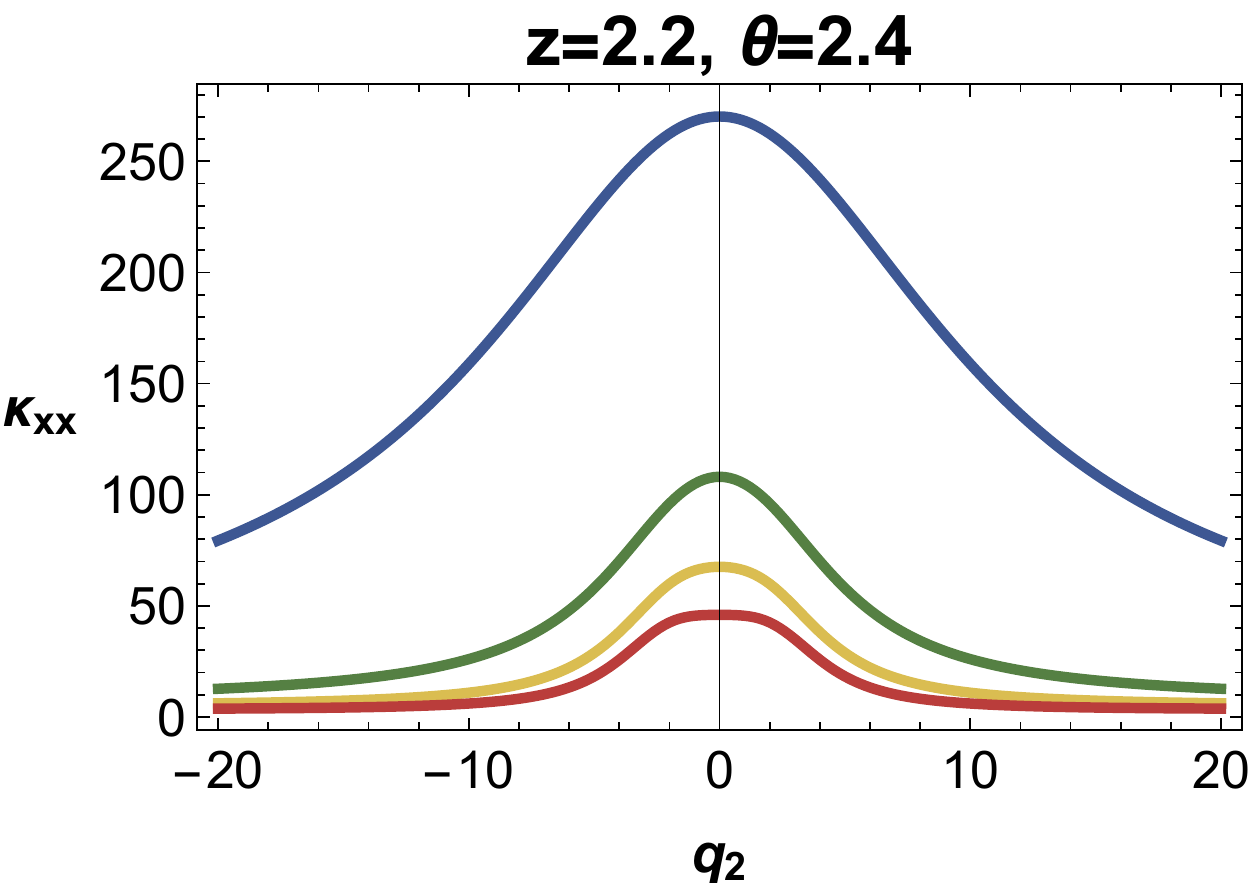} }   
                 \caption{Temperature evolution for $\kappa_{xx}(q_2)$ for different $(z,\theta)$. Each curves corresponds to $T=0.04,0.1,0.160.24$ for blue, green, yellow, and red respectively. In all figure, $q_{\chi}=5$.}   
       \label{fig:kxx}           
\end{figure}
\begin{figure}
\centering
	\subfigure[$q_{\chi}=0.7$]
   {\includegraphics[width=45mm]{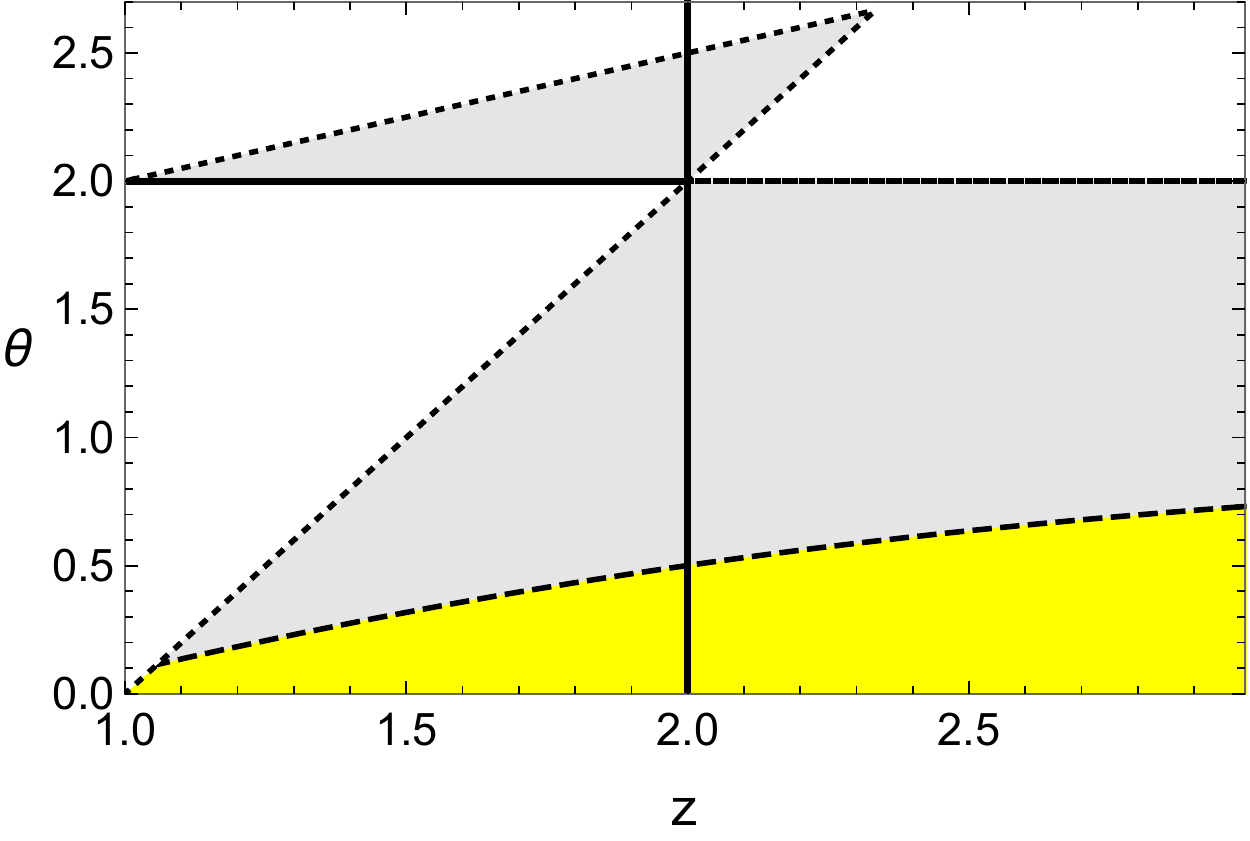} }    
\hskip.2cm
\subfigure[$q_{\chi}=1.4$]
   {\includegraphics[width=45mm]{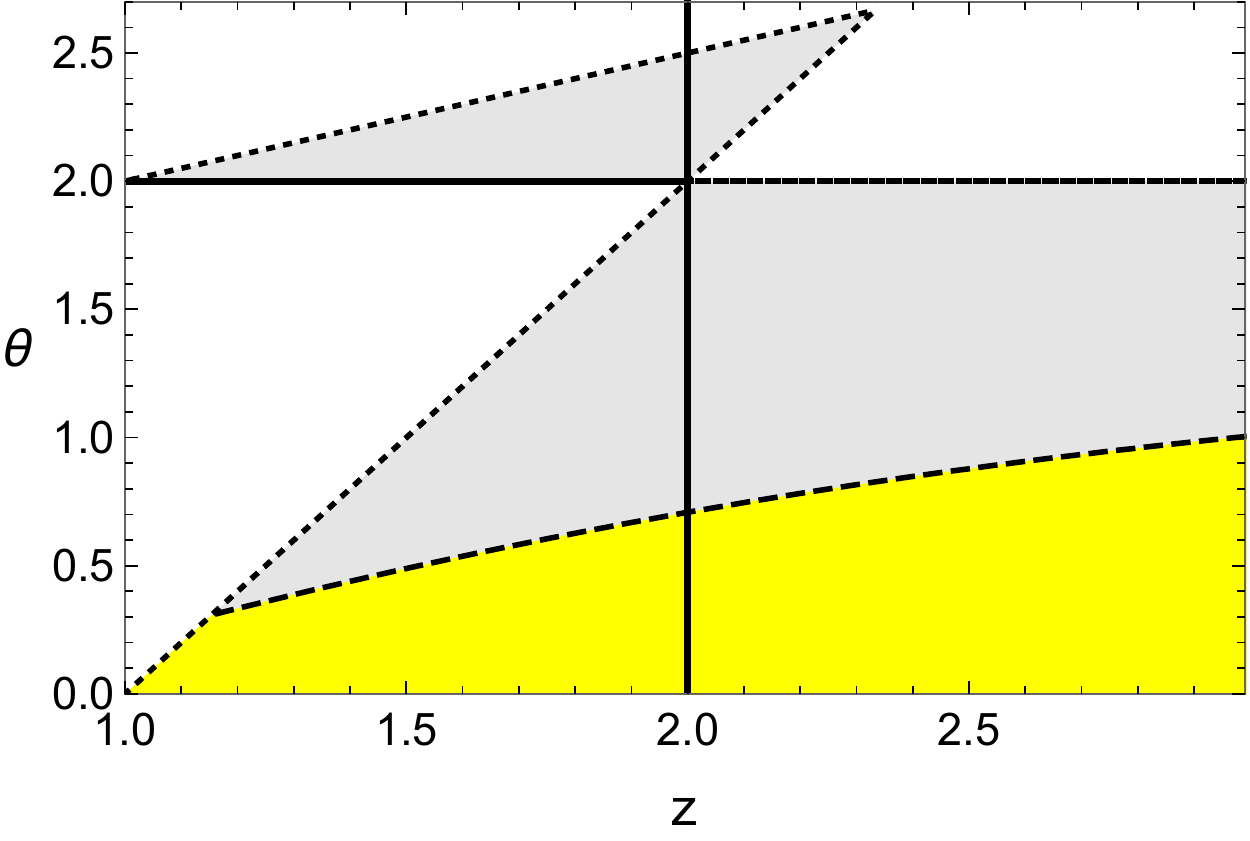} }   
	\caption{Sign of the coefficient of $q_2^2$ of the  $\kappa_{xx}$ near $q_2=0$. Yellow region denotes positive sign(WL) and gray region is for negative sign(WAL). Dotted line is NEC. Here we use the same parameters as Figure \ref{fig:kxx} and $T=0.04$.} \label{fig:kxxq}
\end{figure}
Similarly, transverse thermal conductivity in small density limit is expressed as
\begin{align}\label{kxyexp}
	\kappa_{xy}(q_2)=\frac{16\pi^2T q_{\chi}\lambda^2r_0^{\theta-2}}{\beta^4(1+q_{\chi}^2\lambda^4r_0^{-4(z-\theta)})
	}q_2^2+\cdots ,
\end{align}
Notice that $\kappa_{xy}(q_2=0)=0$. The coefficient of $q_2^2$ in (\ref{kxyexp}) is positive definite for every $(z,\theta)$. See Figure \ref{fig:kxyq}.
\begin{figure}[ht!]
\centering  
 \subfigure[$P_{0A}$    ]
   {\includegraphics[width=45mm]{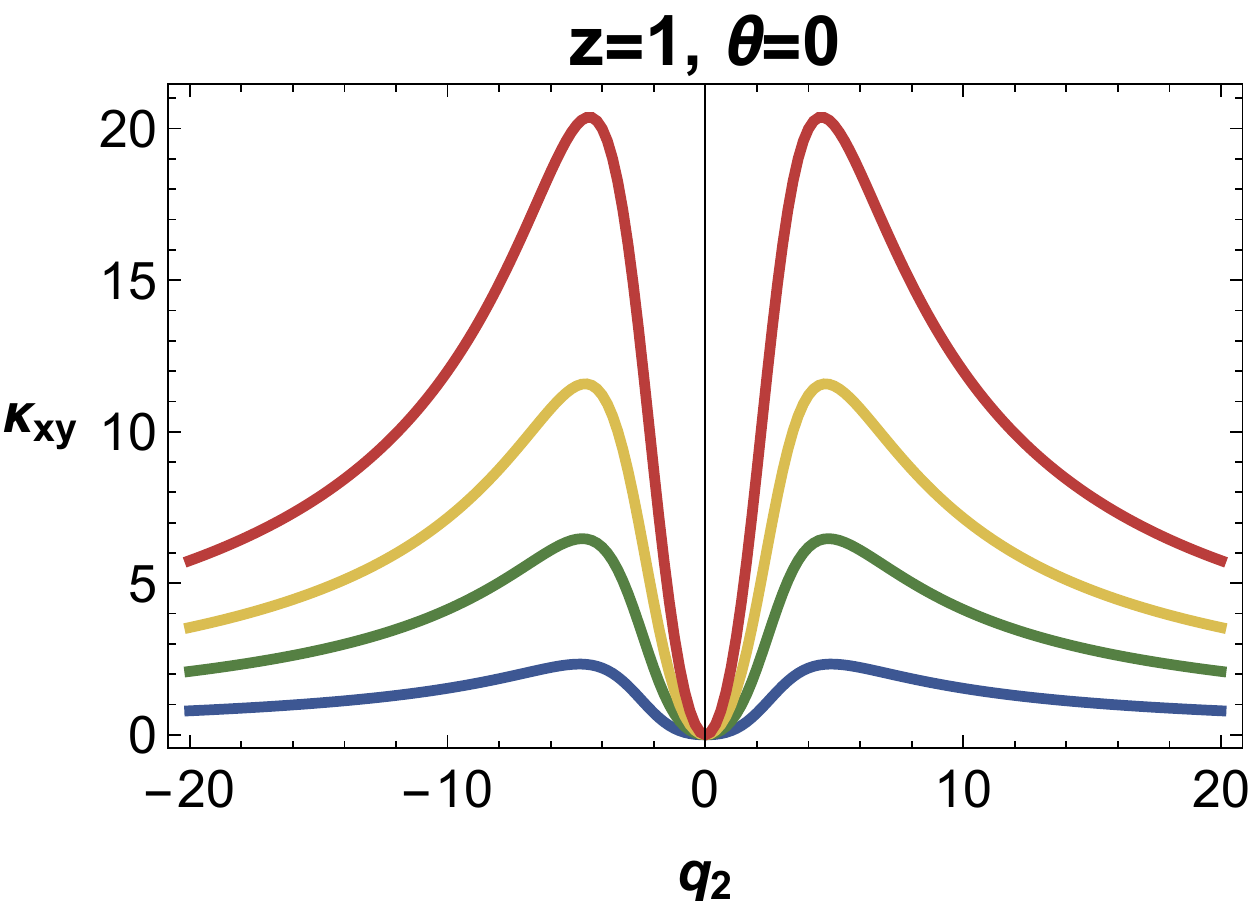} }
         \subfigure[$P_{A}$    ]
   {\includegraphics[width=45mm]{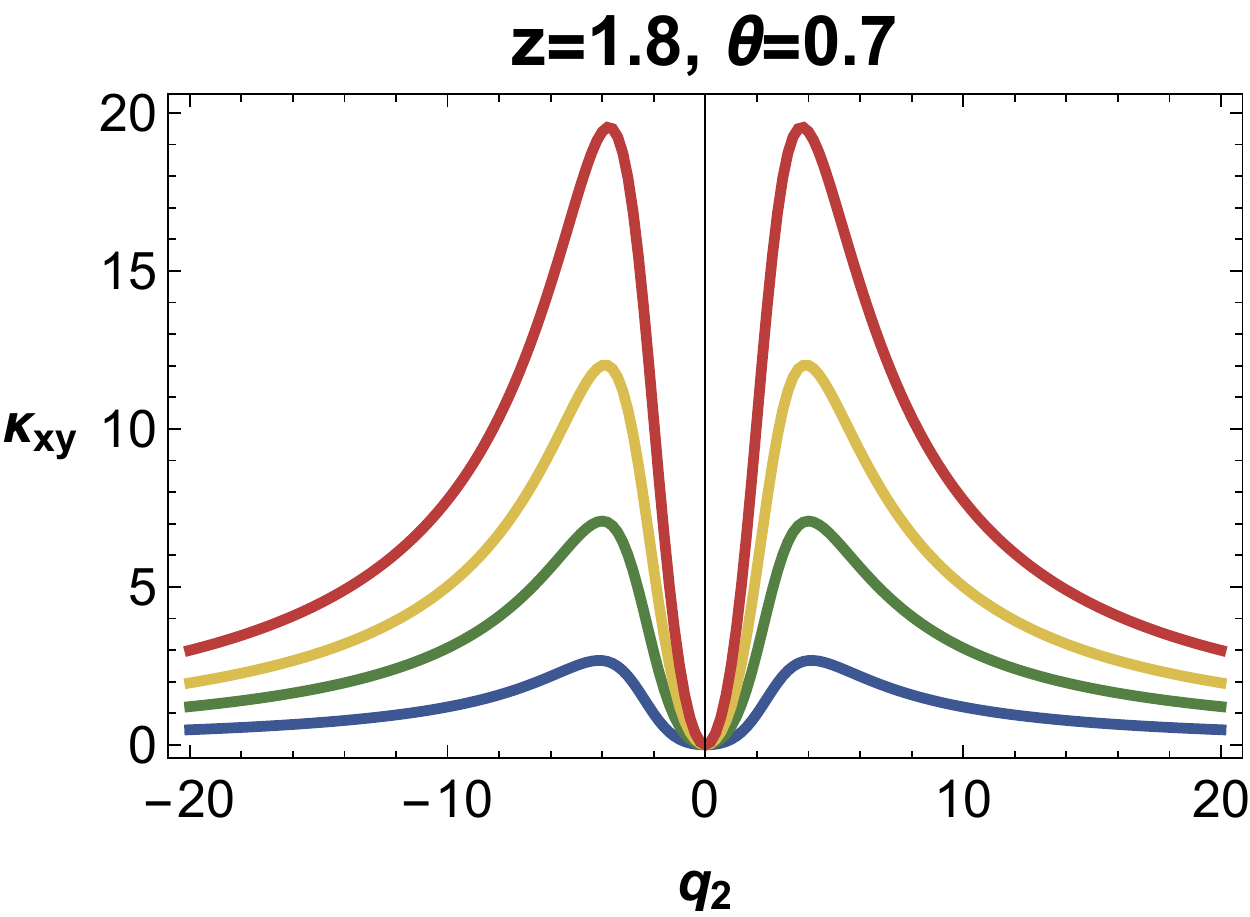} }
      \subfigure[$P_{C}$   ]
   {\includegraphics[width=45mm]{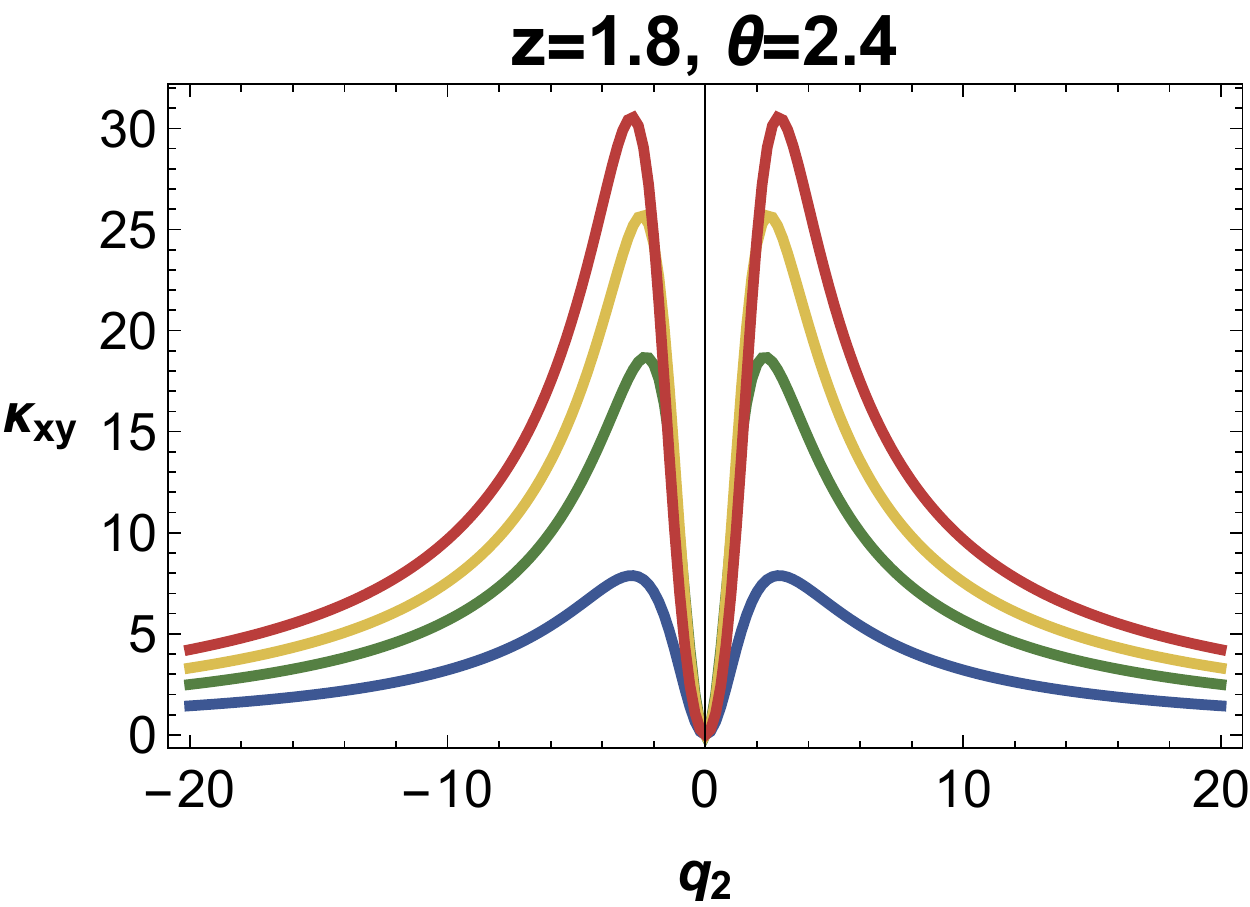} }
    \subfigure[$P_{0B}$   ]
   {\includegraphics[width=45mm]{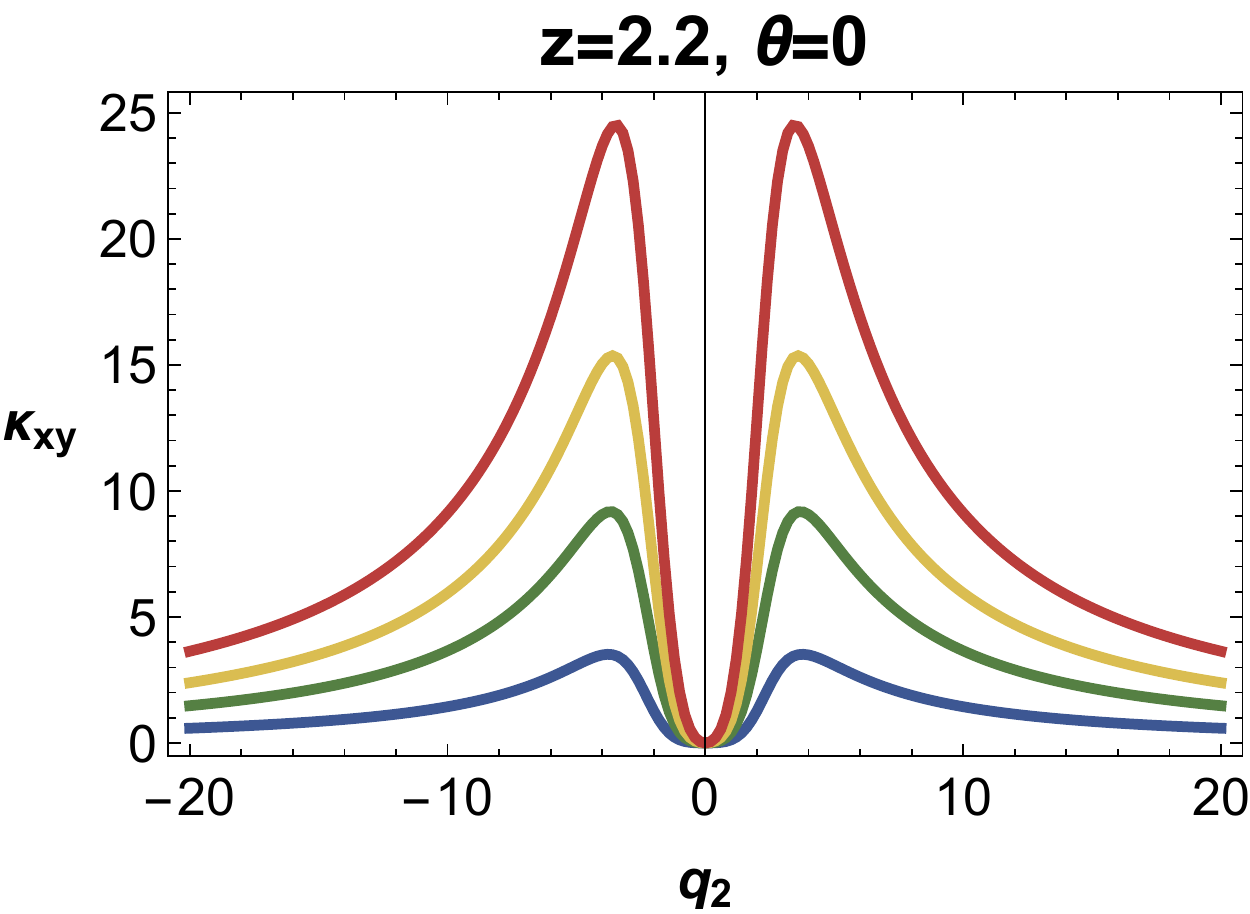} }
      \subfigure[$P_{B}$    ]
   {\includegraphics[width=45mm]{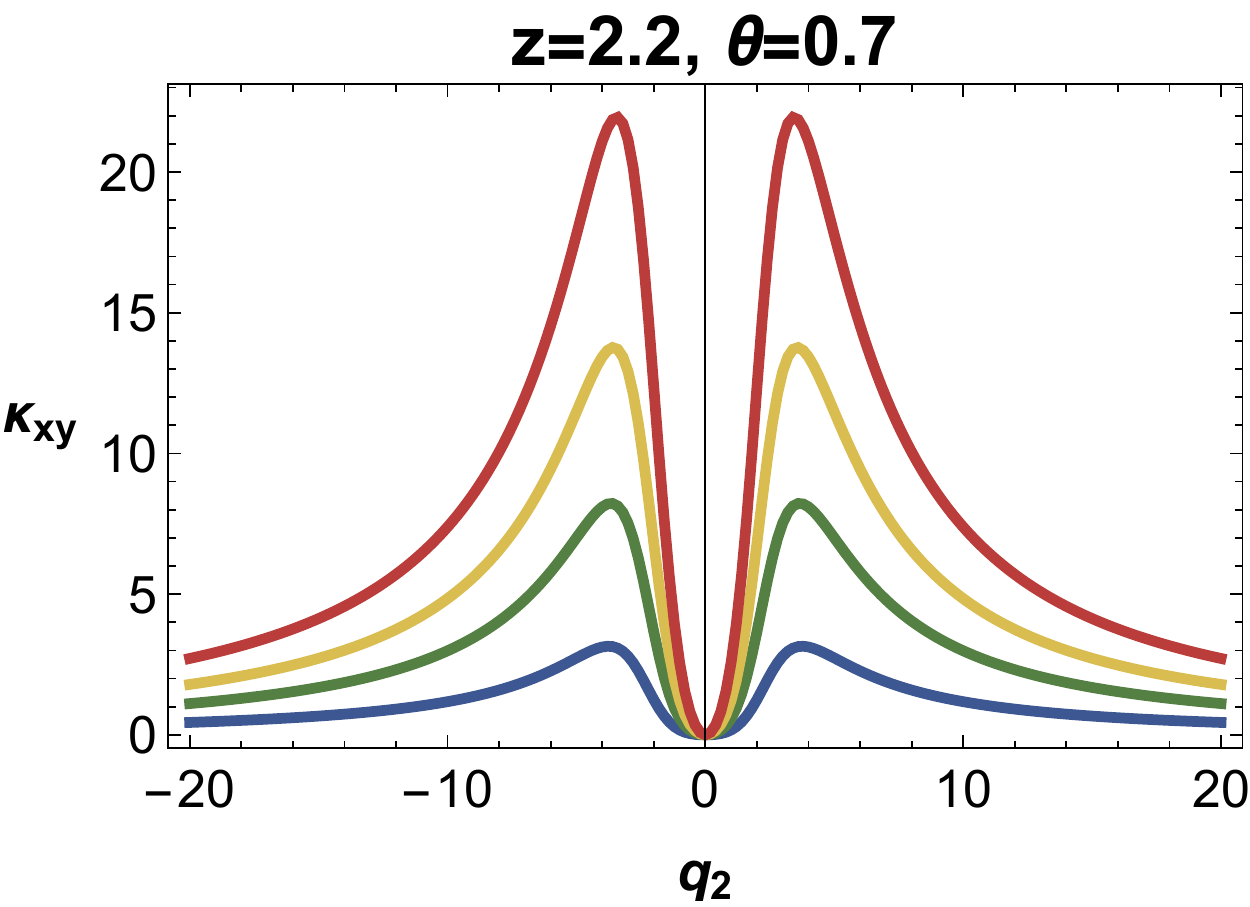} }
      \subfigure[$P_{D}$   ]
   {\includegraphics[width=45mm]{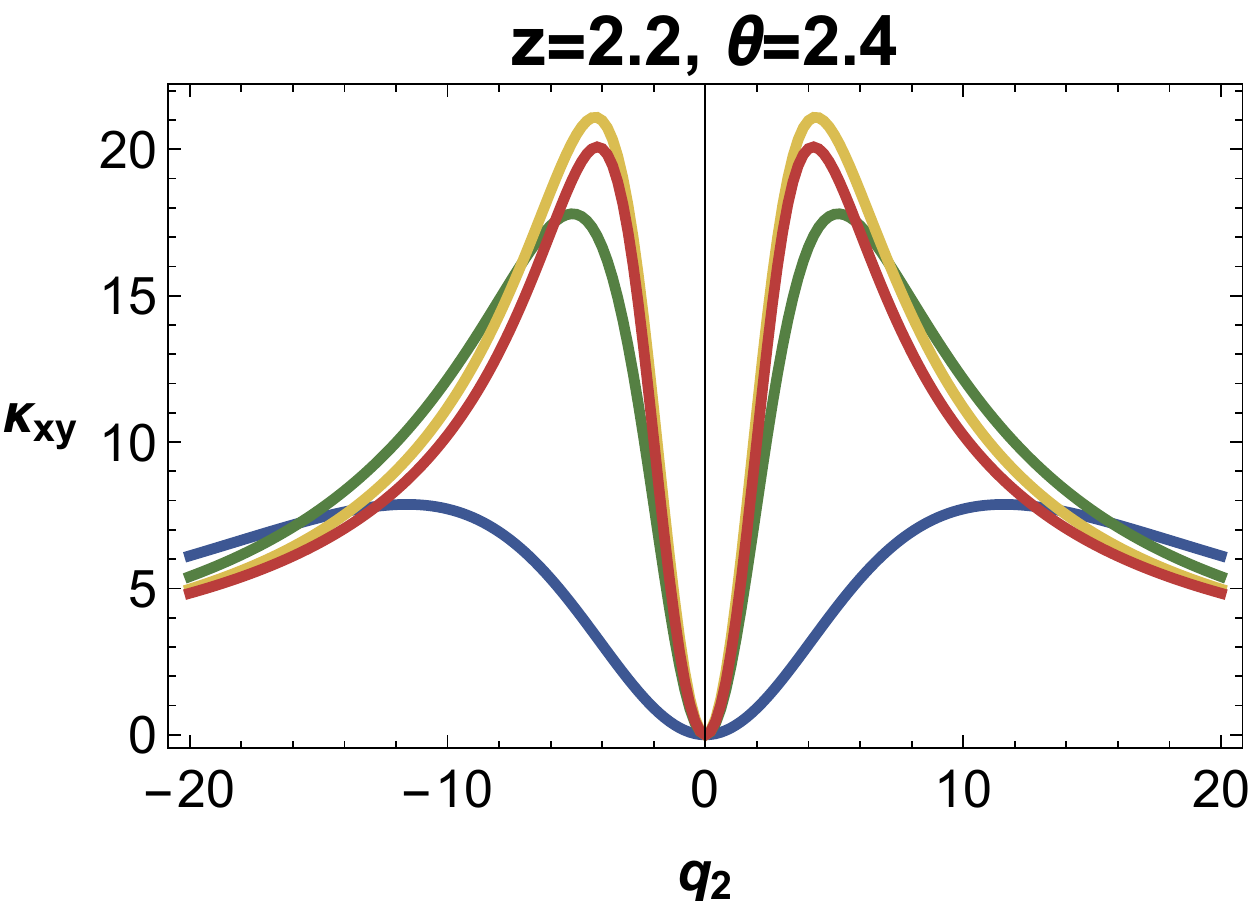} }   
                 \caption{Temperature evolution for $\kappa_{xy}(q_2)$ for different $(z,\theta)$. Each curves corresponds to $T=0.04,0.1,0.16,0.24$ for blue, green, yellow, and red respectively.  We used $q_{\chi}=5.$}    \label{fig:kxyq} 
\end{figure}

\begin{figure}[ht!]
\centering
 \subfigure[$P_{0A}$    ]
   {\includegraphics[width=45mm]{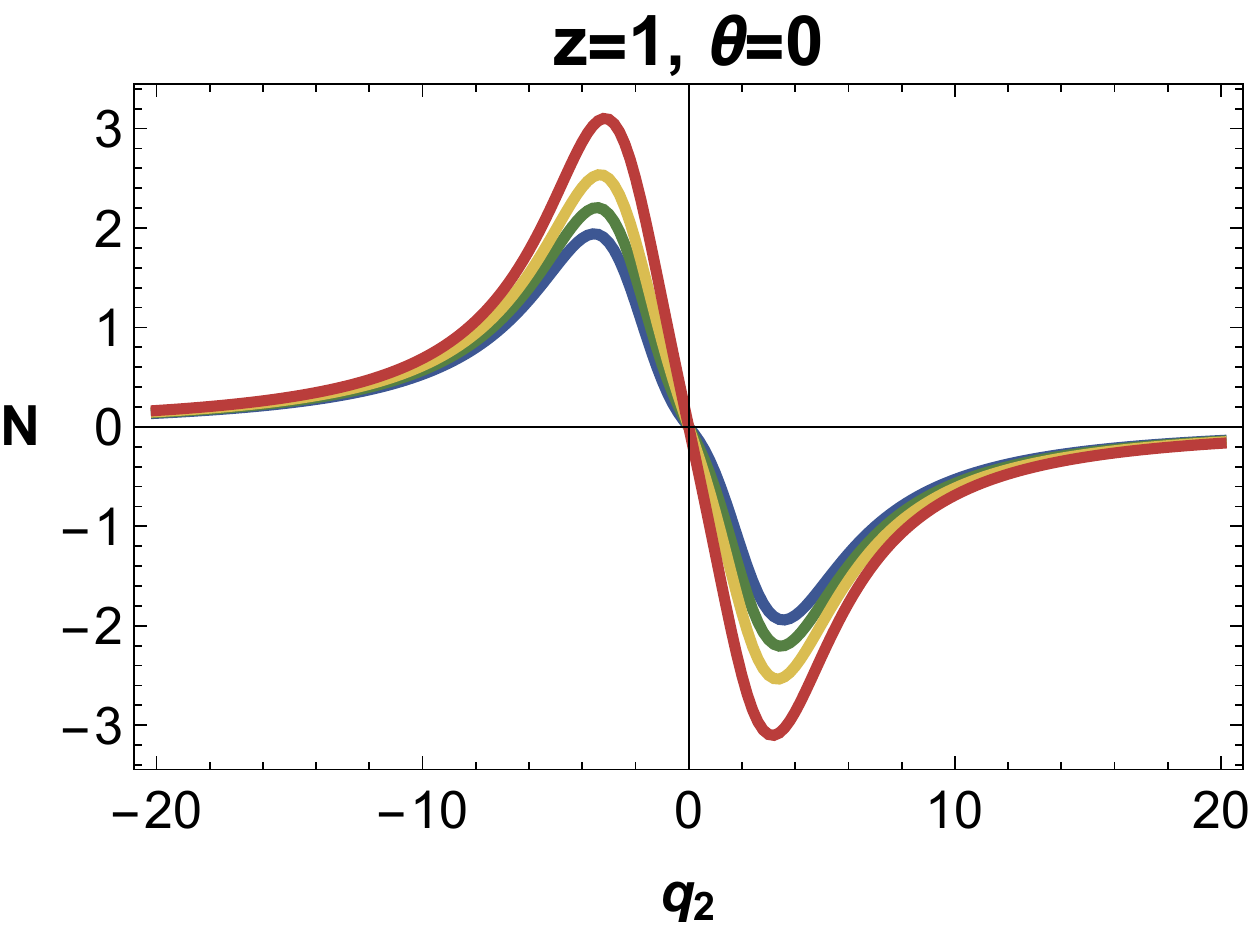} }
         \subfigure[$P_{A}$    ]
   {\includegraphics[width=45mm]{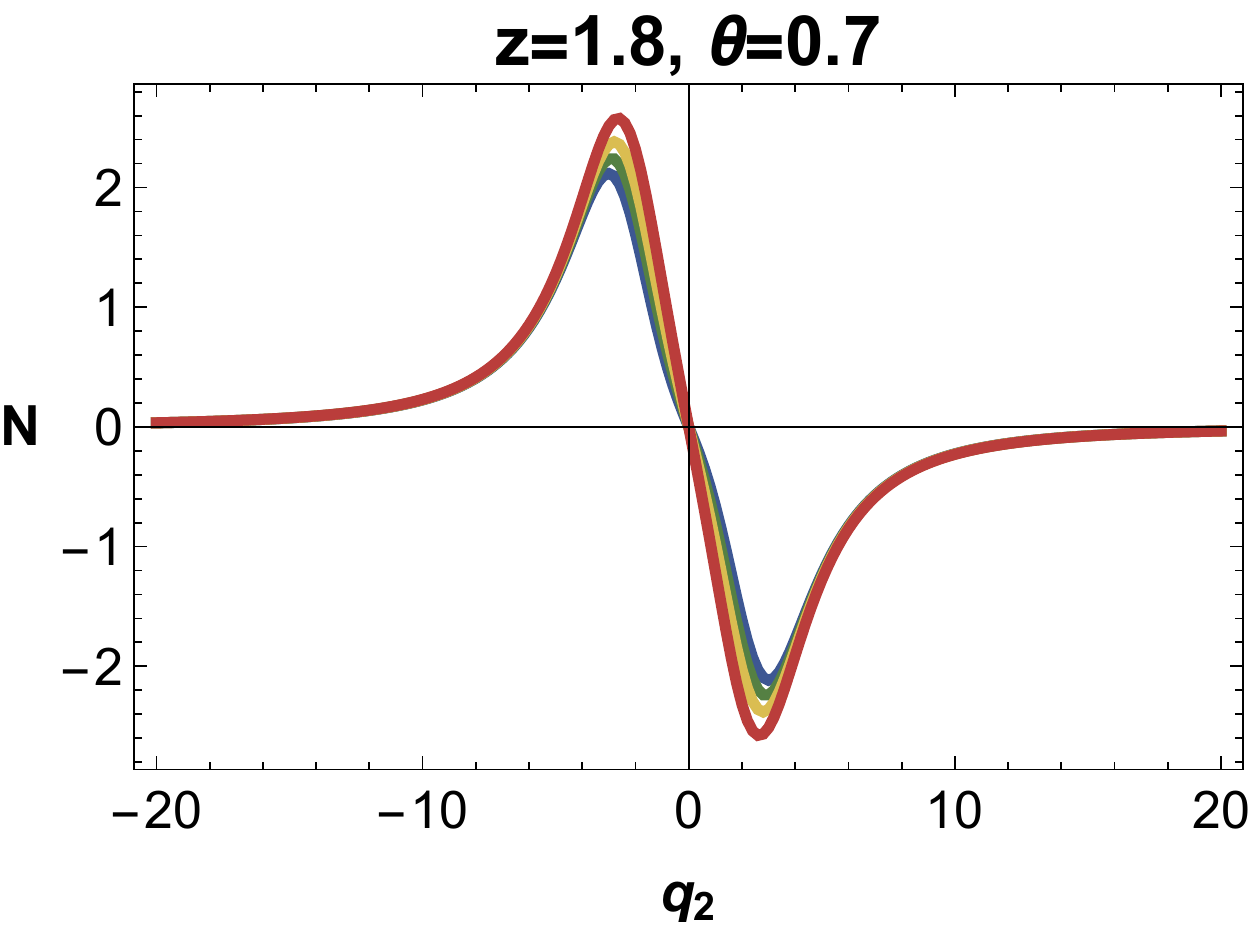} }
      \subfigure[$P_{C}$   ]
   {\includegraphics[width=45mm]{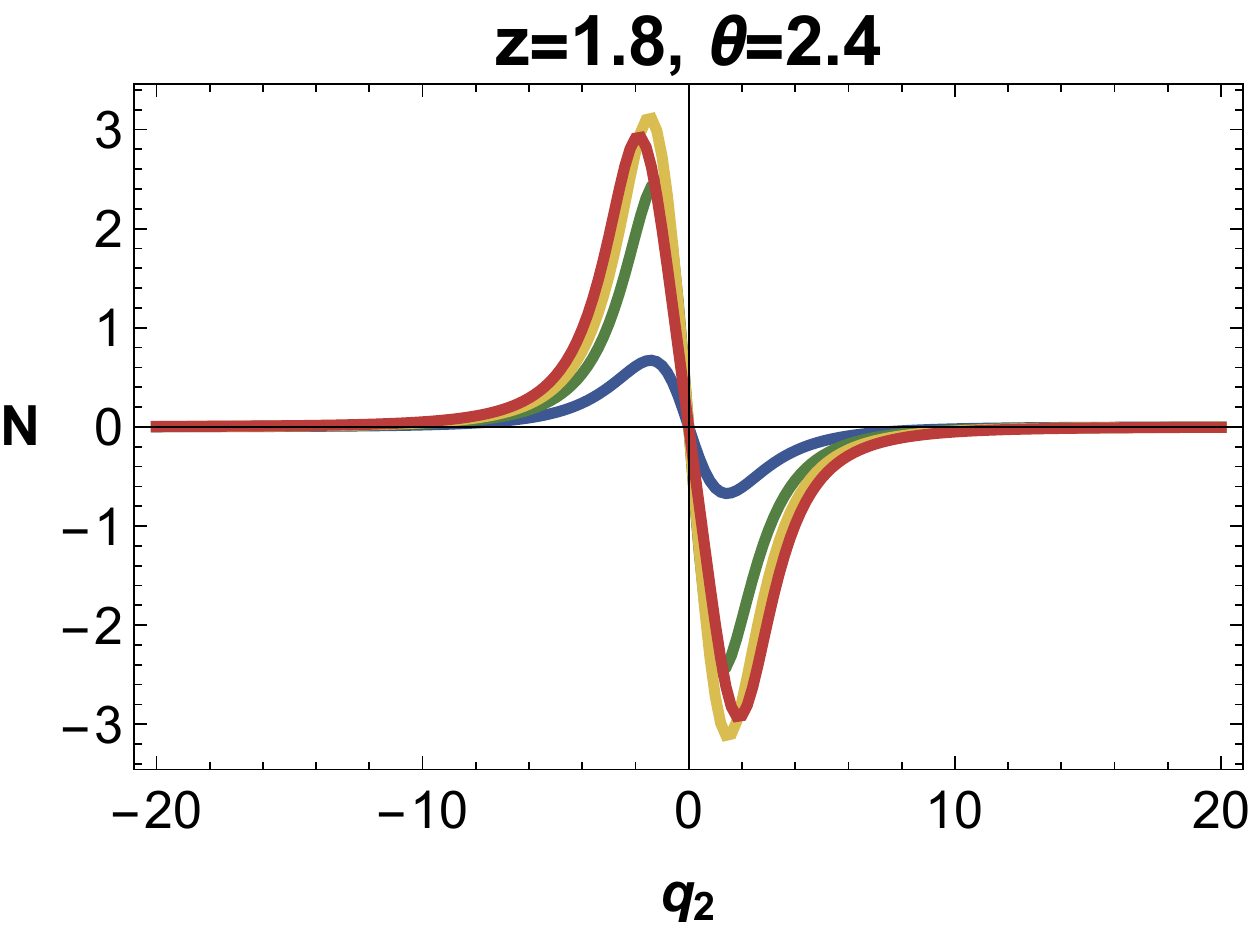} }
    \subfigure[$P_{0B}$   ]
   {\includegraphics[width=45mm]{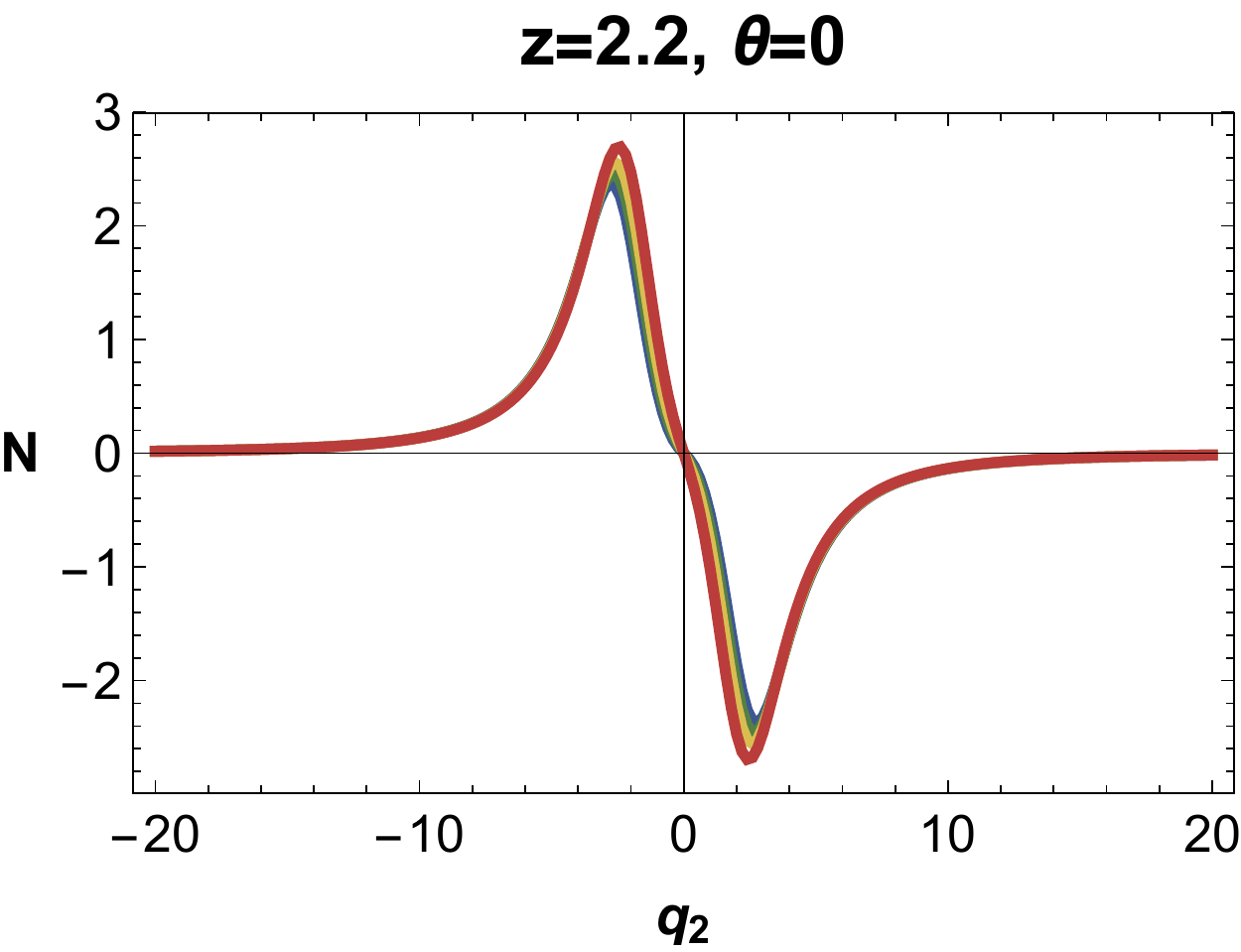} }
      \subfigure[$P_{B}$    ]
   {\includegraphics[width=45mm]{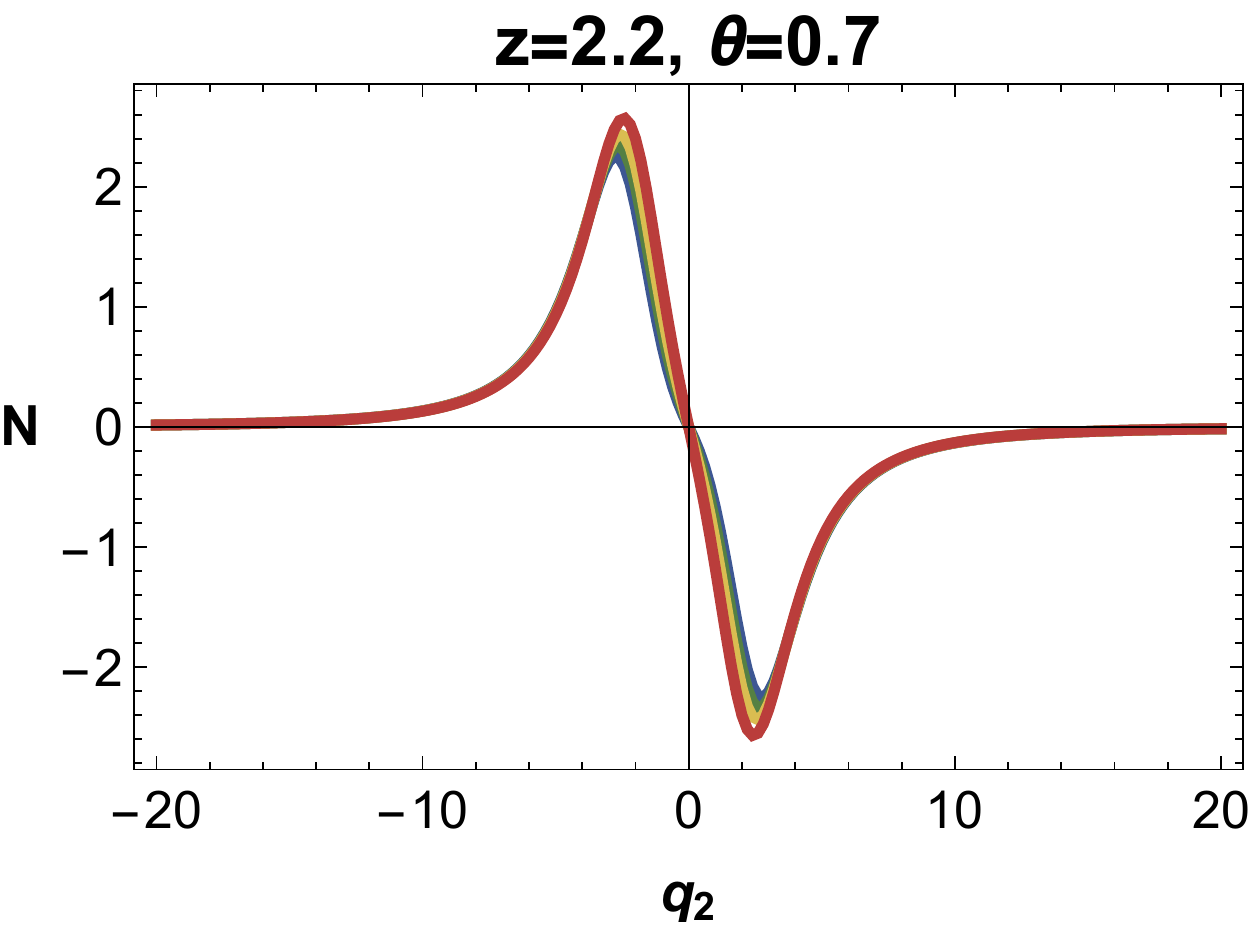} }
      \subfigure[$P_{D}$   ]
   {\includegraphics[width=45mm]{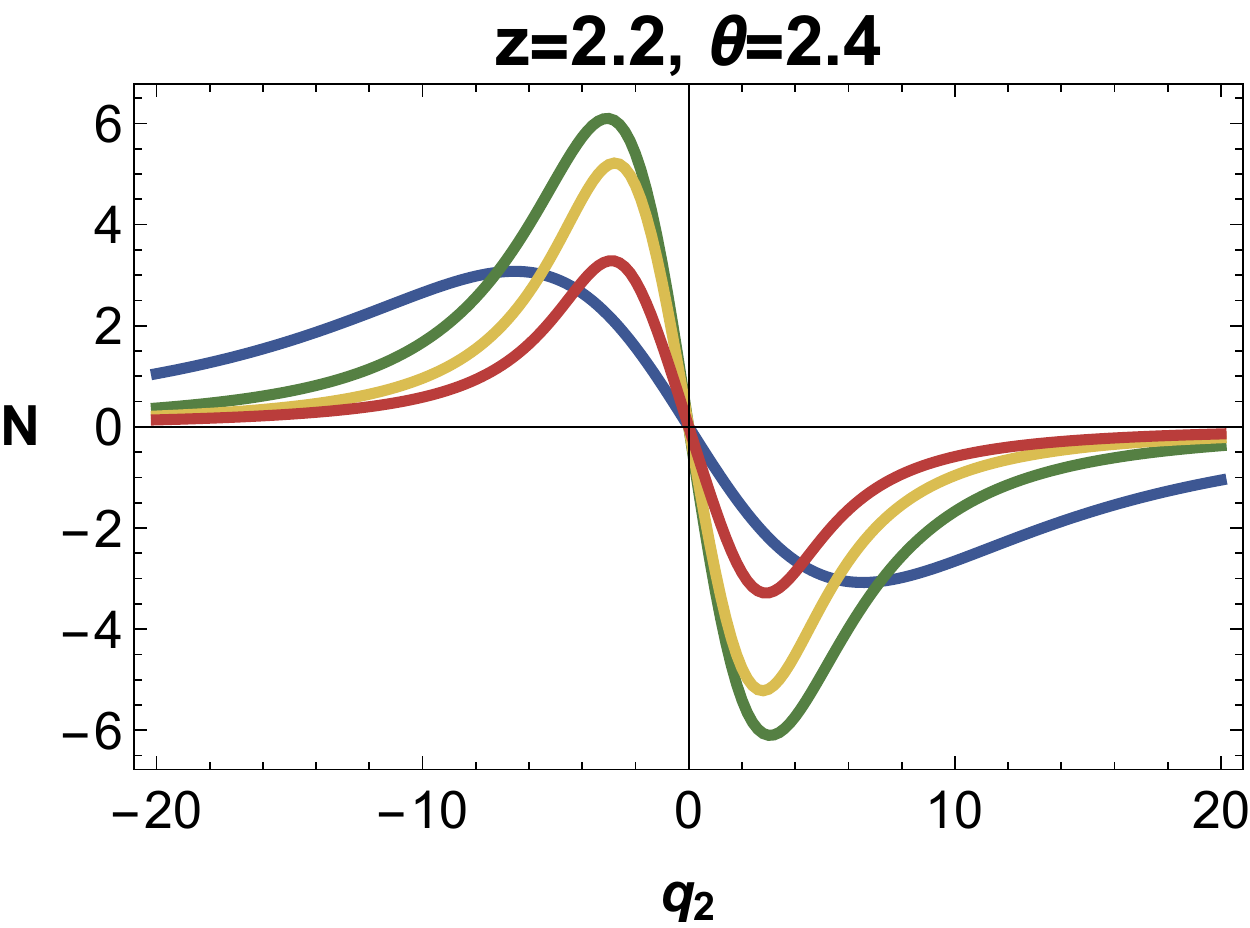} }   
                 \caption{Temperature evolution for $N(q_2)$ for different $(z,\theta)$. Each curves corresponds to $T=0.04,0.1,0.16,0.24$ for blue, green, yellow, and red respectively. We used $q_{\chi}=5.$}    \label{fig:Nq} 
\end{figure}

\begin{figure}[ht!]
\centering
	 \subfigure[$P_{0A}$ at $q_{\chi}=0$]
   {\includegraphics[width=30mm]{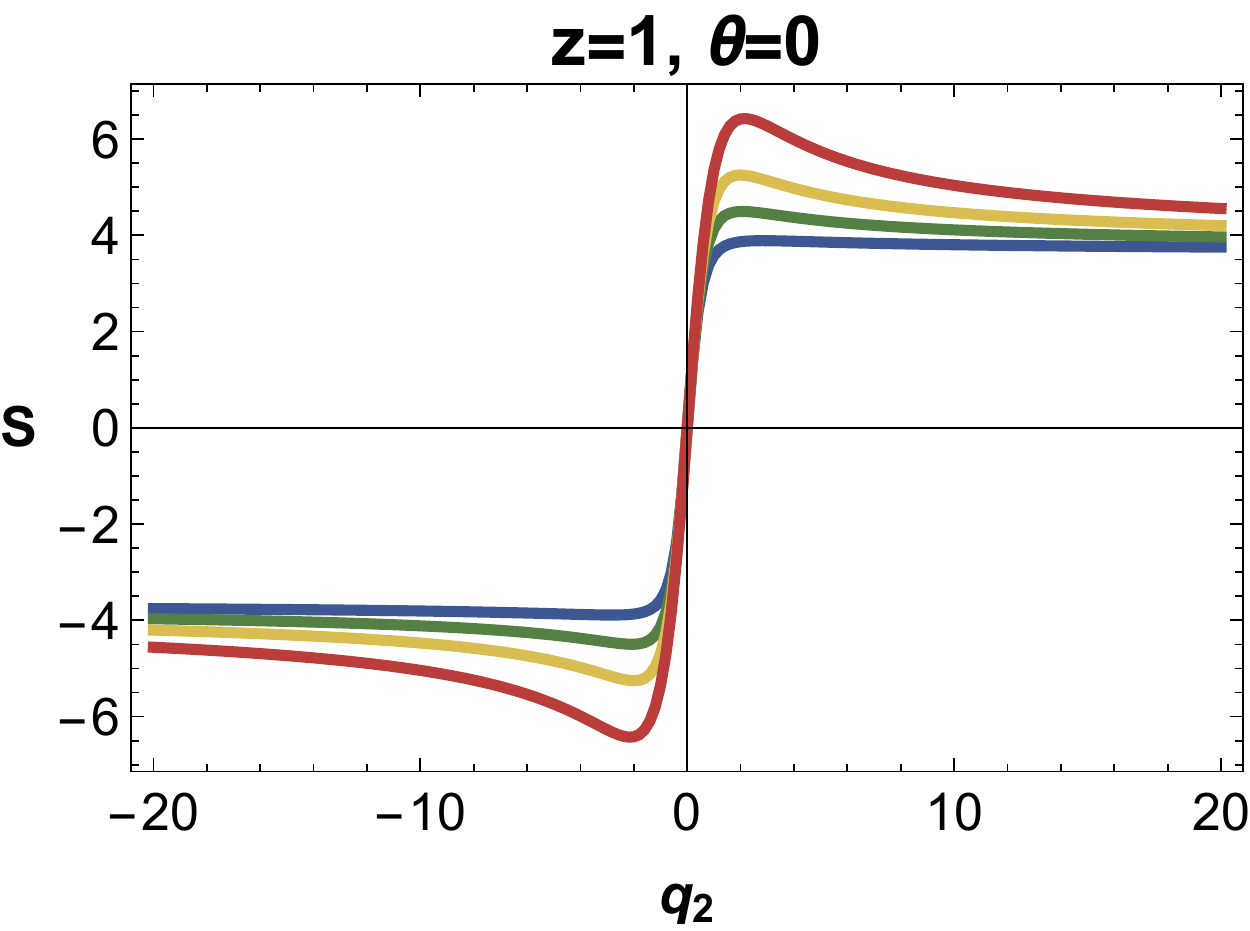} }
    \subfigure[$P_{0A}$    ]
   {\includegraphics[width=37mm]{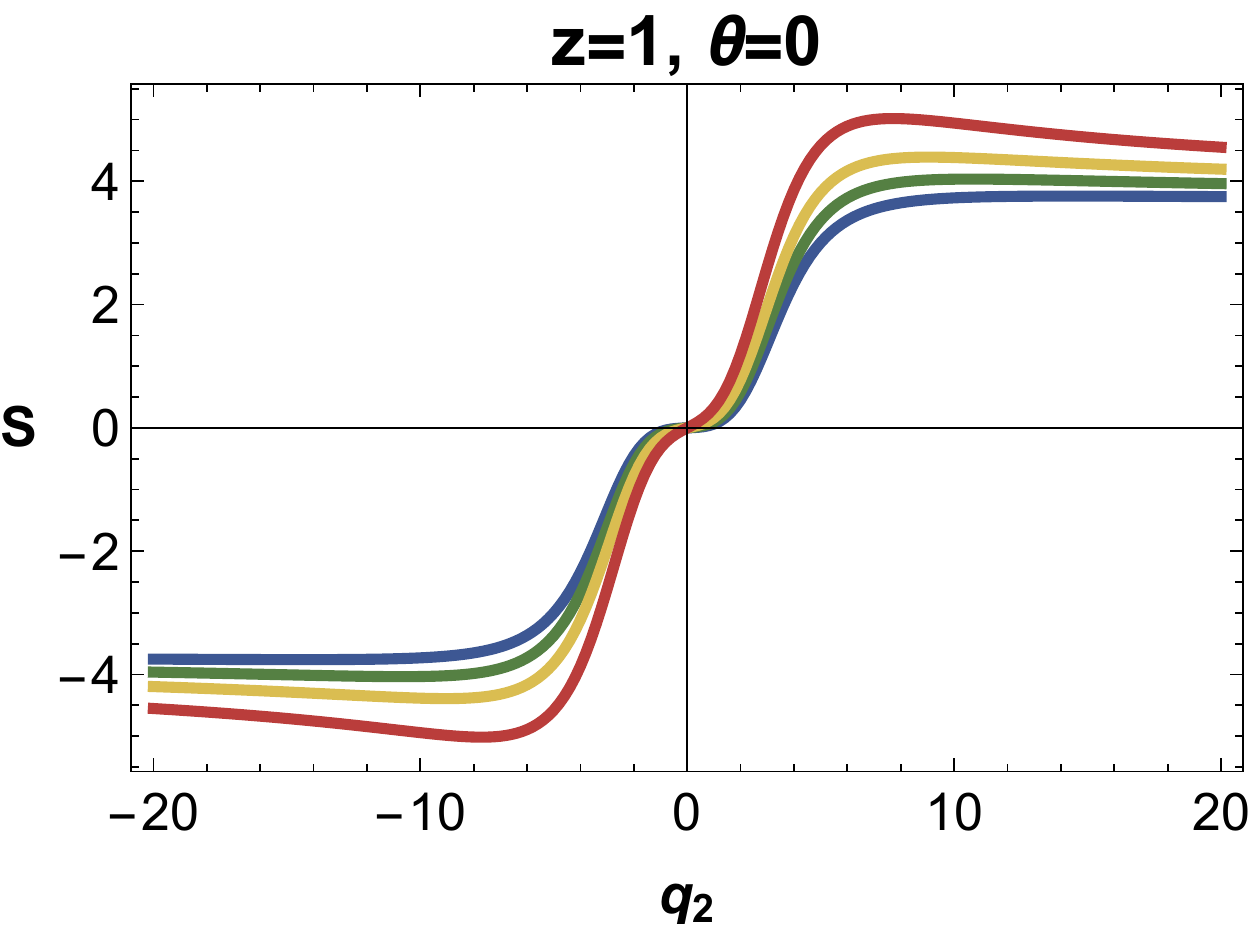} }
  \subfigure[$P_{0B}$ at $q_{\chi}=0$]
   {\includegraphics[width=30mm]{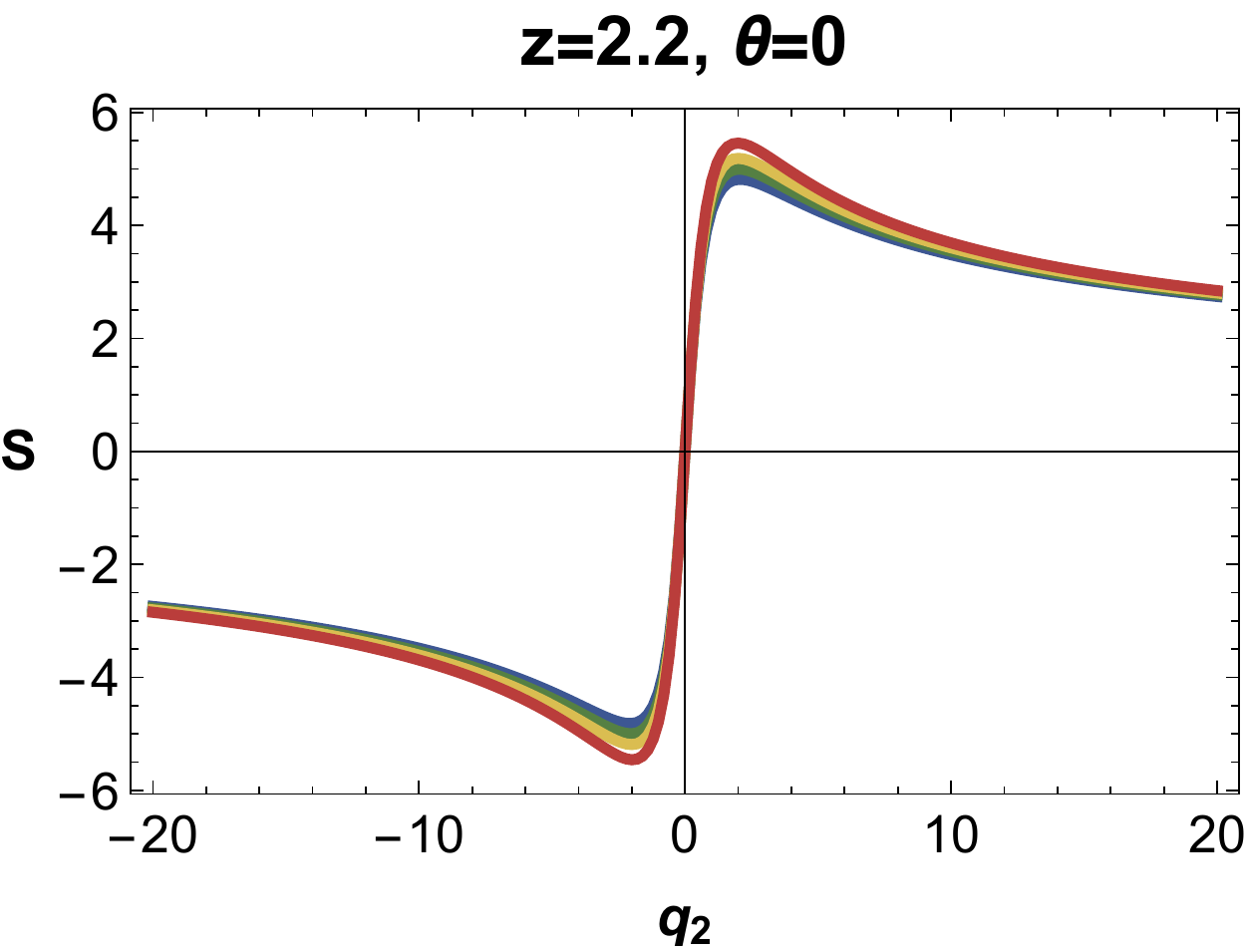} }
    \subfigure[$P_{0B}$   ]
   {\includegraphics[width=37mm]{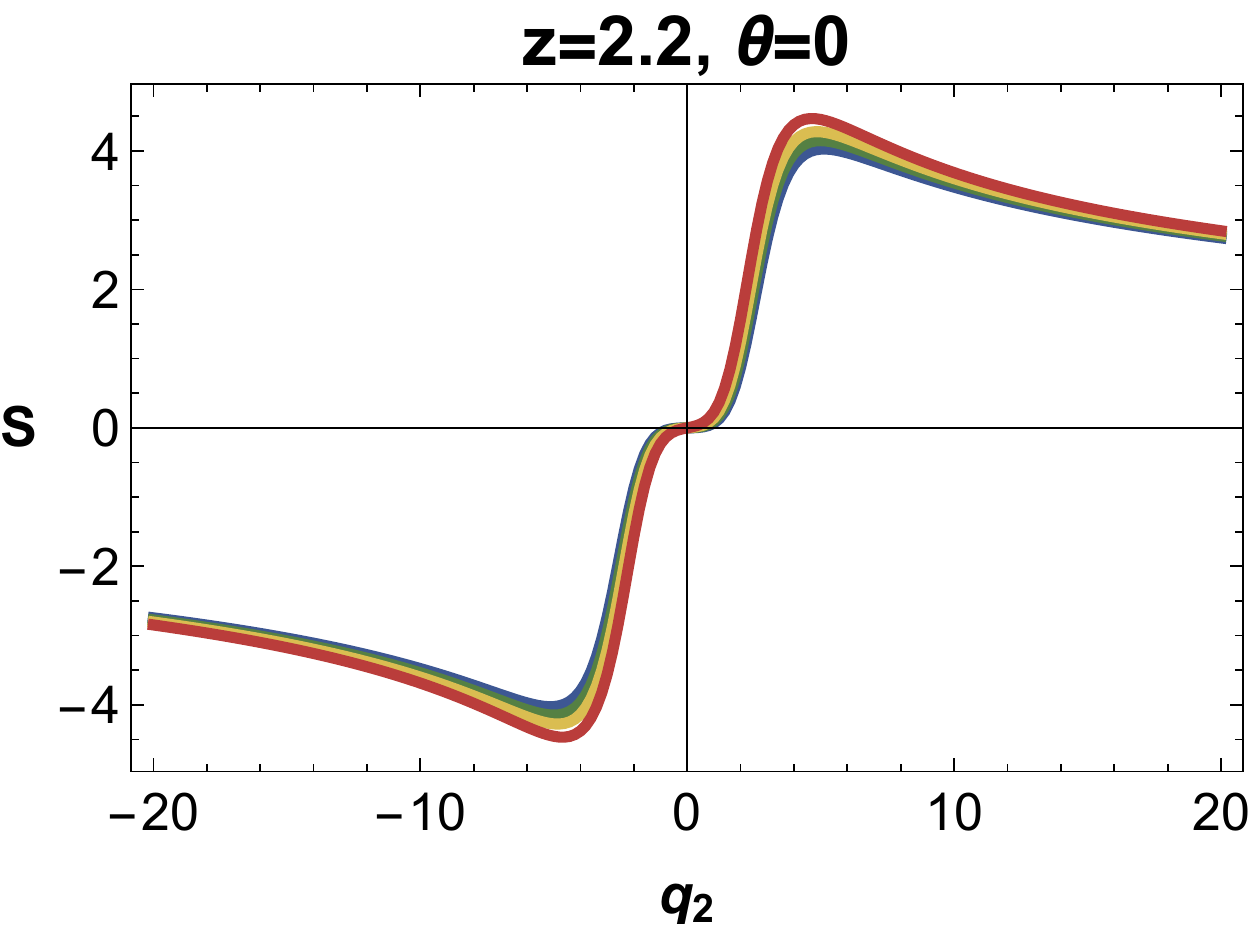} }
      \subfigure[$P_{A}$ at $q_{\chi}=0$]
   {\includegraphics[width=30mm]{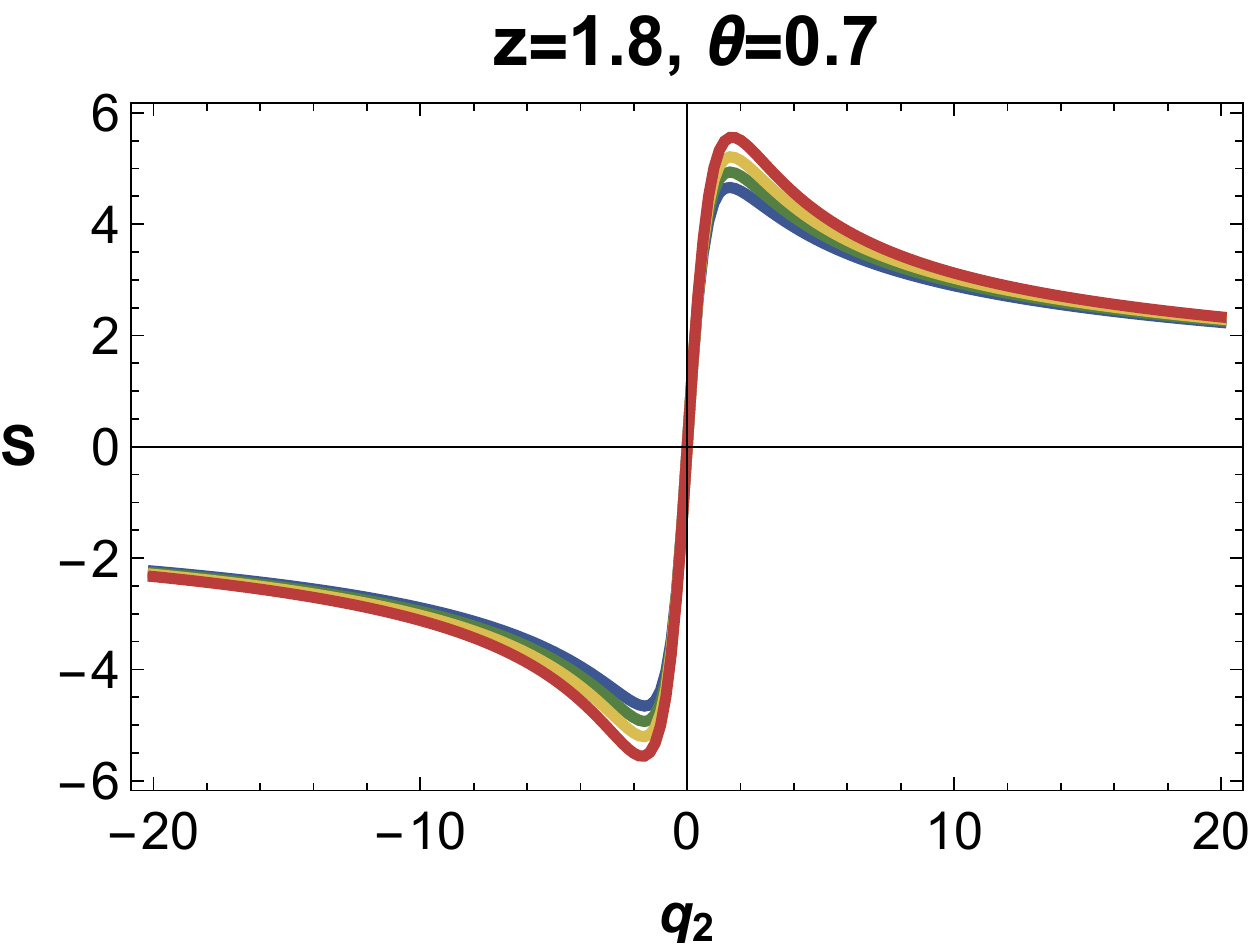} }
    \subfigure[$P_{A}$    ]
   {\includegraphics[width=37mm]{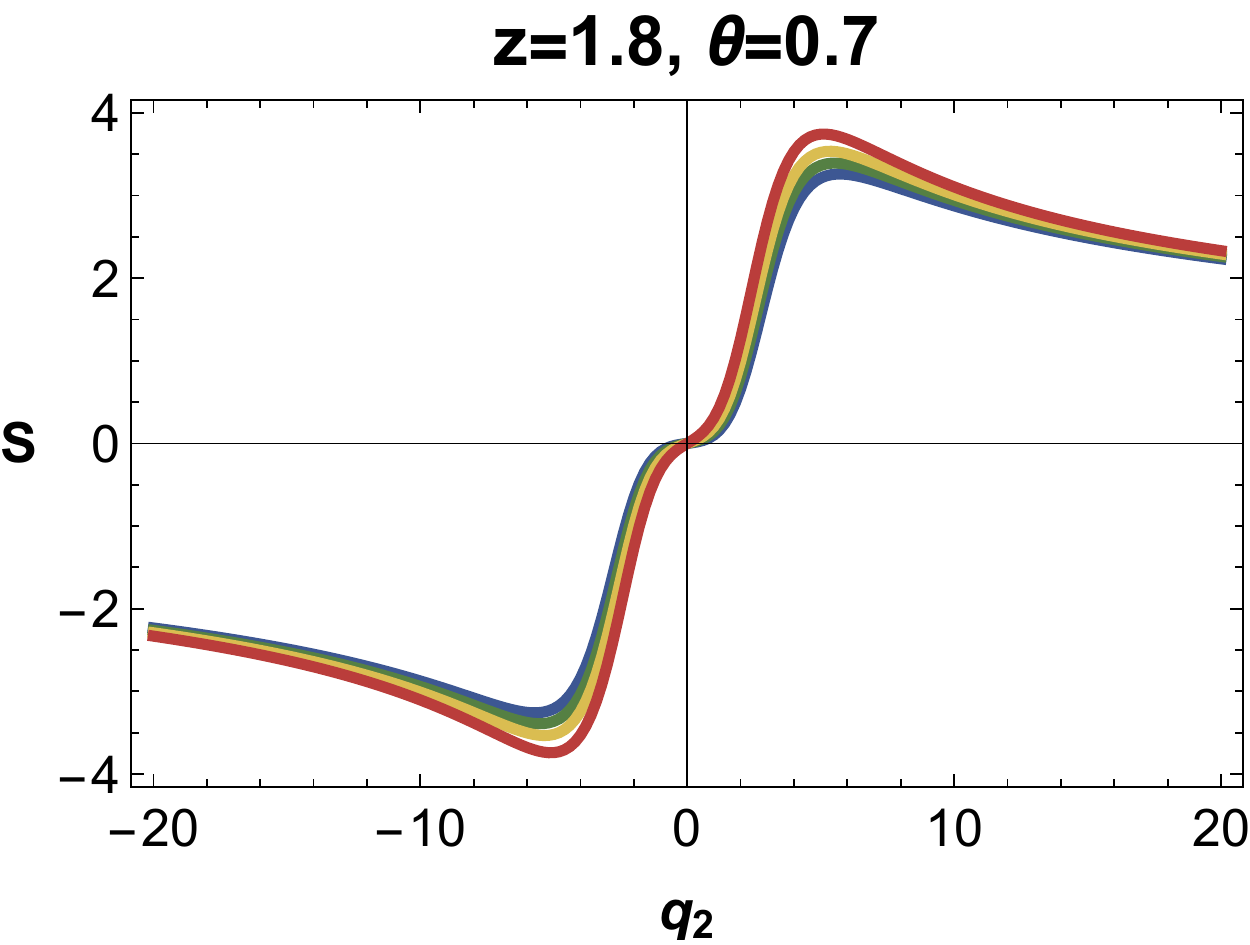} }
  \subfigure[$P_{B}$ at $q_{\chi}=0$]
   {\includegraphics[width=30mm]{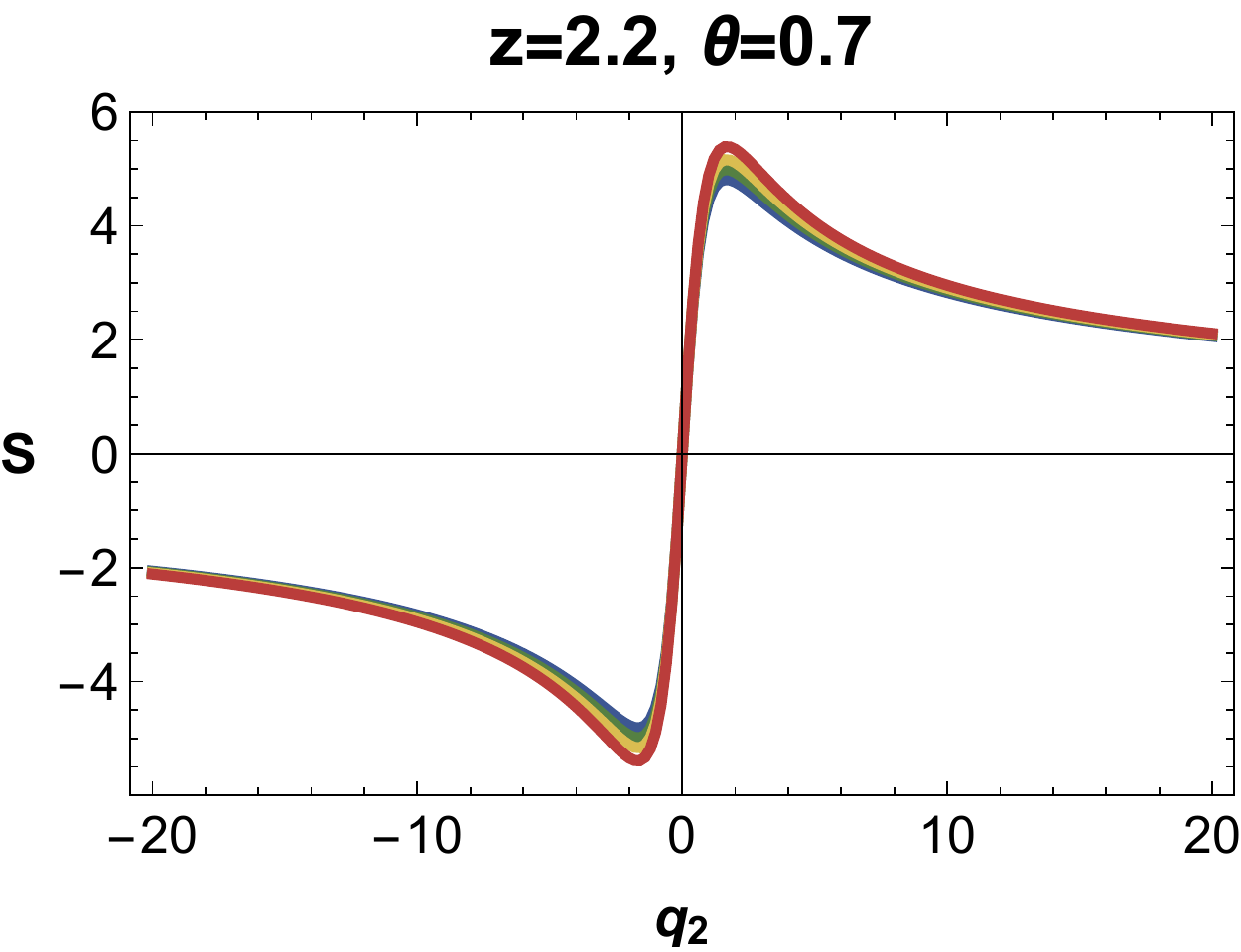} }
    \subfigure[$P_{B}$    ]
   {\includegraphics[width=37mm]{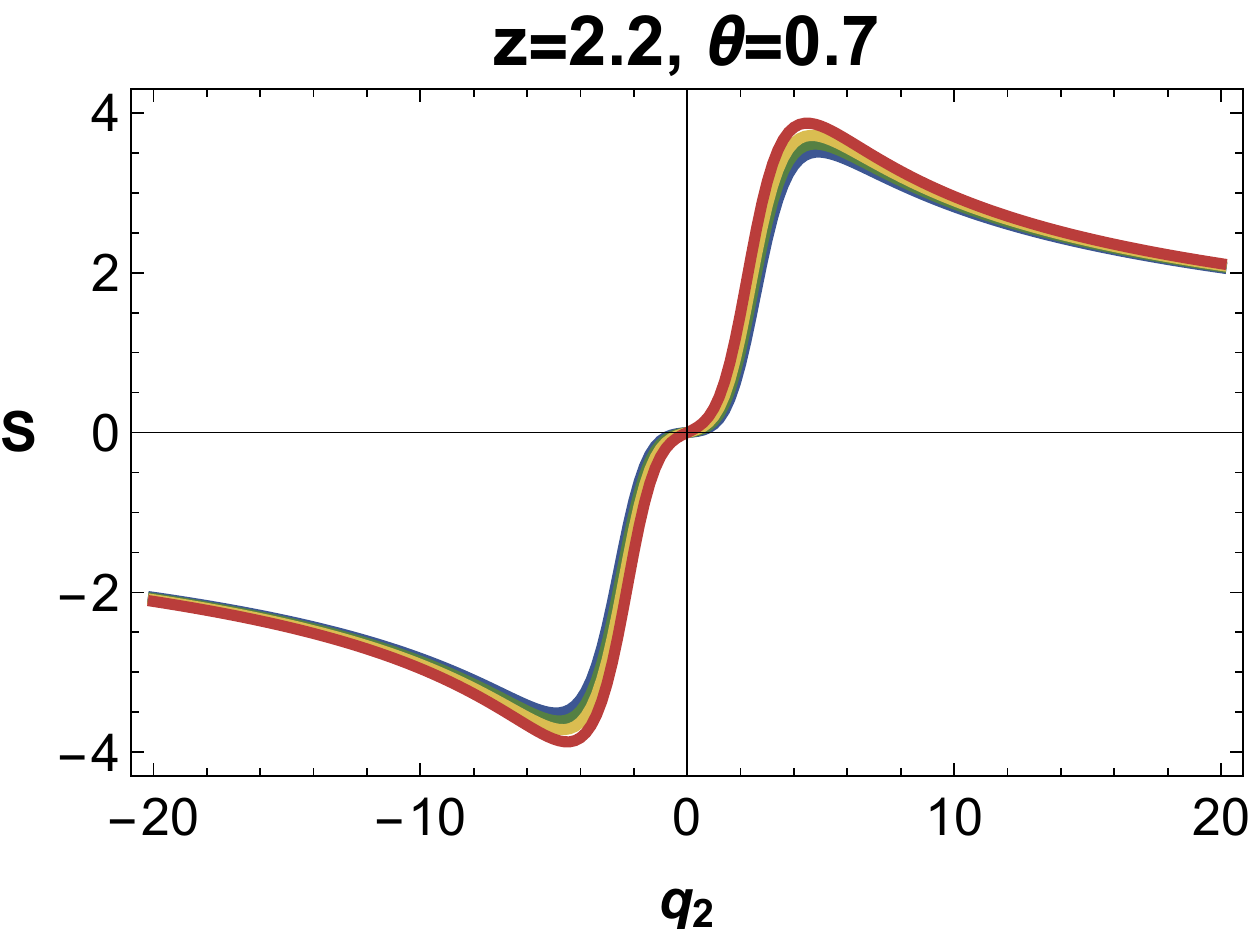} }
    \subfigure[$P_{C}$ at $q_{\chi}=0$]
   {\includegraphics[width=30mm]{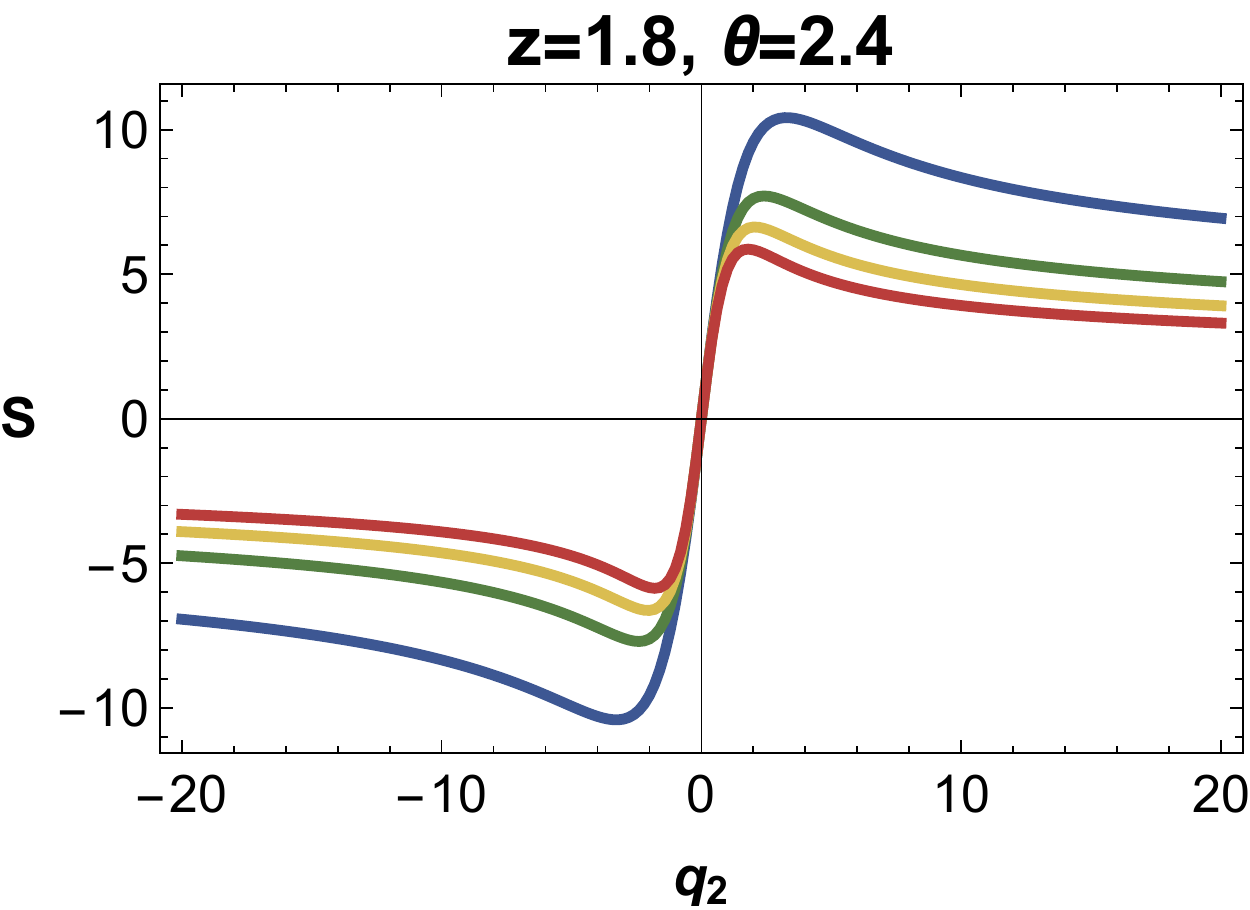} }
    \subfigure[$P_{C}$   ]
   {\includegraphics[width=37mm]{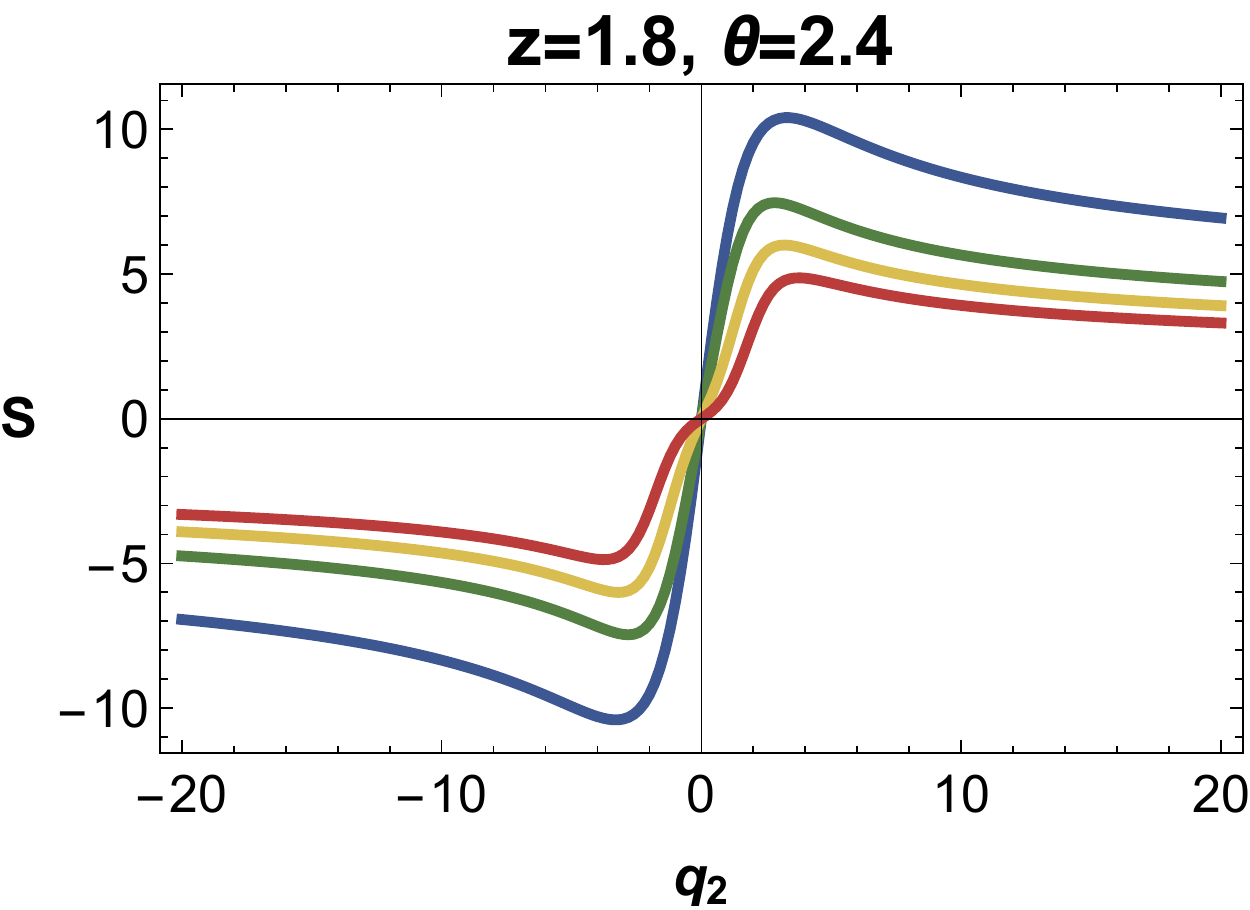} }
         \subfigure[$P_{D}$ at $q_{\chi}=0$]
   {\includegraphics[width=30mm]{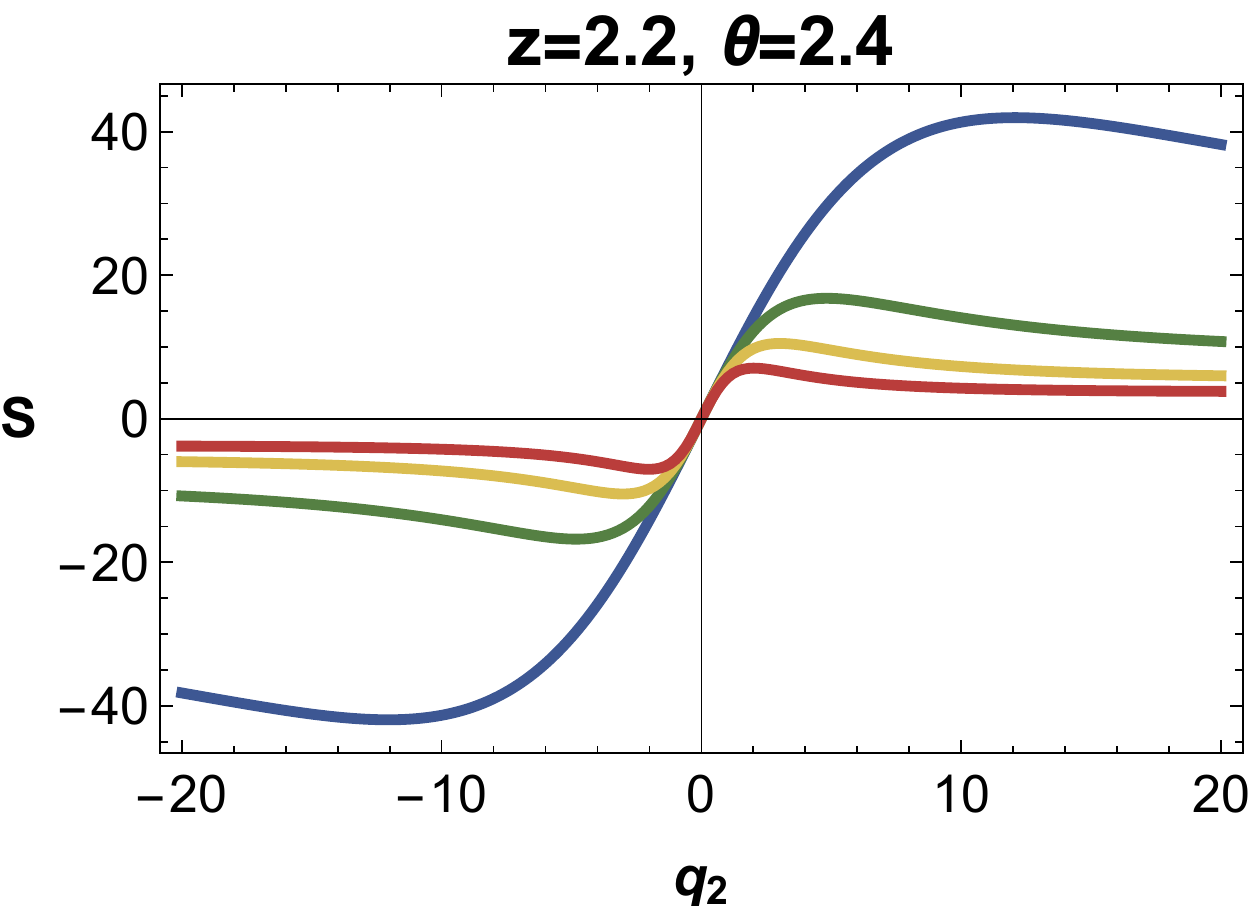} }
    \subfigure[$P_{D}$   ]
   {\includegraphics[width=37mm]{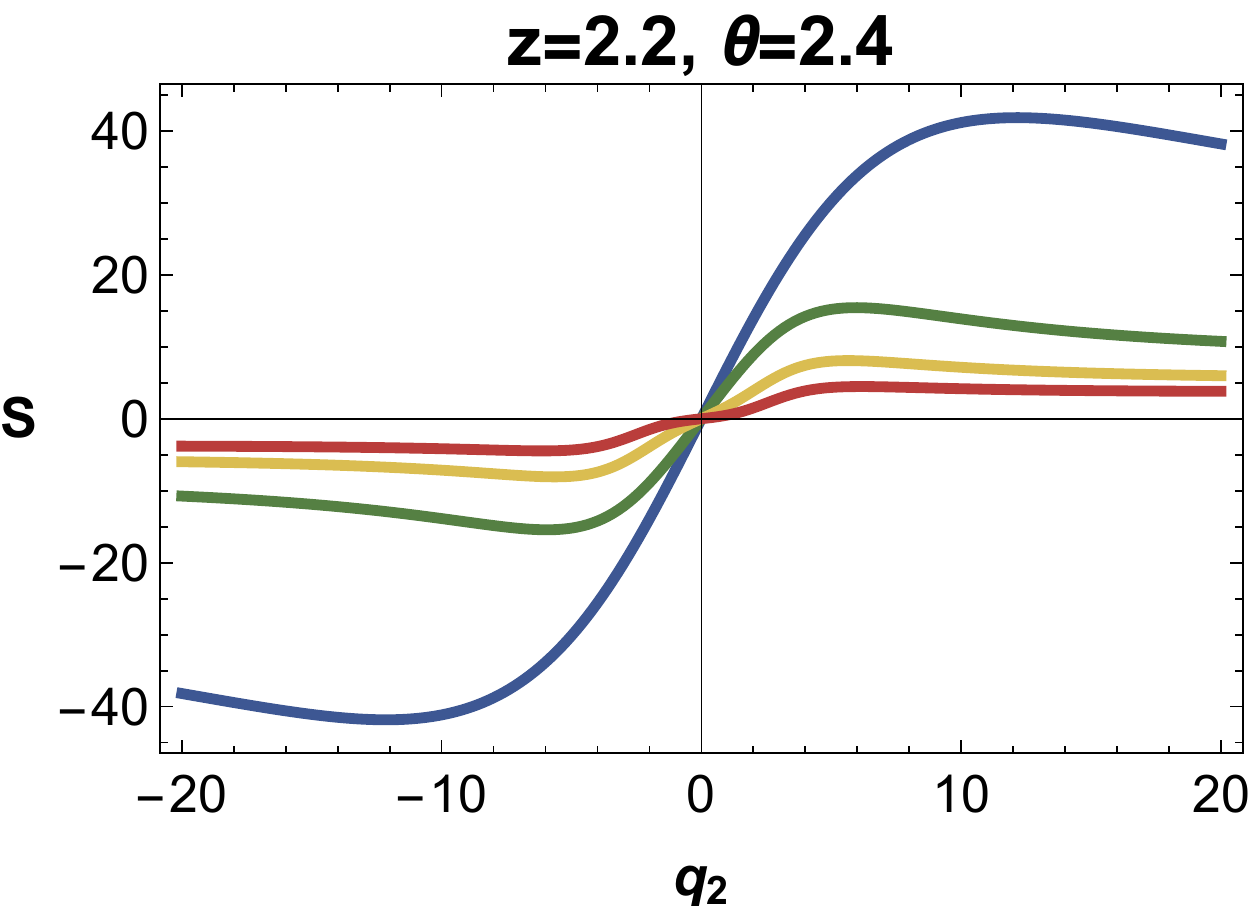} }   
                 \caption{Temperature evolution for $S(q_2)$ for different $(z,\theta)$. Each curves corresponds to $T=0.04,0.1,0.160.24$ for blue, green, yellow, and red respectively.  $q_{\chi}=5$ if $q_{\chi}\neq 0$. }    \label{fig:Sq} 
\end{figure}

Figure \ref{fig:Nq} and Figure \ref{fig:Sq} show the density dependence of Nernst signal and Seebeck coefficient respectively, which are related to the thermoelectric conductivity $\alpha$. We can expand $S(q_2)$ and $N(q_2)$ in the small density limit:
\begin{align}
	N&\sim -\frac{4\pi q_{\chi}\lambda^2r_0^{2(z+\theta)}}{\beta^2 \mathcal{D}}q_2+\cdots , \nonumber\\
	S&\sim \frac{4\pi r_0^{4z}}{\beta^2\mathcal{D}}q_2+\cdots ,
\end{align}
 where $\mathcal{D}=r_0^{4z}+q_{\chi}^2\lambda^4r_0^{4\theta}$. 
 The stepping feature near the zero $q_2$  in the figures \ref{fig:Sq}  for nonzero $q_\chi$  case  is due to the suppression of the linear term by the presence of $\cal D$ that contains $q_\chi^2$ which is chosen to be 5, which is  rather large.

\section{Null energy condition for Lifshitz black branes}
The null energy condition (NEC) requires that, at every point in spacetime $T_{\mu\nu}v^{\mu}v^{\nu}\geq 0$, where $v^{\mu}$ is a light-like vector.  $T_{\mu\nu}$ is the energy-momentum tensor of matter, suggesting that the NEC should be a property of matter. The NEC is used in deriving the second law of thermodynamics for black holes and in prohibiting the traversability of wormholes and the building of time machines.  Einstein's equations imply that the NEC can be reformulated in a form as
\be
R_{\mu\nu}v^{\mu}v^{\nu}\geq 0,
\ee

 However, the NEC can be violated by quantum fields. There are several examples of effective theories that violate NEC, but without manifest confliction with the principles of quantum field theory.
At the quantum level, the NEC is violated by the Hawking effect. In this case, the quantum null energy condition is proposed as \cite{Bousso:2015wca}
\be
\langle T_{\nu\nu}\rangle\geq \frac{h}{2\pi}S'',
\ee
where $S''$ is the local piece of the second null derivative of the entropy, evaluated on either side of the null surface.

The null energy condition for Lifshitz black branes with hyperscaling violating factors are $(z-1)(d+z-\theta)\geq 0$ and $(d-\theta)[d(z-1)-\theta]\geq 0$, where $d$ is the number of the spatial dimension.

The combination of the $tt$ and $rr$ components of Einstein equation is given by
\be\label{null1}
R^t_t-R^r_r=\frac{1}{2}g^{rr}(\partial_r \phi)^2.
\ee
On the other hand, after evaluating  the Ricci tensor for our ansatz, we obtain
\be\label{null2}
R^t_t-R^r_r=-(2-\theta)(z-1-\frac{\theta}{2})r^{-\theta}f(r).
\ee
From (\ref{null1}) and (\ref{null2}), we obtain the solution for the dilaton field
\be
e^{\phi}=r^{\sqrt{(2-\theta)(2z-2-\theta)}}.
\ee
To guarantee our black brane solution satisfies the Null energy condition, we require
\be\label{nec}
\sqrt{(2-\theta)(2z-2-\theta)}\geq 0.
\ee

\bibliographystyle{JHEP}
\bibliography{Refs_HSV}

\end{document}